# RAVENS CONCEPT STUDY

# FINAL REPORT

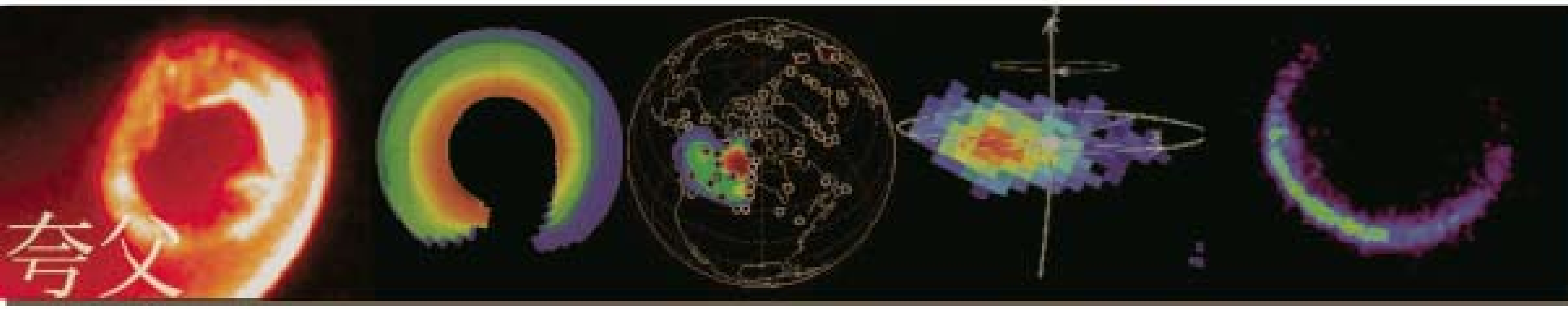

The whole world wide, every day,
Fly Hugin and Munin;
I worry lest Hugin should fall in flight,
Yet more I fear for Munin.

***From Grímnismál***

# RAVENS CONCEPT STUDY

# – FINAL REPORT -

## ERIC DONOVAN – UNIVERSITY OF CALGARY

**SUBMITTED TO PWGSC & CSA DECEMBER 4, 2006**

**With contributions from**

| | |
|---|---|
| Bonnell, John | University of California, Berkeley |
| Brandt, Pontus Carlson | Johns Hopkins University |
| Cogger, Leroy | University of Calgary |
| Fazakerly, Andrew | Mullard Space Sciences Laboratoty |
| Freeman, Mervyn | British Antarctic Survey |
| Gérard , Jean-Claude | Université de Liège |
| Gordon, Blair | Routes Astroengineering |
| Hackett, John | COMDEV |
| Henderson, Michael | Los Alamos National Laboratories |
| Immel, Tom | University of California, Berkeley |
| Jackel, Brian | University of Calgary |
| Jamar, Claude | Centre Spatial de Liège |
| Kabin, Konstantin | University of Alberta |
| Lester, Mark | University of Leicester |
| McKenna-Lawlor, Susan | National University of Ireland |
| Meurant, Matthiue | University of Calgary |
| Phan, Tai | University of California, Berkeley |
| Pulkkinen, Tuija | Finnish Meteorological Institute |
| Rankin, Robert | University of Alberta |
| Ridley, Aaron | University of Michigan |
| Rochus, Pierre | Université de Liège |
| Spann, James | NASA, Huntsville |
| Sofko, George | University of Saskatchewan |
| Spanswick, Emma | University of Calgary |
| Syrjäsuo, Mikko | University of Calgary |
| Trondsen, Trond | University of Calgary |
| Tu, Chuanyi | Peking University |
| Wang, Jing-Song | Chinese National Meteorological Administration |
| Xia, Li-Dong | University of Science and Technology of China |
| Østgard, Nikolai | University of Bergen |

# Table of Contents

Cover: The five images are (left to right) an LBH global electron aurora from WIC, equatorial plasma densities inferred from magnetoseismology, a substorm injection recorded in the CANOPUS/NORSTAR riometer data, an ENA image, and an IMAGE SI12 global proton auroral image. The figures are courtesy of Stephen Mende, Konstantin Kabin, Emma Spanswick, Pontus Carlson Brandt, and Harald Frey, respectively. The logo was designed by Jo-Anne Brown, and the cover layout by Emma Spanswick.

# 1. Context and Overview

Space physics research began with early discoveries that the aurora is a consequence of the precipitation of charged particles from outer space [*Vegard,* 1932], and the causal relationship between transient ejections of mass from the sun and magnetic storms [*Chapman and Ferraro,* 1930]. In the late 1950s and the 1960s, space flight brought with it a period of accelerated discoveries that included the existence of the radiation belts, the plasmasphere, and the general topology of the terrestrial magnetosphere. In the same time period, the International Geophysical Year saw an explosion of ground-based observations that led to discoveries that included the auroral oval and the relationship between negative H-bays and the breakdown of the magnetotail current sheet in substorms, and the first true inklings that GeoSpace[1] behaves as a large system [*Feldstein,* 1960; *Atkinson*, 1967]. These and other early discoveries prompted research into the fundamental physical processes at work in the GeoSpace environment. Just one example was the early exploration of the role of wave-particle interactions in the acceleration and loss of radiation belt electrons [see for example*, Lyons*, 1973, 1974].

As our understanding of the morphology of GeoSpace and the variety of physical process at work within it grew, we also grew to understand how extensively the various components of geospace interact with each other, and that these larger-scale processes are powered by the solar wind and solar radiation. These concepts were first brought to the fore by the ground-breaking works of *Dungey* [1961], and *Axford and Hines* [1961], who proposed competing theories of global convection and currents powered by dayside and tail reconnection and flank viscous interactions, respectively. With these pioneering studies, space physics had reached maturity, wherein our field recognized the importance of exploration and discovery, fundamental physics at work on the small scale, and the concept of GeoSpace as a complex coupled system. It is clear that continued work on all three fronts is essential to understanding how the near-Earth space environment works. This is an enormous endeavor and an important one. Further progress requires multiple strategically placed satellite missions and extensive ground-based programs to combine simultaneous synoptic, MHD-scale, together with kinetic-scale observations, associated data analysis, theoretical and simulation work. The benefits are significant. From the practical standpoint, near-Earth space is part of our environment and therefore something we simply have to understand. More importantly, the near-Earth environment is host to plasma physical processes that are at work throughout the cosmos, but that (for obvious reasons) can be best observed in GeoSpace. By exploring GeoSpace, we are exploring the Universe.

Understanding GeoSpace as a complex coupled system is a concept that has emerged as an overarching umbrella for our field. In order to accomplish this we have to understand the morphology of GeoSpace, the fundamental physical processes at work on the small scale, and the global dynamics. How are Central Plasma Sheet (CPS) electrons and ions energized to become ring current and radiation belt particles? What controls the evolution of storms and substorms? What are the physical processes at work in the electron and proton aurora? What are the system-level consequences of reconnection? These and other key questions are interrelated. Their answers will require knowledge of the physical processes that are relevant in geospace, have direct observations of those processes, and simultaneous synoptic observations of the system-level dynamics.

[1] For our purposes, we define GeoSpace as the near-Earth space environment, including the ionosphere, thermosphere, magnetosphere, the latter including the various plasma regimes such as the plasmasphere, radiation belts, ion and electron central plasma sheets, mantle, lobe, low latitude boundary layer, cleft, etc.

The International Solar Terrestrial Physics (ISTP) program clearly recognized the necessity of simultaneous system-level and simultaneous direct observations of the relevant physical processes. The overarching objectives of ISTP included (and include to this day) obtaining quantitative measurements of the transport of mass, energy and momentum through GeoSpace, carrying out detailed studies of the physical processes that affect this flow, and the study of GeoSpace processes control the deposition of energy into the Earth's atmosphere. These objectives were the science rationale for the WIND, Geotail, Polar, SOHO, and finally Cluster missions. With this constellation our field had the opportunity for simultaneous observations of the Sun, solar wind, and depending on where the remaining satellites were, some fraction of the magnetopause, magnetosheath, bowshock and foreshock, CPS, Low Latitude Boundary Layer (LLBL), and inner magnetosphere. Polar's UVI and VIS auroral imagers provided global observations of energy deposition, and coordinated ground-observing programs such as SuperDARN, CANOPUS, and MIRACLE rounded out our ability to track the continent-scale and even global ionospheric dynamics that are so closely coupled to the rest of GeoSpace.

ISTP is an ongoing program, and its constellation of satellites have for more than a decade been systematically exploring various regions of GeoSpace. The ISTP satellites have provided observations the interpretation of which have significantly advanced our fundamental knowledge of virtually every aspect of the GeoSpace system. Examples are plentiful, and understanding of the consequences of reconnection via direct observations on the magnetopause and in the mid-tail, the role of Bursty Bulk Flows (BBFs) and turbulence in the evolution of the CPS during substorms and other times, the role of heavy ionospheric ions in CPS and inner magnetospheric dynamics, the growth of flow driven instabilities on the magnetopause and their relationship to radial plasma transport in the CPS and inner magnetosphere, the association between Alfven waves and aurora, and the substorm expansive phase onset. The IMAGE satellite mission in particular made tremendous strides towards true systems-level science (see the review "Magnetospheric Imaging: Promise to Reality" by *Burch* [2005]), providing contemporaneous images of the global proton and electron aurora, the plasmasphere, and ring current, and has without question produced the first system-level comprehensive global images of the magnetospheric plasma populations. IMAGE allowed us to see, in many ways for the first time, the interplay between the different large plasma populations in response to the solar driver.

On one hand, there is very little in GeoSpace that we do not know something concrete about. On the other hand, however, the long sought after system level understanding has largely eluded our grasp. This is not at all surprising, for even the ISTP constellation never really allowed for simultaneous observations all the way along the energy/mass transport path. Further, the ISTP studies clarified *what we need to do*. For example, we now understand that the substorm expansive phase onset unfolds in tens of seconds, and involves both current limiting instabilities in the inner CPS and magnetic reconnection in the mid-tail [*see Lui*, 1996; *Baker et al.,* 1996], but we do not have the data to determine which comes first or in other words what the macroscale instability is that is responsible for substorm onset. For another example, we have now direct observations of the consequences of magnetic reconnection via observations by Polar, Geotail, and Cluster, however we do not have the kinetic scale observations that allow us to explore the causes of reconnection. The need to resolve the order of events in substorm onset led to THEMIS. The need for kinetic scale observations in the reconnection region led directly to the Magnetospheric Multi-Scale (MMS) mission. In other words, ISTP moved us forward, but in doing so uncovered deeper mysteries that have in turn led to more sophisticated missions targeted at exploration of specific physical processes. In fact, this push to explore fundamental physical processes in GeoSpace to understand what drives them and what they affect has led not only to THEMIS, and MMS, but also planned and proposed missions including ePOP, the RBSPs, ORBITALS, ERG, the ITSPs, and in the longer term Cross Scale.

At the same time we are also understanding that system-level GeoSpace science has far to go. In one study, for example, IMAGE observations were used to explore the interaction between a substorm related Earthward convection pulse and dynamics of the plasmapause [*Goldstein et al.,* 2005a,2005b]. Now, with THEMIS, we are on the eve of launching a five satellite mission to provide in depth observations of the substorm associated dynamics of the inner CPS. *What physics-based system-level science needs is for us to carry out the global imaging and the targeted in situ observations at the same time.* This is voiced in the Report of the NASA Living With a Star Geospace Mission Definition Team [2002], and is a central theme of the International Living With a Star (ILWS) Program. Although overarching objectives of ILWS are at once sweeping and not entirely well-defined, the spirit of the endeavor is clear:

> **From Launch of ILWS Press Release**[2]: *"The objective of ILWS is to study the connected Sun-Earth system as an integrated entity and the effects that influence life and society; to stimulate collaboration among potential partners in solar-terrestrial space missions; to coordinate international research in solar-terrestrial studies, including all relevant data sources as well as theory and modeling; and to provide open access to all data and results."*

> **Hermann Opgenoorth**[3]**:** *"The overarching objective is to explore how solar variability affects the Earth environment in the short and long term. ILWS will explore physical processes in the sun-Earth system, focusing on those with planetary-scale effects and will quantify the geoefficiency of coupling processes."*

[2] http://ilws.gsfc.nasa.gov/ilws_pressrelease_022003.pdf
[3] personal communication to Eric Donovan in support of CGSM proposal (2003) to CSA

ILWS can be viewed as a natural extension of the International Solar Terrestrial Physics Program. There is an impressive mission line up for the ILWS timeframe, which includes the US LWS RBSPs, and projects as well as SWARM, THEMIS, MMS, and possibly ORBITALS and ERG. We will be at our best ever in many ways in terms of direct observation of the physical processes at work in GeoSpace. On the other hand, with the recent failure of IMAGE, the inevitable demise of Polar, and the fact that there is not a single global imaging mission approved for the future, we will *arguably* be at our worst in over a decade in terms of our ability to specify the temporally evolving state of the global magnetosphere. In no uncertain terms, *system level science is global science*. In order to study global interactions, and in order to understand the roles of the physical processes that the new *in situ* missions are targeting in those interactions, global imaging is essential.

The concept for the Ravens satellite mission was proposed in response to a CSA AO for potential Canadian mission contributions to the International Living With a Star (ILWS) program. Ravens was conceived of to fill an important gap in the ILWS program: global imaging. Ravens will build on the heritage of world-class global imaging carried out in Canada. It would do much more than provide global observations to complete the system level capabilities of ILWS. Ravens would be comprised of two satellites on elliptical polar orbits, relatively phased on those orbits to provide the first-ever continuous (ie., 24 hours per day 7 days per week or “24/7”) global imaging of the northern hemisphere auroral and polar cap regions. This would provide the first-ever unbroken sequences of global images of the auroral response during long duration geomagnetic processes like storms and steady magnetospheric convection events. The name was chosen in reference to the Norse legend of Odin’s ravens Hugin and Munin who were sent out at dawn every day to gather information for their master.

Ravens will carry an instrument complement that provides global electron and proton auroral imaging via a Canadian FUV imager and a Belgian FUV spectrographic imager. The Canadian FUV imager will provide better spatial resolution than has ever been obtained globally, and the best (to date) isolation of two passbands in the N2+ Lyman-Birge-Hopfield (LBH) band-system. Therefore, Ravens could track the spatio-temporal evolution of the global electron and proton auroral distribution, and would obtain maps of characteristic energy and energy flux of the precipitating electrons.

Ravens would be a strategically important mission. As initially envisaged, it would accomplish the following:

- The first ever continuous global proton and electron auroral imaging spanning long duration geomagnetic processes such as the magnetic storm.
- Better spatial resolution than has ever been obtained in global electron auroral images.
- Better global maps of precipitating electron characteristic energy and energy flux than has ever been obtained previously.

As originally conceived, Ravens would provide better time resolution and energy characterization of the electron aurora than did IMAGE, and a fundamentally new opportunity to follow large geomagnetic events like storms from beginning to end. Further, it would fly at the same time as missions designed to carry out the next generation of *in situ* exploration of the physics of the processes that move energy and mass around in GeoSpace. Finally, it would be the only global imaging mission of its era. We assert that it would be a cornerstone in the set of tools used top carry out system-level science, and would have significant impact for Canada in the international GeoSpace Community.

This is the context in which the Ravens Concept Study was carried out. In what follows we provide the following:

- An outline of how we broadened the planned instrument complement to include ENA and conjugate imaging, so that together with global networks of magnetometers, GPS receivers, and riometers Ravens could credibly claim to be tracking the global evolution of the ion and electron CPSs, the ring current, the plasmasphere, and high energy (for example substorm injected) electrons and protons.

- That the Ravens concept has evolved into the imaging component of KuaFu-B, which is a planned Chinese GeoSpace mission comprised of two satellites on orbits determined by the Ravens Concept study, and which will, if it flies, carry the entire Ravens imaging instrument complement.

- An outline of our mission design, including the basis for selecting the 1.8 Re X 8 Re (geocentric) polar orbits for the Ravens satellites.

- Our rationale for including a wide field of view electron auroral imager to provide tens of minutes of conjugate imaging every orbit.

- An analysis of the amount of mid-latitude auroral imaging capability of Ravens.

- A description of how Ravens would synergize with ground-based programs to provide the first-ever simultaneous imaging of the aurora across all relevant spatial scales.

- An argument that it is essential to include *in situ* observations on KuaFu-B because it would be the only geospace mission on a high-altitude high-latitude orbit when and if it flies.

- A brief description of the instruments that will form Ravens imaging package, including issues that we feel must be dealt with in phase A.

- A more thorough description of our study of the feasibility of Canada providing the FUV LBH imagers for Ravens, including a discussion of work sharing in the development and building of those imagers (with our partners in Canadian industry, China, Finland, and Norway).

- A brief discussion of the mission-level issues related to KuaFu-B.

We also include in a series of appendices the following:

- Three publications that arose from the Ravens Concept Study Work. These are papers first-authored by Trondsen, Donovan, and Tu, on the Canadian UV imager, global imaging during ILWS, and the KuaFu mission, respectively.
- A Strawman Phase A plan
- Relevant letters of support
- A proposal for a Canadian UV imager (simpler than the one proposed for KuaFu) on a proposed NASA "Molniya" mission (note that this is not a GeoSpace mission, but this opportunity arose directly as a consequence of the Ravens Concept Study work).
- The report on the "KuaFu Assessment Study" (Chuanyi Tu, PI) that was submitted to the Chinese National Space Administration for a competition for flight opportunities in China and that led to the KuaFu-B opportunity for the Ravens team.
- A web-tool for browsing the results of the orbit simulator we wrote to explore the implications for continuous viewing and mission lifetime of different orbits.

At the time of completion of this report, partners in KuaFu-B in Europe, and China are making arrangements for Phase A activity. This is the Canadian Ravens team's next objective. Ravens should from now forward be viewed as an element of the larger KuaFu mission.

# 2. Executive Summary

Ravens was proposed as a two satellite mission to provide 24/7 global electron and proton auroral imaging. Initial work on Ravens led to a partnership involving an instrument team in Canada for the Ultraviolet Auroral Monitoring Camera (UVAMC) and an instrument team in Belgium for the Far Ultraviolet Spectrographic Imager (FUVSI). UVAMC was to deliver images in two isolated parts of the LBH, providing better spatial resolution and characteristic energies than previously obtained. The UVSI was to deliver global proton auroral images similar to those provided by IMAGE SI12, but at a higher cadence (30 seconds compared to 2 minutes on IMAGE).

As part of the Ravens Concept Study, the Canadian Ravens team was to identify opportunities for launch of the Ravens imaging package. As part of that effort, the Ravens PI traveled to China as part of a large contingent of Canadian space scientists in March 2004. Ravens was presented by the PI to Chinese space science counterparts as one avenue of possible future collaboration between Canada and China. Five months later, the PI of the Chinese KuaFu mission approached the Ravens PI to explore merging the two missions into one. KuaFu emerged as a three satellite mission, one at the L1 point (KuaFu-A) providing solar imaging and solar wind measurements, and two in elliptical polar orbits specified by the Ravens team (KuaFu-B1 and –B2). KuaFu-B1 and –B2 would be identically instrumented with the UVAMC and FUVSI. Both the Canadian and Belgian teams agreed (enthusiastically) to this evolution. Furthermore, the KuaFu-B payload was expanded to include ENA imaging and a wide field of view "perigee" electron auroral imager to provide ring current and episodic conjugate imaging, respectively. As well, because KuaFu-B would be the only geospace mission on a high-altitude high-latitude polar orbit in its era, it was agreed to include some *in situ* observations, while at the same time respecting payload mass restrictions. KuaFu is described in more detail in the KuaFu Assessment Study Report (included as Appendix E), and the paper by the mission PI (Chuanyi Tu) that is currently under review for publication in Advances in Space Research (included as Appendix C).

KuaFu-B will consist of two satellites (B1 & B2) on 90 degree inclination (polar) 1.8 Re X 8 Re (geocentric) orbits. The KuaFu-B satellites will provide the first-ever continuous (24X7) global hemispheric auroral imaging, and the only auroral imaging and high-altitude polar *in situ* observations currently planned for a key period during ILWS.

KuaFu launch is planned for 2012, with a nominal mission lifetime in excess of 2.5 years. If it proceeds, KuaFu-B1 and B2 will fly with the following instrument complement:

KauFu B1 & B2

1. Ultraviolet Auroral Monitoring Camera (UVAMC; Canada: PIs Donovan & Trondsen)
2. Far Ultraviolet Spectrographic Imager or "FUVSI" (Belgium; Pre-launch PI: Rochus)
3. Neutral Atom Imager on KuaFu or "NAIK" (Ireland; PI: McKenna-Lawlor)
4. Wide Field Auroral Imager or "WFAI" (UK; PI: Lester)
5. Fluxgate Magnetometer

KuaFu B1 only

6. Imaging Energetic Electron and Proton Instrument
7. KuaFu Agile Plasma Pitch Angle Instrument and Ion Mass Spectrometer

KuaFu B2 Only

8. High Energy Electron Detector
9. High Energy Proton Detector
10. Linear Energy Transfer Experiment
11. Tri-Band Beacon

During that period, KuaFu-B will provide

- the first-ever 24 hour per day 7 day per week (24/7) global auroral electron (FUV) and proton (Doppler shifted Lyman-alpha) imaging.
- global LBH observations with the best to-date isolation of the LBH-long and short for determining energy deposition and characteristic energy.
- the first systematic program of conjugate auroral observations.
- the best spatial resolution (35 km over much of the auroral zone) ever obtained for global electron auroral images.
- the first global auroral imaging program carried out in conjunction with the operation of networks of higher-resolution ground-based imagers in Scandinavia, North America, and Antarctica.
- in conjunction with the ground-based imaging programs, the first-ever simultaneous imaging of the aurora across all relevant spatial scales.
- 24X7 ENA imaging of the ring current ion population (this will have been accomplished by TWINS in the immediate future).
- the only high-latitude high-altitude platform for *in situ* geospace observations at that time.
- *In situ* **B**-field & particle as well as radio-beacon tropospheric observations.

With these observational capabilities, KuaFu-B will play an essential role in the overall KuaFu program of geospace science and space-weather exploration. This component of the mission will quantify energy and mass transport in the inner magnetosphere and energy deposition in the ionosphere and thermosphere, thus studying the consequences of the varying inputs to the geospace system that will be studied by KuaFu-A. As well, KuaFu-B will address exciting stand-alone science questions related to MI-coupling, mass and energy transport in the CPS and inner magnetosphere, multiscale processes in geospace plasmas, and natural complexity. KuaFu-B will also provide global contextual information for *all other LWS and ILWS observations that will be carried out during its mission lifetime.*

# 3. Ravens (KuaFu-B) Mission Design

## *3.1 Ravens Mission Objectives*

In this section, we describe our motivations, which fall under two themes. First, we begin with more overarching motivation, which is to provide continuous “24/7” observations of major magnetospheric plasma regions such as the ion and electron CPSs, the higher energy (ie., injected) ions and electrons, the ring current, and plasmasphere to address key “system level” geospace questions. This honestly really followed from the second theme, but resonates with KuaFu and ILWS, and manifests synergy with the LWS mission and the ground-based theme that is so prevalent in Canadian space science. In the second theme, (which was our initial motivation), which reflects the fact that it was an imaging group that proposed Ravens and so Ravens was viewed as a mission built around UV auroral imaging. From this perspective, we wanted to see what we could do that was new in imaging that would facilitate new science. One of those new things was the 24/7 capability afforded by the two satellite configuration, but as we describe below, there is more.

### Ravens System-Level Observational Objectives

Our objective is to present a high-level view of what we might do, as an international community, in future global auroral imaging projects, particularly within the framework of ILWS. In order to do this, we start with a brief review of the technical accomplishments of the imaging community, woven together with a very brief review of how those technical accomplishments have facilitated ground-breaking science. We then

Numerous plasma processes mediate energy and mass transfer from the solar wind to the magnetosphere, and subsequent transport through the system and deposition in the sinks. These include magnetic reconnection, plasma waves, wave-particle interactions, magnetohydrodynamic (MHD) instabilities, and parallel electric fields. Understanding these geospace processes is important because they are of obvious significance for both expanding our fundamental knowledge of cosmic physical processes and applied reasons. For a “fundamental knowledge” example, the Sun-Earth system is the only astrophysical object in which magnetic reconnection and its consequences can be directly observed. For examples of real-world applications with societal and economic effects, magnetic storms and MHD waves enhance radiation belt fluxes of charged particles that pose a danger to satellites and astronauts, magnetic storms have knocked out power grids, and there are possible links between geospace phenomena and climate.

Previous satellite and ground-based missions have mapped out the overall magnetospheric topology, and identified interesting plasma physical processes. We are now moving into what can reasonably be called the quantitative era of geospace science. The objectives are to go beyond phenomenology, and now quantify effects of key geospace processes and develop predictive capabilities that both test our understanding against observation, and bring this developing knowledge to real world applications. The International Living With a Star (ILWS) satellite and ground-based observational programs will be designed to complement each other and facilitate quantitative global studies. Key players in ILWS must build on their historical strengths, and undertake bold new initiatives. The end result will represent a true leap forward in terms of physical understanding and benefit to society. One of the overarching themes of ILWS is to develop the capacity to study geospace as a complex coupled system.

The schematic diagram in Figure 3.1 illustrates our view of geospace as a complex coupled system. All the energy and much of the mass that populates geospace comes from the Sun, represented by the box on the left (ie., sources). Ultimately, energy and mass end up in a number of sinks listed in the box on the right. Physical processes affect how energy and mass are transported from the source to the sinks. These operate in geospace, and are numerous (we have listed just a few at the center of the diagram). Many of these processes are understood only empirically, and some are understood only through simulation and theory. As well, we do not know right now which of these processes are dominant in terms of global mass and energy budget. In other words, while some of these processes occur in the system, it may be that if they were excised, there would not be any measurable effect on the global dynamic. For example, Kelvin-Helmholtz and other surface waves certainly exist on the magnetopause, but if the boundary was somehow changed so that it could not support their existence, would the ULF wave power distribution in the magnetosphere be measurably different? Would the radiation belt evolution be different? Would the MI coupling be different in its overall effect? Moreover, these processes interact with each other, there is interaction between the sinks and these processes, and interaction between the sinks (the latter in some sense being the mission deliverable of IMAGE). So it is not as simple as source, process, and sink. Geospace is a complex coupled system that is a more than a passive recipient of solar energy - it is an active participant and source of much of the dynamics that are observed.

This “systems level approach” to geospace science is the natural next step in the evolution of our field. Key missions under the ILWS banner will target physical process such as reconnection (ie., Magnetospheric Multiscale), wave-particle interactions (ie., the LWS Radiation Belt Storm Probes and ORBITALS), ionospheric-thermospheric variability (LWS Ionosphere Thermosphere Storm Probes), or Alfvenic acceleration (ie., Geospace Electrodynamic Connections and Swarm), with the objective of physics-based understanding. At the same time, greatly enhanced and ever more integrated global networks of ground-based instruments, and contemporaneous global imaging of the ring current, plasmasphere, aurora, and even the CPS [*Ergun et al.,* 2000] will provide the quantitative observations of the source and sinks necessary for complete specification of the geospace system

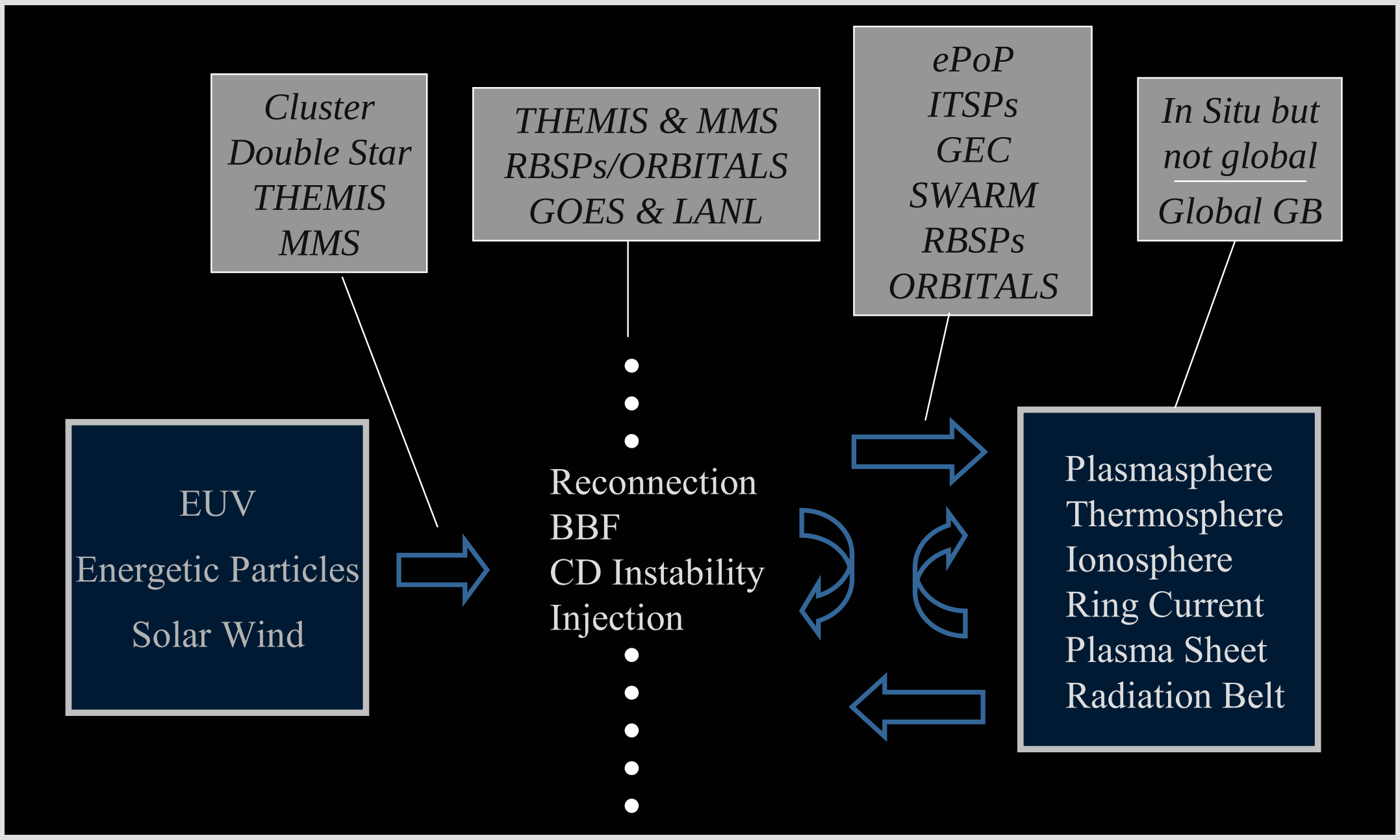

Figure 3.1: Geospace at the systems level. Energy and mass comes from the sun (the "source" at left), physical processes in geospace mediate mass and energy transport within geospace (center), and there are consequent effects on various important "sinks" (at right). There is cross-coupling between the sinks, and even between the physical processes. Virtually every planned geospace mission during the ILWS timeframe except Ravens (or KuaFu) targets *in situ* observation

.

The upcoming NASA Time History of Events and Macroscale Interactions During Substorms (THEMIS) mission is a microcosm of this systems level approach. THEMIS will involve five satellites in a constellation on the nightside magnetosphere, arranged to provide observations of current disruption, mid-tail reconnection, and communication between those two processes via bursty bulk flows and rarefaction waves. THEMIS also includes a ground-based network of all-sky imagers and fluxgate magnetometers which will specify the azimuthal evolution of the substorm disturbance via its effect on the aurora and ionosphere [see *Donovan et al.,* 2006]. This combination of targeted *in situ* and coordinated ground-based observations is necessary to close the question of what causes substorm expansive phase onset.

Although it has not always been so strategically targeted as it is in the case of THEMIS, the idea of monitoring the global spatio-temporal evolution of geospace via its impact on the above-mentioned sinks has been the driver behind SuperDARN [*Greenwald,* 1995] 1995, MIRACLE, Canadian Geospace Monitoring [see cgsm.ca], and global imaging of the aurora, ring current, and plasmasphere [*Williams.,* 1992]. It is the motivation for current activities designed to create virtual observatory networks of magnetometers, riometers, and all-sky imagers that span the globe. This includes the planned United States Distributed Arrays of Small Instruments (DASI) initiative, which if it goes forward will involve the deployment of a worldwide array of ground-based geospace instrumentation, with the stated objective being to facilitate system level science [*DASI Reports,* 2004, 2006]. It was certainly the motivation for the Geospace component of the proposed Chinese KuaFu satellite program [see Tu paper on KuaFu included as Appendix C].

ILWS must promote space missions and ground-based observations that target the solar energy and mass sources, that explore the physics of and interrelationship between geospace processes that affect energy and mass transfer, and finally that provide quantitative synoptic observations of every mass and energy sink in the Geospace system. This is a grand challenge, by any means. It is also, however, the logical next step in the over forty year exploration of geospace. Furthermore, it is fitting of an international effort that draws on the collective experiences of the last forty years and more. As we point out in the Context and Overview Section, there are a number of upcoming missions that are been flown to create a next generation of *in situ* observations of physical processes in geospace (ie., that are targeted at the middle part of Figure 3.1). At the same time, there is not one approved mission that will provide the global observations of the sinks on the right side of the figure. *Thus, while we will be at our best in terms of probing the mediating physical processes, if we are not careful we will be at our worst in terms of quantitative observations of their effects on global scales. This was the motivation for Ravens, and remains the motivation for the geospace component of the Chinese KuaFu mission which has absorbed the Ravens concept and instruments, and adopted the Ravens orbits.*

Our objective with Ravens (and now KuaFu-B) is to facilitate continuous “24/7” observations of the major plasma regions in geospace. As we state elsewhere, this comprehensive magnetospheric imaging was done to a great extent by IMAGE, but not in conjunction with the new set of missions targeting the physical processes that mediate energy and mass transport in geospace, and not 24/7. The proposed approach for Ravens (and KuaFu) is predicated on the assertion that this 24/7 global imaging will by itself lead to new insights in the study of, for example, storms. In addition, global imaging, in conjunction with the new *in situ* missions such as THEMIS, MMS, RBSPs, and SWARM, will make true system-level science possible, wherein we are able to observe the interplay between the processes and the larger-scale plasma populations.

From the outset, Ravens was to be a simple mission with a limited complement of imaging instruments. While teaming up with China on KuaFu broadens our horizons somewhat, we are still limited in payload mass, and still have a strong desire to keep the mission as simple as possible. With that in mind, we want to briefly explore what large-scale plasma regions we hope to be able to observe the global behavior of as part of ILWS. Obviously, we cannot see everything about this system, so the objective is to think in broad terms, much the same as was done with IMAGE. Further, to keep mass down, and to simplify the mission, we make compromises.

Our broad-brush view of systems-level geospace observations is outlined in Figure 3.2. As in other programs, we think of subsystems that comprise the system. These are the ion and electron CPSs, higher energy (injected) ions and electrons, the ring current, plasmasphere, global field-aligned current system, and radiation belts. Below each subsystem, we point out what the optimal solution for "imaging" that subsystem would be, given infinite resource. In light of the previous paragraph, we assert the following. The evolution of the plasmasphere, and the high-energy injected electrons can be less directly but *adequately* (note what is meant by adequate is something that needs to be explored as the mission evolves) observed via networks of ground-based instruments such as magnetometers, GPS systems, and riometers. Low altitude satellites such as SWARM and possibly AMPERE can provide latitudinal cuts through global field-aligned currents (FACs). The RBSPs will provide similar near-equatorial cuts through the radiation belts. Global UV imaging in the LBH and Lyman-alpha is the best way we have to remote sense the electron and proton CPSs (at least their projection into the ionosphere). Further, Lyman-alpha imaging can give information about injected protons. Finally, the best information about the global ring current is provided by ENA imaging.

These ideas are developed further in the subsections on what instruments we need, below. For now we summarize up front where we get to in the end. The imaging objective of KuaFu-B is for global electron and proton auroral imaging to follow electron and ion CPS dynamics, and ENA imaging to track the evolving ring current ion population. This will be complemented by global networks of ground-based instruments, low-altitude satellites such as SWARM, and the RBSPs. The combination will be global specification of the subsystems identified in Figure 3.2. This combination will be available on a continuous 24/7 basis.

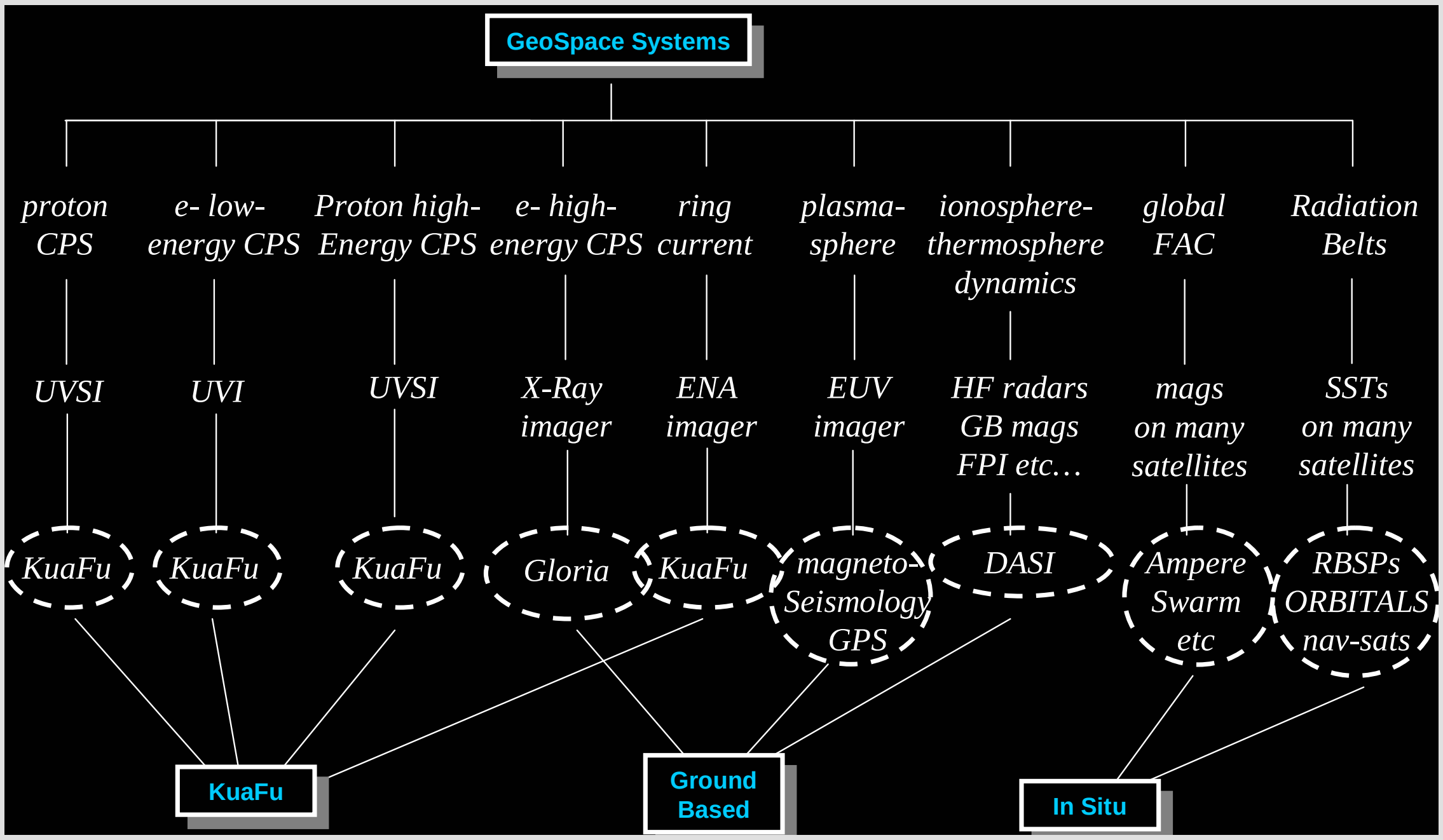

Figure 3.2: The main geospace plasma regions that we would image won KuaFu-B are the proton and electron CPS, high energy injected protons and electrons, and the ring. Other subsystems would be observed via global arrays of ground-based instruments or SWARM and the RBSPs.

## New Frontiers in Global Auroral Imaging

A brief history of global auroral imaging is spread through Sections 5 and 6 in the introductions to the specific imagers for the mission (particularly the UVAMC). There is a concise history of UV auroral imaging presented in the Donovan Advances in Space Research paper on global imaging for ILWS that is included as Appendix B. A very concise statement that reflects what shaped our goals is as follows.

By the time IMAGE was flown, there had been great advances, and great achievements, in global auroral imaging. Still, after 35 years of global imaging from space, there is a surprising list of technical challenges that have either not been met, or have been done to a degree that could be improved upon. These include the following:

1. *Simultaneous imaging across all relevant scales*: Spatial resolution of global images has to date been limited to no better than ~100 km. As well, it is only in recent years that ground-based arrays of all-sky imagers (ASIs) are capable of providing continent-scale images with ~km resolution. Imagers with resolution down to tens of meters [see *Trondsen et al.*, 1997] have not been operated with larger-scale imagers (ie., with spatial coverage larger than that of one ASI). Consequently, there has never been a systematic program for simultaneous auroral imaging across all important spatial scales (ie., from global down to tens of meters). All relevant spatial scales can be observed with existing technologies (see Figure 5.2). Provided that global observations are available in the future, simultaneous imaging across all relevant scales will be a reality. There is an important caveat, however, in that kilometer scale resolution images will be available for only a part of the global auroral distribution, and even higher resolution images for even smaller regions. This scenario will provide a complete picture of the cross-scale distribution only if fine scale structure is not globally organized. While this is a logical assumption, it has not been proven.

2. *Spectral Resolution of sufficient quality to allow for global maps of characteristic energy and total energy flux of incoming electrons*: The most direct way of deriving the auroral energetics is to compare large scale images in two spectral regions of the far-ultraviolet LBH band-system; 140-160 nm (``LBH short" or ``LBH-S") where O2 absorption at auroral altitudes is significant and 160-190 (``LBH long" or ``LBH-L") where the emissions are relatively free of absorption [*Strickland et al.,* 1983; *Germany et al.,* 1990, 1994]. While there are difficulties with this approach due to factors such as O2 upwelling and other factors that obfuscate atmospheric composition, it is the best technique available. Previous global auroral imaging experiments have obtained nearly simultaneous images from different parts of the electron auroral spectrum (ie., see Polar UVI above and studies utilizing Polar-PIXIE data [*Imhoff et al.,* 1995; *Anderson et al.,* 2000]. The UV data, however, has not yet met the technical requirements necessary to quantify the spatial dynamics of the average energy and total energy flux of precipitating particles on a global scale, and as such, time evolving maps of the global distribution of both characteristic energy and total energy flux have not been obtained. Instrument development programs are currently underway in several institutions to achieve the technical requirements stated here. Strategies that are being explored include the use of newly developed materials for reflective coatings and filters for better out of bandpass rejection.

3. *Long duration continuous global auroral imaging*: Previous global imaging experiments have been flown on single satellite missions, and hence have provided continuous imaging (at whatever cadence the imager is working at) for only a fraction of the satellite orbit. The UV imagers on Polar, for example, provide continuous complete hemispheric auroral coverage for less than half of Polar's 18 hour orbit. As a consequence, there are no unbroken sequences of images that span the relatively long time-scale geospace processes such as magnetic storms. Continuous global auroral imaging of one hemisphere can be obtained with two satellites on elliptical polar orbits, relatively phased on those orbits so that when one satellite is at apogee the other is at perigee. As is pointed out in the subsection on Orbital Design, below, it is straightforward to generate several years or more of continuous imaging with such an orbital configuration (ie., this is subject to precession of the line of apsides).

4. *Simultaneous global imaging of both hemispheres*: The northern and southern hemisphere auroral ovals have been simultaneously imaged during serendipitous events when two UV imaging satellites had a clear view of the two hemispheres [*Craven et al.,* 1991; *Østgaard, et al.*, 2004; *Stubbs et al.,* 2005], or when one satellite had an oblique view of the two hemispheres. To date, there has never been a systematic program for interhemispheric conjugate auroral observation on the global scale. As in the case of continuous global auroral observations, interhemispheric conjugate observations can be achieved with well planned orbital configurations in multi-satellite missions. In a KuaFu-B type scenario, in which two satellites are are used to provide continuous global observations in one hemisphere, a ``perigee-imager'' on the satellites would provide some conjugate observations during every orbit of both satellites. As the continuous imaging can only be achieved with two satellites if they are on highly elliptical orbits, the perigee imager would have to have a wide field of view. As we point out in the subsection on Conjugate Imaging, below, a design for an imager with a wide field of view (based on microchannel plate properties) which could be flown on a low earth orbiting spacecraft has recently been developed [*Hamilton et al.,* 2005], utilizing a proposal for an all sky X-ray camera [*Fraser et al.,* 2002].

All of the above technical objectives could be met within the next decade. Over and above meeting these one by one, there will be synergy between missions that are carried out at the same time. There are in general several geospace missions and the world-wide array of ground instrumentation operating simultaneously. This was the essence of ISTP, and will be an objective in ILWS. With synergy between missions and ground-based observations it is possible to carry out new science with instruments that have been flown before, but not previously in a given combination. Synergy is not a new tool in space physics, but new synergies will bring new opportunities.

## Science Objectives

We therefore have two sets of operational objectives. The first, which is very broad-brush and which can only be met in combination with global networks of ground-based instruments[4], is to provide continuous 24/7 global specification of the subsystems that make up geospace. The second is to push the envelope on what has been done with global auroral imaging. These are clearly interrelated as, for example, the ability to provide 24/7 global auroral imaging is an advancement of auroral imaging capabilities, but is necessary for the system-level geospace studies.

A mission that would provide 24/7 global auroral and ring current imaging, involving better spatial and spectral resolution than ever before achieved in the electron aurora would provide data that would pervade virtually every aspect of geospace research. This broadness is in some ways a strategic weakness for global auroral imaging. It is sometimes perceived of as providing nothing more than context for more "physics-based" missions (ie., MMS, THEMIS, RBSPs, etc) that target *in situ* processes. In other words, global auroral imaging can be viewed as playing a not-very-sexy supporting role, being something for everyone but nothing special in particular. This, we believe, is an unfair characterization.

In another view, "imaging is everything". It is our only way to observe the global system we are studying, and the only way to quantify the effects of physical processes of interest. System-level science is simply not possible without imaging. Further, new technical capabilities in imaging would support fundamental new science on its own (or in conjunction with ground-based imaging systems). We (particularly the PI) thought hard to find that key isolated and interesting problem[5] that Ravens or KuaFu-B would go after. We (at least the PI) could not accomplish this. The fact is, however, this is not why we conceived of Ravens and ultimately KuaFu-B. This mission arises out three desires on our part. First is our belief that system-level geospace science is where our field is going and that global auroral imaging is inextricably linked to it. The second is that a next-generation electron and proton auroral imaging system would open up new avenues of research in areas that we (the Ravens or KuaFu-B team) are fundamentally interested in, related to, for example, the effects of auroral type on MI coupling, and the relative importance of Alfven wave acceleration in the overall auroral process). The third is we believe that global imaging with new capabilities will play an essential role in studies of Natural Complexity and Universal Processes which have recently entered the lexicon of our field. We believe that natural complexity will lead to some of the most fundamental geospace science, addressing how structure arise spontaneously in cosmic plasmas and perhaps shattering our reductionistic views of geospace processes such as the substorm. An honest statement is *this is why we do what we do.* In this subsection, we provide just three examples (one from each of the three themes mentioned here) of research directions that Ravens (and KuaFu-B) would make possible.

[4] It is virtually certain that global networks of magnetometers, GPS, riometers, and HF-radars will grow over the next decade expanding in density and extent. Further, the advent of Virtual Observatories means that these data will be available in an integrated fashion (this is already true for SuperDARN and the GPS systems), and will be true for the magnetometers and riometers.
[5] For example – THEMIS is motivated by a simply stated question: "what comes first, current disruption or the near-Earth neutral line?"

At the system-level, the interactions between the ion and electron CPS, the ring current, and the ionosphere/thermosphere in storms and substorms are only understood in part. This is an overarching theme in geospace that encompasses many problems. Just a few include: How important is precipitation in the thermospheric energy budget? How important is precipitation and Joule heating in modifying the mid-latitude ionosphere? What is the role of substorms in storms? How does the CPS feed plasma into the ring current? How does substorm injected plasma interact with pre-existing inner magnetospheric populations? Are sawtooth events substorms? There are many more. Moreover, these interactions take place over days or even weeks, and "system level" understanding of geospace phenomena, which is the overarching objective of the ILWS geospace program, will never be possible without global observations that span long duration geospace phenomena such as magnetic storms. By way of example, we focus on one topic which is "how do dynamic processes stage CPS particles for transport to the inner magnetosphere, and what are the effects of those injected populations?"

In figure 3.3, we show the evolution of the ion plasma sheet as inferred from FAST Electrostatic Analyzer (ESA) ion data from successive evening sector passes through the northern hemisphere auroral zone over 130 hours that brackets a large geomagnetic storm. There is a detailed description of the plot in the figure caption. The geomagnetic activity unfolds over a time period longer than three days. From the FAST ESA data, we can see the ion plasma sheet undergoing three successive "injections" into the inner magnetosphere, starting from poleward of 70° magnetic latitude (reflecting a very quiet magnetosphere), ultimately reaching equatorward of 55° magnetic latitude, with clear evidence of significant energization in the ion spectra. These inward steps of the ion plasma sheet transport the ion population across the outer radiation belt (as evidenced by the band of penetrating radiation). Furthermore these inward steps are not rapid, unfolding over 10 or more hours in each case.

This is a classic situation. The very quiet magnetosphere prior to the beginning of the storm undoubtedly means an expanded plasmasphere. The inward transport of the ions undoubtedly corresponds to increased electric fields in the inner magnetosphere which would in turn lead to erosion of the plasmasphere. The overall consequence is spatial overlap of the plasmaspheric, ring current, radiation belt, and CPS plasma in an extremely dynamic electric and magnetic field. These plasma populations will interact, leading to enhanced precipitation, changes in the radiation belts, and an enhanced ring current. Furthermore, these interactions happen on global scales, and unfold over tens of hours. These interesting dynamics have been exposed by the ground-braking IMAGE observations as interpreted by *Goldstein et al.* [2005a, 2005b], but to date we cannot track these subsystems for events of this duration (in other words – for the duration of any magnetic storms of note).

KuaFu-B (in conjunction with the above mentioned ground-based networks) will provide us with just this capability. We will be able to follow the global ionospheric projection of the ion and electron CPSs throughout such an event. This will enable us to explore how the CPS ion and electron populations are staged to interact with the inner magnetospheric ring current, radiation belts, and plasmaspheric populations, and quantify the effects on the inner magnetospheric populations and the ionosphere-thermosphere via storm-time precipitation for entire storms. These data will be invaluable when used to truth global models that simulate such events, pointing the way to gaps in our understanding of these global interactions.

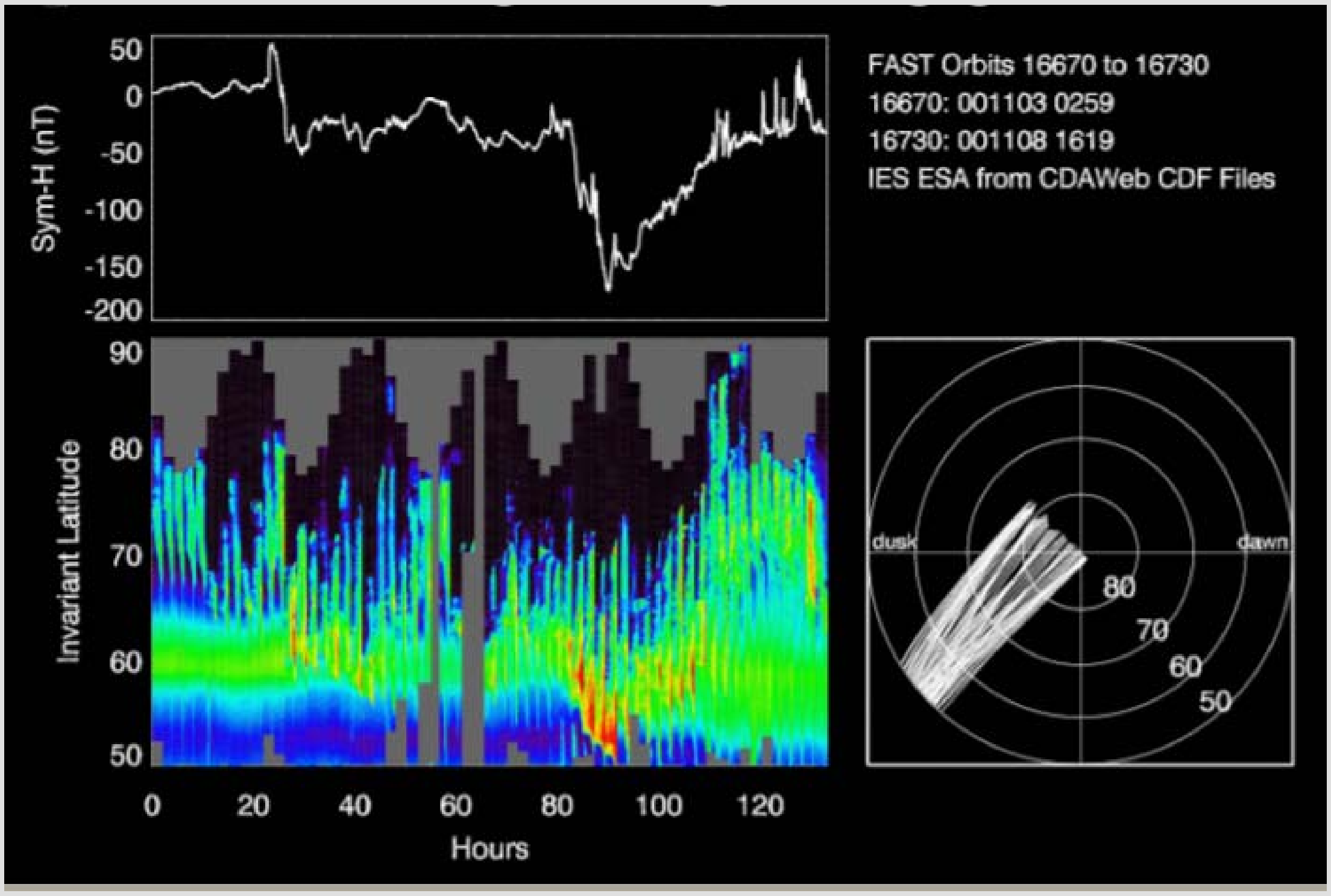

Figure 3.3: Fourteen days of in situ ion data from the FAST ESA ion instrument. The lower left panel shows successive electron spectra for evening sector orbits that span an active period during which there are several large geomagnetic storms. During these events, the equatorward boundary of the ion CPS undergoes significant movement in latitude. More important, that motion caries the ion plasma sheet deep into the inner magnetosphere, across the outer radiation belts. This is an excellent example of how different the effects of storm-time convection can be on different plasma populations. One of the *hot topics* in geospace science is the transport, acceleration and loss of inner magnetospheric particle populations (ring current, radiation belt, plasmasphere). Especially during storm-time, the CPS serves as a source for ring current particles, and potentially interacts with the plasmasphere. Time scales for ring current and plasmaspheric changes are often long compared to typical geospace satellite orbits. Lower left panel) FAST ion ESA energy spectra for 130 hours bracketing two geomagnetic storms (see Sym-H top left). The fast data is from successive passes of the evening sector auroral oval (see footpoints bottom right). In the energy spectra, the outer radiation belt shows up as the diffuse green band straddling 60 degrees PACE invariant. At T=0, the inner edge of the CPS is at ~70 degrees PACE. During the two successive storms, the CPS ion plasma moves across the radiation belt, then retreats. This is a clear example of the intrusion of the ion CPS into the inner magnetosphere on a timescale of tens of hours. Studying the two-dimensional spatio-temporal evolution of the ion and electron CPS is a primary technical objective of KuaFu. KuaFu data would allow for the role of the ion and electron CPS in storms to be explored in fundamentally new ways.

Moving away from the system-level and into more traditional areas of auroral research, spectral resolution of the LBH-S and -L as proposed for KuaFu-B will allow for time evolving maps of the energy deposition and characteristic energy. There are two main sources of the electron aurora, the first being the magnetospheric plasma sheet. The electric fields generated by the large-scale convection of magnetospheric plasma create currents that close through field aligned currents. These currents cannot be supported by the conductivity of the plasma at the ionosphere/magnetosphere boundary and electric potentials develop that accelerate the particles and can divert portions of the plasmasheet population into the loss cone producing the so called inverted-V electron energy distribution that is often observed in satellite transits of the auroral oval. The second source is due to the interaction of the cold electrons in the outer ionosphere, within several Re of the earth, with Alfven waves propagating along Earth's magnetic field. This interaction leads to downward acceleration and precipitation of very high fluxes of low energy beam-like electrons. The Alfven waves are indicative of reconfigurations of magnetic lines of force in the magnetotail that occur during all types of reconnection events including substorms.

Finally, we come to natural complexity. Dramatic auroral displays over Earth's polar regions reveal a range of intricate patterns. Auroral spirals, westward travelling surges, auroral curls, and many other phenomena have been identified on time scales from seconds to hours and spatial scales from 100 metres to thousands of kilometres [Borovsky, 1993]. More recently, attention has begun to focus on whether there may be universal aspects to auroral structure, of the type observed in many natural complex systems [Science, 1999].

One such universal footprint is fractality – statistical similarity when viewed on any scale [Mandelbrot, 2002]. Analyses of the durations, areas and energy of auroral bright patches [Lui et al., 2000; 2002; Uritsky et al., 2002; Kozelov et al., 2004] and of durations and sizes of magnetic fluctuations from associated electrical currents [Freeman et al., 2000] have shown that they have a power law distribution over a wide range of time and space scales. This implies that, in general, auroral fluctuations have no characteristic scale because a power law has no characteristic scale in its functional form. For example, consider the maximum area of an evolving auroral bright patch. For any area from the smallest observed area of x $km^2$ to the largest observed area of y $km^2$, there are always x times as many patches of half this area and y times as many patches of double this area. Thus the auroral landscape looks similar at any scale over x orders of magnitude.

It has been suggested that the ubiquity of scale-free landscapes in Nature may result from Self-Organised Criticality (SOC) – the tendency of complex systems to self-organise into a critical state that is invariant under scale transformations [Bak et al., 1987]. This was first identified in computer experiments mimicking a slowly growing pile of sand, but this abstract model describes a general class of systems whose governing equations have the form of a diffusion equation with a non-linear diffusion coefficient. An analogous equation can be derived from the magnetohydrodynamic induction equation in a simplified 2-D current sheet with a resistivity that depends non-linearly on the current [Klimas et al., 2000]. In this way, scale-free auroral structure has been argued to arise from a projection of dissipative structure in the magnetotail [*Kozelov et al.,* 2004]. Some support for this interpretation is provided by the observation of bursty bulk flows (BBFs) in the magnetotail that are correlated with auroral emissions and have a scale-free distribution of durations [*Klimas et al.,* 2000]].

More generally, from the scaling of different moments of fluctuations in the auroral electrojet indices, evidence is emerging for multi-fractal structure which may be described by a Fokker-Planck equation or fractional kinetic equations. Thus, the identification of scaling in auroral fluctuations has led to increased recognition of the importance of coupling between all time and space scales and to new physical models. But there is still much to be understood, for which the KuaFu auroral imaging offers unique opportunities:

*First, at what scale does scaling break down?* Analysis of auroral electrojet indices in the time domain and of auroral bright patches in the spatial domain reveals a power law distribution of fluctuation duration and area that is broken by an additional population with a characteristic duration of ~2 hours and area ~2 million km2, corresponding to the substorm. However, analysis of evolving auroral bright patches in both time and space simultaneously reveals no such break on scaling. This has led to two different views of the substorm: On one hand, the substorm is a large-scale auroral fluctuation that is distinct from other auroral fluctuations due to either different physics or because it is constrained by the magnetospheric system size. On the other hand, the substorm is just part of a continuum of auroral fluctuations that extends up to the substorm scale and possibly beyond. A resolution of this paradox is clearly important for understanding. One possible explanation is that a sufficient number of long and large evolving patches have not been observed in order to distinguish between these two possibilities. For example, assuming a characteristic substorm duration of 2 hours and average area 2 million km2, the time-integrated area of a substorm may be estimated to be ~10^10 km2 s. It is interesting to note that the distribution of time-integrated sizes of evolving auroral patches appears to deviate from power law behavior above this size, although it is unclear whether this is due to an additional substorm population of characteristic scale or whether it is due to increased statistical noise due to low occurrence. Certainly the evolving patch distribution provides no evidence for scaling beyond the substorm scale because the maximum time-integrated scale is ~10^11 km2 s. This is limited by the duration for which an auroral imager on a single spacecraft can trace a global auroral disturbance, which is typically less than 9 hours. By its continuous monitoring of the whole auroral oval, KuaFu offers the possibility to resolve this substorm paradox by both extending the range of time-intergrated area by orders of magnitude and by observing many more events at ay given scale, compared to a single spacecraft imager.

*Second, what is the origin of scaling?* Analyses of scaling in the magnetosphere have been primarily limited to the auroral oval and to its magnetically conjugate magnetotail plasmasheet. However, a very recent study has found that ionospheric velocity fluctuations scale differently in the polar cap and auroral zone [*Gary Abel and Mervyn Freeman*, private communication], supporting an earlier suggestion based on a more limited analysis of magnetic fluctuations. These differences between fluctuations in the polar cap and auroral oval likely reflect the processing of solar wind fluctuations by the magnetopause and magnetotail, respectively, and thus may provide important constraints to respective models of reconnection or other processes in each region. The continuous 24/7 observations of KuaFu should allow sufficient statistics to be compiled on auroral fluctuations in the polar cap to be compared with those of the auroral region to see whether auroral fluctuations also scale differently in these two regions and to build a more complete electrodynamic model of the relevant processes.

In conclusion, we have argued that KuaFu will be essential to system-level science, will advance our understanding of global auroral electrodynamics, and will facilitate exciting advances in natural complexity in geospace. We reiterate that the system-level argument is predicated on three things: first, global observations are essential to system level science and second, KuaFu-B is the only mission currently planned for the ILWS era, and third, continuous 24/7 global imaging will allow studies of global interactions during entire storms, rather than for just hours at a time. Advances in global auroral electrodynamics will follow naturally from the new capabilities of the electron auroral imager on KuaFu, which will include isolation of LBH-L and –S and better global spatial resolution than ever before achieved. KuaFu's unbroken sequences of auroral imaging, better spatial resolution, and coordination with ground-arrays which will allow for contemporaneous imaging down to much smaller scales still will be important (even essential) for the evolving field of natural complexity. *These were our motivations for proposing Ravens and for KuaFu-B.*

In the remainder of the report we introduce further science material. This relates to mid-latitude imaging, conjugate imaging, and synergy with ground-based programs. *These were not motivating factors for the mission, however we include them because in developing a mission plan it became evident that KuaFu will provide new opportunities in these areas. Furthermore, after analyzing the orbits we needed for 24/7 global auroral imaging of the northern hemisphere, we understood that with the addition of one more imager (the UK Wide Field of View UV Auroral Imager or WFAI) we could accomplish systematic conjugate auroral imaging (albeit for tens of minutes every orbit) for the first time. Relevant science issues are discussed in the appropriate subsections below.*

## 3.2 *What Can We Do From the Ground?*

Auroral imaging from space has and continues to be our only true means of studying the spatio-temporal evolution of the magnetospheric system. In the very early days, global auroral images proved in an instant the existence of both the continuous auroral oval and the diffuse aurora. Global electron auroral images have been used to map the evolution of the substorm expansion, the effects of solar wind pressure pulses on the central plasma sheet, and in some cases even the steady-state nature of dayside reconnection. Because of global auroral imaging, we are able to study the complex interplay between convection, conductivity, and field-aligned currents. Whenever they are available, global auroral images are used for essential context for other studies.

### Tracking the Evolution of the Plasmasphere

The plasmasphere is now understood to be important in the overall geospace dynamics. Overlapping CPS, ring current, and plasmaspheric plasma in the outer plasmasphere, drainage plumes, and separated plasmaspheric patches lead to a host of interesting processes that affect the larger particle populations. Examples of this include pitch angle scattering of radiation belt electrons by plasmaspheric hiss, and injected plasma affecting the plasmapause [see, for example, *Lyons et al.,* 1973, 1974; *Goldstein et al.,* 2005a, 2005b]. While radial profiles of the plasmasphere provided by *in situ* particle flux measurements have been available for decades, it was the advent of global imaging provided by the EUV imager of the IMAGE mission (which images sunlight resonantly scattered by ionized Helium, a significant constituent of the plasmasphere which is optically thin at the instrument bandpass) that unveiled the dynamic global nature of the plasmasphere for the first time [see *Sandel et al.,* 2001]. There are, however, two methods of remote sensing cold plasma from the ground that can, in principle, be used to track plasmaspheric dynamics. We point out that while these techniques show promise (magnetoseismology and differential GPS, as discussed below), and in some cases remarkably good results, global imaging in EUV would be our preferred technique for plasmaspheric imaging.

Using geomagnetic micropulsations as a means to probe the magnetosphere (magnetoseismology) was suggested as far back as the sixties [*Troitsaya and Gulelmi*, 1967], and in the recent years it developed into a mature and commonly used technique [*Waters et al.,* 1996; *Menk et al.,* 1999; *Denton and Gallagher*, 2000; *Dent et al.,* 2003; *Clilverd et al*., 2003; *Rankin et al.,* 2005]. Progress in magnetoseismology stems from several factors, which include improved data analysis techniques, advances in description of the Earth magnetic field, and better numerical models for shear Alfven waves propagating along the closed geomagnetic field lines. Modern data analysis techniques, such as the cross-phase analysis [*Waters et al.,* 1994] allow monitoring of the frequencies of the standing shear Alfven waves (Alfven continuum) almost continuously using the ground-based meridional chains of appropriately spaced magnetometers. Empirical models of Earth magnetic field, such as Tsyganenko models [*Tsyganenko* 1995, 2002] provide a very reasonable description the magnetic field near Earth for typical solar wind conditions and require negligible computer resources to run. In addition, Global Circulation (MHD-based) models of the Earth magnetosphere [*Powell et al.,* 1999; *DeZeeuw et al.,* 2004], although still quite computationally intensive, may now be used to obtain a self-consistent description of the Earth plasma and magnetic field environment and may be also used for magnetoseismology purposes. Finally, recent developments in modeling the shear Alfven waves [*Rankin et al*., 2004] have significantly improved our capabilities to accurately estimate Alfven continuum frequencies in complex magnetic field topology, as compared to older models [*Singer et al.,* 1981], which were previously used for these purposes.

In addition to the important developments in the theoretical foundations of magnetoseismology over the last several years significant upgrades and extensions of the existing ground-based magnetometer arrays are currently under way. For the practical purposes of supporting a space mission, we can rely on almost continuous real-time data stream coming from from the CGSM magnetometers, China/MERIDIAN 210, Scandinavian IMAGE chain, SAMBA, MEASURE, West Greenland array as well as numerous additional magnetometer stations in the US, Europe, and Russia and in some cases magnetically conjugate stations in the southern hemisphere. An example from an application of magnetoseismology to CANOPUS Churchill line magnetic field data is included in Figure 3.4, below (courtesy Konstantin Kabin). Therefore, we have a technical ability for a near global almost continuous coverage of the plasma density content of the inner magnetosphere and plasmasphere.

There is another ground-based technique that can be used to provide additional information about the plasmaspheric dynamics, at least as projected onto the ionosphere. *Vo and Foster* [2001], and *Foster et al.* [2002, 2004] have presented the results of using Global Positioning System (GPS) scintillation observations to produce maps of total electron content, which in turn show the unmistakable signatures of plasmaspheric dynamics (see Figure 3.5, below). With both magnetoseismology and networks of GPS systems there is a reasonable expectation that we will have reliable information about the dynamics of the plasmasphere during the KuaFu mission timeframe, even without EUV imaging from space. These techniques must be developed further, and the necessary steps need to be taken on the international front to integrate these data for convenient use global applications. The integration of these data into global data sets is one of the prime motivations for several Virtual Observatories that are currently funded, which means that by 2012 these data should be readily available in a format that allows creation of global-scale data products.

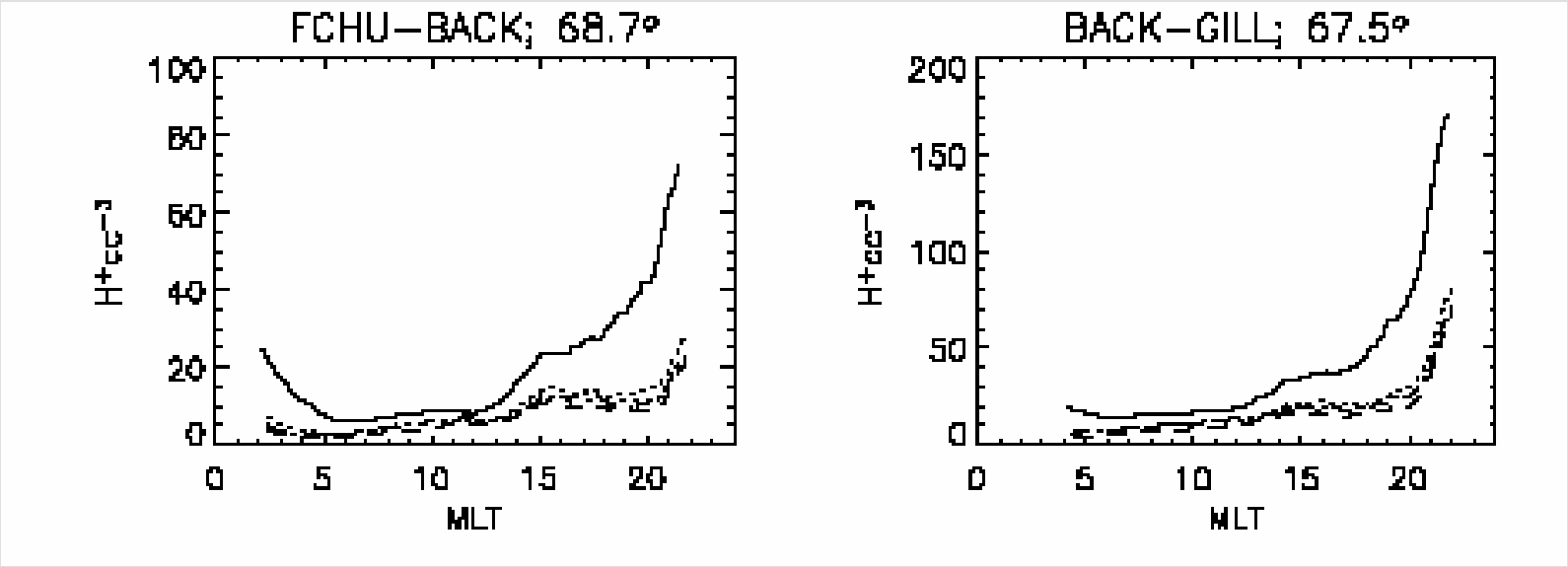

Figure 3.4: Example proton number densities inferred using magnetoseismology along the CANOPUS Churchill line together with three different magnetic field models. Figure courtesy of Konstantin Kabin.

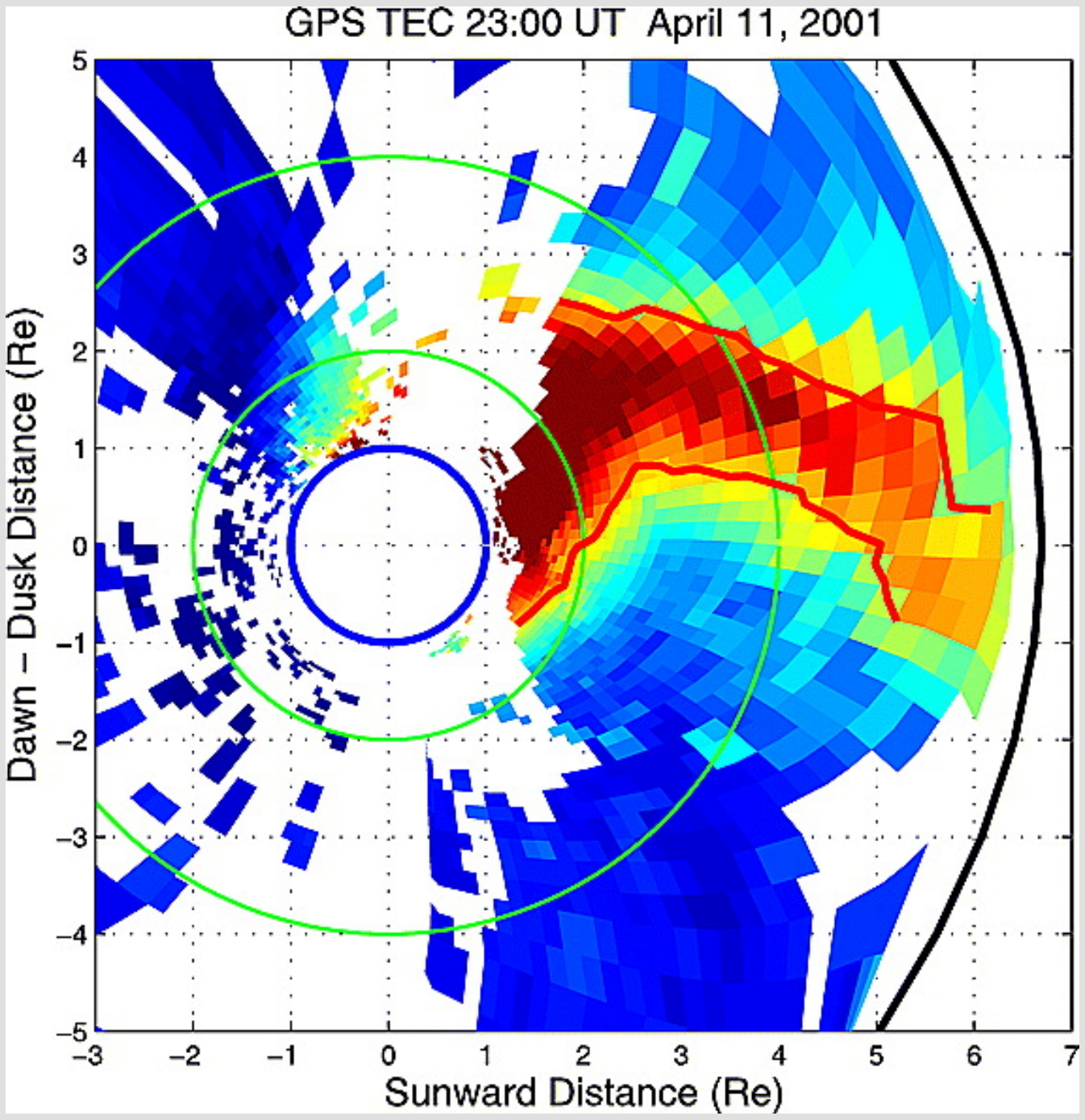

Figure 3.5: GPS TEC observations from the North American network of GPS receivers used by *Foster et al.* [2004] to explore the dynamics of a plasmaspheric plume, mapped to the magnetic equatorial plane via a global magnetic field model. This figure is reproduced directly from *Foster et al.*, [2004], wherein it appeared as their Figure 2.

## Tracking the Evolution of the Global High-Energy Electron Dynamics

The global high energy electron population varies dramatically during geomagnetic activity and as a result of the solar terrestrial interaction. Substorm injections of high energy electrons are present in virtually all substorms, although the nature of the injection process is only partially understood [see for example *Li et al.,* 2003], and extreme solar events have even been shown to create new radiation belts [see for example*, Li et al.,* 1993]. If we are seeking to understand geospace at the system-level, then we want to be able to track the evolution of the magnetospheric high energy electrons. As well, precipitating high energy electrons can significantly modify the ionosphere, and their population is often in strong pitch angle diffusion (for injected plasma) and sometimes subject to precipitation due to wave particle interactions (for radiation belt electrons) during interesting events [see, for example, *Baker et al.,* 1981]. Thus, we can track the magnetospheric high energy electron population global evolution to a reasonable extent by following the high energy electron precipitation on a global scale.

Satellite based X-Ray imaging is currently the only well recognized method of following the evolution of high energy precipitation on a truly global scale. Monitoring auroral X-rays also has some unique advantages over UV and visible imaging. First, they are only produced by high-energy particles via the Bremmstralung effect. The kinetic energy of a particle which produces an auroral Bremmstraulung X-ray is generally >30keV (*see for example Belon et al. 1966*). This means, you really are monitoring the high-energy component of precipitation, allowing studies of, for example, the spatio-temporal evolution of substorm injected electrons [see, for example, *Imhoff et al.,* 1992; *Østgaard* et al., 1999], or loss of relativistic electrons via wave-particle interactions in the inner magnetosphere [see, for example, *Millan et al.,* 2002]. Second, bremmstralung x-ray spectra are extremely predictable which allows inversion techniques for multi-wavelength observation to easily obtain an estimate of the characteristic energy. Third and finally, X-rays have the advantage that they can be observed under both sunlit (day) as well as dark (night) conditions since the solar x-ray flux is not very intense. PIXIE, the Polar Ionospheric X-ray Imaging Experiment, was the first true 2 dimensional satellite borne X-ray imager, and has been widely used in studies of global scale processes that involve the magnetospheric high energy electron population.

Given our stated scientific objectives and the list of parameters that we want to observe, we would either want an X-ray imager in Ravens (KuaFu-B), or an alternative and reasonable proxy measurement for the high energy electron precipitation. There are, in fact, opportunities to use ground based instrumentation as a “stand in” for global x-ray imagers. In general, electrons which possess enough energy to produce X-ray aurora will also penetrate the ionosphere to depths where they can significantly alter the electron density and effect the propagation of HF radio waves. Riometers (*Relative Ionospheric Opacity Meters*) measure the relative attenuation (referred to as absorption) of these HF waves (typically at 20-60MHz) to remote sense the precipitation of high energy particles. Like X-rays, riometer absorption is only produced by the high-energy particle population. In the case of X-rays a particle must have enough kinetic energy to decelerate and produce an X-ray photon, in the case of riometer absorption the particle must have enough speed to reach the ionosphere's D-region where rising electron-neutral collision frequencies make HF radio propagation most affected by changes in electron density. These particle populations, under "normal" ionospheric conditions consist of the >30keV electrons and MeV protons. Indeed, studies that use riometer data as well as global images from PIXIE have confirmed that absorption correlates extremely well with the average x-ray flux seen above the instrument (*see for example Christensen et al. 2003*).

Figure 3.6 shows a sequence of three images derived from the CANOPUS/NORSTAR network of riometers (13 riometers across northern Canada). We use a thin-plate-spline technique in a geographic grid to produce an "absorption image" in which each riometer contributes a pixel to an interpolated image. Also shown are the PIXIE and UVI images corresponding to the same time period. The riometer image which is generated from only 13 instruments does a reasonable job of reproducing part of a PIXIE image. It is also apparent from the UVI image that the morphology of the x-ray and UV aurora are different highlighting the differences in the two particle populations (see Østgaard et al 1999).

The technique for generating absorption images obviously needs more testing (as is evident in Figure 3.7) and will benefit greatly by including denser networks of riometers from around the globe. Effort is currently being made by the worldwide riometer community to integrate data from existing wide beam riometers and produce the Global Riometer Array (GloRiA). GloRiA will consist of potentially 100 riometers (North American and European sector riometers that will contribute to GloRiA by 2012 are shown below). We point out that an expansion of the riometer coverage across Canada, which will increase the number of riometers from 13 to 28, is currently underway. This has significant implications for our ability to produce nearly global images of high-energy precipitation.

From a data standpoint the riometers add temporal resolution and continuity of observations that cannot be achieved with a single satellite borne x-ray imager such as PIXIE. With often 1 second resolution, a network of riometers can produce higher cadence data than is possible using current X-ray imager technology. This is useful, for example, in wave-particle interactions and substorm onset studies where temporal resolution has an impact of the scientific quality and usability of the data.

The major drawback to using riometers as a replacement for an instrument such as PIXIE is the lack of energy information in the riometer signal. Riometer absorption is determined by the integrated energy flux deposited in the ionosphere so there is little hope of obtaining characteristic energy. Studies using multi-wavelength all sky cameras and imaging riometers have, on a case by case basis, been able to invert the precipitation measurements into an energy map (see Kosch et al. 2001), but these techniques are limited to specific events and small regions.

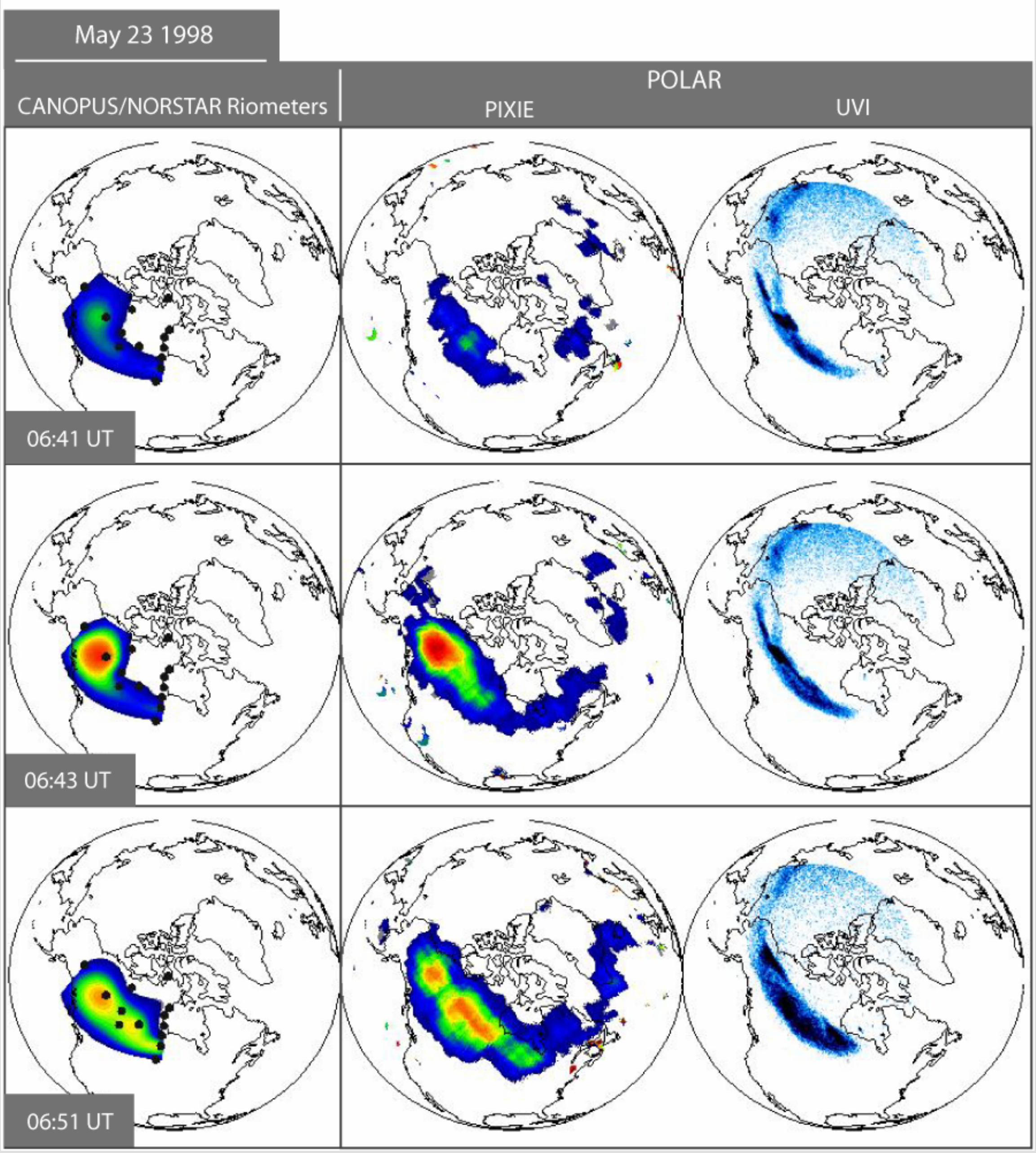

Figure 3.6: Sequence of near-simultaneous images from Polar PIXIE, UVI and those generated from the NORSTAR/CANOPUS riometer network. Riometer images are generated with a thin-plate-spline technique from 1 minute averaged data centred about the PIXIE image timestamp. (PIXIE data courtesy A. Aasnes, UVI data courtesy K. Liou)

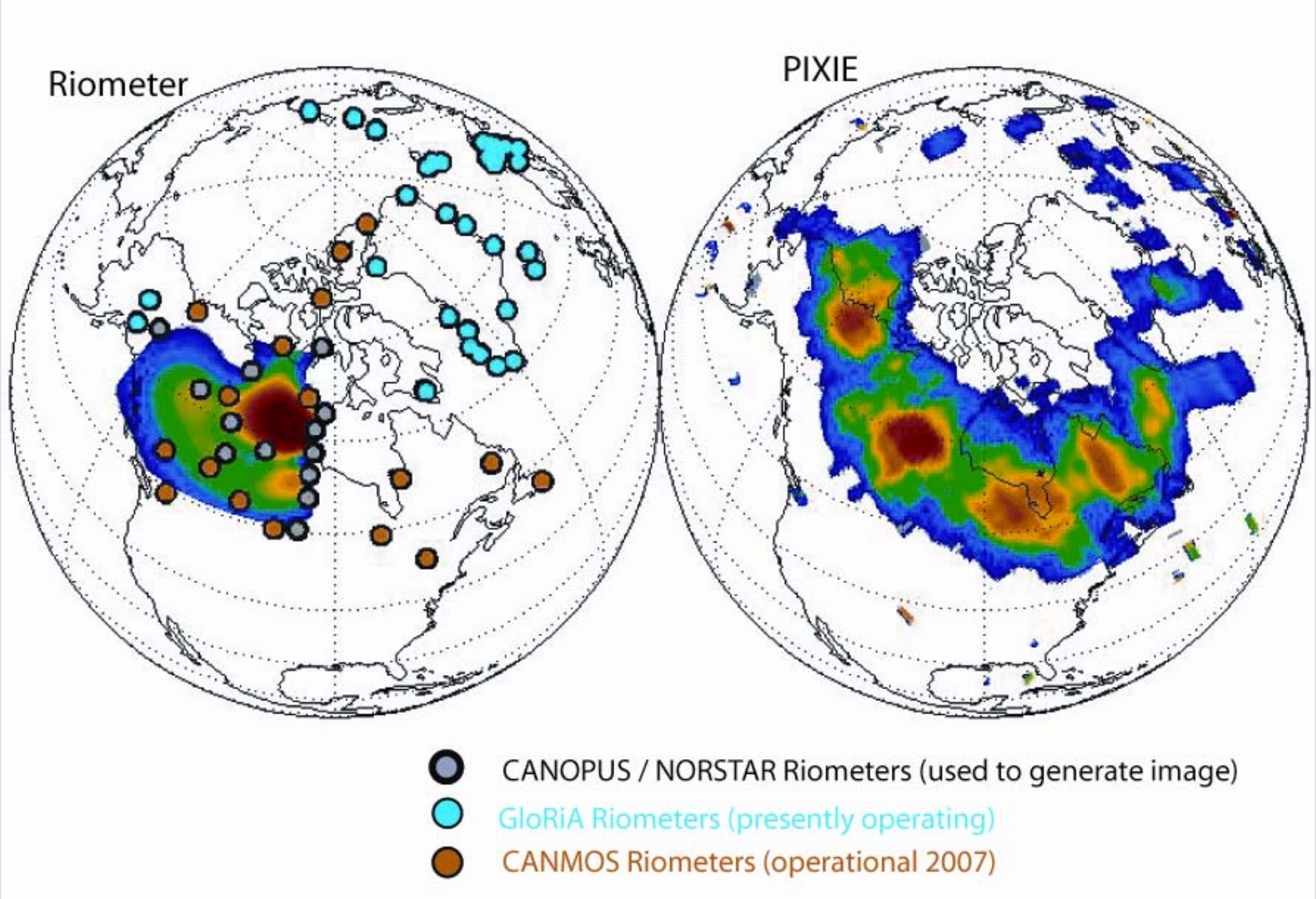

**Figure 3.7: (Right PIXIE X-ray image and Left) riometer absorption over north central Canada at the same time. Overplotted on the left globe are the locations of riometres that will contribute to the GloRiA initiative during the KuaFu timeframe.**

The differences between global X-ray imaging are highlighted in the table below. To recap, the global network of riometers cannot match the combined spatial resolution and global coverage provided by an instrument like PIXIE, nor can they provide estimates of energy of the electrons causing the absorption. However, global X-ray imagers cannot match the temporal resolution afforded by the riometer network. Furthermore, the effective signal to noise of the riometer network in discerning changes in the high energy electron population is superior to that afforded by a global X-ray imager. These last two points are highlighted by the recent work of *Spanswick et al.* [2006], who demonstrated that the a continent-scale network of integrating riometers can follow the evolution of the substorm dispersionless electron injection as projected along magnetic field lines into the ionosphere, with a temporal resolution at least as good as 5 seconds.

Table 3.1: Summary of relevant capabilities of PIXIE and GloRiA.

| | Multi Energy X-Ray Images (PIXIE) | Global Riometer Array (GloRiA) |
|---|---|---|
| Temporal Resolution | ~1-7 minutes (dependent on x-ray flux) | 1 second |
| Energy Resolution | yes | No (>30 keV e- in general) |
| Spatial resolution | 64x64 Pixel<br>1000km at apogee | Variable – depends on density of riometers |

## *3.3 What Imaging Instruments Do We Need?*

From the mission objectives, we have the requirement to obtain *global* information about at least the following:

1. the electron CPS as projected in the ionosphere and obtained by auroral imaging.
2. the ion CPS as projected in the ionosphere and obtained by auroral imaging.
3. the ring current ion population.
4. the plasmasphere.
5. the magnetospheric high energy electron population as projected into the ionosphere by diffuse auroral mechanisms.

This information is to be available continuously (24/7) to support studies of long duration dynamic geospace processes such as storms. Continuity of observations obtained from space will be obtained by utilizing a two satellite configuration as discussed in the next section on Orbital Design. The mission needs to be as simple and low cost as possible while leading to new science. Finally, the mission is to build on the Canadian heritage of world-class global electron auroral imaging.

With these goals, and the likelihood we can obtain sufficient information on the plasmasphere and high energy electron populations using ground-based networks of magnetometers and riometers, respectively, we have come up with a simple imaging instrument complement for Ravens (and ultimately KuaFu-B). Each of the two satellites is to be identically outfitted with the following:

1. A two channel FUV electron auroral imager from Canada capable of providing images in two isolated passbands within the LBH, thus allowing for inference of characteristic energy and energy flux of precipitating electrons.
2. A spectrographic imager from Belgium capable of producing global images of the proton aurora in Doppler shifted Lyman-alpha. It would be advantageous if this instrument could provide even crude information about the proton auroral energies.
3. An Energetic Neutral Atom imager capable of providing global images of the ring current ion population, *preferably with mass resolution.*

In the section on conjugate imaging, below, we argue that a wide field of view UV electron auroral imager on both Ravens (KuaFu-B) satellites would provide tens of minutes of conjugate observations every orbit, and we further argue that we should therefore include a fourth instrument in the imaging package:

4. A wide field of view electron auroral imager to support conjugate observations while one satellite is passing though perigee.

## *3.4 Orbit Design*

All previous global imaging missions have been single satellite efforts. These satellites have been on high inclination orbits, and invariably on highly elliptical orbits. The high ellipticity was chosen to increase the fraction of the orbit during which global imaging was possible. Nevertheless, in such a situation, the satellite spends a not insignificant part of the orbit in locations from which good viewing is impossible. This is due to either bad look angles (i.e., looking across rather than down at the oval), or due to the oval being out of sight altogether. In the figure below, we show the location of the IMAGE satellite at 11 evenly spaced times during one orbit on November 24, 2001, and the WIC images from hose times. As can be seen from this sequence, even with the high ellipticity of the IMAGE orbit, the quality of viewing was poor for almost 5 hours out of the ~13 hour orbit.

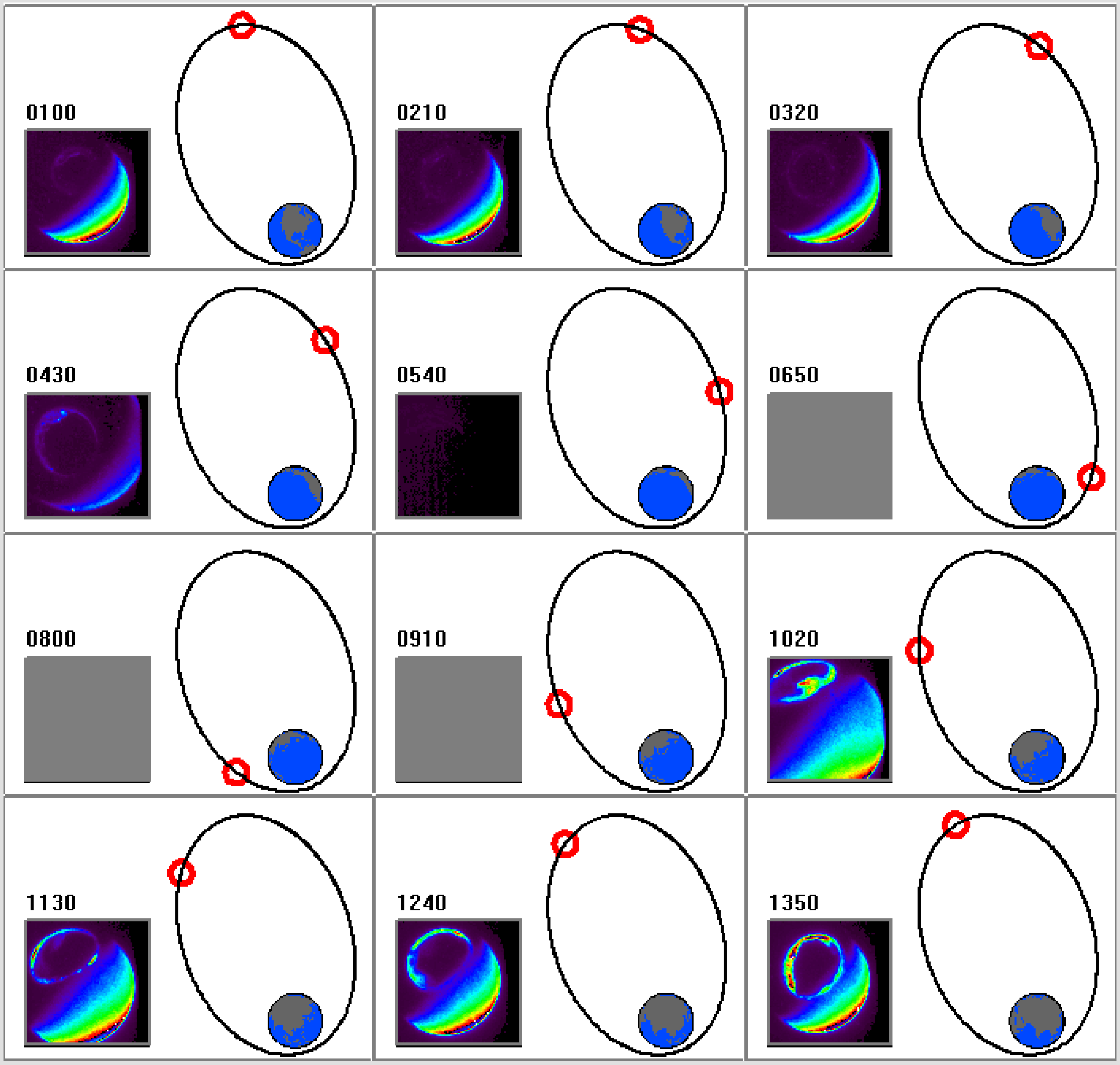

Figure 3.8: IMAGE location at 11 evenly spaced times during one orbit, with the WIC images collected at those times shown in the insets.

The Ravens (and now KuaFu-B) mission scenario is based on the fact that it is possible to achieve continuous imaging of the auroral distribution in one hemisphere using two satellites. One way to accomplish this is with two spacecraft relatively phased on identical (coplanar) elliptical polar (90 degree inclination) orbits so that when one is at apogee, the other is at perigee. The relatively slow passage through apogee, and quick passage through perigee means that at all times at least one of the two spacecraft will be near apogee. We illustrate this concept in Figure 3.9 below.

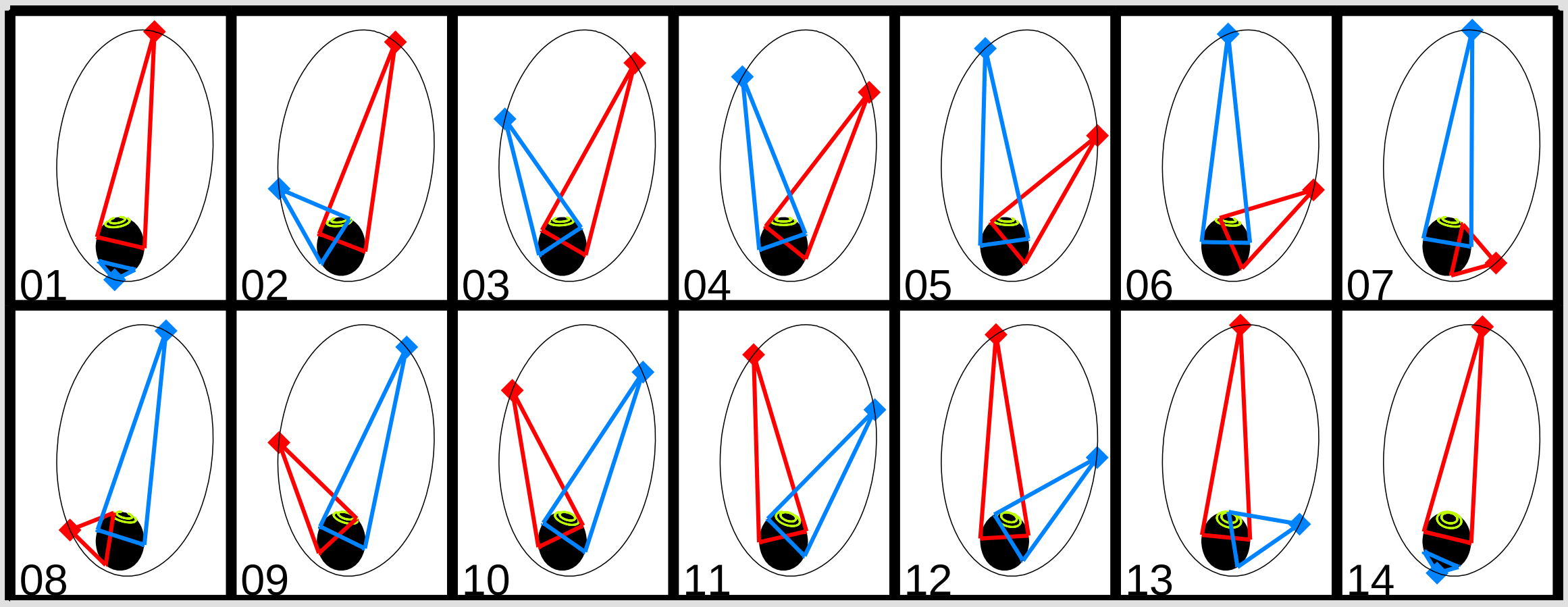

Figure 3.9: 14 snapshots of two satellites relatively phased on identical polar orbits so that when one is at apogee the other is at perigee. An auroral oval, at appropriate latitudes, is indicated in green, and the part of the globe visible to each spacecraft is indicated with the red and blue cones. Note that, at all times, the entire northern hemisphere oval and polar cap are within view of one, the other, or both satellites.

The purpose of this subsection is to describe the work that was carried out as part of the Ravens Concept Study to pick a suitable orbit for the proposed Ravens mission. We point out at the beginning that the orbital parameters we came up with have been accepted by the KuaFu mission design and engineering teams as the orbital parameters for the KuaFu-B satellites. What we present here is a subset of the exploration that we carried out. Further, the work we have carried out is a subset of what we will ultimately need to do. For example, we used only the aspheric Earth terms in tracking the evolution of the orbit, neglecting the moon, and atmospheric drag. While we are sure this is enough at this "feasibility stage", we will need to carry out more in depth calculations to determine what will need to be done to maintain these orbits (if KuaFu proceeds, it is certain that this will be done by the Chinese engineering team in any case). Further, although we explored a number of different inclinations, we elected to stick with the simplifying 90° inclination that was used for IMAGE. Although we did not start out with the objective of carrying out systematic conjugate auroral observations, we have ended up incorporating another imager (the Wide Field of View Auroral Imager from the UK) to accomplish just that. The 90° inclination is not only simplifying, it is optimal for conjugate viewing (as we state in the conjugate viewing subsection below, we explored a Molniya orbit, for example, and found that it supported virtually no conjugate imaging).

Given that we settled on a 90°, there are a number of competing factors that must be considered in choosing an orbit. The orbit must be significantly elliptical so that each satellite spends a large enough fraction of its orbital period near apogee. This is necessary for achieving the 24X7 continuous auroral coverage (consider for instance the limiting case of two satellites corotating on identical, polar, circular orbits: when they are both at the equator, neither can see the pole, so continuous excellent viewing at all auroral and polar latitudes in one hemisphere is impossible with low ellipticity orbits). There is a competing set of needs that perigee should be high enough and apogee low enough to facilitate conjugate imaging with the wide field of view imager, and better resolution near apogee with the UV auroral imager, respectively. For a given apogee and perigee that allows 24X7 viewing, the precession of the line of apsides then limits the duration of continuous viewing. This precession is made faster if either perigee or apogee is lowered, as the effect of the aspheric Earth is increased by anything that lowers the orbit-averaged radial distance of the satellite (see Figure 3.10, below). Thus, subject to the above factors, we would like both apogee and perigee to be as high as possible to maximize the effective mission lifetime. Finally, higher apogee and perigee makes for lower possible payload mass for a specified launcher and bus. For cost reasons, and to maximize the overall payload mass, we would want low apogee and perigee.

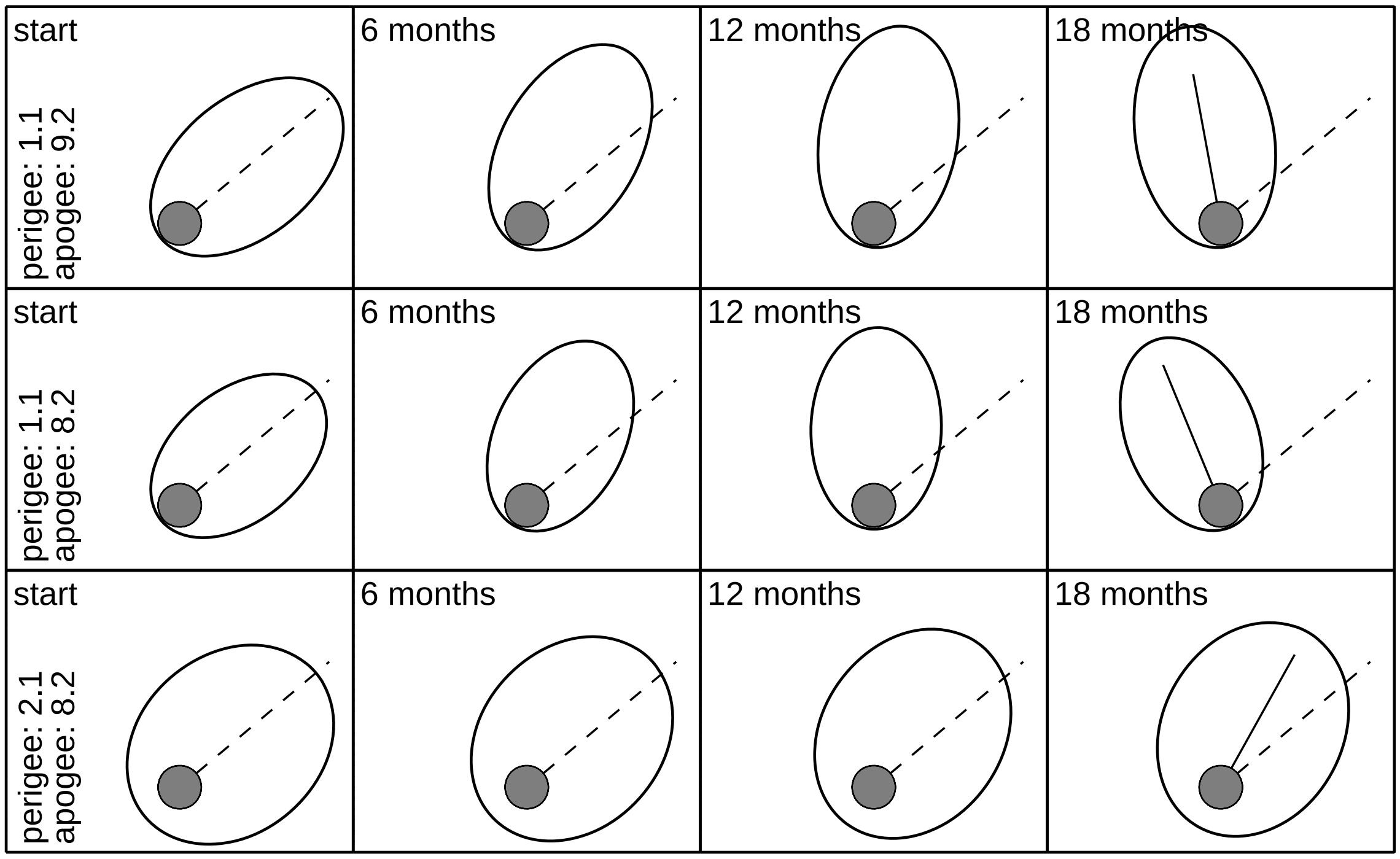

Figure 3.10: The three rows show the precession of the line of apsides over 18 months for 90° inclination orbits with three different sets of apogee and perigee. The middle row has apogee and perigee of 8.2 and 1.1 Re (geocentric), respectively. The top row has higher apogee and the bottom higher perigee. Note that increasing either decreases the rate of precession (increasing the duration of excellent viewing for our proposed mission), but that increasing perigee has a more dramatic effect than does increasing apogee (as would be expected by the relative effects on average radius).

This is a daunting set of competing factors, and it would be difficult to argue that after the Ravens Concept Study we have arrived at the *optimal* solution. We have, however, carried out sufficient analysis to argue convincingly that we have identified a *reasonable* apogee and perigee that everyone involved has agreed meets our scientific needs, allowing for a long enough period of sustained excellent quality viewing of the hemispheric global aurora and enough payload capability on the KuaFu launcher and bus combination to have an interesting and valuable set of instruments on each satellite.

First of all, higher apogee and perigee increases the duration of sustained excellent viewing but is expensive in that for a given launcher and bus this would mean small payload allotment for instruments. We therefore had to consider just how long a period of sustained imaging is required. Our primary scientific/technical objectives are to follow the evolution of the global electron and proton aurora for long durations, to allow us to study the system-level geospace dynamics during long duration geomagnetic processes such as magnetic storms and periods of steady magnetospheric convection (SMC). Obviously, we want to have a period of sustained excellent viewing of the global hemispheric auroral distribution that is long enough that we can expect there to be a significant number of storms and SMC events. Storms typically last 3-4 days, and occur as a result of processes on the Sun that we cannot predict (see, for example, the introduction of the Tu et al., KuaFu SSR paper that is included with this report as Appendix C).

In Figures 3.11 and 3.12, below, we show SYM-H for the years 2001 and 1996, respectively. From Figure 3.11, we can see that near solar maximum, storms come fairly frequently, there having been ~15 significant magnetic storms during the year 2001. On the other hand, at solar minimum, storms *can be* relatively infrequent. From Figure 3.12 we can see that in 1996 there was only two relatively small magnetic storms. At the present time, the KuaFu mission launch is planned for 2012, which would mean that the main phase of the KuaFu mission would be during the declining phase of the solar cycle. Any significant slip in this date (which is not at all unreasonable to expect and which Canada would have little control over if any), would mean that the main phase would likely unfold over solar minimum. If the mission launches as expected, we can reasonably expect a dozen or more magnetic storms during just one year of the mission. If there is a slip, the duration of sustained excellent quality viewing of the hemispheric global aurora would need to be significantly longer than one year in order to *guarantee* (we cannot guarantee but history dictates that storms are an ever present albeit episodic feature of the solar dynamic) that some significant storms would occur during KuaFu. Considering the Sym-H in the years around 1996, we have come up with the target lower limit of two years for sustained excellent quality viewing of the hemispheric global auroral distribution.

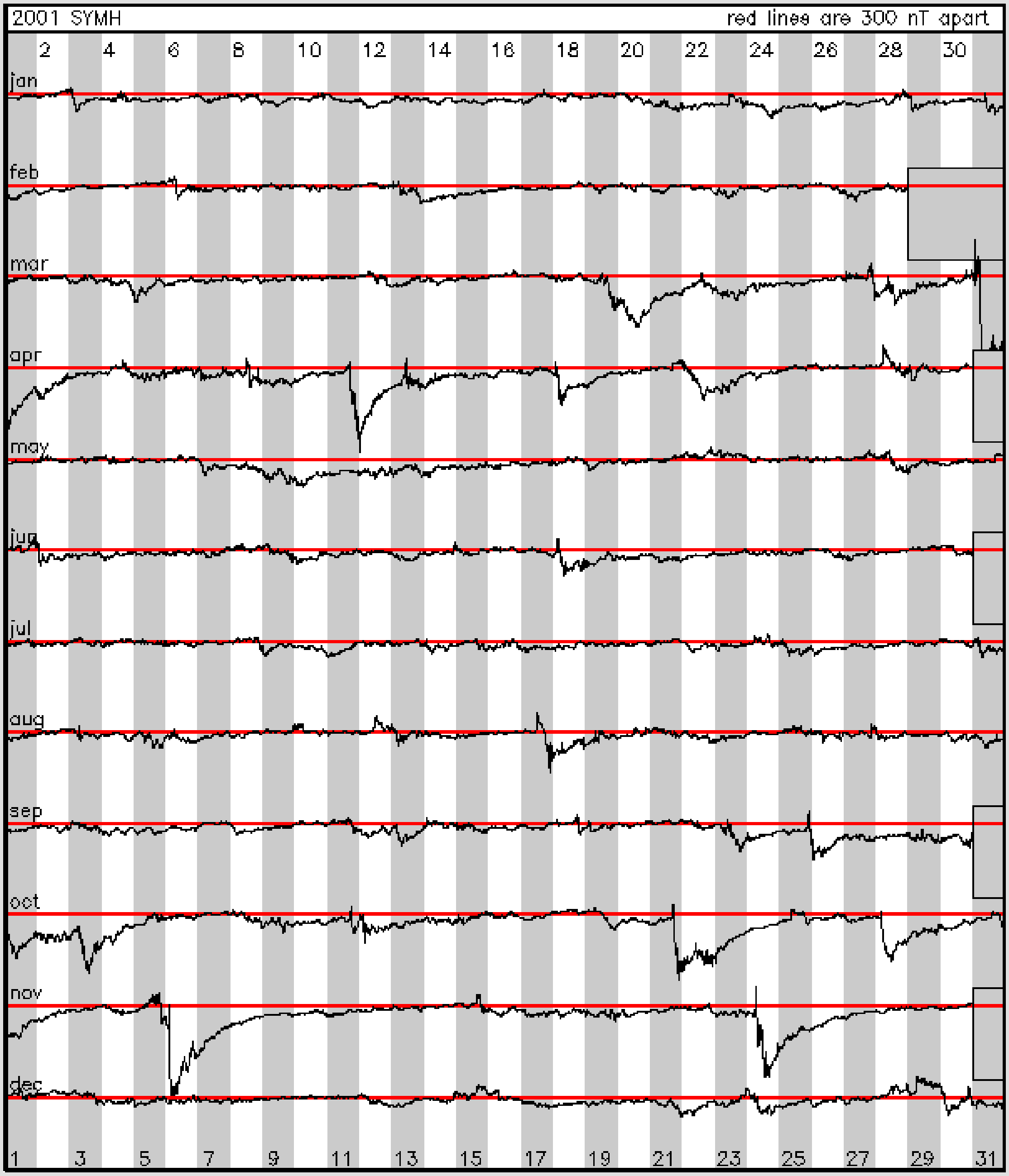

Figure 3.11: The SYM-H index is a one-minute estimate of the average deviation of the northward equatorial magnetic field, and with caveats discussed in [*Wanliss and Showalter*, 2006] can be treated as a high time version of DsT. Large negative SYM-H corresponds to storm-time conditions. This stack plot shows the SYM-H for the year 2001. From this plot, we can see that there were ~15 significant storms during 2001. The data is courtesy the World Data Center in Kyoto.

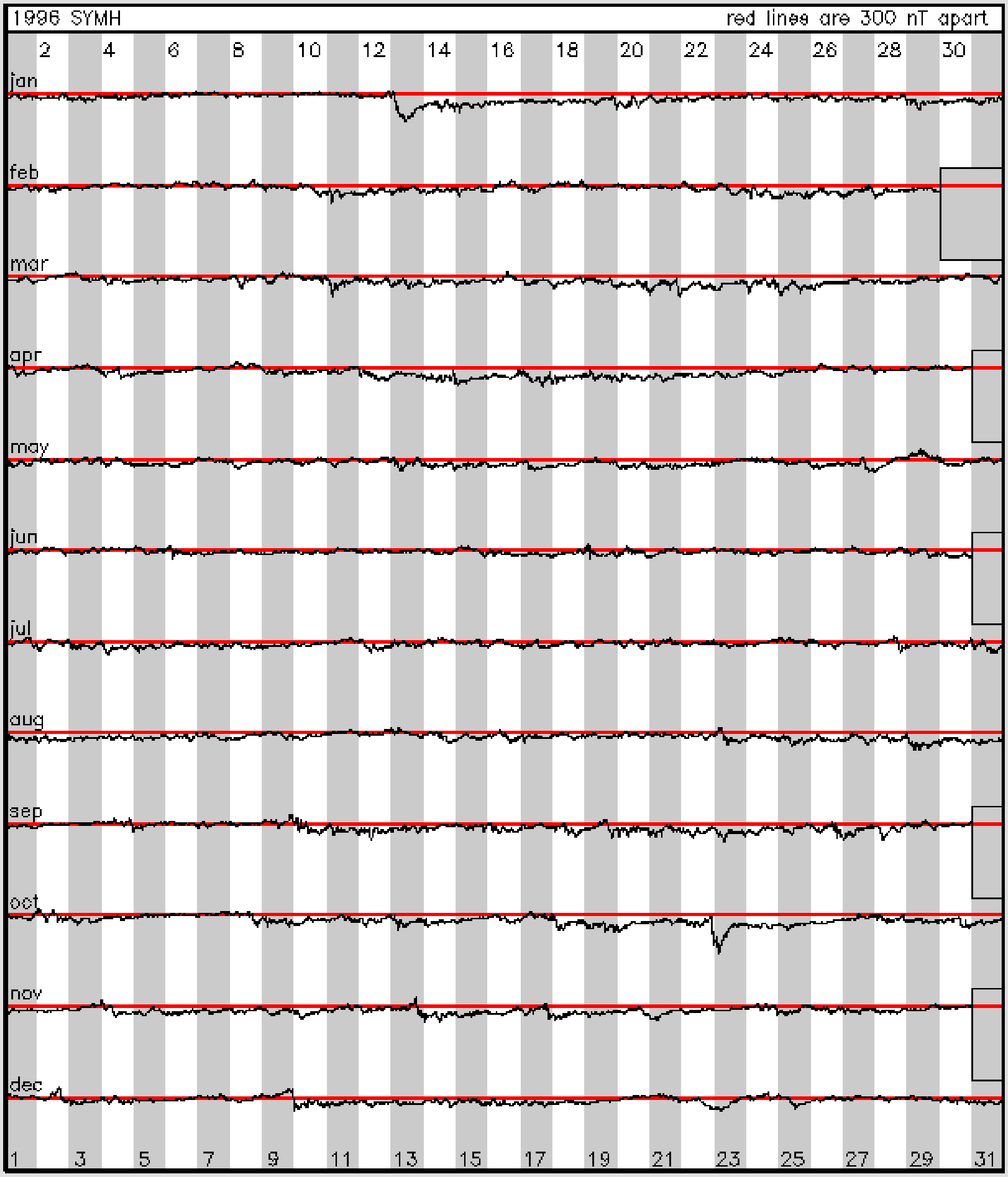

Figure 3.12: This stack plot shows the SYM-H for the year 1996, in the same format as used for the previous plot. There were two small magnetic storms during 1996. The data is courtesy the World Data Center in Kyoto.

On the basis of the above, we are planning to use polar orbits, and we want the period of sustained viewing to be in excess of 2 years to ensure that we have at least several significant storms during the period of sustained excellent quality viewing. With this in mind, it was necessary to explore different perigees apogees to *quantify* how long we could sustain excellent quality viewing as a function of perigee and apogee. We had initially intended to use Satellite Tool Kit (STK) to do this, however after some initial trials we determined that while STK made excellent graphical representations of specific orbital situations, it was not well suited to our task. As a consequence, we had to write a computer simulation of our own to address our questions (this code is available on request by the CSA). This program places two satellites corotating on coplanar identical orbits. The satellites are corotating, and 180° out of phase so that when one is at apogee the other is at perigee. We are interested in how well the pair of satellites can see the northern hemisphere polar and auroral regions. Given that one of our chief objectives is to study magnetic storms, we want to observe the global aurora even during active times. We have therefore assessed how well the satellites as a pair see the region above 50 degrees PACE invariant magnetic latitude.

To make this assessment, we take "pixels" in the region of interest. For a given pixel, we determine which of the two satellites sees it better (i.e., is looking more directly down at that location). Using that satellite, we determine the "look angle", with 90 degrees being "straight down" or nadir. We then average this look angle so determined over the entire region. An average measure of 0 means every pixel is *at best* seen obliquely (note that spatial resolution is best for a look angle of 90 and decreases away from 90. An average measure of 90 means that we are looking straight down on every pixel, which, while being as good as it could be is only achievable with an infinite number of satellites. Our best estimate is that a number of 40 degrees is a "good image" (based on consideration of WIC images from IMAGE and spatial resolution considerations as discussed in the section on FUVAMC below). We are therefore defining successful continuous global imaging (or sustained excellent viewing) to be a situation where the average look angle is greater than 40 degrees on and ongoing basis.

In Figure 3.13 (below), we attempt to illustrate this point. Figure 3.13 is comprised of two panels (top and bottom). Taking the top panel to start, there are two satellites on 1.1X8.2 Re (geocentric) orbits. The image at left shows the Earth from above the North Pole, with the best look angle from either of the two satellites indicated by the shading. The satellite track and instantaneous geographic footpoints of the satellites are also indicated. Note that the look angle is best at nadir, so there are two regions with large (essentially 90°) look angles directly below the satellites. The bottom panel shows the same information, but in that case for a line of apsides that is aligned with the geographic axis, which provides better overall quality of view. In each case, we calculate the average viewing angle (with the minimum being 0°) for each satellite (represented as l_a and l_b in the figure legend) as well as the average of the better of the two viewing angles (l_c in the legend). The objective is to keep the average of the better of the two viewing angles as large as possible, and always over 40° everywhere poleward of 50° geomagnetic if at all possible.

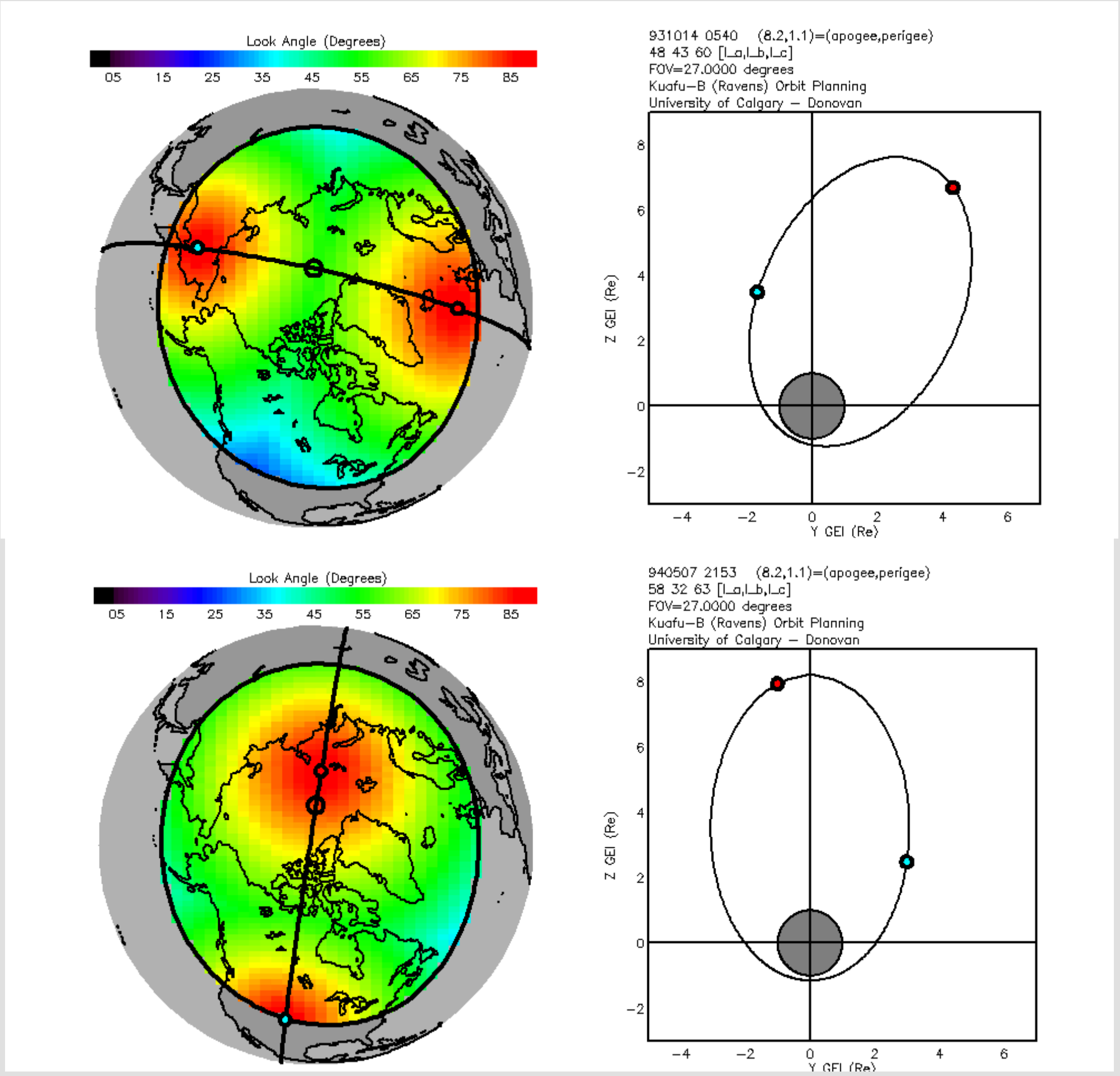

Figure 3.13: Quality of viewing can be determined from an orbital simulation. The two panels show the "look angle" for two different times in a possible two-satellite mission. The look angle for each cell is the better of the two look angles corresponding to the two satellites. Important variables here are argument of perigee and time in specific orbit.

Our simulation then moves the satellites forward in time. We take 11 steps from an initial argument of perigee until the line of apsides is the pole. At each of the 11 steps, we determine the "average look angle" (or "quality of viewing") for 160 evenly spaced steps in time during four orbits (we rotate the Earth in step as well, which is necessary as the geomagnetic and geographic poles are not the same). The seven steps are illustrated in Figure 3.14 for one apogee and perigee pair. The 160 quality of view values for each of the 11 steps form a well defined pattern as a function of time measured as "years into mission". For each step the average viewing angle varies from a minimum to a maximum value. For sustained quality of viewing we are interested in keeping the minimum values above 40°. This translates to a period of time bracketing the time at which the line of apsides passes through the pole. This is illustrated in the two panels in Figure 3.13. If you are having trouble visualizing this, we invite you to look at the web browser of sequences of figures like Figure 3.13. This browser is Appendix I of this report, and can be found at the URL http://aurora.phys.ucalgary.ca/cgi-bin/appendix_i.pl? .

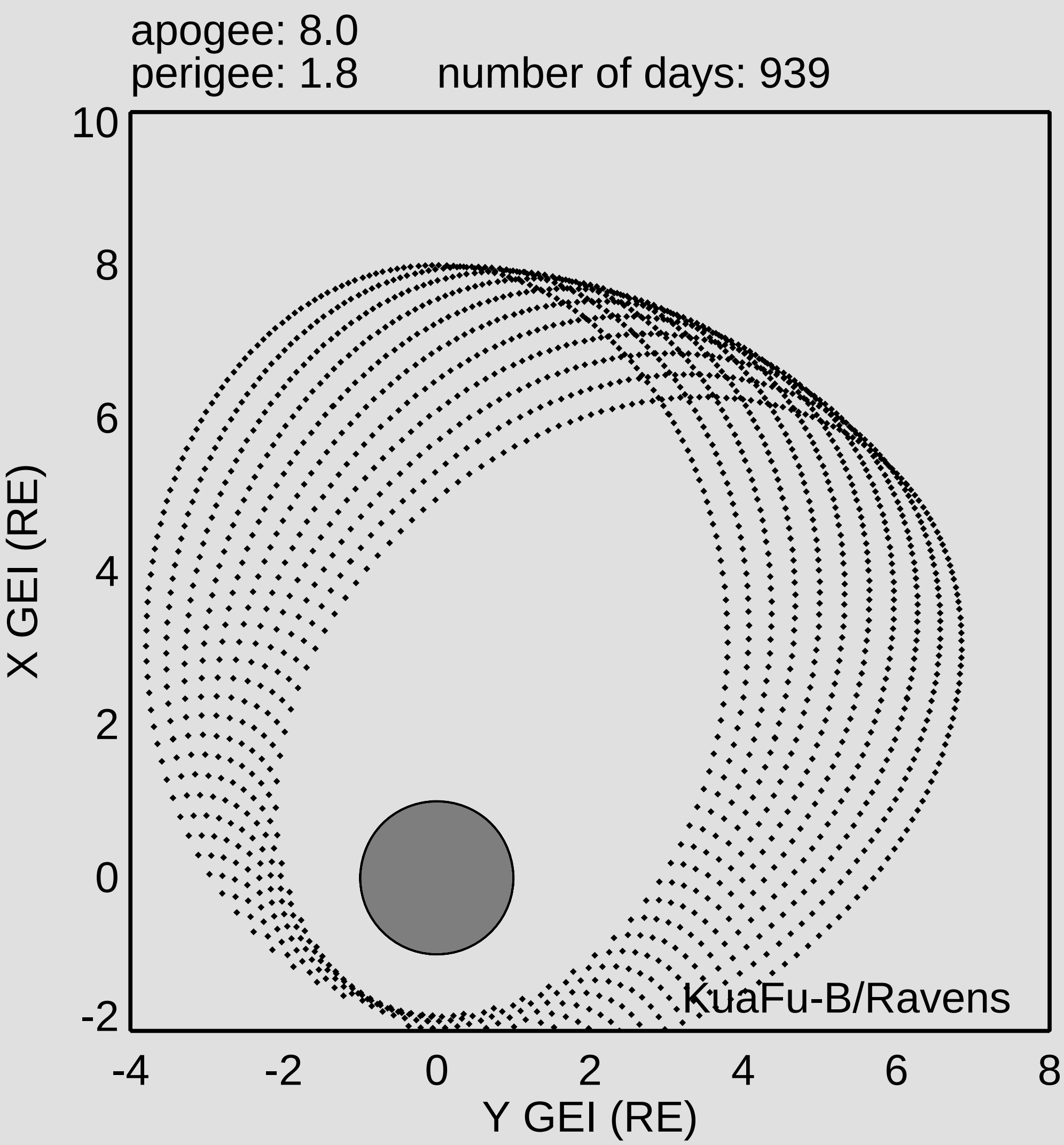

Figure 3.14: For our simulation we “inject” the satellite with an argument of perigee that corresponds to low enough latitudes that the viewing is not excellent throughout one orbital period according to our criterion. As the line of apsides precesses (as a function of course of both apogee and perigee) towards and through the pole, the average, best, and worst quality of view on any given orbit changes systematically, with better and better average viewing as the line of apsides approaches the pole. For our simulation we take 11 sets of four orbits (shown here for one set of apogee and perigee vales). On those 11 sets of orbits (each set consists of four orbits), we calculate the quality of view. The results for two sets of apogee and perigee are shown in the next figure.

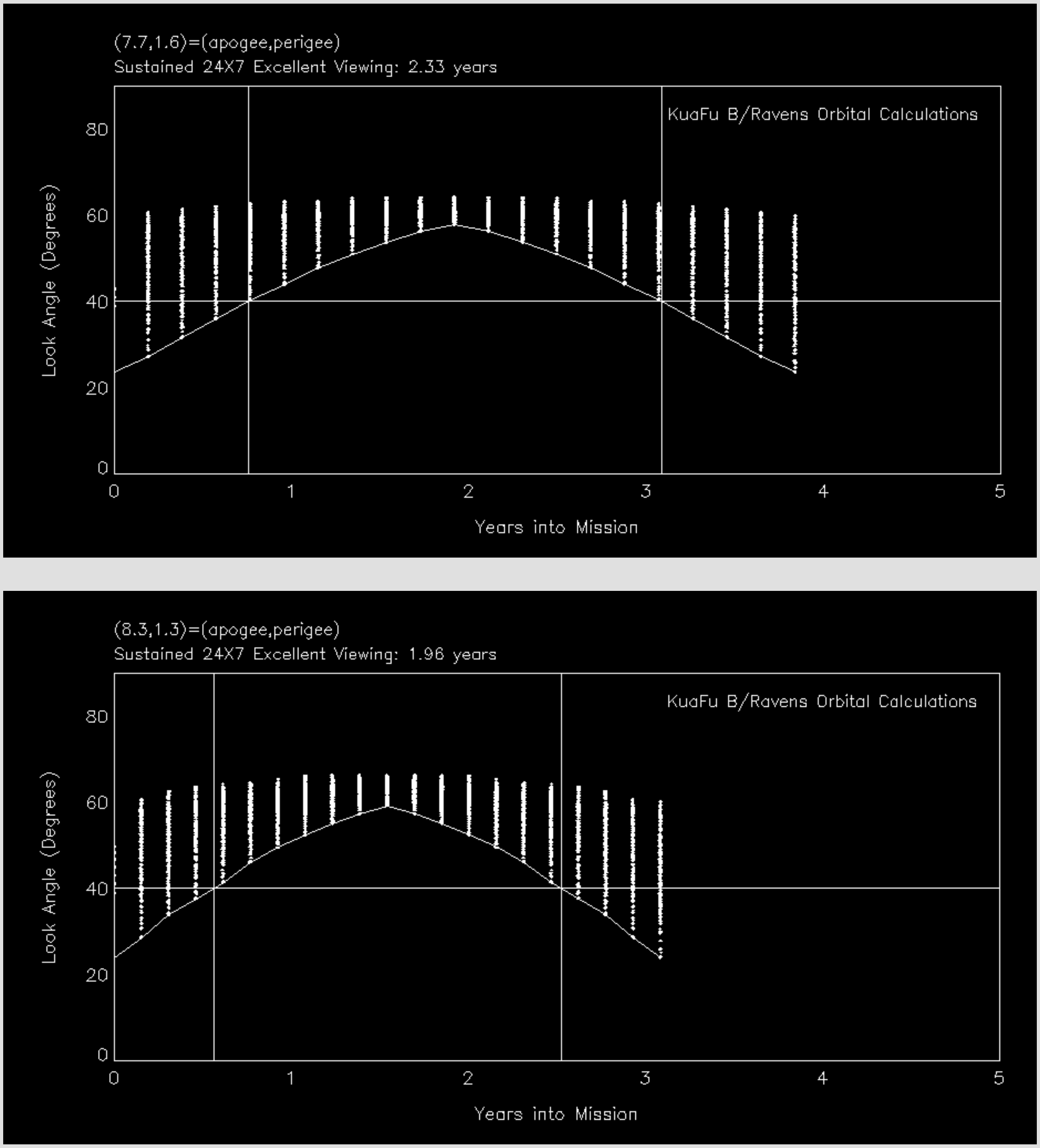

Figure 3.15: Average look angle for different times in a simulated mission. Each vertical set of points corresponds to the quality of view for four successive orbits at a particular argument of perigee. The argument of perigee changes with precession of the line of apsides as a function of the apogee and perigee. From this simulation, we can estimate the duration of “excellent viewing”, which we define as average quality of view better than 40°. For the two cases shown here, the duration of sustained excellent viewing is 2.33 (top) and 1.96 (bottom) years.

As stated above, the simulation of quality of view for the 11 steps for a given apogee and perigee pair allows us to quantitatively determine the length of time the quality of viewing is better than 40° for a given apogee and perigee. This is done automatically in the simulation. First, the results for the 11 steps are mirrored about the last step (line of apsides aligned with the geographic pole). The lower bound from each step is determined, through which a curve is fit, which is our estimate of the time evolving lowest quality of view. The two intersections of that curve with the line corresponding to 40° are determined. The separation in time between those two points is the length of time for which we have sustained excellent imaging. For the two examples shown in Figure 3.15 above, the durations so determined are 2.33 and 1.96 years. We carried out this simulation for orbits with apogees ranging from 7 to 9 Re and perigees ranging from 1.2 to 2 Re (geocentric for both). The results are shown below in the form of a contour plot of duration of sustained excellent viewing.

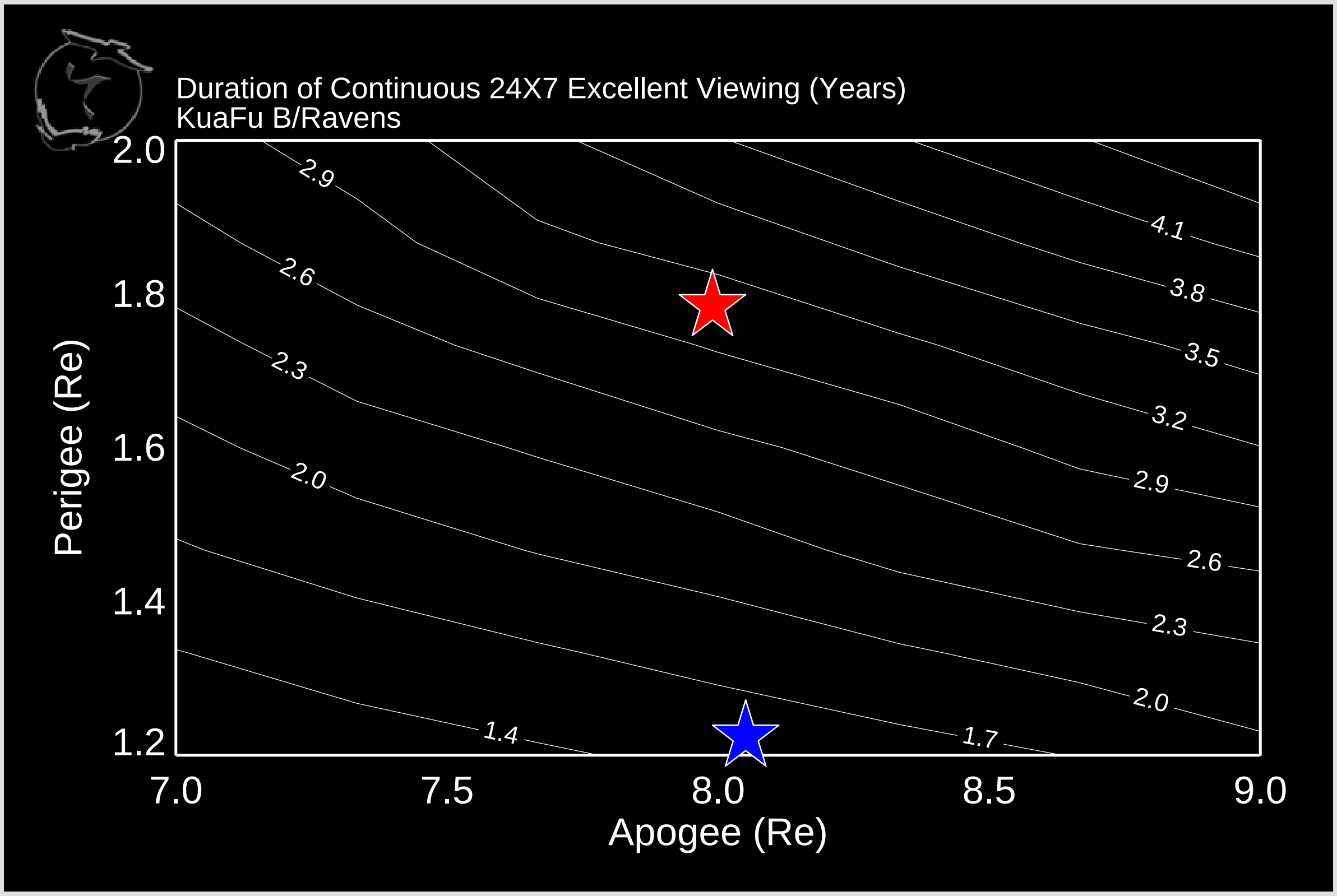

Figure 3.16: We carried out the simulation of quality of view as a function of time for apogees ragnging from 9 to 9 (geocentric) and perigees ranging from 1.2 to 2 (geocentric). For each pair of apogee and perigee, we determined the duration of continuous "excellent viewing" as determined by the duration of time during which the quality of view was above 40. The above contour plot shows duration of "excellent viewing" as a function of apogee and perigee. For reference we have indicated the orbital parameters of IMAGE and the proposed orbital parameters for Ravens (KuaFu B) with the blue and red stars, respectively.

To check that our simulation is giving reasonable results, we went to the NASA Satellite Situation Center (http://sscweb.gsfc.nasa.gov/ ) and obtained several years of locations of the IMAGE satellite ( a data set comprised of 50000 points). We placed two satellites on orbits identical to the IMAGE orbit, one leading the other be exactly ½ of an orbital period. We used these locations of IMAGE and its imaginary counterpart to determine the quality of view as a function of time for the two satellites. This simulation uses actual positions of a satellite on an orbit within the parameter space we are exploring with our simulation. The results are shown in Figure 3.17, plotted in the same format as are the results from our simulation. From this, we determine that two Ravens-type satellites with orbits identical to that of IMAGE would have had sustained "excellent viewing" (ie., average look angle always above 40°), for ~1.6 years. Based on the results shown in Figure 3.16, above (see the blue star indicating the IMAGE orbital parameters), we see that at least for the IMAGE parameters, our simulation is giving an excellent estimate of the duration of sustained excellent quality viewing.

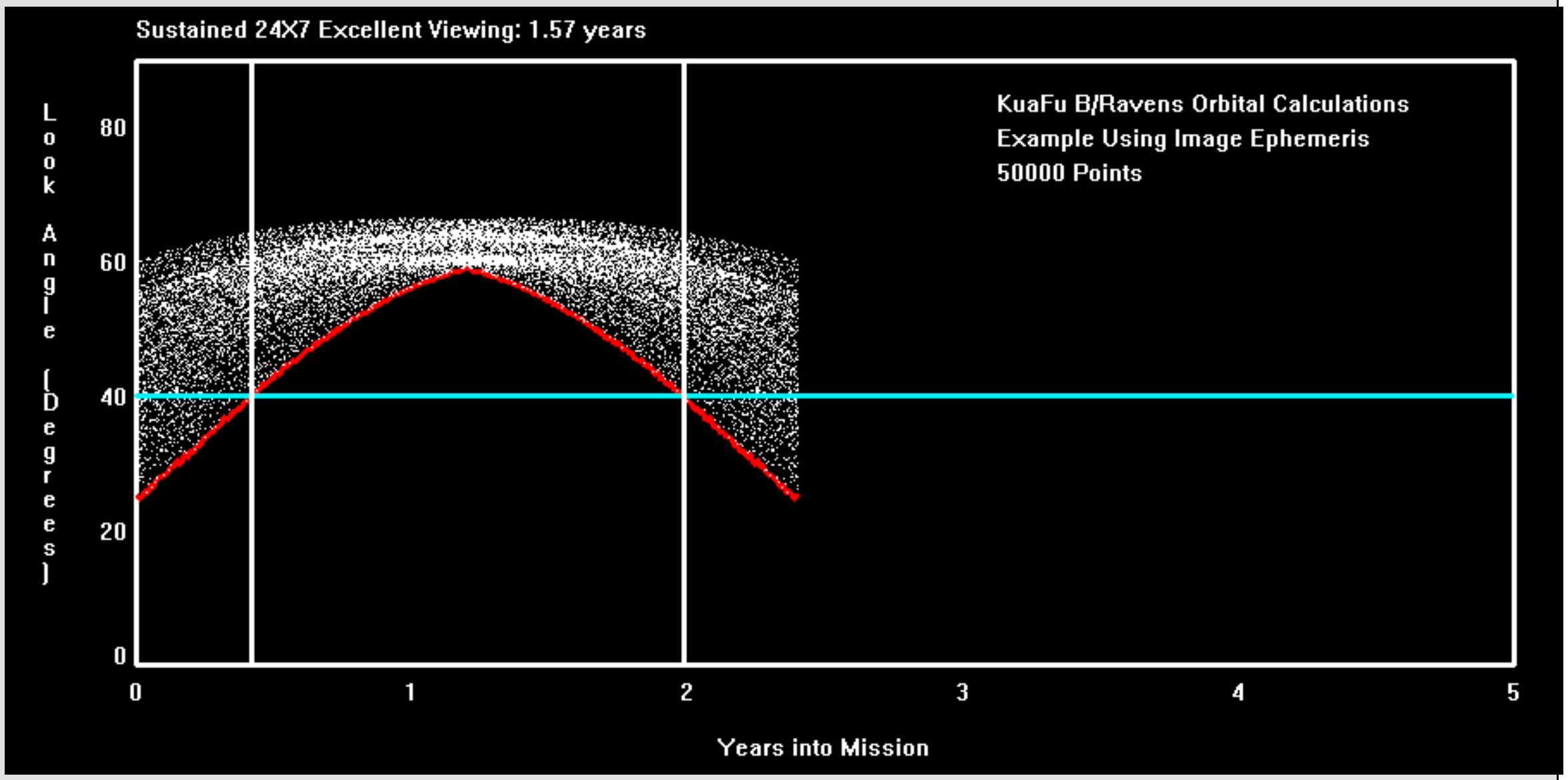

Figure 3.17: Duration of sustained excellent viewing of the northern hemisphere auroral and polar regions by a pair of satellites with the same orbit as IMAGE, determined using 50000 IMAGE locations.

With the exercise described here, we have determined that there is a wide range of apogee and perigee values for which we will have more than two years sustained excellent viewing of the northern hemisphere viewing (everywhere above the 2 year contour in Figure 3.16). With this information in hand, we considered the additional issue of how well a specific orbit would support systematic conjugate observations (see below), the impact of increasing apogee on spatial resolution, and negotiated with the KuaFu-B engineering team and KuaFu-B science team about payload implications. In the end, we settled on a 8 Re X 1.8 Re (geocentric) orbit, which will allow for roughly 2.6 years of sustained excellent quality viewing.

For completeness, we include the following figure, which shows contour plots of precession of line of apsides (indicated in red) and orbital periods as a function of apogee and perigee (note these curves are symmetric about the line apogee=perigee). On this plot, we indicate the parameters of the proposed Ravens (KuaFu) and the IMAGE orbits.

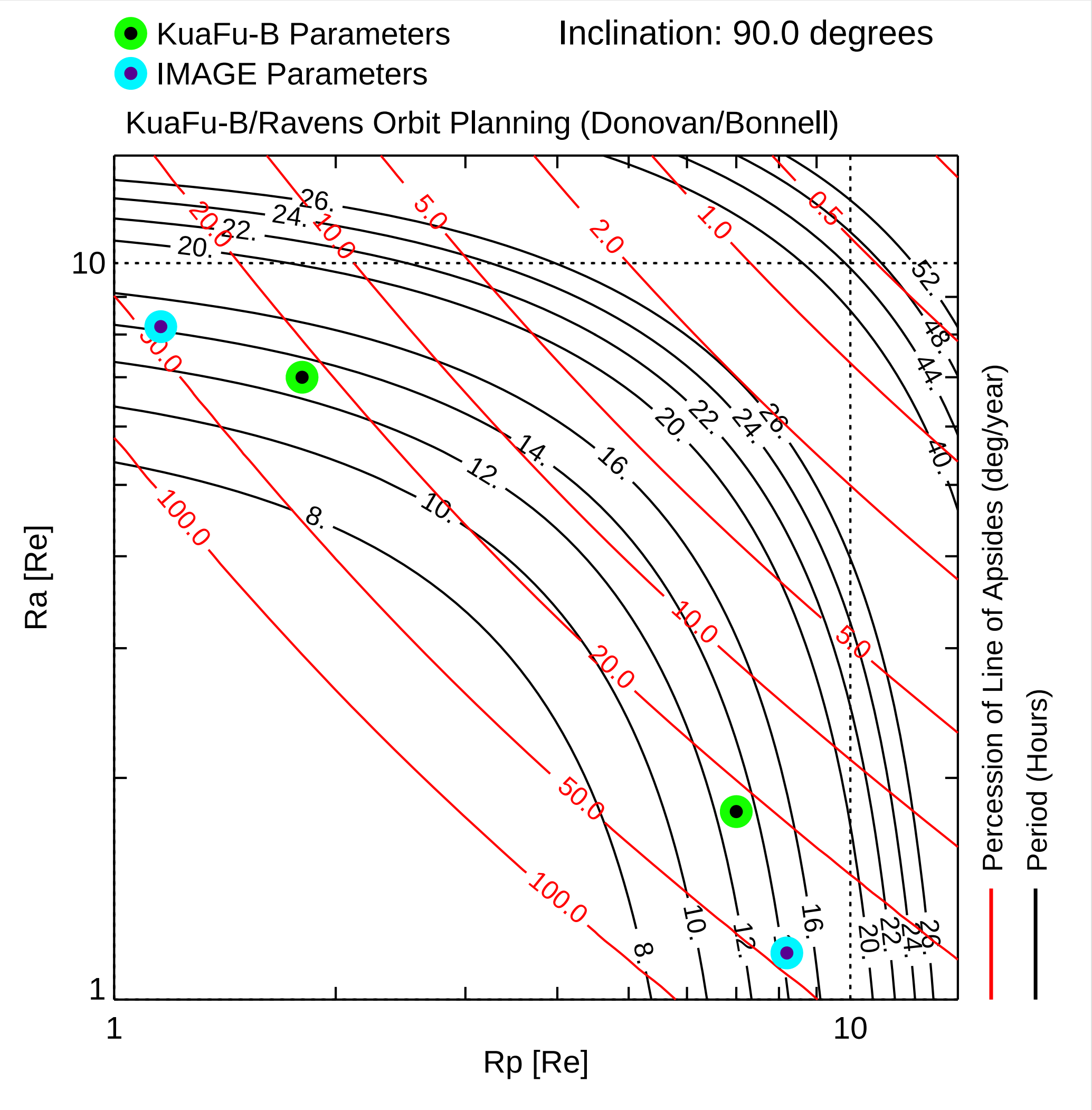

Figure 3.18: The proposed Ravens (KuaFu-B) orbits are 90 degree inclination, with 7 and 0.8 Re apogee and perigee altitudes, respectively. The orbit parameters are shown here, together with those of IMAGE, on a plot of orbital period and precession of the line of apsides.

We point out that we ignored the effects of the Moon and other perturbing factors on the evolution of the orbit (although our comparison using the real evolution of the IMAGE orbit described above gives us complete confidence that what we propose is feasible, and the Chinese engineering team has agreed to these orbits). During the Phase A, should we proceed, we would work with the Chinese engineering team to explore more fully the consequences of the selected orbit for the feasibility of the mission.

## 3.5 Serendipitous Opportunities

Our objectives in designing the Ravens (and ultimately KuaFu-B) global imaging mission were relatively straightforward. As stated above, this was primarily motivated by the desire to provide long duration continuous global imaging of the electron and proton aurora, as well as (if possible) ENA imaging of the ring current. This would be a significant first, allowing the first-ever imaging of long duration geospace processes such as magnetic storms, and supporting system-level science within the context of the ILWS and LWS mission lineups (ie., THEMIS, MMS, RBSPs, GOES, Swarm, etc). Further, it builds on the successful Canadian heritage of global auroral imaging, and would provide Canada with a high-profile contribution to ILWS. We believe there is no doubt that we will be able to achieve these objectives if the Chinese KuaFu mission goes ahead.

For reasons that we explore in this subsection, KuaFu-B would provide some additional significant opportunities that were not envisioned when we initially proposed the Ravens mission idea. Subject to the practical objective of keeping the mission simple, it has proven very interesting to explore these “serendipitous opportunities”, and two of these opportunities (systematic conjugate imaging and *in situ* observations on KuaFu-B) have materialized in the KuaFu-B program. In our view, these opportunities derive directly from the Ravens mission concept, and we spent considerable time during the concept study exploring their implications. For those reasons, we report briefly on these opportunities here.

### Systematic Conjugate Auroral Imaging

The expectation of observing similar auroral features simultaneously in the conjugate hemispheres arises from the fact that charged particles are “tied” to magnetic field lines, due to their gyro motion on the time scale of their mirroring period from one hemisphere to the other. If the magnetic field lines connecting the two hemispheres were symmetric, as in a simple dipole field, any disturbance/instability in the magnetosphere that causes particles to precipitate in the ionosphere would result in auroral features that could be observed at exactly the same geomagnetic locations in the southern and northern hemispheres. On the other hand, if the magnetotail is twisted (i.e., rotated around the Sun-Earth line) or the topology is changing on time scales short compared to the time it takes information and particles to propagate to the ionosphere, then we can reasonably expect non-conjugate effects. As conjugate imaging was not an initial target of the Ravens program, we present a brief review of recent work on conjugacy before we address how well KuaFu-B will be able to support systematic conjugate auroral observations.

Early studies with all-sky cameras reported similarities in the spatial and temporal development in both hemisphere [e.g., Belon et al., 1969]. However, we now have sufficient evidence that this is not always the case. It has been clearly demonstrated that auroral nightside features may be largely displaced in the two hemispheres [Sato et al., 1986; Burns et al., 1990; Stenbaek-Nielsen and Otto 1997; Frank and Sigwarth 2003]. The displacements have been found both in latitude and longitude [Stenbaek-Nielsen and Otto 1997] although the longitudinal displacement seems to be the most pronounced and range from a few hundreds of km [Sato et al., 1986; Sato et al., 1998; Frank and Sigwarth, 2003] up to 1-2 MLT sectors [Burns et al., 1990].

During 2001 and 2002, the Polar and IMAGE spacecraft offered a unique opportunity to study the aurora simultaneously in the conjugate hemispheres. Due to the apsidal precession of the Polar spacecraft orbit and the large field of view of the Polar VIS Earth camera and the IMAGE - FUV instruments, substorms and auroral features were imaged on a global scale from the southern (VIS Earth camera) and the northern (IMAGE-FUV) hemispheres simultaneously. Examining 13 events where the locations of substorm onsets and auroral features were unambiguously determined in the conjugate hemispheres Østgaard et al. [2004a] presented the first quantitative results of how much the magnetotail is twisted around the Sun-Earth line and how this is controlled by the IMF orientation. They explained these findings as being the result of magnetic tension force acting on newly opened (reconnected) field lines as they are convected from the dayside to the nightside. Under the influence of a strong IMF By component only field lines with asymmetric foot points will "find" each other in the magnetotail and be able to reconnect. One may also think of this as the penetration of IMF By [e.g., Cowley, 1991; Stenbaek-Nielsen and Otto,1997]. Although these asymmetries are predicted by theory, model predictions [Tsyganenko, 1995; 2002] show asymmetries an order of magnitude smaller than the observations [Østgaard et al., 2004c].

In a recent paper Østgaard et al. [2003b] identified two events where a theta aurora was observed in one hemisphere, but not in the other. In one example, the theta was observed in the northern hemisphere (IMAGE/FUV - 135.6 nm). At the same time the VIS Earth camera images (130.4 nm) from the southern hemisphere did not show any theta aurora. In another event, they observed the opposite scenario. A clear transpolar arc was seen in the southern hemisphere, while no theta was seen in the northern hemisphere. They attributed these two non-conjugate occurrences of theta aurora to the IMF Bx control of the different rates of lobe reconnection in the two hemispheres, which is the driver of the plasma convection and shear flows, producing the electric fields that causes the theta aurora. These results apparently contradict the conjugate observations of theta aurora presented by Craven et al. [1991]. However, during their observations, the IMF Bx component was close to zero in the reference frame of the Earth's magnetic dipole axis, and consequently did not favour reconnection in a specific hemisphere. The papers by Craven et al. [1991] and [Østgaard et al., 2003b] are the only ones so far to describe conjugate and non-conjugate occurrences of theta aurora.

From the previous three paragraphs, we see that there has been some work carried out on auroral conjugacy. That research has been limited due to the relatively infrequent opportunities to observe conjugate phenomena. It has led to some interesting questions. Specifically from the work discussed above, the azimuthal separation of the auroral breakup as seen in the two hemispheres is larger than can be accounted for in models, and the nonconjugacy of theta aurora is a mystery. There are additional questions related directly to conjugacy (or nonconjugacy as the case may be). For example, due to the nature of the source of diffuse aurora, we assume its interhemispheric conjugacy, but it certainly has not been established. We do not understand how conductivity affects MI coupling. By examining, for example, the consequences of differential conductivity at opposite ends of flux tubes one could assess the role of conductivity in some dynamic processes such as expansive phase onset. How does reconnection occur during northward IMF? By examining F-region patches in both polar caps simultaneously one could assess this important question. As well, reconnection at the distant neutral line may be the ultimate source of poleward boundary intensifications. Evidence of conjugacy of PBIs would shed light on the possible role of reconnection in CPS dynamics.

In fact, our collective understanding of conjugate phenomena are based on such limited data that any systematic observational program aimed at conjugacy would undoubtedly lead to new discoveries as well as giving us an excellent shot at answering the above questions. The proposed KuaFu-B orbits are not ideal for conjugate viewing. The orbits are highly elliptical, with perigee only 0.8 Re in altitude. The field of view of the electron auroral imager (which will be 25° as discussed in the section describing UVAMC in depth), will not provide anything near a global image from that altitude. As well, the transit through perigee is relatively rapid, so any conjugate imaging would be possible for only tens of minutes. Still, as one satellite passes through perigee, the other will be near apogee, so the potential for some conjugate imaging clearly exists. Further, if conjugate imaging is possible, it would happen every orbital period, making KuaFu-B the first ever satellite mission that could support systematic conjugate imaging. The field of view issue could be addressed by adding another imager with a significantly wider field of view. These issues led to the addition of the Wide Field FUV Auroral Imager (WFAI) from the University of Lancaster. This is discussed further in Section 4.3.

We finish by pointing out that investigation of a Molniya type orbit demonstrated that conjugate imaging could not be supported if that type of orbit was adopted. At various times during the definition of KuaFu-B, there was pressure from a number of fronts to move to a Molniya type orbit. Our calculations showing that the Molniya orbit would make conjugate imaging all but impossible were an important factor in rejecting that orbit.

## Mid-Latitude Imaging

Again, as stated above, our primary objective has been to provide a platform for continuous global auroral imaging in the northern hemisphere. This was motivated by system-level science objectives, and the opportunity to complement in situ missions such as MMS, THEMIS, the RBSPs, ORBITALS, ERG, and Swarm. In fact, NASA's recent geospace thrust (the Living With a Star program – see http://lws.gsfc.nasa.gov/ ) is heavily focused on the inner magnetosphere (ie., the radiation belt acceleration, transport and loss), and space weather effects on the mid-latitude atmosphere.

Particle precipitation at typically sub-auroral latitudes due to inner magnetospheric processes has been a topic of interest for decades. Such precipitation is a natural consequence of extremely expanded auroral ovals during storms, as well as a consequence of ultra-low frequency pulsation induced loss of high energy electrons, wave-particle interactions leading to precipitation of protons, and so-called SAR arcs. There are many articles on these mid-latitude auroral phenomena, but we refer the reader to, for example, *Baker et al.,* [2002], *Spanswick et al.,* [2005], *Thorne* [1971], and *Brace et al.* [1988] for each of the above processes in turn.

While there is no high-latitude global auroral imaging mission currently planned except for KuaFu, it is clearly and widely recognized that mid-latitude auroral imaging is essential to the NASA LWS science mission [see, for example, section 4.3.3 of the Report of the LWS Geospace Mission Definition Team (GMDT), 2002]. As part of the recently announced RBSP Missions of Opportunity program, NASA has funded a global disk and limb UV imager (called GOLD for Global-scale Observations of the Limb and Disk please see the GOLD factsheet at http://fsi.ucf.edu/GOLD/Gold_Fact_Sheet.htm ). This is not certain to fly, but communicates serious interest in mid-latitude imaging as an integral part of LWS and a powerful complement to the RBSPs and ORBITALS.

During the Ravens Concept Study, we were approached by the NASA LWS Geospace Project Scientist (David Sibeck), who asked us to provide an idea of how well Ravens (and now KuaFu by default) would do at mid-latitude imaging. Given the importance of inner magnetospheric processes at the systems-level in geospace, and the increasing interest in the international community in mid-latitude ionospheric effects during large geomagnetic events (owing in no small part to NASA's LWS and Canada's ORBITALS programs), this was clearly an avenue worth investigating.

We carried out a brief study to assess this for Sibeck, shown in Figure 3.19, below. These results indicated that, not surprisingly, Ravens (and KuaFu) would do very well at providing long-term coverage of the mid-latitude ionosphere. To do this we modified the programs we developed to assist us in the design of the Ravens (now KuaFu-B orbits). We carried out the calculations for only the orbital parameters selected for KuaFu-B (ie., 90° inclination, 1.8 Re X 8 Re (geocentric) orbits), and only for an *argument of perigee* of 270° and for the plane of the orbits being the noon-midnight meridian. We then assessed the quality of view (in other words the look angle as described above) as a function of Magnetic Local Time (MLT) and magnetic latitude. This is illustrated in Figure 3.19, below, and described in the following paragraphs.

Figure 3.19 illustrates assessment of quality of viewing as a function of MLT at 35° magnetic latitude, throughout one orbital period. The panels at left show the locations of the two satellites at 8 evenly spaced times during the orbit. Each satellite is identified by a different color symbol. The bottom two panels at right show the quality of view as a function of time during orbit and as a function of MLT. Each of those two panels corresponds to quality of view provided by one satellite, as indicated by the symbol at lower left. The top panel shows the quality of view in the same format provided by the combination of the two satellites, where for each MLT and time we take the better of the two views. We remind the reader that 90° quality of view corresponds to "perfect" (or nadir), and 0° to oblique (ie., looking tangent to the ionosphere) viewing, and negative values correspond to the location being over the horizon.

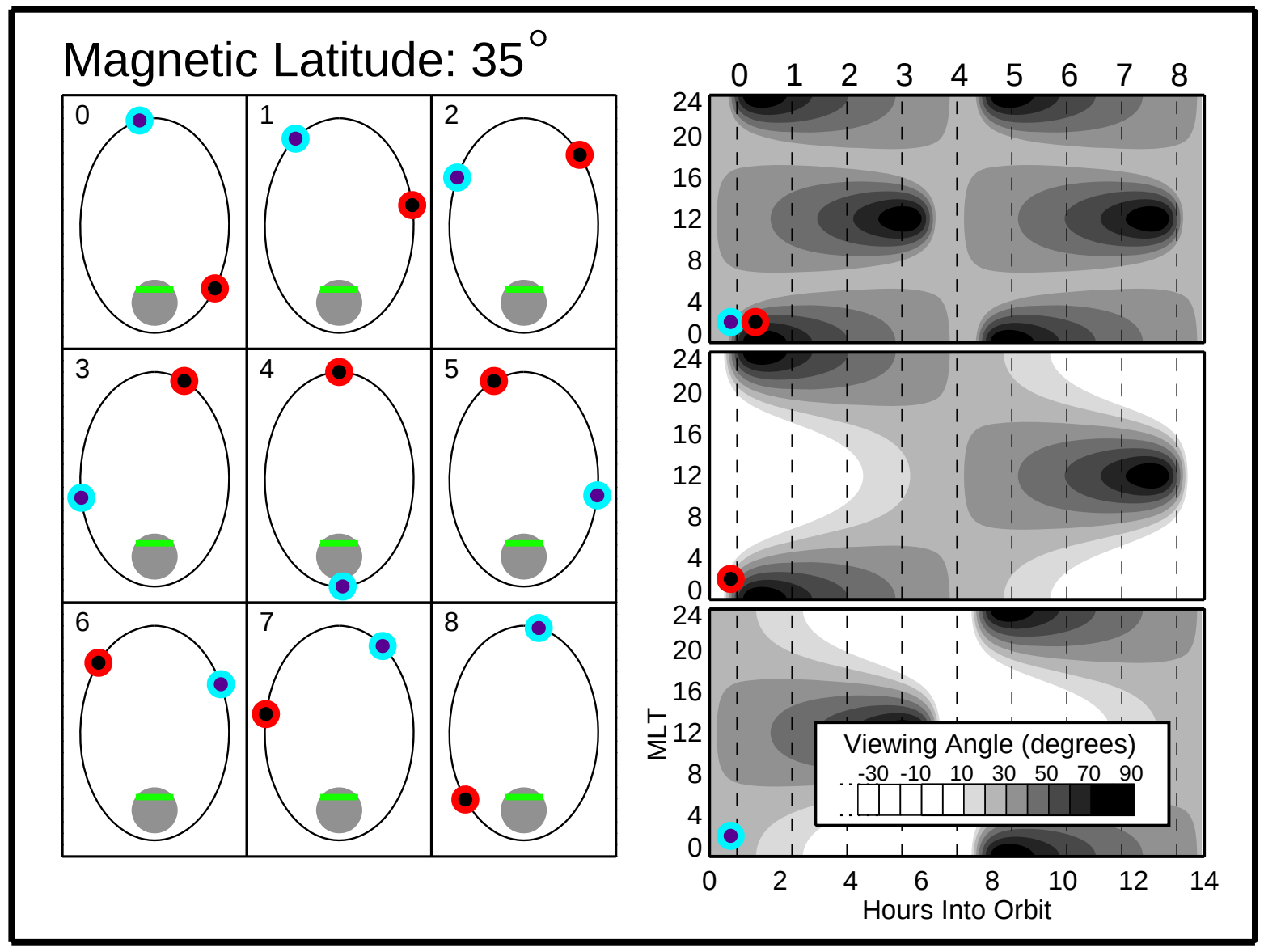

Figure 3.19: Mid-latitude imaging coverage afforded by two satellites in the proposed Ravens (KuaFu-B) orbits as a function of MLT and time during orbit at 35° magnetic latitude (see text for description of this figure). Note that the green ring on each globe in the panels at left indicates the typical auroral oval, and not the locus of points at 35°.

From Figure 3.19, we see that at 35° magnetic latitude, there are two extended periods during which there is excellent viewing of a several hours of MLT wide region centered on local noon and also on local midnight (ie., the plane of the orbit). In Figure 3.20, we show the results presented in the top right panel of Figure 3.19, but for magnetic latitudes ranging from 5° to 65°. For 5° (essentially equatorial), there is are two very narrow in MLT and brief periods of excellent viewing per orbit period. At 65°, of course, the viewing is excellent at essentially all MLT for the entire orbit (the orbits were chose to accomplish this at latitudes above around 50° - see discussion above).

From Figure 3.20, we can see that KuaFu-B would provide excellent sustained coverage of MLT extended mid-latitude regions for a significant fraction of the orbital period. In an attempt to quantify this, we asked the question "for what fraction of the orbital period is the quality of view at a given magnetic latitude above a specified value for a specified range of MLTs?" The results of our analysis to address this question are shown in Figure 3.21. Looking at the top pane, for example, we can see that 25% of the time the quality of view is above 44° across 8 hours of MLT at 33° magnetic latitude. The contour plots allow the reader to examine the tradeoff between magnetic latitude, MLT extent of region to be viewed, and quality of view.

The argument of perigee we chose is best for high latitude imaging. Earlier or later in the mission, the line of apsides will be at lower latitudes, and this will make for even better mid-latitude imaging. Furthermore, there is a possibility that the above mentioned GOLD mission will move forward. KuaFu-B mid-latitude imaging would complement images obtained by GOLD, extending coverage in MLT and latitude, and enhancing the GOLD program.

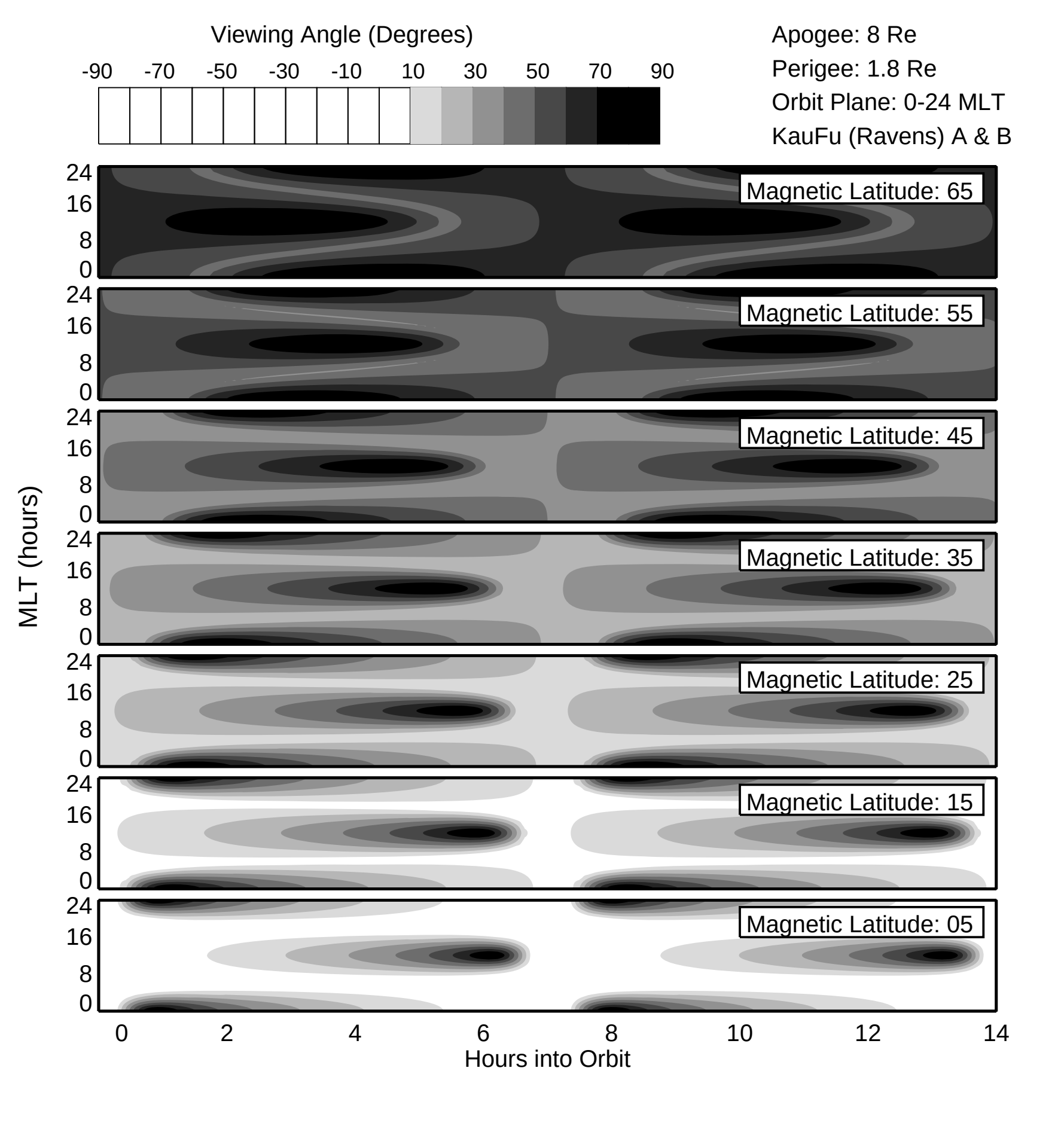

Figure 3.20: Quality of view from both KuaFu B1 and B2 as a function of MLT for magnetic latitudes ranging from 5° to 65°. Note that, for example, the top panel corresponding to 65° magnetic latitude shows excellent quality view at all local times and throughout the orbit, consistent with the design of the orbits as discussed above. See text for a more complete description.

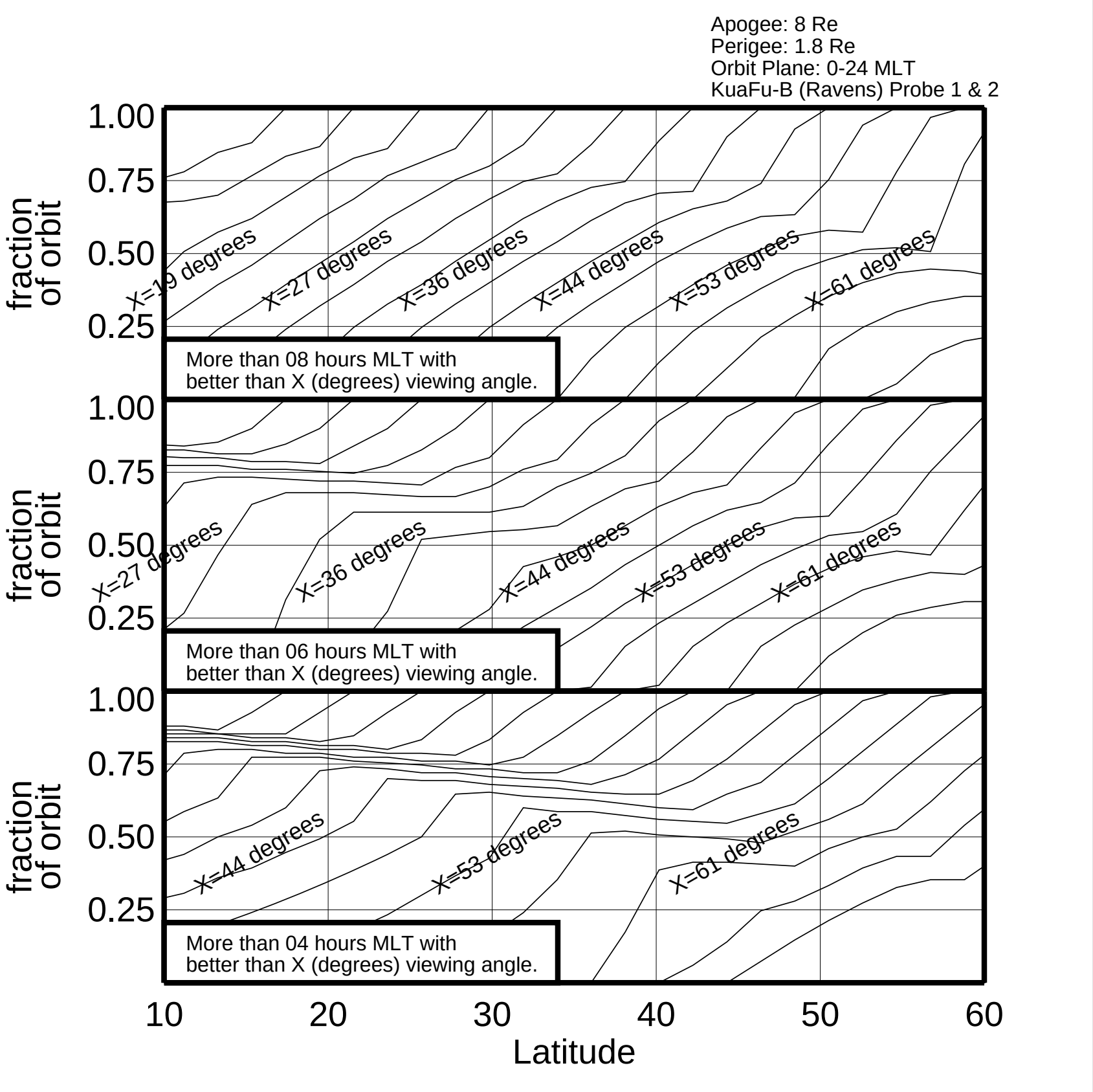

Figure 3.21: As discussed in the text, we asked the question "for what fraction of the orbit at a particular magnetic latitude is the quality of view above a specified value for a given range of MLT. The top pane, for example, shows the results assessing quality of view on 8 hours of MLT (4 centered on local noon and 4 centered on local midnight, keeping in mind the orbital plane for this simulation was the noon-midnight meridian). For example, for 40° magnetic latitude, the quality of view is above 44° across 8 hours of MLT (in two 4 hour segments) for roughly 65% of the orbit. The results show that KuaFu-B will not support continuous global mid-latitude imaging, but will support high quality mid-latitude imaging across a significant MLT sector for significant parts of the orbit.

Of relevance to the proposed Canadian Phase A for UVAMC and KuaFu is the question of what one needs to observe in terms of mid-latitude imaging. Simply imaging the LBH-L and LBH-S will be valuable in studying mid-latitude geospace effects, but there was considerable interest expressed by the LWS Geospace Mission Definition Team [see report of the GMDT, 2002] in studying effects on thermospheric composition via ratios of O emissions to those of N2. In the Phase A we propose to work with researchers who are studying mid latitude space weather effects, as well as the science teams of the other KuaFu imaging instruments, in order to explore ways to maximize the mid-latitude scientific output of our imaging program. As well, during Phase A, we would want to establish a connection with the GOLD effort, and develop an understanding of how those observations would synergize with the KuaFu-B imaging.

## In Situ Observations on Ravens (KuaFu)

Global remote sensing of the auroral distribution and ring currents and are the primary drivers of the orbits and technical specifications of the satellites. Such global observations are essential to studying the interaction between various geospace plasma populations and their response to external inputs, both of which are at the heart of the large-scale geophysical processes that we wish to study. These observations require two satellites on elliptical high altitude high latitude (90° inclination orbits).

KuaFu, with a planned launch in 2012, would fit into a larger international geospace and solar mission line up. This includes magnetospheric missions such as MMS, THEMIS, the RBSPs (and possibly ORBITALS and ERG), GOES, and LANL, *all of which are equatorial or nearly so*, as well as SWARM, NPOES, and possibly the ITSPs, *all of which are high inclination but low altitude*. Thus, at the time of KuaFu there are no high-altitude polar orbiting geospace satellites planned during this phase of ILWS. Thus, at the present time, excluding KuaFu-B, there is no plan for any high latitude high altitude *in situ* observations during the timeframe 2010 onwards (Cluster and Polar, both of which have been extended well beyond their nominal mission lifetimes, will almost certainly not be operating by that time). This is a serious problem. Without this class of *in situ* observations, the ILWS mission lineup (including missions flown as part of NASA's LWS) will be limited in its capacity to address questions of how the energy, mass and momentum are dispersed within the magnetosphere-ionosphere-thermosphere system.

For example, the Summer 2005 KuaFu Assessment Report highlights (p.30) that KuaFu-B will provide global time evolving maps of the characteristic energy and energy flux of the electron aurora, global time evolving maps of the proton aurora and systematic conjugate images. The scientific contribution these could make would be greater if simultaneous information about the changing conditions in the auroral particle acceleration regions were available from in situ measurements – for example in distinguishing Alvenic vs electrostatic (inverted-V) aurora. Ion outflow and precipitation of CPS ions as seen at higher latitudes is important for understanding the formation of and response of the plasma sheet and ring current during storms and substorms. Further, satellites on KuaFu-B type orbits equipped with higher energy particle detectors would provide latitudinal cuts through the radiation belts and ring current a number of times every day. This would provide important complementary observations to those provided by the RBSPs.

Further, dramatic changes in the solar wind that affect the global geospace system are the ultimate target of our investigations. The information from these changes is communicated through the magnetosphere via fast and shear mode Aflven waves. We refer to, for example, to one event in which Polar recorded the passage of a shear Alfven wave communicating the arrival of a sudden pressure increase in the solar wind. This is illustrated in Figure 3.22, below, which is drawn from an ongoing event study about the effect of a sudden increase in solar wind pressure on the ion CPS [*manuscript in preparation, Donovan et al*]. Our point here is that *in situ* high latitude high altitude observations convey important information about *how* information and mass is communicated between the ionosphere and higher altitude magnetosphere. In Figure 3.23, we reproduce a figure from *Fowler and Russell* [2001] which shows the Solar Magnetic (SM) YZ coordinates of Polar when it observed the passage of 25 sudden impulse disturbances (including the one illustrated in Figure 3.22). Such observations are essential to our ability to unravel the chain of events from the solar wind, through the magnetosphere to the ionosphere. This is absolutely not to downplay the crucial importance of lower latitude near-equatorial and lower altitude high latitude *in situ* observations. *The point is the lack of high latitude high altitude observations.*

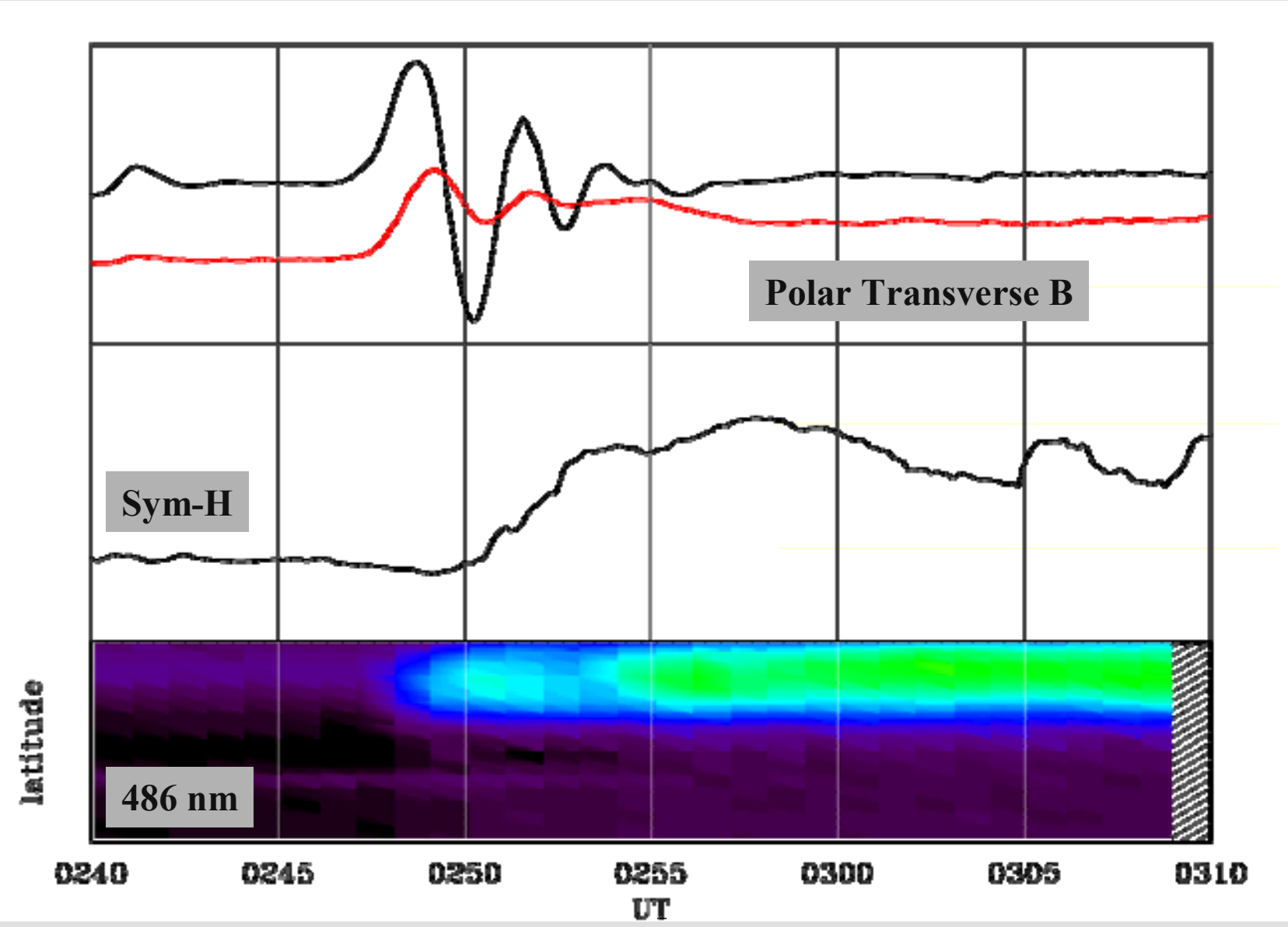

Figure 3.22: (Top) Transverse Polar magnetic field measurements in the high latitude CPS at the time of arrival of a pressure pulse in the solar wind on February 18, 1999 (Data courtesy of Chris Russell and obtained from CDAWeb); Middle) SYM-H for the same time period showing a 40 nT increase with the arrival of the increase in pressure (storm initial phase signature); Bottom) Hydrogen Balmer Beta emissions from the CANOPUS Gillam MSP showing the response of the ion (proton) CPS to the step increase in pressure of the solar wind.

KuaFu-B1 and –B2 will be in excellent orbits to provide high latitude high altitude *in situ* observations. As stated above (and illustrated in these figures), KuaFu-B will likely be the *only* geospace mission providing such a capability at that time (note the Molniya Mission described in Appendix G would be in such an orbit, but it is not a geospace mission and would not have room for any *in situ* instruments).

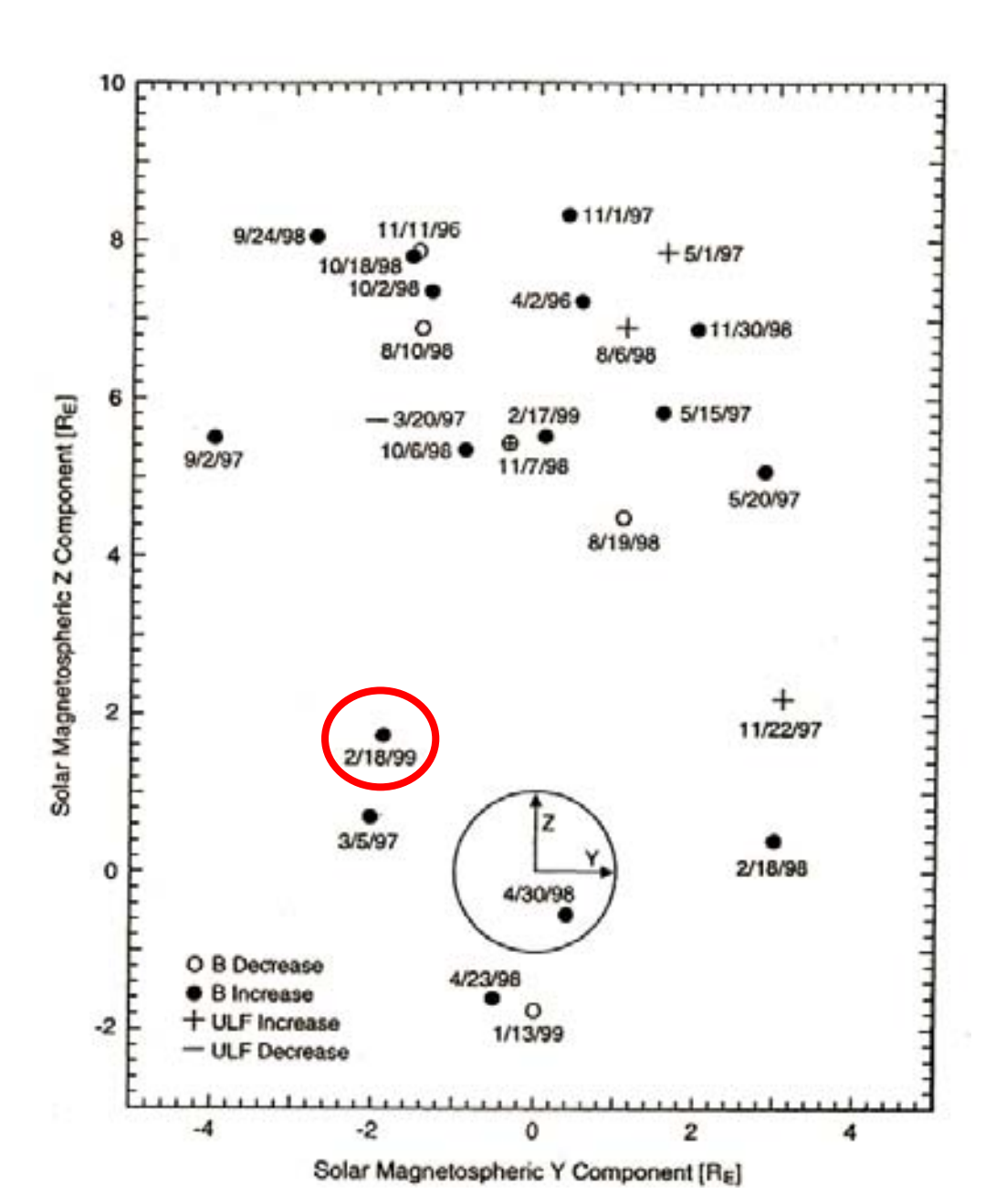

Figure 3.23: Reproduction of Figure 5 from *Fowler and Russell* [2001] which shows the SM YZ coordinates of Polar at the time of 25 different Sudden Impulses (their study was of the "geomagnetic field response along the Polar orbit to rapid changes in the solar wind dynamic pressure." The event highlighted in Figure 3.22, above, occurred on February 18, 1999 (when by the way Polar was roughly magnetically conjugate to Iceland). The location of Polar for that event is circled by us in red.

KuaFu-B was conceived of as an imaging mission. Very early in the mission concept development, the mission definition team acknowledged that the argument for *in situ* instruments in addition to imaging instruments was compelling for the reasons pointed out above. It was agreed by all parties that, subject to mass limitations and with the understanding that everyone wants to keep the mission simple to reduce overall risk, KuaFu-B would include a small complement of *in situ* instruments. While the imaging complement must be the same on both satellites (to facilitate 24/7 coverage), it was not essential that *in situ* instruments be identical on both satellites. The complement we arrived at fits within the 65 kg mass budget we have for payload on each of KuaFu-B1 and –B2, and will allow us to observe important quantities in the high latitude high altitude magnetosphere. That *in situ* instrument complement is listed in the introduction to Section 4.

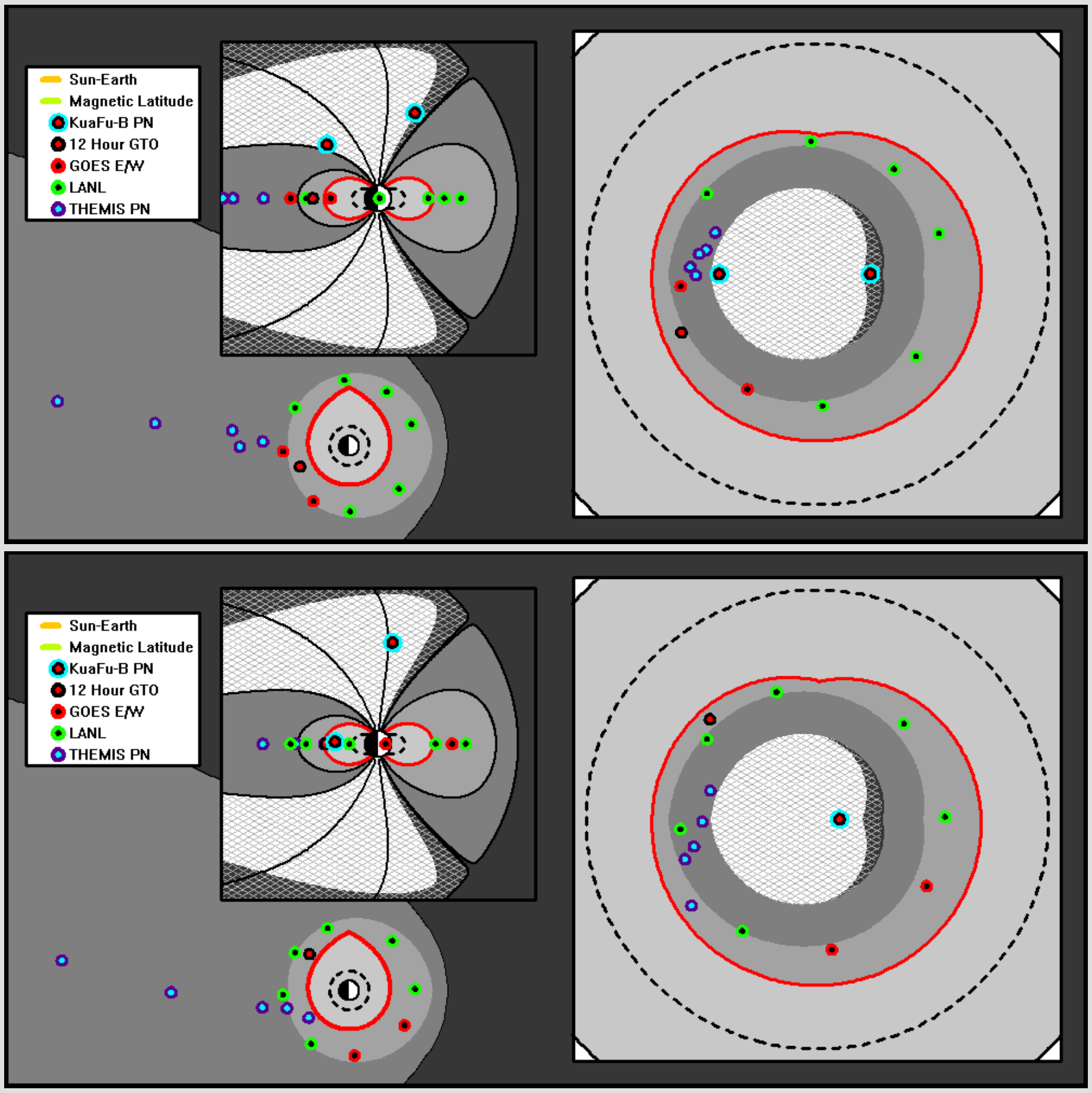

Figure 3.24: (Top Panel) Locations of geospace probes that will likely be flying during KuaFu. Lower left in the panel is an ionospheric projection of the locations of the satellites with roughly equatorial orbits. The LANL and GOES satellites will almost certainly still be in operation with a constellation roughly the same as now. THEMIS will also likely still be operating early in the KuaFu mission. The "GTO" satellite is meant to be one Radiation Belt Storm Probe (all of the NASA RBSPs, ORBITALS, and ERG would, if flown, have roughly a Geosynchronous Transfer Orbit (GTO) – so there are certainly two and possibly four satellites that might be on such an orbit during KuaFu). KuaFu-B1 and –B2 are indicated by the light blue symbols. The inset plot at upper left shows the satellites locations projected in the GSM XZ plane. The inset plot at right shows the locations of all spacecraft mapped along a simple magnetic field model to the ionosphere. In all three plots the sun is to the right, the red curve indicates a reasonable plasmapause, the dashed curve indicates L=2, the white area indicates the lobe/polar cap, and shades of grey indicate the plasmasphere, cusp/cleft, and Central Plasma Sheet. Bottom Panel) Locations of the satellites at a different time. In the top panel, one probe is in the vicinity of the cleft and descending in latitude, while the second is ascending through the PSBL. By the time the ascending satellite is approaching the cusp region, the descending satellite has passed through perigee and is transitting the night-side plasmasphere before making a latitudinal cut through the CPS and back into the lobe. Considering the nature of the orbits, the KuaFu-B pair will provide an almost continuous observing platform in the lobe, and frequent transits of the PSBL, cleft, and CPS.

## 3.6 Coordinated Ravens (KuaFu) and Ground-Based Observations

Up to now, the only techniques we had to track the global time evolving geospace system for long periods (ie., spanning storms and other long duration events) have involved the use of networks of ground-based instruments. Furthermore, we will need to rely on ground-based magnetometry and riometry to track the spatio-temporal evolution of the plasmasphere and high energy electron populations (magnetoseismology and radio absorption due to >30 keV electron precipitation, respectively). It is clear that there will be synergy and in some ways even symbiosis between KuaFu-B and global networks of ground-based instruments. In section 3.2, above, we dealt with what we could do towards meeting our mission objectives with networks of ground-based instruments. In this section, we briefly outline some other ways in which synergy between KuaFu and networks of ground-based instruments will contribute to our scientific goals.

To begin with, imaging across all relevant scales is an important objective in geospace science. Previous work has allowed for statistics of the distributions of scales from tens of meters to thousands of kilometers but has not, in general, allowed an instantaneous let alone a time-evolving picture of the spectrum of fluctuations. Although KuaFu will have better spatial resolution in the electron aurora than has ever before been achieved (see section entitled UVAMC in Depth), this resolution will not be adequate to get down to the very relevant tens of kilometer and smaller scales. This is important, as it would support studies of natural complexity in geospace, the evolution of structure in geospace plasmas, and the effect of structure on magnetosphere-ionosphere coupling. We are fortunate that at the time of KuaFu, existing large networks of ground-based All-Sky Imagers (ASIs) such as NORSTAR, THEMIS-ASI, and MIRACLE, will likely still be in operation and possibly expanded. Furthermore, it will not be difficult to deploy several narrow field of view imagers capable of resolving structures down to tens of meters embedded in the networks of ASIs. The continuous coverage afforded by KuaFu-B imaging, coupled with the networks of ASIs with ~1km resolution, and the narrow field of view instruments will provide the first ever simultaneous observations of the spectrum of spatial structures from global down to local scales. In Figure 3.25. we show the fields of view of the THEMIS and MIRACLE ASIs overplotted on a color-level plot of the spatial resolution of KuaFu-B electron auroral imaging. The concept of probing the full spectrum of fluctuations with nested instruments of different field of view and resolution is illustrated in Figure 5.2.

From the planned spectrally resolved FUV electron auroral images, we can obtain a good approximation of the ion and electron energy deposition, and hence the effects of precipitation on ionospheric conductivity. By using forward modeling, we can estimate the ionization rates from the ion and electron particle precipitation, and estimate the 3D ionospheric density and composition. Because of the complexity of the feedback between the composition and the energy deposition, this is a difficult process, and requires modeling of the coupled thermosphere and ionosphere. Even with the difficulties involved, imaging with appropriate spectral resolution is the best hope for determining the large-scale particle input into the atmosphere.

The SuperDARN radars measure the doppler shift of electromagnetic waves scattering from ionospheric irregularities. This is the best technique for inferring the global distribution of ionospheric electric field, although it is at times limited by absorption or lack of irregularities. Ground-based magnetometer measurements can be used to estimate the distribution of ionospheric current. Combining the current with the conductance pattern determined in part from global auroral observations as described in the previous paragraph, an electric field can be estimated, thus providing an important complement to SuperDARN convection observations.

By determining the state of the ionospheric electric field and particle precipitation, one can start modeling the thermospheric and ionospheric dynamics with much more accuracy. The electric field and particle precipitation are of primary importance when considering the temperature structure, composition, winds, and energy balance in the thermosphere. In addition, the magnetosphere is in part controlled by the ionosphere, so specifying the conditions in this region would allow a more accurate representation of the global magnetosphere.

*Sofko* [1995] showed that the upward field-aligned current (FAC) in the ionosphere can be written as

$$\mathbf{J}_{\parallel} = \Sigma_P \, \mathbf{B} \cdot \nabla \times \mathbf{v} - \mathbf{E} \cdot \nabla\Sigma_P + \mathbf{B} \, \mathbf{v} \cdot \nabla\Sigma_H$$

where $\sum P$ and $\sum H$ are the height integrated Pederson and Hall conductivities, respectively (this is equation 3 in their paper). Based on this, if the conductivities and the conductivity gradients can be determined, then the complete solution to the FAC distribution can be computed. The complementarity of the SuperDARN radars and optical instruments is obvious - provided there are adequate irregularities for obtaining good radar echoes, and that global images with adequate spectral resolution in the FUV spectral band are available, the combined radar and global auroral images will provide us with the ability to specify the global distribution of upward FAC.

In Figure 3.26, we show the fields of view of the global network of SuperDARN radars overplotted on a color-level plot of the spatial resolution of KuaFu-B electron auroral imaging. KuaFu-B will provide a powerful complement to SuperDARN, as the wavelength separation in the LBH should allow for estimation of energy and energy flux and hence give a better idea of conductivity maps than have been available up to now. Further, the continuous nature of the imaging, in conjunction with the continuous nature of the SuperDARN radar observations means that we will be able to track the global electrodynamics (such as the FAC distribution) through long duration events such as storms.

On another front, EISCAT and the Sondrestrom Incoherent Scatter Radars (ISRs) will continue operation on Svalbard for the foreseeable future. The new Advanced Modular Incoherent Scatter Radar (AMISR) is currently under construction in both Alaska (at Poker Flat) and Resolute Bay Canada. These radars probe the upper atmosphere with narrow UHF radar beams measuring details of both the ionospheric ion and electron distributions. These radars will provide detailed properties of the plasma and convection velocities in key regions such as the cusp/cleft, the open closed boundary (particularly the south facing AMISR face at Resolute Bay), and the polar cap. Taken together with the KuaFu-B electron and proton auroral images which will be continuously available, we will be able to address questions related to dayside reconnection (utilizing proton and electron auroral observations in the region of the cusp, particularly around the winter solstice in Svalbard when the cusp region is dark), the relationship between reconnection at the distant neutral line and poleward boundary intensifications and bursty bulk flows [see, for example, *de la Beaujardièreet al.,* 1991; *Zesta et al.,* 2000; *Lam et al.,* 2006]. As well, the polar cap ISRs together with the KuaFu electron auroral imaging will allow for detailed studies of polar cap patches and theta aurora. In Figure 3.27, we show the fields of view of ISRs that we believe will be operating in 2012, together with a color-level plot showing the spatial resolution of KuaFu-B electron auroral imaging.

There is a long list of research thrusts which would be facilitated by coordination between KuaFu-B imaging and ground-based observations. *Without question, KuaFu-B will provide context for and fundamental new science based on coordination with essentially every component of the ILWS program, including programs which are of strategic importance to Canada (such as SuperDARN, CGSM, THEMIS-ASI, and AMISR), and to China (such as SuperDARN, MERIDIAN, and EISCAT).*

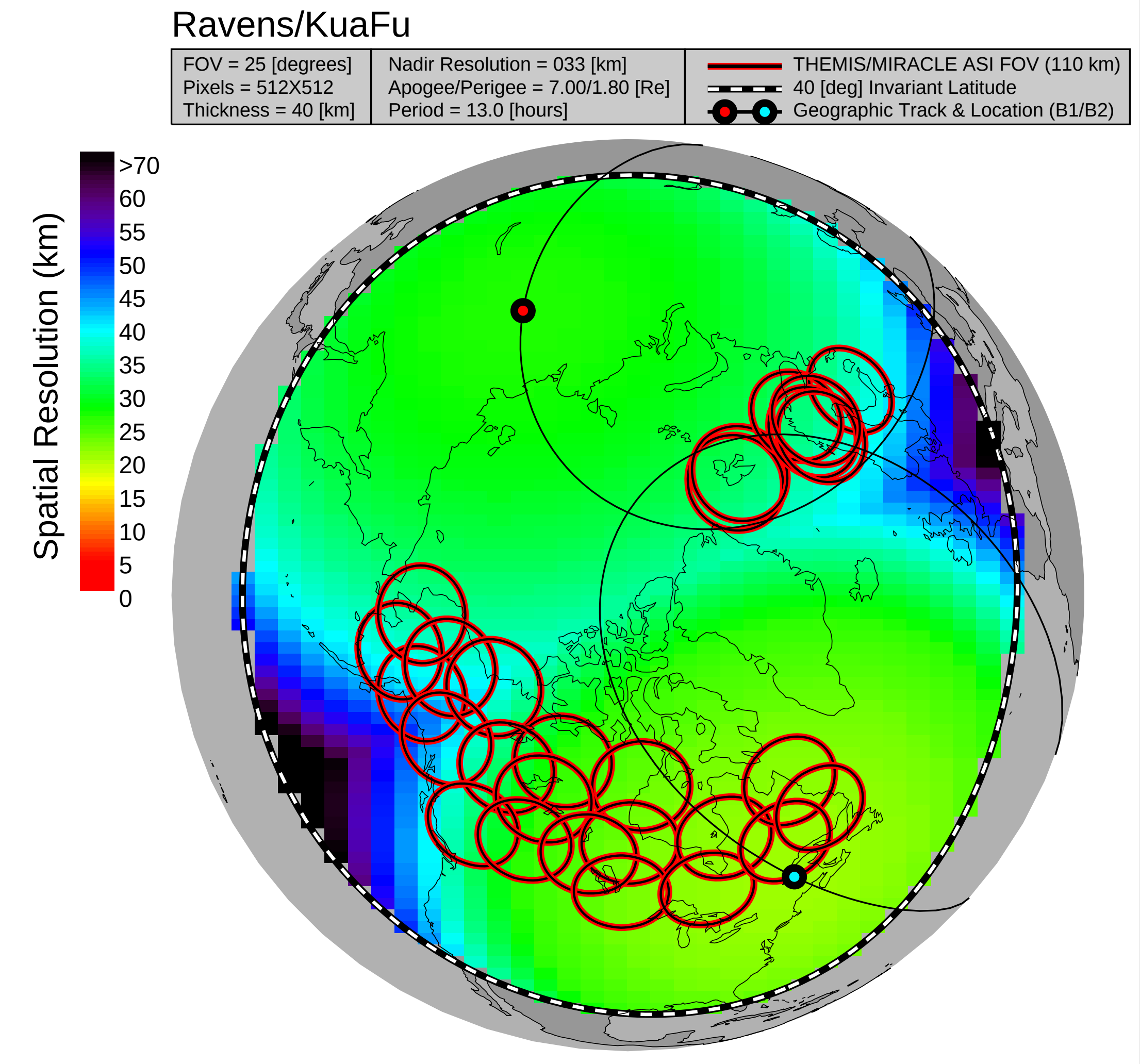

Figure 3.25: The spatial resolution for locations poleward of 40 degrees invariant latitude afforded by the UV electron auroral imagers on the two KuaFu-B satellites. Note the ground-tracks and ground locations of the KuaFu-B satellites. This is for one time during the orbit; however such global scale spatial resolution would be sustained for over 2 years. A key objective of KuaFu-B is to achieve simultaneous imaging of the electron aurora across all relevant spatial scales (note that due to charge-exchange collisions the small scale structure is not present in the proton aurora). Even given modern telemetry, CCDs, and optical materials and designs, this is simply not achievable from an orbiting platform. It is also not achievable from the ground because of daylight, clouds, ocean, and inaccessible polar land. We must utilize ground-based and space-based imaging with overlapping spatial resolutions. KuaFu-B would provide global images with resolution of ~30km. Ground ASI arrays in Canada and the US and northern Europe will provide observations spanning spatial scales from hundreds of kilometres down to ~1 km at 1 image every 3 seconds. Telescopic imagers such as Trond Trondsen's Portable Auroral imager will provide observations down to tens of meters and up to 30 Hz. We have included the fields of view of the current MIRACLE array and the planned (and funded) THEMIS array.

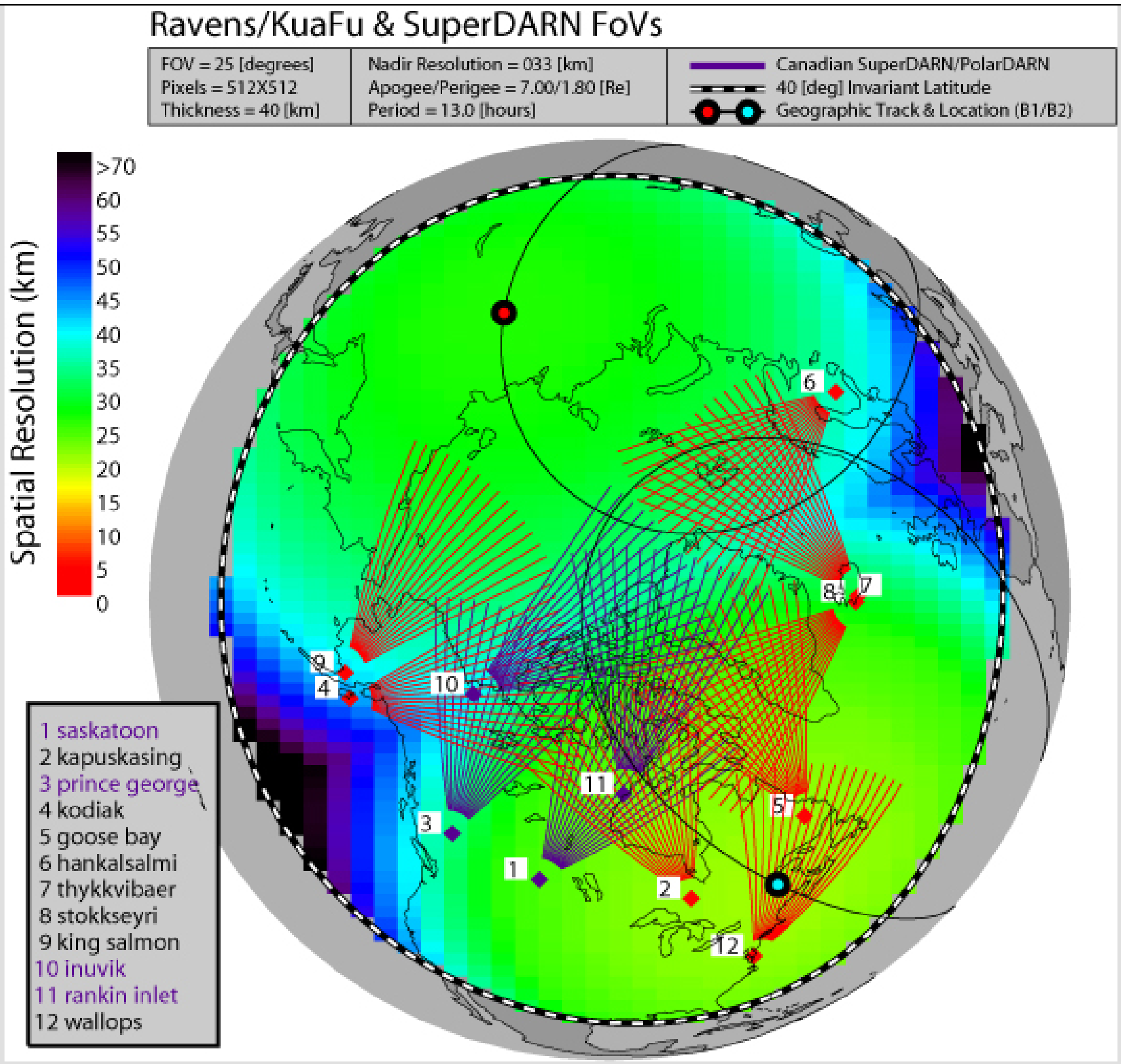

Figure 3.26: The spatial resolution for locations poleward of 40 degrees invariant latitude afforded by the UV electron auroral imagers on the two KuaFu-B satellites. (see caption for B2.2 for more details). As discussed above, KuaFu-B optical data would provide an important complement for global convection maps from the SuperDARN HF-radar array. The spatial resolution of the global images would be comparable to or better than the spatial resolution of SuperDARN convection measurements under normal operating modes. We have over-plotted the fields of view of the currently operating and planned northern hemisphere SuperDARN radars. Note that the two planned PolarDARN radars (10 and 11) would fill in the gap in coverage over the polar cap that has plagued SuperDARN to date. SuperDARN and KuaFu-B would be able to, for example, provide quantitative global maps of convection electric field and conductance, from which one can derive the global distribution of field-aligned currents. This combined observation would be a powerful tool for studying the coupled ionosphere-thermosphere-magnetosphere system and how it reacts to variations in solar wind and solar radiant inputs. Just one of many important questions that could be explored is how convection relates to substorms as a response and as a driver.

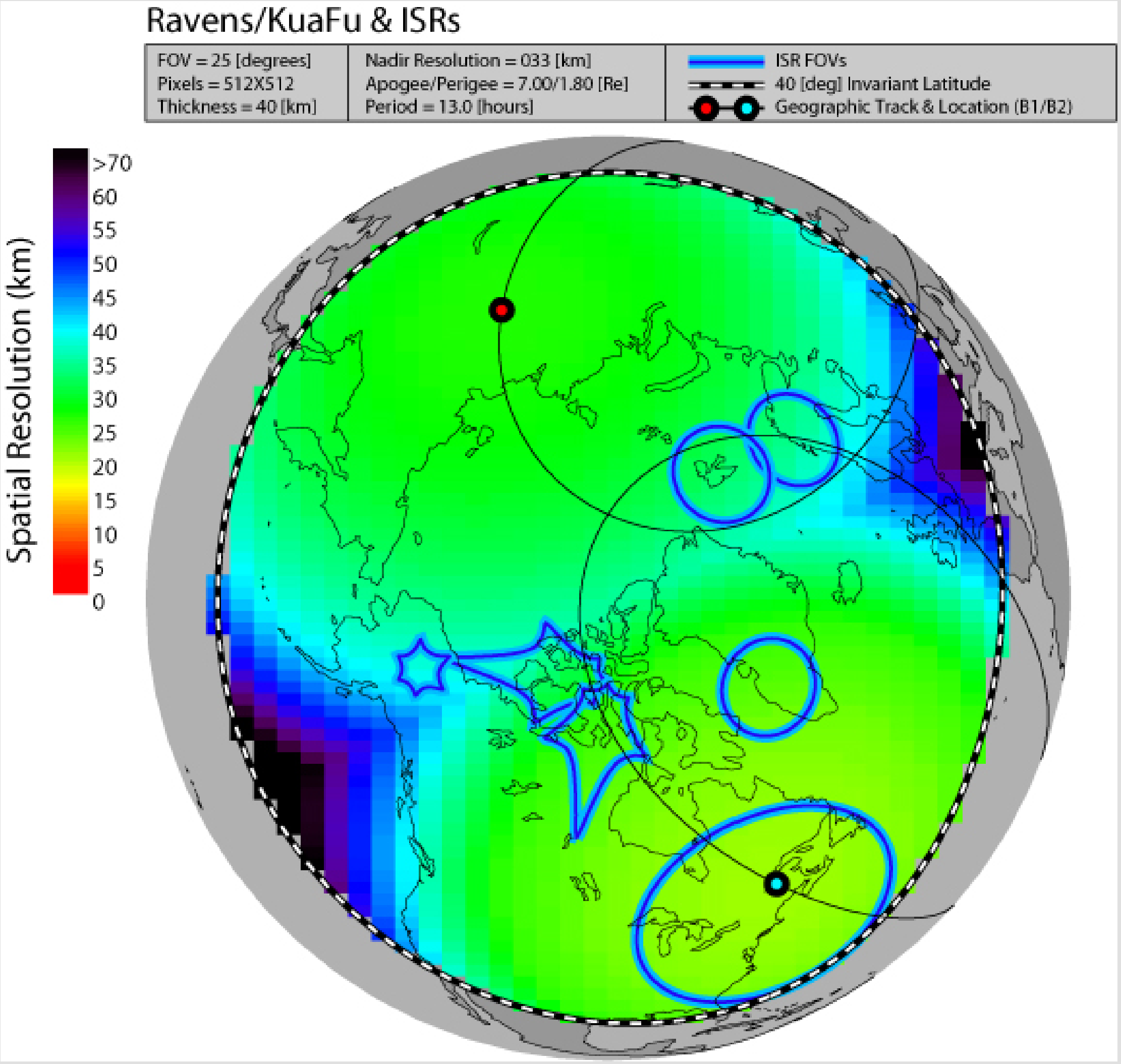

Figure 3.27: spatial resolution for locations poleward of 40 degrees invariant latitude afforded by the UV electron auroral imagers on the two KuaFu-B satellites. (see caption for B2.2 for more details). KuaFu-B optical data would provide important context for measurements from incoherent scatter radars. For example, global electron and proton auroral measurements would support detailed studies of the cusp dynamics. As Svalbard dark for 24 hours per day during the month or so around winter solstice, it is ideally located for cusp observations that are supported by optical data. Proton auroral signatures of dayside reconnection have been identified in IMAGE SI-12 data (SI-12 is the heritage instrument for the proton auroral imager on KuaFu-B). As well, one face of AMISR (Resolute Bay) will be looking south into central Canada (leaf-shaped FOV in central Canada). That field of view will span the open closed boundary. AMISR observations of convection at the open/closed (polar cap) boundary, in conjunction with global electron auroral observations, will provide an exciting opportunity to study the relationship between reconnection at the distant neutral line (as inferred from plasma transport across the open closed boundary) and poleward boundary intensifications (as inferred from the electron auroral images).

## 3.7 Conclusions

We proposed Ravens to facilitate system-level science, investigations of global auroral electrodynamics, and exploration of natural complexity in geospace. We argued that we want to track the global evolution of the electron and ion CPSs, the injected electron and ion populations, the ring current, and plasmasphere on a continuous 24/7 basis for long enough to observe several large magnetic storms. We argued that we could follow the plasmaspheric and injected electron populations with global networks of magnetometers, GPSs, and riometers. Our mission proposal called for two satellites on 1.8 Re X 8 Re (geocentric) elliptical polar orbits, each carrying a two-channel global imager for LBH-L and LBH-S, a spectrographic imager for global proton auroral images in Lyman-alpha, and an ENA imager for ring current imaging. With these orbits, this program would provide excellent viewing of the entire auroral and polar region in the northern hemisphere for more than 2.5 years, more than long enough to guarantee several storms. We want to make advances in global auroral electrodynamics, to ensure we could create global maps of electron characteristic energy and energy flux for system-level science, and to obtain better spatial resolution than any previous global imager to facilitate multi-scale observations. The electron auroral (LBH) imager from Canada should thus have out of bandpass rejection sufficient for energy estimation based on LBH-L to LBH-S ratios and ~30 km spatial resolution.

We went on to consider some of the ancillary benefits of this mission design. We demonstrated that the Ravens orbits would allow for tens of minutes of conjugate imaging every orbit. This motivated inclusion of the wide field of view auroral imager in the KuaFu mission plan. Further, we showed that KuaFu would have significant synergy with ground-based arrays of All-Sky Imagers, the SuperDARN global HF radar network, and the northern hemisphere Incoherent Scatter Radars (particularly the EISCAT Svalbard and soon to be operating Resolute Bay AMISR radar).

The KuaFu-B orbits were chosen based entirely on the work presented in this chapter. Furthermore, the KuaFu-B payload includes the four imagers specified here: these are the LBH (Canada), Lyman-alpha (Belgium), ENA (Ireland), and WFAI (UK) imagers which are described in more detail in Sections 4 and 5.

There are unresolved issues related to the work presented in this chapter. Phase A work on KuaFu and the Canadian LBH imagers should include at least the following:

- more detailed orbital analysis including significant perturbing factors such as the moon, to be carried out in conjunction with the Chinese KuaFu engineering team.

- Exploration of potential synergies between GOLD and KuaFu for mid-latitude imaging.

# 4. Ravens (KuaFu-B) Instrument Payload Overview

In the section on mission design, we outlined our rationale for an identical complement of four imagers on each KuaFu-B satellite. These imagers combined with global ground-based networks of magnetometers and riometers would come together to address the primary scientific objectives of the mission which have to do with system-level In addition, because *in situ* observations are so important in the high latitude high altitude region where KuaFu-B1 and -B2 will be spending the most time (ie., ascending through and descending from apogee) and there is no other mission planned for this timeframe with an orbit that could provide those measurements, we have added several *in situ* instruments. The number of additional *in situ* instruments is quite limited by overall payload mass restrictions (65 kg of total instrument payload for each of KuaFu -B1 and –B2 given the chosen 0.8 X 7 Re orbit).

The KuaFu-B instrument payload is as follows

B1 & B2

12. Ultraviolet Auroral Monitoring Camera (UVAMC)
13. Far Ultraviolet Spectrographic Imager (FUVSI)
14. Neutral Atom Imager on KuaFu (NAIK)
15. Wide Field Auroral Imager (WFAI)
16. Fluxgate Magnetometer

B1 only

17. Imaging Energetic Electron and Proton Instrument
18. KuaFu Agile Plasma Pitch Angle Instrument and Ion Mass Spectrometer

B2 Only

19. High Energy Electron Detector
20. High Energy Proton Detector
21. Linear Energy Transfer Experiment
22. Tri-Band Beacon

The Ravens Concept Study was initially envisioned as an investigation of a two satellite 24/7 global imaging mission. In the spirit of the original concept study, we restrict our attention in this report on the technical end only to the four imaging instruments, which as this came together were viewed as the “Ravens Integrated Imaging Package”. In this section, we provide a *brief overview* of the four imaging instruments, concentrating on the observations to be delivered, and significant issues that need to be dealt with in the Phase-A period (note – we anticipate a Canadian KuaFu and UVAMC Phase-A, which would address issues related to UVAMC directly, ensuring the combined imaging package addresses our scientific requirements, technical issues related to the integration of UVAMC with the other auroral imagers and the auroral imaging package with the overall instrument package, and the mission level science plan). In the following section, we deal with UVAMC in depth. Note that a small amount of the introductory material for UVAMC included in this section is repeated in "UVAMC In Depth". The reason for this is that both this section and UVAMC In Depth are meant to be stand alone documents.

## 4.1 UVAMC

The Ultraviolet Auroral Monitoring Cameras would provide global observations of the electron auroral emissions, with adequate spectral isolation of two parts of the LBH to allow for estimates of characteristic energy and energy flux. The science requirements that would drive the design of the UVAMC are as follows:

1. determination of characteristic energy and energy flux of precipitating electrons on a global scale (motivates isolation of LBH-L and LBH-S).
2. global images with ~30 km spatial resolution to facilitate studies of Multiscale processes and natural complexity (motivates large CCD and selection of field of view).
3. imaging of dim features (places limitations on CCD size, aperture, and field of view, and places strong requirements on optical throughput).

The UVAMC instruments would each be comprised of two FUV imagers and a common electronics box. The two component imagers on each obtain global images at a 30 second cadence (corresponding to the satellite spin period which is in turn determined by the requirements of the FUVSI). They would deliver quantitative intensities from well separated parts of the LBH band, with the objective of using relative and absolute intensities for determination of characteristic energy and energy flux of precipitating electrons on a global scale. Our technical objective is $10^{10}$ out of bandpass rejection. In order to achieve the spatial resolution that we are aiming for from near apogee, the instrument must have at least a 512X512 pixel CCD and a FOV that is wide enough to span the entire auroral zone and polar cap for a significant fraction of the orbit. Finally, the aperture must be as small as possible (for compactness and lower mass) while admitting enough light for reasonable signal to noise for dim aurora (ie., in the tens of Rayleighs integrated across the appropriate part of the spectrum).

The heritage for these imagers are the past University of Calgary FUV instruments (Viking, Freja, InterBall, and IMAGE-WIC); however, the design of this new instrument will need to be different if we are to meet our technical objectives, and hence our scientific objectives. In particular, we will be moving to a 4 mirror design, which allows us better out of bandpass rejection. This new instrument will provide wavelength separated LBH electron auroral observations with ~30 km resolution over the global auroral distribution. Furthermore, it will image even dim aurora (~30 Rayleigh features in LBH-L or LBH-S). The UVAMC spacecraft resource requirements are included in Table 4.1 (which is also in the UVAMC In Depth section).

Table 4.1: Spacecraft resource requirements (preliminary) for new 4-mirror design.

| | **UVAMC-0** | **UVAMC-1** | **UVAMC-E** |
|---|---|---|---|
| **Mass** [kg] | 9.0 | 9.0 | 3.0 |
| **Dimensions** [$cm^3$] | 25 x 24 x 15 | 25 x 24 x 15 | 20 x 20 x 10 |
| **Power** [W] | 4 | 4 | 3 |
| **Data Rate** [Mbits/s] | 0.1 | 0.1 | N/A |
| **Temperature Limits** | -40 to +60 C (surv.)<br>-25 to +20 C (ops.) | -40 to +60 C (surv.)<br>-25 to +20 C (ops.) | -40 to +60 C (surv.)<br>-25 to +40 C (ops.) |

The UVAMC concept is described in detail in the UVAMC In Depth section of this report. We provide somewhat more detail about the other three imagers, although we point out that this material was provided by collaborators in Belgium (FUVSI), the UK (WFAI), and Ireland (NAIK). In each case we attempt to highlight issues that relate directly to our design of UVAMC, or affect the overall science objectives.

## 4.2 FUVSI

Scattering of solar Lyman-alpha by the geocoronal hydrogen surrounding the Earth produces the Ly-alpha emission line at 121.6 nm. Part of the auroral Ly-alpha emission of neutral hydrogen is due to charge exchange of energetic protons that cascade into the atmosphere. One objective of the FUVSI instrument is the imaging of this type of Lyman-alpha emission. As the protons penetrate into the atmosphere, they capture electrons from the major constituents by charge exchange reactions. They become fast neutral hydrogen atoms radiating Doppler-shifted Lyman-alpha emission.

As we point out below, proton aurora comes from pitch angle scattering which in turn fills the loss cone (in other words diffuse aurora). With proton auroral observations, therefore, we are remote sensing the ion (proton) CPS. *Baseline scientific requirements driving the FUVSI design are*:

1. to image the entire auroral oval from a spinning spacecraft (30 second spin rate) at 7 Re apogee altitude.
2. to separate spectrally the hot proton precipitation from the statistical noise of the intense, cold geocorona.
3. to provide information on the predicted contribution to separately obtained (ie., UVAMC) electron auroral signals due to secondary electrons produced by primary proton precipitation.

In addition, if it is at all possible, we would want the following additional requirement to be met:

4. provide some information about the characteristic energy of the precipitating protons.

The first requirement is met by a 15 degree field of view. The second requirement needs a high spectral resolution, better than 0.2 nm, in order to separate the Doppler shifted emission resulting from proton precipitation from the typically 100 times brighter but cold (hence narrow in spectral width) Lyman-alpha geocoronal emission. With better spectral resolution, one can shift the primary part of the response profile closer to the Lyman-alpha central (cold) wavelength. This is desirable, in that one wants the response to encompass as much as possible of the complete profile of Lyman-alpha of proton auroral origin. This means that anything that decreases spectral resolution pushes the response curve to longer wavelengths with in turn means the instrument will not be measuring photons emitted as a result of the precipitation of lower energy (say less than ~2 keV which is important if one considers the typical precipitating spectrum as measured by FAST or DMSP). The spectral resolution, however, comes at the expense of mass, and so the ability to include the signal due to ~keV precipitating protons is fundamentally limited by how large the instrument can be. In addition, Nitrogen emissions near 120 nm (triplet lines at 119.955, 120.022 and 120.071 nm) also have to be filtered out. The third requirement can be met to some extent provided the first two are met

The fourth requirement can obviously be better met provided there is any information about the global distribution of the characteristic energy of the precipitating protons. Furthermore, there are obvious scientific motivations for achieving the fourth requirement. In short, the proton aurora is virtually always diffuse aurora, and the characteristic energy of the protons would be a very valuable and direct measure of the characteristic energy of the CPS proton population that is being scattered into the loss cone by non-adiabatic processes [see for example *Tsyganenko,* 1982; *Gvozdevsky and Sergeyev*, 1995]. One way that an SI-type instrument could meet the fourth requirement would be to spectrally resolve two parts of the Doppler shifted Lyman-alpha. This would require two separate channels dedicated to Lyman-alpha imaging, each with comparable or better spectral resolution than IMAGE SI12. The basic idea can be discussed with the help of Figure 4.1. This shows the simulated Doppler shifted Lyman-alpha for a 2 keV (characteristic) precipitating proton distribution, together with the response of the SI12 [note, taken from *Gérard et al.,* 2000]. The Doppler shifted profile will shift to longer wavelengths with larger energy. Two channels with shifted and possibly narrower spectral response would allow one to crudely track the shifting peak, with the caveat that the wings of the spectral response will introduce complications in interpretation.

The heritage for the FUVSI on KuaFu is the Spectroscopic Imager on the NASA IMAGE mission [*Mende et al.,* 2000a,b]. The IMAGE FUV proton auroral data is delivered via the IMAGE SI12 channel. The satellite had a two minute spin period, which was the cadence at which the images were collected. This was a revolutionary step in our ability to study global processes in geospace. Proton aurora had been observed via ground-based photometers, spectrometers, and more recently All-Sky Imagers for years now [see for example from the ground (see for example *Galand* [2001] and *Sakaguchi et al.* [2006]).

These observations have provided an excellent understanding of the temporal evolution of the proton aurora in localized regions, such as in the case of substorm events observed with the CANOPUS Meridian Scanning Photometers [*Samson et al.*, 1992], or the statistical behaviour of the global proton auroral distribution on the night side inferred from long-term observations from isolated ground stations [*Creutzberg et al.,* 1988].

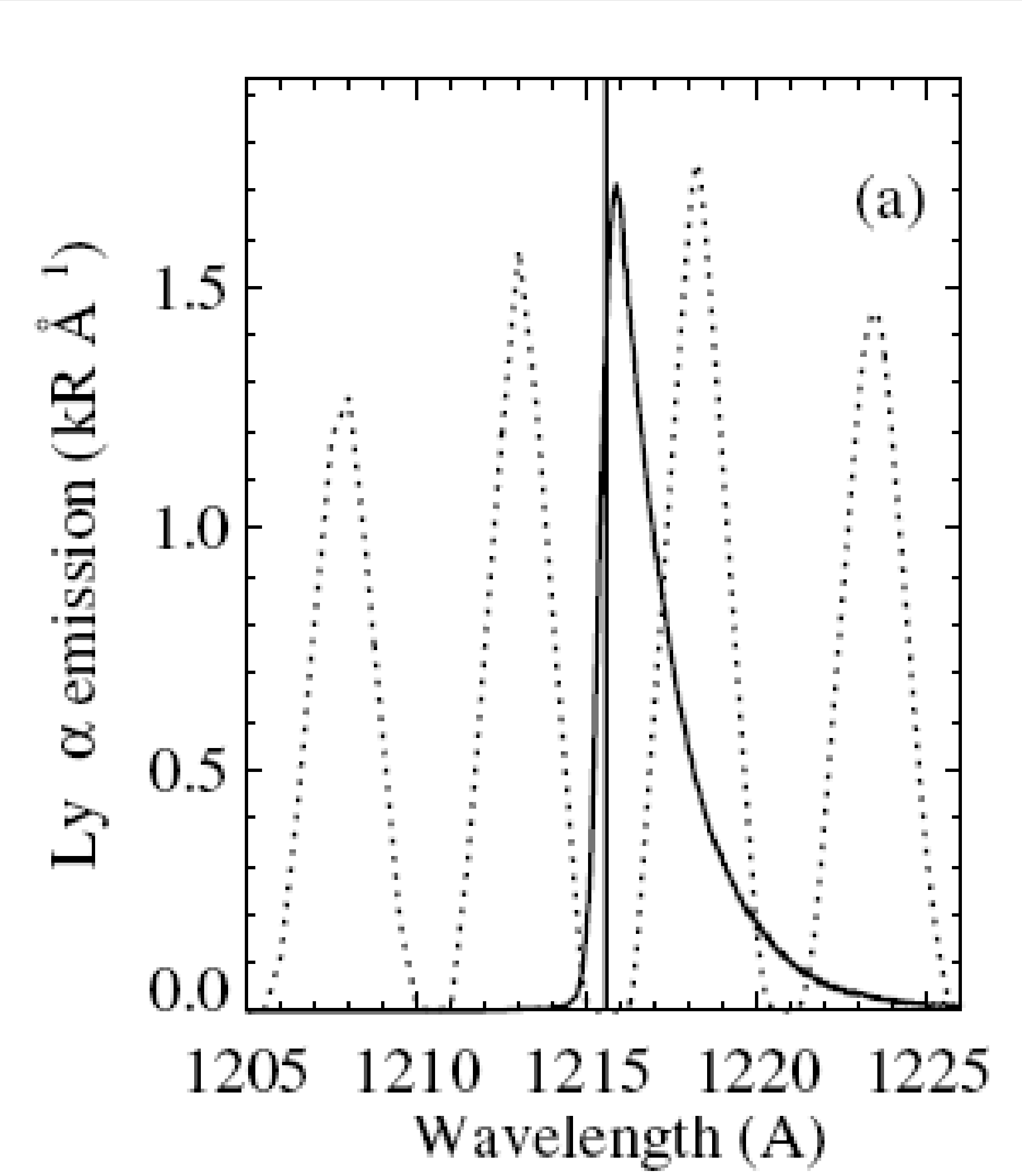

Figure 4.1: Reproduction of Figure 1 panel a from Gérard et al. [2001]. This shows a simulated Doppler shifted Lyman-alpha profile (solid) and the SI12 response (dashed). The emission profile is for nadir viewing and a precipitating 2 KeV Kappa (K=3.5; 1 erg $cm^2$ $s^{-1}$) proton spectrum. This is discussed in more details in section 5.3.

With the SI12 observations from IMAGE, our community got its first true global images of the proton aurora, and consequently the time evolving ion plasma sheet as projected along magnetic field lines into the ionosphere by diffuse auroral processes. This has led to a wealth of scientific firsts, just one example of which was the observations of the ionospheric signature of continuous dayside reconnection [*Phan et al.,* 2003].

The quality of the data shows that the requirements which drove the development of the instrument were justified. Very early on in IMAGE, *Frey et al.* [2001] used FAST and DMSP ion precipitation observations to show that there was good agreement between the SI12 intensities and the emissions expected from the *in situ* precipitation observations. In addition, comparisons between ground MSP Hydrogen Balmer Beta observations and SI-12 intensities show general agreement between important boundaries such as the "optical-b2i" as inferred from the ground and space and, taking appropriate ratios into account agreement between MSP Hydrogen Balmer Beta and SI12 Lyman alpha intensities [*Trondsen et al., manuscript in preparation*; *Meurant et al.*, 2006]. An example qualitative comparison between Hydrogen Balmer Beta measured by the CANOPUS MSP at Gillam Canada, and a keograms for the MSP meridian extracted from the SI12 images is included in Figure 4.2. An example global image from SI12 is shown in Figure 4.3.

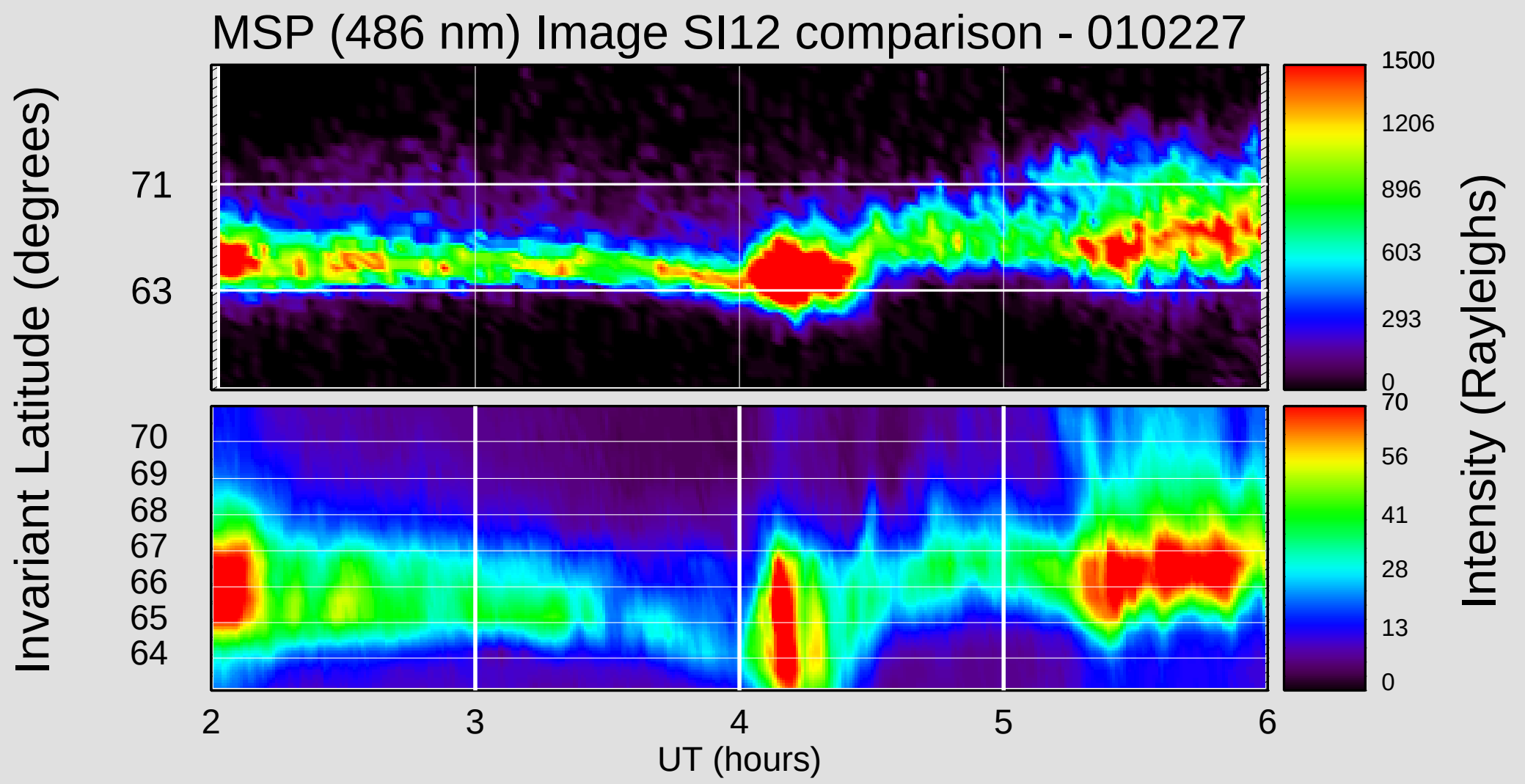

Figure 4.2: Comparison of proton auroral observations obtained in Lyman-alpha by SI12 on IMAGE (top), and in Balmer-beta by the CANOPUS MSP at Gillam (bottom). To facilitate visual comparison, a keogram has been constructed from the SI12 data.

In Section 3.2, we illustrated how the global network of riometers is capable of tracking the evolution of the high magnetospheric energy electron population, provided that population is subject to strong pitch angle scattering. This capability has been refined to the point where we can follow the spatio-temporal evolution of the dispersionless electron injection with 5 second temporal resolution, albeit with no energy resolution. Recent work using SI12 data from imager has shown that the FUVSI should be able to provide the same information about the high energy substorm injected protons, subject to the limitations of the FUVSI special resolution and the proposed 30 second image cadence imposed by the satellite spin. This was demonstrated for several events by comparison of LANL SOPA *in situ* geosynchronous particle fluxes and SI12 data from IMAGE [*Chi et al.*, 2006].

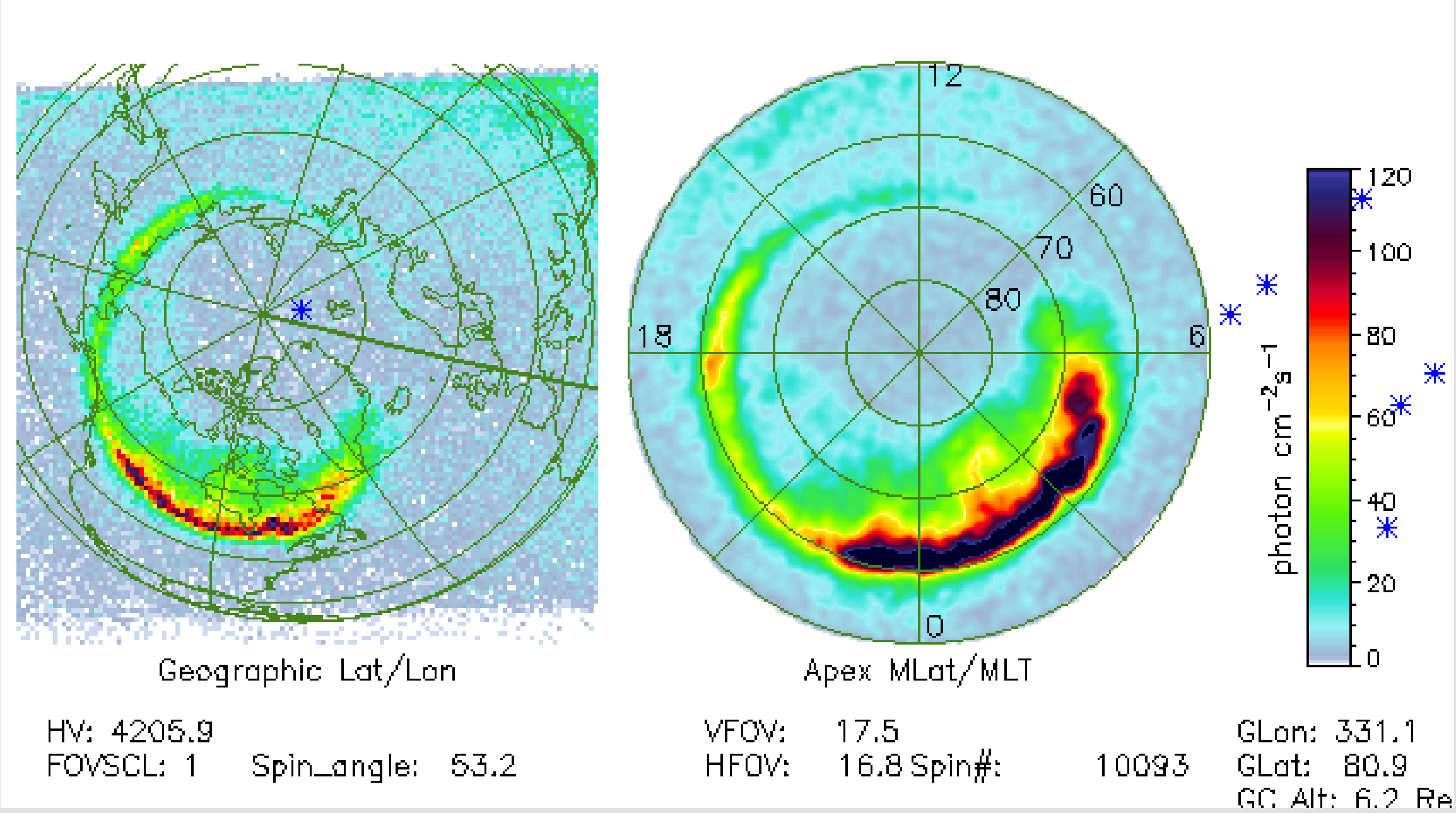

Figure 4.3: Example of IMAGE SI-12 global proton auroral image in geographic (left) and geomagnetic (right) coordinates. Image courtesy Harald Frey.

As stated above, the IMAGE Spectroscopic Imager is the heritage for FUVSI on KuaFu. A technical description of the IMAGE instrument is included in section 4.2.2 in the KuaFu Assessment Study report, which is attached to this report as Appendix E. There are several points that we extract from that report and include here because they are likely relevant to work under the anticipated Canadian KuaFu Phase A. These are:

1. "Two wavelengths are selected with two separate exit slits. One for imaging the Doppler-shifted Lyman-alpha, the other one for the OI 135.6 nm emission." … and as well… "The monochromator was designed to produce the largest dispersion, in order to reach the required high spectral resolution. Focal length is 500 mm and the grating ruling is 3600 lp/mm (holographic recording). The grating curvature is spherical; the collimator curvature is hyperbolic. The linear spectral dispersion is about 1820 µm/nm in the exit slit plane. As depicted in figure 1, a first image of the Earth is produced approximately at the grating plane. The aim of the back-imager is to relay this image after diffraction by the grating, in order to obtain two spectrally separated images. Each sub-assembly of the back-imager is a two-mirror optical system. The first mirror is spherical. The second one is conical (elliptical) and off-axis. A flat folding mirror is added to the 135.6 nm imager in order to allow fitting the two detectors in close proximity." (both from page 91)

2. “The detectors use a KBr photocathode on a $MgF_2$ window image tube with a triple stack microchannel plate (MCP) intensification. The intensified image is detected by a crossed delay-line type detector with two 128 x 128 pixels, each active areas covering 20x20 mm. This kind of detector presents a QE of 17% at 121.6 nm and of 14% at 135.6 nm. The detectors could probably also be CMOS covered by some type of phosphor transferring FUV to visible. The selection of the best technology is still TDB and has to be studied during the design phase of the instrument. The FUVSI instrument is in orbit on the IMAGE NASA spacecraft since March 25, 2000. Therefore, the instrument is fully qualified with MCP detectors. If a CMOS detector is chosen, it should be similar to the SWAP/PROBA2 detector and would be qualified before 2007.” (from page 92)

3. “The absolute response of FUVSI is about 1% at both wavelengths of interest:121.8 nm and 135.6 nm. The entrance aperture slit grill is about 1 sq.cm.. Therefore, assuming an input of 1.6 photon per Rayleigh, the overall efficiency of the instrument is about 0.017 counts per Rayleigh for a 5s exposure.” (from page 93)

Further, we include Figure 4.10 from the Assessment Study Report (from page 92), which shows the design of the IMAGE SI, which is the starting point of the design of FUVSI

Our reason for including these three points and the figure is that each relates to work that must be done by our team (and/or Canadian industry) during a phase A. For example, the first point and the figure have implications for the UVAMC team. The Spectrographic Imager on IMAGE provided global images in two wavelengths. One is the Doppler shifted Lyman Alpha (ie., the SI12 discussed above) while the other is the OI 1356 Å emission. The latter is potentially useful in conjunction with WIC images as ratios should be able to give characteristic electron energy. This has not found much use which *may* indicate it is not useful (note that lack of use does not necessarily mean the information is not useful – it may simply be that the community has found enough interesting things to go after with just the WIC and SI12 auroral images). It is worth mentioning converting the 1356/LBH ratio to electron energy is dependent on a neutral atmosphere model. Observations have shown that the O/N2 ratio is affected by auroral heating (particle + Joule) in a way which is difficult to predict. Therefore, using 1356/LBH might never be a very accurate method, even if the two signals are strong and the emissions spectrally well isolated locally and separated from each other.

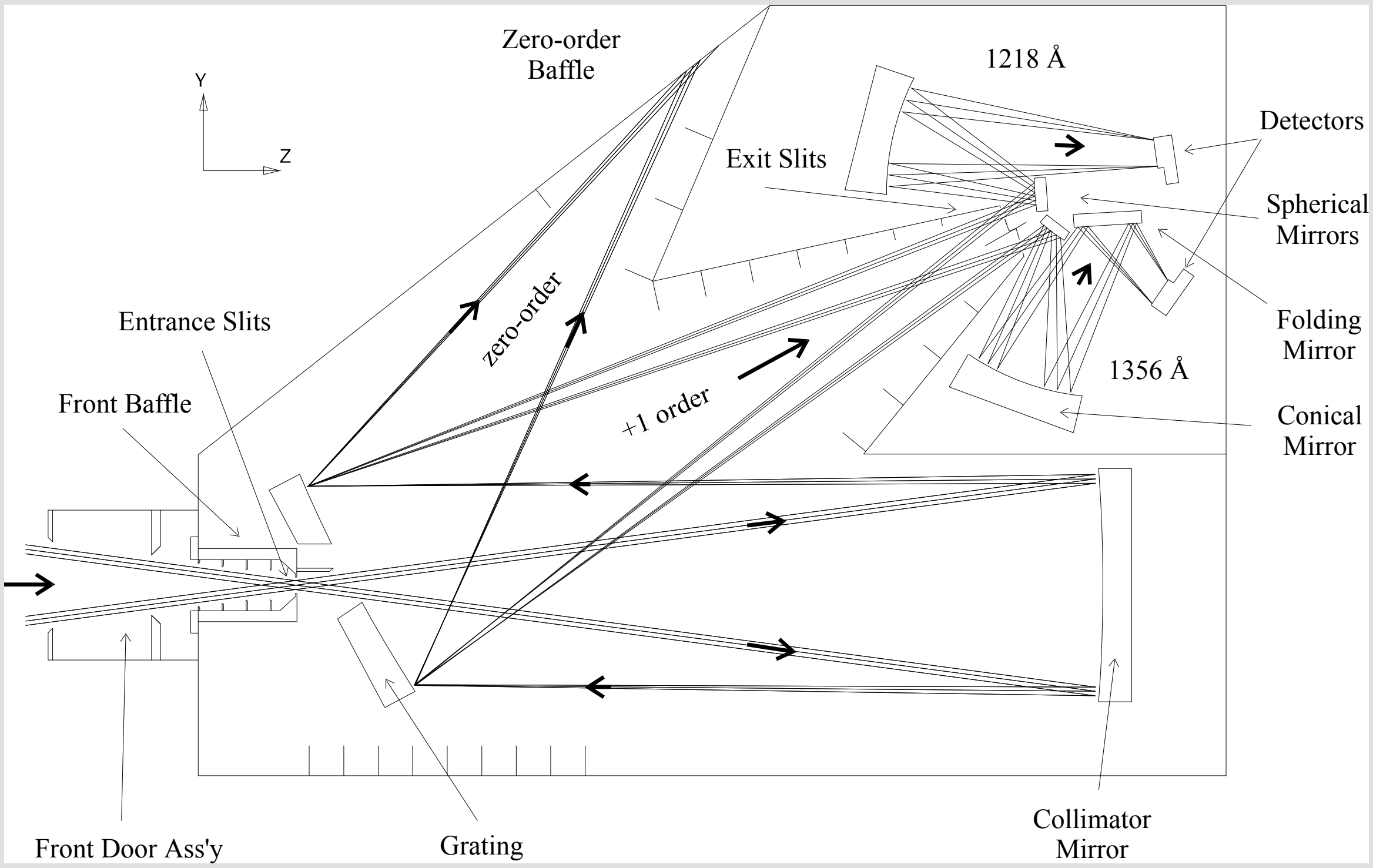

Figure 4.4: Initial design of FUVSI for KuaFu (reproduction of Figure 4.10 from the KuaFu Assessment Study Report). This illustrates the different optical paths for 1218 and 1356 Å, leading to images in the two different wavelengths on two separate detectors.

For KuaFu, our intention is to deliver LBH-L and LBH-S with UVAMC enabling estimates of characteristic energy and energy flux in the electrons, so there may be no need for 1356 even if it is useful. We will also explore the possibility of splitting the Lyman-alpha signal into "long" and "short, possibly in successive 30 second spins, possibly in the same spin. This opens up the possibility of creating time-evolving maps of very crude (but very powerful) estimates of the CPS proton characteristic energy, but obviously would demand something extra of the instrument. This in turn is likely only possible if (among other things) the FUVSI does not image in 1356 Å. To pursue this, we (the Belgian and Canadian teams) would have to address the following questions early in the Phase A:

I. Given what we plan to do on UVAMC, is 1356 Å necessary to the overall success of the mission?

II. If FUVSI does not do 1356 Å, should the Belgian team try to split the Lyman-alpha?

III. If FUVSI splits the Lyman-alpha is split, can it be done with no moving parts?

IV. If FUVSI splits the Lyman-alpha is split, is it necessary (or even possible) to image in both Lyman-alpha "long" and Lyman-alpha "short" during every spin, or would it be sufficient, for example, to image in one passband during one spin, then the other during the next?

The UVAMC team should contribute to providing answers early in our proposed phase A. Question I requires consideration of the usefulness of ratios, and the reasonable expectation of what we can deliver on UVAMC. Question IV requires thinking about how quickly the proton aurora really evolves on these global scales. The CANOPUS/NORSTAR MSP data might be the best way to go after that, owing to the two minute time resolution of SI on IMAGE (in global processes there is an enormous difference between 2 minutes and 30 seconds, and 30 seconds is the inherent temporal resolution of the CANOPUS (now NORSTAR) MSP. Questions II and III must be dealt with directly by the Belgian team.

Points 2 and 3 relate to the detector and the sensitivity requirement. Regarding point 2, our understanding is that the Belgian team cannot acquire a suitable detector in Belgium. Further, they are legally bound *not* to acquire certain types of hardware above a specified price limit with Belgian funds from a non-Belgian company. For IMAGE, the detector was acquired by UCBerkeley through a NASA contract, giving the US, and Berkeley an important role in the Spectrographic Imager. This is potentially an equivalent opportunity for Canadian scientists and industry. We suggest that as part of the Phase A for the Canadian contribution to KuaFu, the industry partners explore whether detectors are available from Canadian firms, or international firms based in countries where such hardware could be provided for a Chinese mission (likely an impediment for repeating what was done on this point for IMAGE).

Point 3 highlights the fact that the sensitivity of the IMAGE SI was adequate for 5 second exposures, which translates to how long a feature stays within the 15 Å field of view during a two minute spin. For KuaFu, the current plan is for a 30 second spin rate. This has been set by the competing requirements of the Chinese engineering side in terms of stability, the desire for as high as possible time resolution (driving down the spin period), and the desire for good signal to noise (driving up the spin period). Regardless, the spin period being 25% of that on IMAGE means less photons for a given image. This in turn needs to be explored by the Belgian team in terms of whether this 30 second spin rate is viable for the Lyman-alpha viewing. If it is not, then that couples back to all other instruments on KuaFu-B, and most importantly (we think) the UVAMC. Assessing the viability of imaging at this cadence is a fundamentally important objective for Phase A for both the FUVSI and UVAMC teams.

## *4.3 WFAI*

The KuaFu mission will enable the continuous imaging of the global auroral oval in the northern hemisphere. In turn this continuous global imaging will provide crucial information on the response of the auroral oval during magnetic storms which tend to have time scales of days and represent the major large scale manifestation of “space weather” events in the ionosphere. However, imaging of one hemisphere alone does not provide full information on the auroral response to large scale energy input if the southern auroral oval is not likewise being imaged. *The scientific purpose of the Wide Field Auroral Imager is to provide such systematic coverage.*

The science case for systematic conjugate imaging is very easy to make (see Section 3.5). There is a dearth of such observations from past missions and ground-based programs because conjugate imaging was never a primary motivator of those programs (ie., the recent work published by Østgaard and colleagues pointed out in Section 3.5 was carried out with data from a few serendipitous relative alignments of Polar and IMAGE. We know that conjugate imaging is important. With such imaging, we can explore when diffuse and discrete auroral forms are conjugate and when they are not. Such work will immediately uncover flaws in our understanding of interhemispheric mapping, situations where parallel electric fields decouple the two hemispheres, and either confirm or refute our understanding of various forms of diffuse aurora. Conjugate imaging will be particularly valuable when brought to bear one substorm optical onset, PBIs, cusp aurora during northward IMF, trans-polar arcs, and the global auroral response to storms.

Systematic conjugate imaging will be provided by observations of the southern auroral oval as one satellite passes through perigee. In principle, we could use the UVAMC imagers to carry out this task, however their FOVs will be selected to facilitate viewing of the whole oval from above 4 Re. This FOV will be too narrow for optimum conjugate viewing from perigee, so it is desirable to carry a light weight wide FOV imager designed specifically for qualitative global imaging from near perigee. In other words, this imager would not meet the strict wavelength isolation, spatial resolution, and SNR requirements that the FUVAMC would be held to. The WFAI would be based on an imager originally designed for operation on low Earth orbiting spacecraft that incorporates a square pore slumped MCP optic that provides wide field of view imaging [see *Fraser et al.,* 2002; *Hamilton et al.,* 2006]. This instrument has not been flown for this purpose before and is under development at Leicester University under the leadership of Steve Milan and Mark Lester. The instrument is described in some detail in section 4.2.3 of the KuaFu Assessment Study report which is included with this report as Appendix E. The instrument has low technical requirements, demanding only 10 cm X 10 cm X 10 cm volume, 1.5 kg of mass, and 10 W (again, please see the Assessment Study Report).

Thus, our intention is to carry out a program of systematic conjugate imaging on KuaFu-B via the UK Wide Field Auroral Imager, used in conjunction with the UVAMC instrument (ie., the UVAMC will provide the conjugate image in the Northern hemisphere). This is clearly of relevance, then, to the ongoing development of UVAMC.

As pointed out in the Mission Design section, the conjugate imaging would take place only while one satellite passes through perigee (the other will definitely be in the vicinity of apogee at that time, as the orbits are relatively phased so that when KuaFu-B1 is at perigee, KuaFu-B2 will be at apogee and vice versa). Thus any conjugate imaging would occur for only tens of minutes every orbit. This is not a fatal limitation. Referring to the conjugate processes that we would want to study that are discussed in the previous paragraphs, these processes are all effectively transient (onsets, PBIs, etc) or could be studied with snapshots (as in the case of nonconjugate theta aurora). Interesting new science could be accomplished within a time limited conjugate observing period.

Given that conjugate auroral imaging with a wide field of view UV auroral imager was a possibility, in our capacity on the KuaFu mission definition team, we checked that the orbits selected for KuaFu-B support high-quality conjugate imaging for a reasonable fraction of the orbital period. To do this, we used a modified version of the code developed to establish the Ravens orbits. We picked pixels in both the northern and southern hemisphere. For each pixel we carry out the look angle calculation. For each southern hemisphere pixel there is a corresponding northern hemisphere pixel, where that correspondence is determined using the PACE invariant geomagnetic coordinates. In other words, for this calculation, we are assuming that conjugacy is determined entirely by the internal field of the Earth, an assumption that is valid for simply estimating whether or not the orbits sustain conjugate viewing, especially at auroral latitudes.

The quality of conjugate viewing for that pair of pixels *is determined by the poorest quality view of the two pixels*. In other words, if one is seen well, and the other poorly, the conjugate viewing is poor. In Figure 4.5, we show an example output of this calculation. The map at top left is the quality of view for the northern hemisphere only, while the map at bottom left is the conjugate quality of view (in this case excellent). The plot at top right shows direct (around 60) and conjugate (spikes at bottom) quality of view for four successive orbital periods. The red vertical line indicates the specific time that corresponds to the plots at left. The plot at bottom shows the locations of KuaFu-B1 and -B2 for that same time (in this case, excellent conjugate viewing is of course consistent with one satellite near apogee and the other near perigee). We also limited our attention to pixels above 60 degrees PACE latitude, with the idea that conjugate imaging will not be restricted to storm time, and in fact will likely be most valuable scientifically during non-storm time.

The calculation illustrated in Figure 4.5 assumes a 25° field of view UVAMC (ie., the Canadian electron auroral imager) imaging the northern hemisphere and a 40° field of view wide field UV camera imaging the southern hemisphere. The orbits are conjugate in geographic (GEI) coordinates, however the auroral conjugacy is ordered by magnetic coordinates. The eccentricity of the dipole terms in the internal field shift the conjugate (southern) field of view relative to the direct (northern) field of view. Regardless, our calculations show that we can expect reasonable conjugate viewing (comparable to or better than that shown above), for roughly 8% of the orbit. Another way of looking at this is that we will have good quality conjugate imaging for around 30 minutes twice per orbital period.

An important issue that needs to be addressed by the UVAMC and WFAI teams is the compatibility of UVAMC images with WFAI images for the purpose of conjugate imaging. In other words, conjugate images would be most useful if both imagers have identical properties. In flight, we will be able to inter-calibrate the two by simultaneous imaging on the southern aurora within the narrower field of view of UVAMC. In advance, we should at least consider the spectral response of WFAI in comparison to that of both UVAMC channels, and consider any implications of those differences in terms of conjugate imaging.

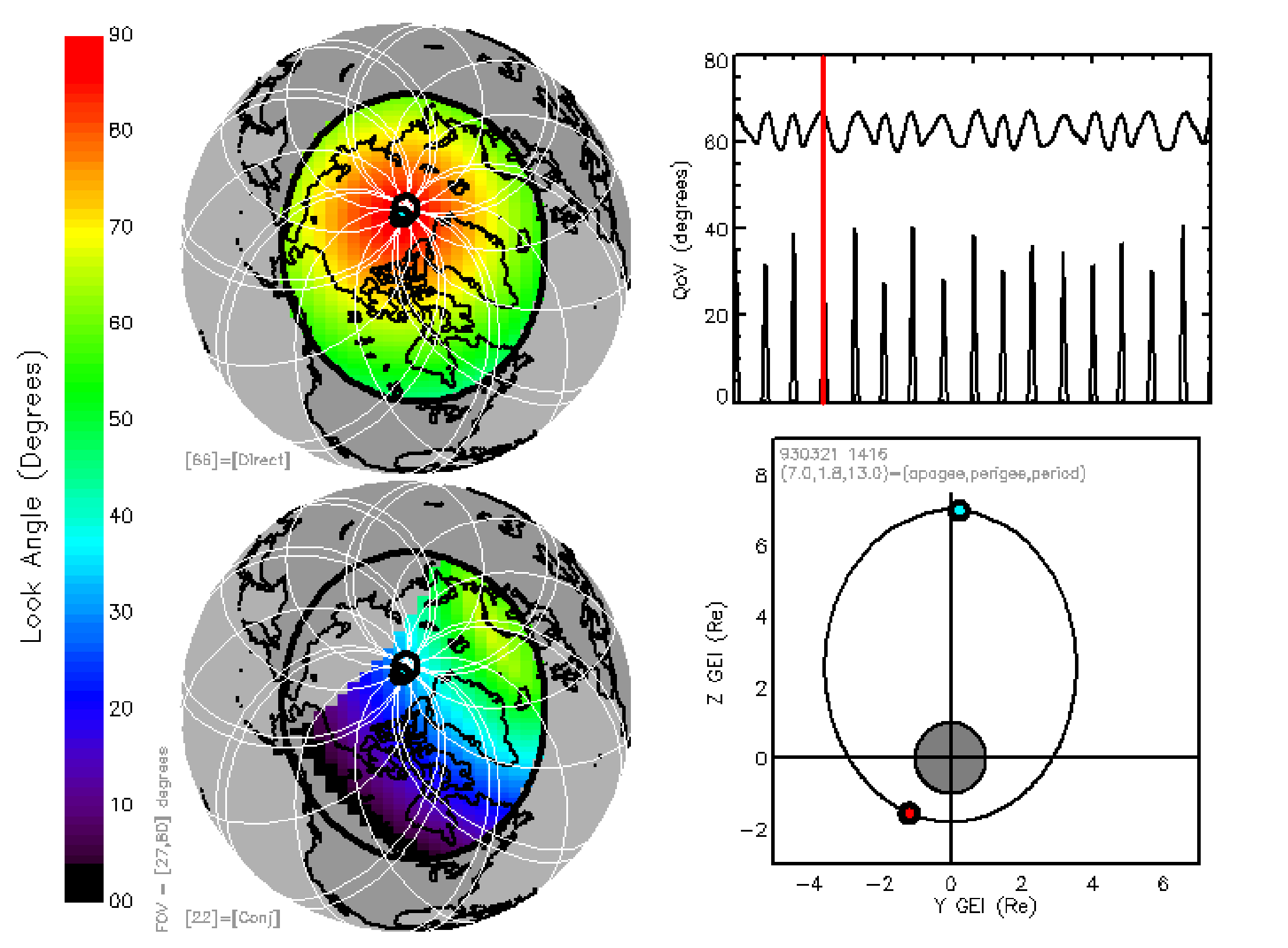

Figure 4.5: Quality of direct and conjugate viewing for KuaFu-B using the proposed orbital parameters.

## 4.4 NAIK

Energetic Neutral Atom (ENA) imaging is a novel technique to remotely image space plasma. *In particular, as we discuss in this section, the motivation for including ENA imaging is to track the spatio-temporal evolution of the ring current ion population.*

ENAs are produced by charge exchange between a singly charged ion and a neutral atom or molecule. In the terrestrial ring current and plasmasheet, singly charged ions are trapped in the magnetic field and charge exchange with the neutral atoms of the Earth's exosphere and upper atmosphere. Within the ring current energy range (>1 keV) the resulting ENAs travel in straight trajectories, unaffected by the electric, magnetic and gravitational fields. The energy loss of the charge-exchange reaction is on the order of a few 10's eV, so that the energy spectral information of the plasma is conserved. In addition, the mass spectral information of the plasma is also maintained. At Earth, ENA imaging has been applied to the ring current, plasmasheet, low-energy ion outflow and the solar-wind interaction with the magnetosphere [Roelof et al., 1987; Brandt et al. 2002; McComas et al. 2002; Collier et al., 2005].

ENA imaging has recently also been proven to be an invaluable tool for the large-scale dynamics of the magnetospheres of Jupiter and Saturn (see Figure 1) [Mitchell et al., 2004; Krimigis et al., 2005], whose rotational speeds are of such magnitude that the plasma of the inner magnetosphere is corotation dominated. One of the non-magnetized moons of Saturn – Titan – sits in this corotational flow and displays a complex interaction between the hot, ambient magnetospheric plasma and the dense upper atmosphere that have been revealed by ENA imaging from the INCA instrument on board the Cassini spacecraft [Mitchell et al., 2005]. The interaction between Titan and the magnetospheric plasma of Saturn, resembles, in many ways, the interaction between Mars and the solar wind. At Mars, the solar wind interacts directly with the upper atmosphere, resulting in the escape of atmospheric ions, atoms, and molecules [Lundin et al., 2004]. A significant fraction of the atmospheric escape is carried by energetic neutrals. The determination of the global escape rates from ENA measurements on board Mars Express is an ongoing effort that holds great potential [Brinkfeldt et al., 2005; Futaana et al., 2005; Gunell et al., 2005].

While the discovery of the fundamental large-scale behavior dominates the objectives of the extra-terrestrial measurements, terrestrial ENA imaging is moving towards quantitative understanding and specification of the inner magnetosphere. At Earth, the dominating fraction of the plasma pressure of the inner magnetosphere consists of ions contained in the ring current energy range (1-300 keV), which is targeted by ENA imaging. This is important since the plasma pressure distribution in this energy range is believed to drive most of the electrical currents in the inner magnetosphere that close through the ionosphere. This magnetosphere-ionosphere coupling is the source of a range of phenomena that only with the recent advances of imaging missions have been explained on a global scale [Brandt et al., 2005; DeMajistre et al., 2005, Goldstein et al., 2005a,b].

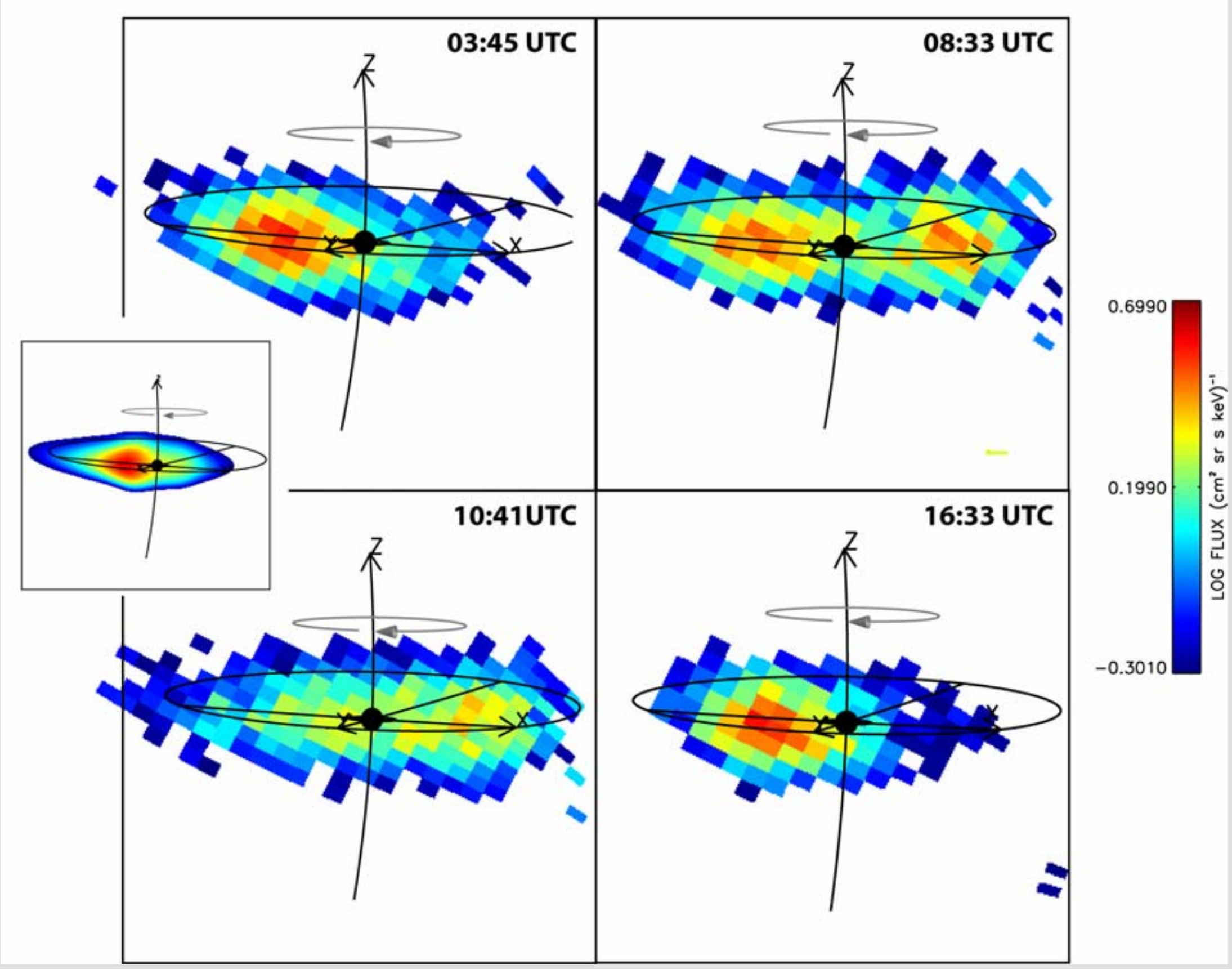

Figure 4.6: Hydrogen ENA images of the inner magnetosphere of Saturn obtained by INCA in the 20-50 keV range. The X-axis points towards the Sun, Z-axis is parallel to the rotation axis of Saturn, and the Y-axis completes the right hand system. The rotation direction of the observed ion cloud is indicated by the dashed curved arrow. The outer ring marks the radial distance of 20 $R_S$ and the inner ring (barely visible) marks the extent of the visible rings. The inset shows simulated ENA image of the first ENA image at 03:45 UTC. The required energetic proton distribution extends beyond the orbit of Titan.

Therefore, one of the most important scientific objectives of terrestrial ENA imaging is the specification of the global pressure distribution of the inner magnetosphere. An inversion technique has been developed to retrieve the ion intensities from ENA images [DeMajistre et al., 2004]. It is based on expanding the ENA line-of-sight (LOS) integral into a linear equation system that then can be solved using constrained linear inversion methods. A similar technique has also been developed using expansions into spline functions [Perez et al., 2004]. Figure 2 shows a validation of the result from the constrained linear inversion technique [DeMajistre et al., 2004], against in-situ measurements from Cluster.

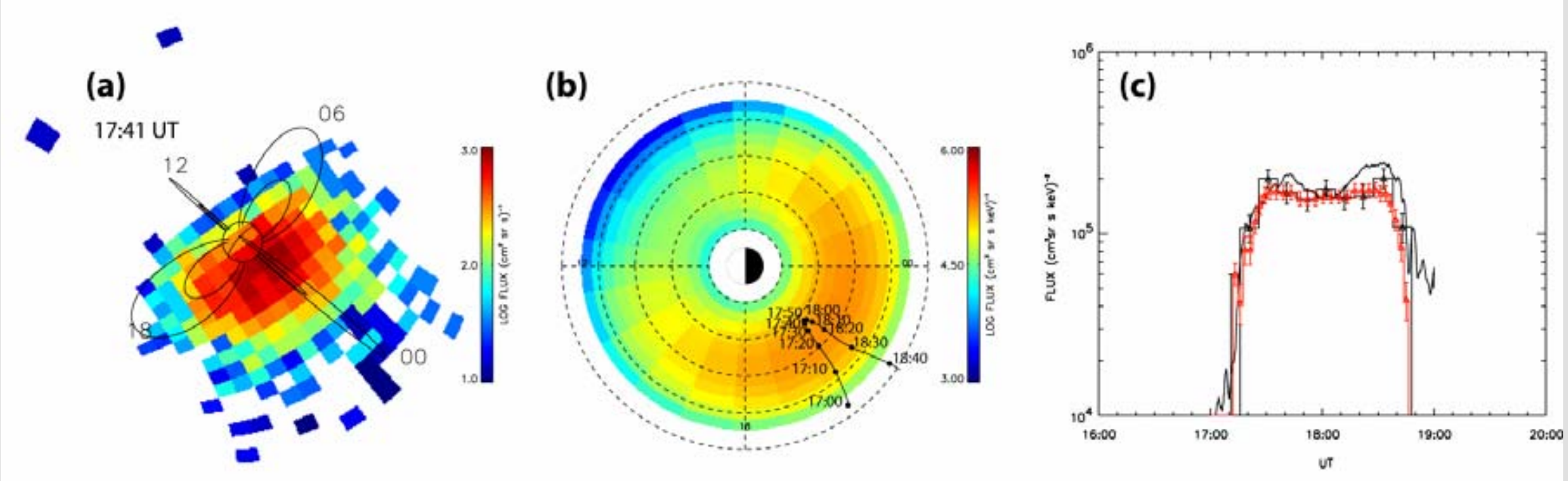

Figure 4.7: Hydrogen HENA image in the 27-39 keV range during the the 20 April 2002 storm. (b) Resulting equatorial proton distribution (perpendicular pitch-angle component) from an inversion of the HENA image. The orbit track of Cluster-4 is shown in black. (c) The perpendicular pitch-angle component from the HENA inversion and the Cluster/CIS/CODIF data. The proton distribution at each Cluster position has been mapped to the SM equator using the T01-S model, where it is compared with the corresponding proton flux obtained from the inversion. Black line represents the CODIF proton flux in the 27-39 keV range and the red line is the corresponding proton flux from the inversion.

Given the global pressure distribution it is possible to derive the associated current density ***J*** by assuming current continuity and using the force balance equation (***JxB=del P***, neglecting the inertial term in the inner magnetosphere). The current density is conveniently computed by using an Euler potential formalism for ***J*** itself [Roelof, 1989]. Figure 3 shows a part of the resulting 3D current system.

The intensity of high-energy (>MeV) ions and electrons in the radiation belt experience a dramatic decrease during mainphases of magnetic storms. During the recovery phase the intensities recover to a level that can be higher, lower or the same as the pre-storm value. Theoretical modeling shows that this effect is due to the increase of the ring current pressure that severely distorts the magnetic field of the inner magnetosphere so that the high-energy drift trajectories moves outward. While some modeling suggests that this is a fully adiabatic process, modeling taking into account the morphology of the ring current and the fluctuations in solar-wind pressure show that this process can be highly stochastic and result in a significant loss of the radiation belts through the magnetopause. Continuous specification of the ring current pressure during all phases of the storm is required to determine how the radiation belts are lost and energized. ENA imaging is so far the only technique capable of meeting these requirements.

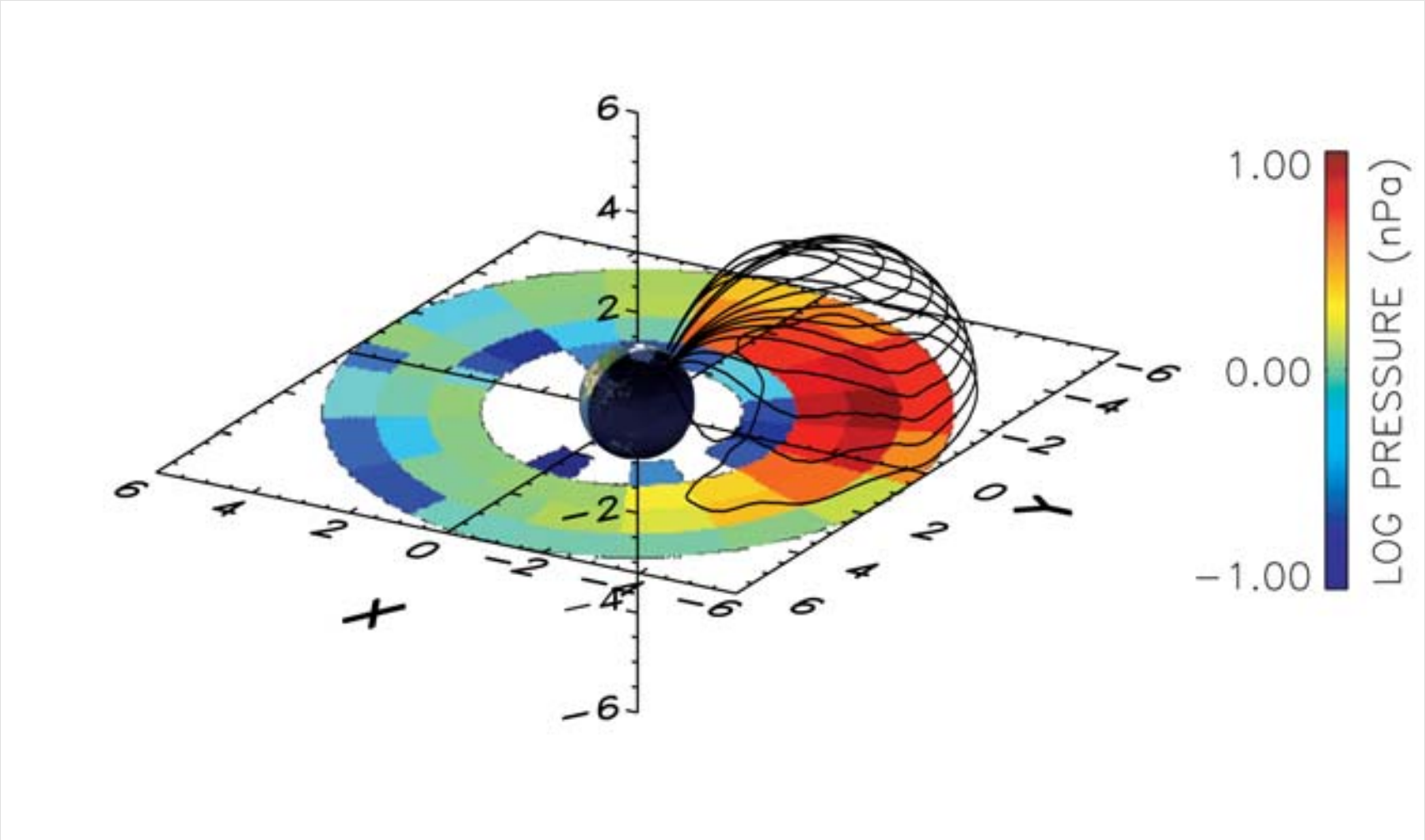

Figure 4.8: The global proton pressure derived from HENA/IMAGE observations (energy range 10-200 keV) by applying a constrained linear inversion technique to the ENA image. Isotropy and a dipolar magnetic field are assumed. The black lines represent the electrical current flow lines computed on one iso-pressure contour of the derived pressure. These currents are the core part of the Region 2 current systems that play the major role in the MI-coupling process.

One of the more severe limitations of existing terrestrial missions carrying ENA instrumentation is their lack of continuous coverage of the ring current through a complete storm. *By launching two, or more, satellites into suitable orbits this insufficiency can be overcome. In addition, multiple vantage points enable more accurate retrievals of the 3D ring current distribution.* For KuaFu-B, global perspective will be also needed for other fundamental aspects of the magnetospheric physics, for example, particle origins, loss processes, transport mechanisms and their time dependence. These are all embedded in the two of the most important dynamic features of the magnetosphere are geomagnetic storms and substorms (Alexeev, 2003). Although numerous single-point in-situ measurements are made and result in significant achievements in the comprehension of both events, global perspectives are still needed to change our understanding from the picture in a statistical average sense (space climatology) to an instantaneous view of large-spatial-scale magnetospheric dynamics (space weather). In-situ magnetic field observations with high spatial and temporal resolutions will describe the real-time magnetic environment. The ring current as an indicator of magnetic storm(s) will be remote sensed by a neutral atom imager onboard each of KuaFu-B1 and KuaFu-B2.

The Neutral Atom Imager on KuaFu (NAIK) instrument may, conveniently, be described in terms first of its Sensor Head and then of its Electronics Box, similar to NAUDU onboard Double Star. In fact, NAUDU is the heritage instrument for NAIK. The fact that Double Star P1 and P2 are not spinning and KuaFu-B1 and B2 will be means that some redesign is essential. The cylindrical Sensor Head protrudes beyond the skin of the spacecraft so as to provide a free field of view for each of its sensors. The Sensor Head contains 16 solid-state detectors, an electrostatic defector/collimator system and a circular front-end electronics board on which is mounted a set of charge-sensitive preamplifiers and calibrated pulse amplifiers..

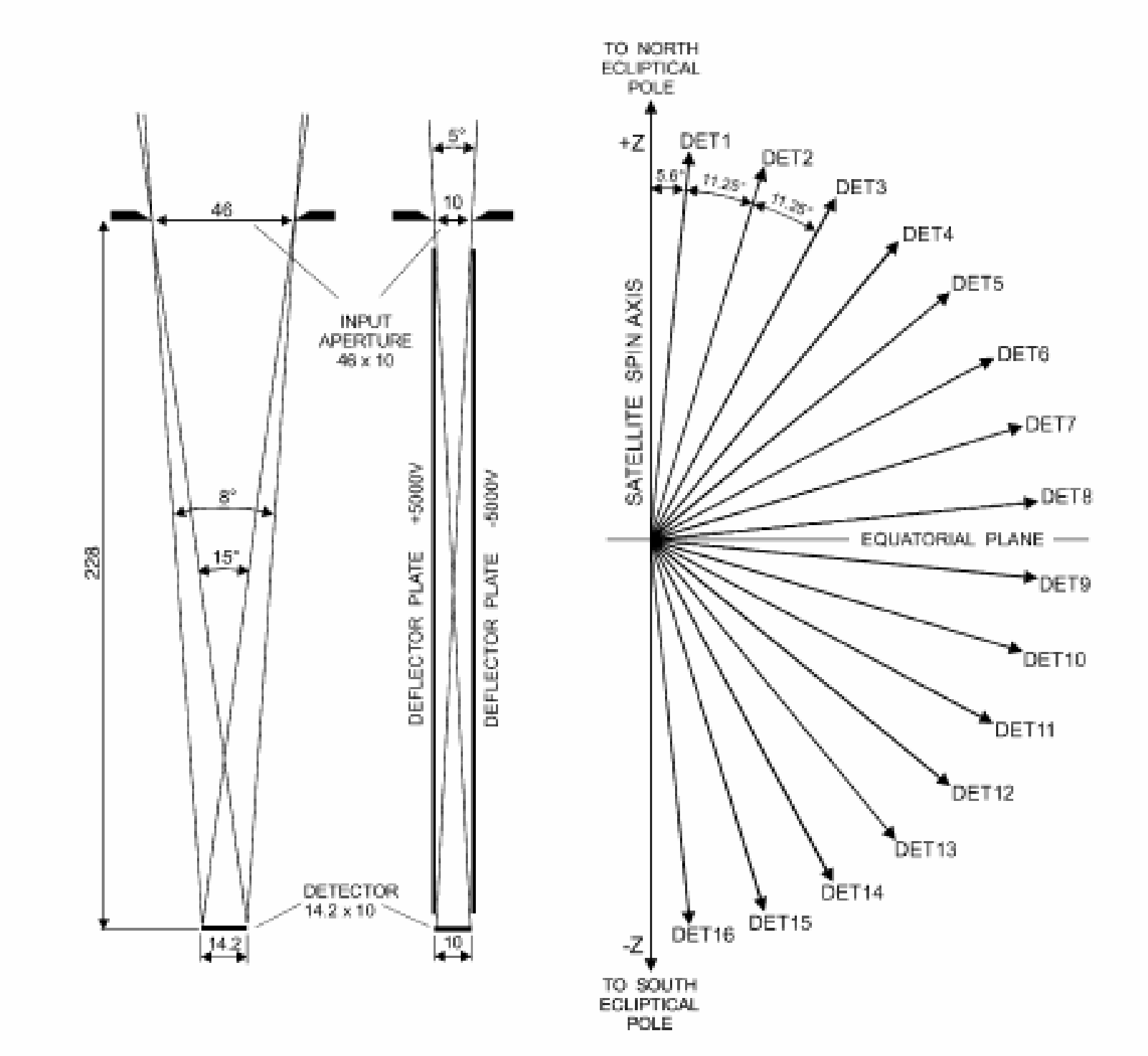

Figure (Left) Definition of the field of view of a single detector; (Right) pointing directions of the 16 detectors.

Sixteen custom designed Passivated Ion Implanted Silicon (PIPS) detectors are mounted at four different heights in the Sensor Head. The sensors have, individually, equal fields of view $11.5^0 \times 2.5^0$ (fwhm), regularly distributed over an $180^0$ angle in the elevation plane (Fig. 3). An important feature of the sensor system is its large geometric factor (0.0125 $cm^2$ sr per $11.5^0 \times 2.5^0$ pixel). Since mechanical collimation is employed, the pixel size does not depend on particle energy and mass. Because of the deposition on each detector of a metalization layer (200nm of Al), the Sensor Head is impervious to Lyman-alpha radiation from the geocorona. The spacecraft spin allows the azimuthal plane to be divided into 128 equal sectors through counting the pulses of the onboard clock. The full 4π solid angle is divided into 16*128=2048 pixels and, given the spacecraft spin period (Ts=4s), the integration time for one pixel is 4s/128=31.25 ms. NAIK can record a complete 4pi image on the completion of each spacecraft spin, in synchronisation with the spacecraft's Sun Reference Pulse. The electrostatic detection system is designed to prevent charged particles with energies >300 keV from entering the detectors. The basic specifications of NAIK are provided in the table, below.

Table 4.2: Main Parameters of NAIK

| | |
|---|---|
| Weight | 6 kg |
| Maximum dimension (LWH) | 290 mm*239 mm*239 mm |
| Power consumption | 5W (from 28V direct current) |
| Number of detectors | 16 |
| Detector type | PIPS , light-tight, custom design |
| Single detector active area | 142 mm2 (14.2 mm 10mm) |
| Discrimination levels | 20 keV, 50 keV, 100 keV, 300 keV |
| Field of view (each detector) | $11.5^0 \times 2.5^0$ (fwhm, elev. × azimuth) |
| Geometrical factor (each detector) | 0.0126 $cm^2$ sr |
| Resolution over 4π solid angle | 2048 pixels |
| Image synchronization | by Sun reference pulse |
| Continuous data rate | 0.5 kbps - 16 kbps |
| Charged particle cutoff energy | 300 keV |
| Deflection high voltage | 10 kV |
| The optimum operating condition range | $-20^0$C-$0^0$C |

The main scientific objective ENA imaging of the Earth's ring current in the energy range 45-300keV. The data obtained will provide insights into the dynamic of the ring current responsible for stimulating disturbed magnetospheric condition that underline many aspects of space weather. NAIK will fly on KuaFu-B, at the same time as a number of other missions (RBSPs, THEMIS, and possibly ORBITALS and ERG) in equatorial or near-equatorial orbits that will provide multiple transits of the ring current region, and valuable *in situ* measurements of radial profiles of relevant particle fluxes. These profiles can be used with the NAIK observations and models to constrain ring current particle distribution inferred from NAIK. Furthermore, the NAIK observations will be available often from both satellites simultaneously, allowing for stereoscopic viewing of the ring current <300 keV ion population. For these reasons, and the two satellite TWINS-like configuration of the KuaFu-B pair, the multi-point ENA sampling obtained through combining with NAIK data made aboard other spacecrafts can provide the potential for hitherto unprecedented, integrated studies of global, dynamic, magnetospheric process.

There is at least one concern with NAIK that the UVAMC team would like addressed in the design phase of the mission. NAIK data will be used together with the UVAMC and FUVSI data to specify the global time-evolving ring current population and the ion and electron CPSs (at least as projected in the ionosphere via diffuse auroral precipitation). It is now very well understood that ion composition in the inner magnetosphere changes significantly during strongly varying geomagnetic conditions [see for example, *Hamilton et al.,* 1988; *Nose et al.,* 2003, and references therein]. The current design of NAIK cannot resolve mass and hence species. We encourage the NAIK team to explore options for adding this capacity. As well, during Phase A, the UVAMC, FUVSI, and NAIK science teams should carry out a study establishing what limits a lack of ENA mass resolution would place on our science objectives. Please note that we do not regard this as a fatal flaw, it is just something we want to explore given the importance of changes in ion composition during storms.

## 4.5 Conclusion

The Ravens Integrated Imaging Package began with UVAMC and FUVSI. With incorporation into KuaFu this has expanded to include the WFAI and NAIK. In conjunction with ground-based networks of magnetometers (from which magnetoseismology can give information about the time evolving plasmasphere) and riometers (which provide a nearly global proxy for X-Ray imaging and hence higher energy electron precipitation), the imaging instruments will allow us to continuously track the global ring current (NAIK), high energy electron precipitation (riometers), ionospheric projection of the electron and proton CPSs (UVAMC and FUVSI), the global distribution of discrete aurora (UVAMC), and the plasmasphere (magnetoseismology). The WFAI and UVAMC together provide the opportunity for systematic conjugate auroral imaging, albeit for limited periods (~30 minutes) during each orbit. This is a powerful combination that will facilitate cutting edge system-level geospace science, and will provide an equally powerful complement to virtually all other ground and satellite based geospace observations.

The UVAMC is described in depth in the following section. Phase A considerations that pertain specifically to UVAMC are derived from that section. The implications related to other instruments of the material presented in this section for the proposed Canadian UVAMC and KuaFu Phase A study are as follows:

- establishment of whether the FUVSI should dedicate a channel to 1356 Å imaging.
- determination of temporal requirements for global proton auroral imaging with FUVSI.
- determination if Canadian industry can play a role in providing and/or acquiring a detector for FUVSI.
- reiteration of the viability of FUVSI imaging on a satellite with a 30 second spin rate and exploration of consequences for UVAMC if a longer spin period is required.
- initial investigation of use of UVAMC and WFAI images for conjugate observations.
- exploration of the implications of a lack of mass resolution by NAIK on the overall science objectives of the imaging component of KuaFu-B.

These issues will be dealt with by the relevant teams in conjunction with the UVAMC team (if we go forward with a Phase A study).

# 5. UVAMC – In Depth

*Ravens* is a planned multi-satellite mission aimed at providing continuous imaging of the global auroral oval in the northern hemisphere. In turn this continuous global imaging will provide crucial information on the response of the auroral oval during magnetic storms which tend to have time scales of days and represent the major large scale manifestation of space weather events in the ionosphere. Ravens would consist of two satellites on identical polar elliptical orbits with an inclination of ~90 degrees, and relatively phased on that orbit so that when one is at apogee the other is at perigee. The Ravens Mission will provide the first-ever 24 hour per day 7 day per week (24/7) global auroral electron (FUV) and proton (Doppler shifted Lyman-alpha) imaging. **FUV measurements will be provided by the University of Calgary Ultra-Violet Auroral Monitoring Cameras (UVAMC)** and Lyman Alpha measurements will be provided by the University of Liege's FUV Imaging Spectrograph (UVSI). The focus of this chapter is on the UVAMC instrument.

The heritage for the UVAMC imagers are the past University of Calgary FUV instruments (Viking, Freja, Interball, and IMAGE-WIC); however, the design of this new instrument (essentially a "next generation" Canadian FUV imager) will need to be significantly different in order to meet the much more demanding scientific objectives.

## *5.1 Brief History of Auroral Imaging – Setting the Stage*

Auroral imaging from space has and continues to be our only true means of studying the spatio-temporal evolution of the magnetospheric system. In the very early days, global auroral images proved in an instant the existence of both the continuous auroral oval and the diffuse aurora. Global electron auroral images have been used to map the evolution of the substorm expansion, the effects of solar wind pressure pulses on the central plasma sheet, and in some cases even the steady-state nature of dayside reconnection. Because of global auroral imaging, we are able to study the complex interplay between convection, conductivity, and field-aligned currents. Whenever they are available, global auroral images are used for essential context for other studies.

Global auroral imaging began in the early 1970s with the Auroral Scanning Photometer (ASP) on ISIS2 satellite (launched April 1, 1971). ASP used the satellite spin and its orbital motion to provide the very first global auroral images (obtained in both 557.7 nm and 130.4 nm once per orbit). The ISIS2 system was operated for almost 10 years and yielded a number of discoveries, including the diffuse auroral oval [Lui et al., J. Geophys. Res., volume 80, page 1795, 1975].

Since ISIS2, global auroral imaging has evolved significantly. Imagers on Kyokko, DE 1, HILAT, and Polar Bear operated in the UV, allowing for the first time imaging of the aurora on the dayside. Global imagers on Dynamics explorer and Viking allowed more than one image per pass, with the Viking UV instrument providing an impressive cadence of one image every 20 seconds. In fact, the Viking instrument utilized a combination of UV filters, UV-reflective coatings, and a UV-sensitive image intensifier fiber-optically coupled to a 256 by 256 pixel CCD to obtain short-exposure (1 second) simultaneous global UV images in the LBH-short band and the LBH plus an extension into the part of the spectrum including the 1304 Å OI line. Viking provided an exciting new picture of the global time-evolving auroral distribution and new insights into the substorm dynamic [see *Henderson et al.*, *Geophys. Res. Lett.,* volume 25, pages 3737-3740, 1998]. Images from the two filters could be used to estimate energy deposition and the mean electron energy, however resonant scattering and a requirement of knowing the O/N2 ratio seriously limited the accuracy of these derived quantities.

The next significant leap forward in global imaging came with the ISTP Polar satellite, which carried two imaging packages. The Visible Imaging System [VIS; see *Frank et al., Space Sci. Rev.,* volume 71, pages 297-328, 1995] consists of two cameras for nightside auroral imaging in the visible wavelengths. The two cameras deliver images with different spatial resolutions (~10 and ~20 km at apogee), and have produced images that have been widely used in studies of magnetospheric dynamics [see, eg., *McWilliams et al., Annales Geophysicae,* volume 19, pages 707-721, 2001; *Nakamura et al., Annales Geophysicae,* volume 17, pages 1602-1610, 1999]. In addition to VIS, Polar carries a UV imaging package (UVI for Ultraviolet Auroral Imager), which provides images of the entire auroral oval for roughly half of the 18 hour Polar orbit. Polar UVI has also been widely used in studies of global magnetospheric dynamics [see, for example, *Sergeev et al., Geophys. Res. Lett.,* volume 26, pages 417-420, 1999]. The nominal UVI spatial resolution is 50 km at apogee, with an integration time of ~30 seconds and sensitivity of 50 R. This sensitivity is still the best ever achieved with a global imager.

In 2000, NASA launched the IMAGE satellite. This was the first space science satellite with an instrument complement entirely devoted to remote sensing via optical and radio means. IMAGE carried with it a global imaging "FUV" package consisting of three instruments. One of the FUV instruments onboard IMAGE is the Spectrographic Imager, one channel of which (the SI-12 channel) provides global proton auroral images every two minutes (the satellite spin period).

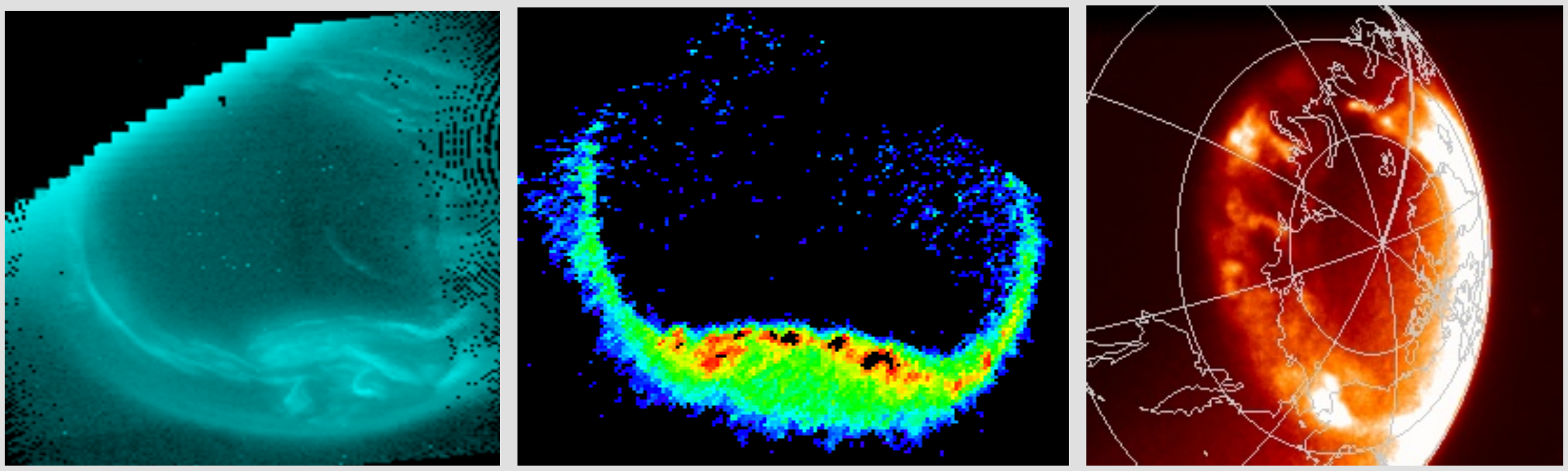

Figure 5.1: (left) ISIS II image of the global aurora obtained in 1971 (once per orbit); middle) Viking image from April 1, 1986 (once every 20 seconds); right) IMAGE FUV WIC.

Brief summary of operating parameters of previous global auroral imaging experiments

| Platform (instrument) | Year of launch | Orbit | Wavelengths | Exp. Time | Frame rate | Spatial resolution |
|---|---|---|---|---|---|---|
| ISIS 2 | 1971 | 1400 km | 557.7 nm, 391.4 nm | 15 min | 2 h | 10 km |
| ISIS 2 (redline) | 1971 | 1400 km | 630 nm | 15 min | 2 h | 70 km |
| DMSP | 1971 | 850 km | 400-1100 nm | 15 min | 1.5 h | 5 km |
| KYOKKO | 1976 | 650-4000 km | 120-140 nm | 12 s | 2 min | 4-21 km |
| DE-1 | 1981 | 675-25,000 km | 12 narrow bands from 130 to 630 nm | 12 min | 12 min | 24-130 km |
| HILAT | 1983 | 830 km | 115-200 nm with 2 nm resolution | 25 min | 100 min | 20 x 4 km |
| Viking | 1986 | 800-13,500 km | 125-160, 135-190 nm | 1 s | 20 s | 14-20 km (1 x 1), 27-40 km (2 x 2) |
| Polar bear | 1986 | 800 km | 115-200 nm with 2 nm resolution | | | |
| EXOS-D | 1989 | 300-8000 km | 115-150 nm & 557.7 | 1 s | 8 s | 1 km |
| Freja | 1992 | 650-1700 km | 125-180 nm | 0.4 s | 6 s | 5 km (2 x 2) |
| Interball | 1996 | 475-20,000 km | 125-160 nm | 6.5 s | 120 s | 27 km (1 x 1) 55 km (2 x 2) |
| Polar (VIS) | 1996 | 5100-51,000 km | 732.0, 656.3, 630.0, 589.0, 557.7, 391.4, 308.5 nm | s | 54 s | 20 km |
| Polar (VIS Earth Camera) | 1996 | 5100-51,000 km | 125-150 nm | s | 54 s | 80 km |
| Polar (UVI) | 1996 | 5100-51,000 km | 128-133, 132-139, 140-158, and 166-174 nm | s | 38.6 s | 40 km |
| IMAGE (WIC) | 2000 | 1000-44,500 km | 140-190 nm | 10 s | 120 s | 100 km |
| IMAGE (SI-12) | 2000 | 1000-44,500 km | 121.8 nm | 5s | 120 s | 100 km |
| IMAGE (SI-13) | 2000 | 1000-44,500 km | 121.6, 135.6 nm | 3 Hz | 120 s | 100 km |

Note that we have simplified the information in some cases for clarity. The table is ordered chronologically, from top to bottom. The imagers on Polar are still operating.

Table 5.1: properties of all global imagers flown previously (from Donovan et al. ASR paper included in Appendix B).

## *5.2 Objectives Related Specifically to UVAMC*

By the time IMAGE was flown, there had been great advances, and great achievements, in global auroral imaging. Still, after 35 years of global imaging from space, there is a surprising list of technical challenges that have either not been met, or have been done to a degree that could be improved upon. Surpassing previous accomplishments or striking out in new ways would facilitate new science. Two examples of technological improvements that would facilitate exciting new science include better spatial resolution, and better wavelength isolation.

Better spatial resolution in global images would facilitate scientific advancement in the study of cross-scale coupling and multi-scale processes. The aurora shows significant structure on scales ranging from global (ie., thousands of kilometers) down to tens of meters and even smaller. Some of this structure is a manifestation of the global structure of the magnetosphere and its significant plasma regions such as the Central Plasma Sheet (CPS), lobe, and cusp. As well, structure also arises as a consequence of electrodynamics involved in auroral acceleration. It is clear that the scales are related, as are the processes, but it is not clear how. Furthermore, temporal evolution of features of various scale sizes arises as a consequence of magnetospheric dynamics and the auroral electrodynamics. An imaging program to systematically explore the spatio-temporal structure and evolution of the aurora would allow for exciting studies of multi-scale processes, cross-scale coupling, and even the emerging field of natural complexity. As illustrated in the right hand panel of the figure below, such a program would involve ground-based observations of the smallest scales (tens of meters to tens of kilometers) and fastest processes (< 30 second time scales), and would not be possible without global auroral images with better spatial resolution than previously obtained.

Isolation of parts of the auroral spectrum is an absolute necessity for the ultimate quantification of the flux and characteristic energy of auroral electrons, which are in turn essential to our ability to truly understand magnetosphere-ionosphere coupling, and magnetospheric dynamics. This is discussed below (separation of LBH-L and LBH-S), but here we state that this requires implementation of new technologies, and it would be advantageous for the contribution to the LBH (L and S) from secondary electrons due primary proton precipitation (also see below).

The instrument-specific objectives for UVAMC are to achieve appropriate spatial resolution and suppression of out of band contributions to two isolated parts of the LBH spectrum (LBH-L and LBH-S) on a satellite in the Ravens 1.8X7 Re orbit with a spin period of 30 seconds.

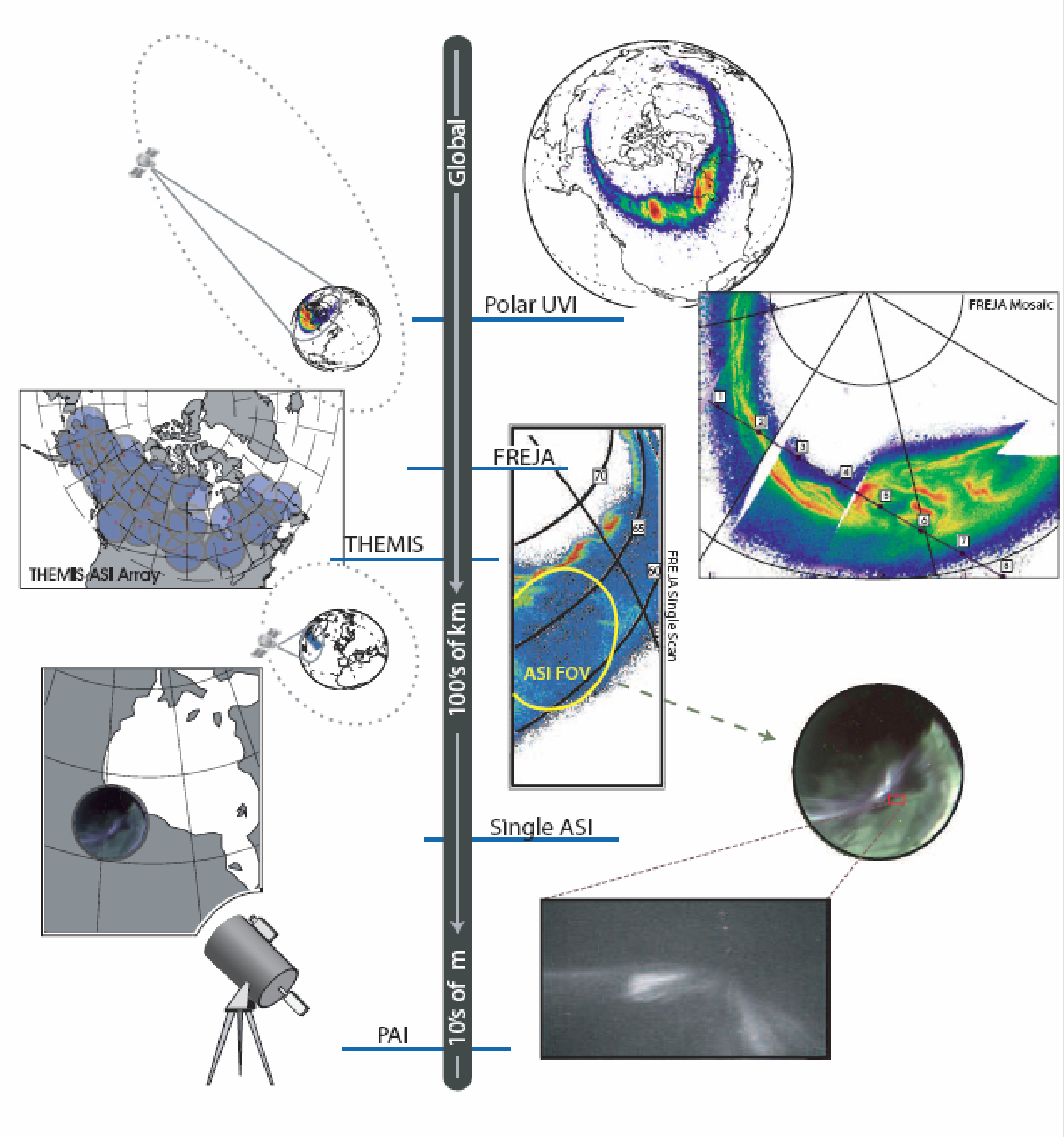

Figure 5.2: Cross-scale coupling, multi-scale processes, and natural complexity are evident in the aurora. Structure arises spontaneously in this system, and understanding how that structure arises is of fundamental importance to future progress in space physics. Observations to support theoretical studies of the distribution of scales and its origins must involve global imaging from space and simultaneous meso- and small-scale imaging from the ground. Ideally, networks of ground instruments would provide sub-second observations of tens of meters scales, and several second observations of kilometer to tens of kilometer scales. The ground-based observations would mesh together with global imaging with ~30 kilometer resolution and 30 second cadence (from Ravens/KuaFu). The effectiveness of this strategy is contingent that the assumption that small-scale fast processes are not organized on global scales. (a reasonable assumption though difficult to substantiate with current data).

### 5.3 Subtraction of contribution to LBH-S/L due to secondary electrons

Proton precipitation into ionosphere induces interactions with the main constituents of the thermosphere ($O_2$, $N_2$ and O) [Gérard et al., 2000, Hubert et al., 2001]. By collision, the incident proton may capture an electron from the main constituent, leaving a fast hydrogen atom, possibly excited in the upper state of the Lyman-alpha transition. As the hydrogen atoms move away from the observer (the spacecraft), a photon is emitted with a Doppler shift to the red of the unshifted line center. The line profile of the emitted radiation that contributes to the signal observed by (for example) the IMAGE SI12 camera is the result of the integration of the contributions of all velocity vectors projected on the line of sight. The Doppler shift creates a profile that varies (slightly) as a function of look angle. That is the Doppler shifts are on average smaller at nadir than they are away from Nadir, but the effect is not very significant. Further, the Doppler shift obviously depends fairly significantly on the characteristic energy as well as the shape of the energy spectrum. Much of this has been worked out to help with initial interpretation of the IMAGE FUV SI12 channel data, and has been presented by Gérard et al. [2001]. We include their Figure 1 below (with permission of the author who is a co-investigator on Ravens), which shows the overlap of a model Doppler shifted emission profile with the SI12 spectral response. We come back to this later when we discuss the requirements for both the UVAMC and the UVSI on Ravens.

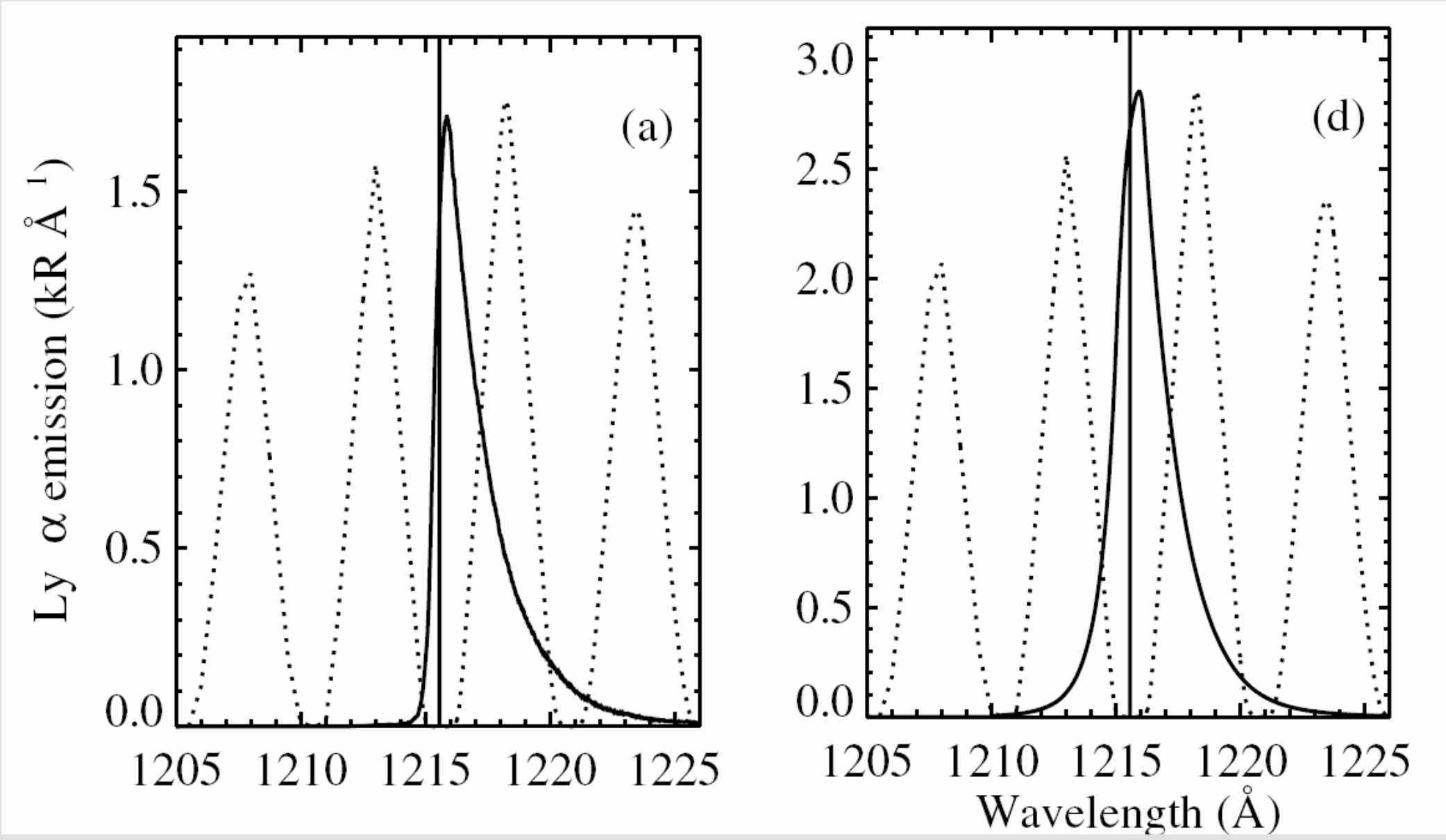

Figure 5.3: Reproduced from Gérard et al. [2001]. Left panel (labeled "a") and right panel (labeled "d") show profiles of Doppler shifted Lyman-alpha proton auroral emission profiles (solid curves), with a precipitating 2 KeV Kappa (K=3.5; 1 erg $cm^2$ $s^{-1}$) precipitating proton spectrum, for nadir and θ=60° (ie., away from nadir). Note that the emission profile for 60° away from nadir is broader, has more blue-shifted contribution, and the emission intensity is brighter by about 50%. The dotted curves are the SI12 response, which is specifically designed to reject the cold (ie., non-Doppler shifted) Lyman-alpha "geocorona".

In addition to direct proton auroral emissions, collisions between incident protons and atmospheric constituents may also induce the ionization of the target, producing secondary electrons which interact with the upper atmosphere constituents in the same way than electron flux coming from magnetospheric sources. This produces what is referred to as “secondary electron aurora”. The standard 557 nm Oxygen green-line is comprised of emissions caused by precipitating electrons (the primary signal) and a secondary signal caused by proton precipitation (a rule of thumb which is obtained by analysis of CANOPUS MSP data is that the secondary signal in 557 nm is ~10 times the brightness of the 486 nm Balmer “H-beta” proton aurora. Thus, research that involves quantitative analysis of, for example, the 557 nm electron aurora requires quantitative knowledge of the contribution to that signal by the precipitating protons. Ideally, one infers information about the precipitating electrons *after* subtracting the contribution due to the protons which would in turn be estimated from knowledge of the precipitating proton spectrum and appropriate transport codes. In reality, at least for the ground-based observations, most studies do not take account of this, or at best apply something like the rule of thumb mentioned above.

The same constraints apply to inferring information about the precipitating electron spectrum from global auroral observations. In LBH, which has been imaged from above by Viking, Freja, InterBall, and IMAGE, and that we hope will be imaged by UVAMC instruments on KuaFu B1 & B2, secondary electrons produce auroral emissions in the LBH spectrum. These emissions will contribute to the signals recorded by the LBHS and LBHL cameras. This excess contribution has to be estimated in some way, and then subtracted before the LBHL and LBHS signal can be used to provide quantitative information about the electron aurora. This effect can be quite significant. Statistical studies of ground-based photometer data [see e.g. Creutzberg et. al., 1988] and straightforward consideration of proton motion in the CPS (which leads to more energetic protons in the evening sector CPS) mean that the proton contribution to the overall precipitating energy often dominates in the lower latitude evening sector aurora. Consequently, in the evening sector auroral oval the emissions due to secondary electrons resulting from primary proton precipitation at times will be the dominant contribution to standard electron auroral signals, and it usually represents a significant fraction of those signals. This point has been made numerous times in the literature (although it is often completely ignored). In particular, this point was made in the abstract of Gérard et al. [2000].

Our scientific objectives demand knowledge of the auroral electron and proton precipitation. The Lyman-alpha emissions accurately capture the global proton auroral distribution, with of course the caveat that the current (ie., SI12 on IMAGE) incarnation of the spectroscopic imager for proton auroral observations cannot discern anything about the characteristic energy of the precipitating protons. The electron auroral emissions are definitely contaminated by signal from secondary electrons due to proton precipitation. That is, the proton auroral observations provide a direct measure of the proton precipitation, although the information that can be inferred is somewhat limited. On the other hand, the electron auroral emissions are in general better imaged (better spatial resolution and with LBHL and LBHS separated some idea about the characteristic energy ad energy flux), but particularly in the evening sector the signal contains significant contribution due to precipitating protons. Based on the above, to infer the "pure" or primary electron signal provided by the proposed LBHS and LBHL cameras, it will be important to be able to be able to estimate and subtract the contamination due to protons. This subtraction can be easily performed via the efficiency curves. These curves provide the count rate of LBHS and LBHL camera in response to a mono-energetic beam of energy flux 1 erg $cm^2$ $s^{-1}$ precipitating protons, as a function of the characteristic (given this is a mono-energetic beam this is just "the energy") energy. Assuming a value of the proton mean energy, the effective proton flux may be deduced from *SI12* data. The combination of the effective proton flux and the count rate for each unitary flux gives the total count rate due to protons for each LBH camera. As shown in the figure below, this method has been already been used for the IMAGE SI12 cameras in conjunction with the WIC imager. It was tested with a combination of data from FAST (ion precipitation), DMSP (ion precipitation), and IMAGE WIC (electron aurora and secondary electron aurora) and SI12 (proton aurora) data.

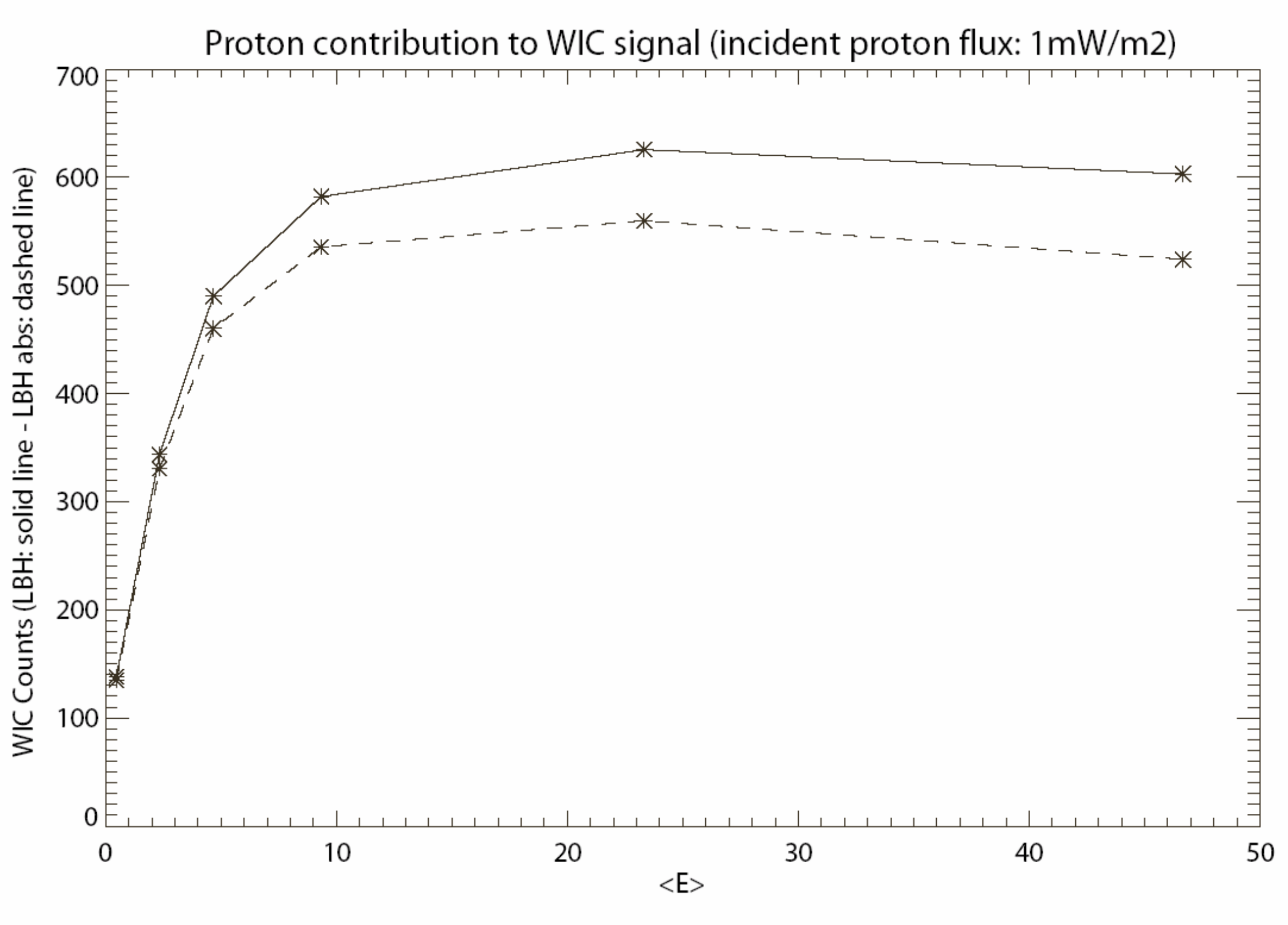

Figure 5.4: Given an incident flux (1 erg $cm^2$ $s^{-1}$) of precipitating protons, secondary electrons will be produced that in turn produce electron aurora. These curves estimate the contribution to the WIC signal due to those secondary electrons (the dashed curve is the estimate with some effort made to account for absorption of emitted photons by atmospheric constituents above where the emission takes place. These curves were provided by Benoit Hubert and Matthieu Meurant

Several steps are necessary to deduce efficiency curves. Using to a Monte-Carlo model, it is possible to simulate the proton aurora by the calculation of the Lyman-α excitation and the ionization rate. The ionization rate is then used as an input of a model providing the total LBH emission. This emission rate has to be integrated along the line of sight, taking in account the relative contribution of each band of the LBH system and the $O_2$ absorption. The outgoing spectrum is then integrated on the pass-band of the camera, taking in account the response of the instrument. The use of transport codes and atmospheric models was shown to be fairly satisfactory [see Gérard et al., 2001]. With this method, the proton contribution may be subtracted from LBHS and LBHL data, assuming a value of the energy of incident protons. This energy value may be fixed *a priori* (thanks to the Hardy model for example) or may be provided by *in situ* spacecraft crossing the oval in the region of interest. Example of this has been provided by *Frey et al.* [2001], and *Gérard et al.* [2001].

However successful this was for previously flown WIC and SI12 instruments, the estimated contributions from proton aurora to the signal in the electron auroral imager depends on the imager itself. For the UVAMC imagers, we will need to estimate the contributions to each of the proposed channels (LBH-L and LBH-S) due to emissions caused by secondary electrons due to primary proton precipitation. This depends on knowledge of the precipitating protons, and in the ideal situation knowledge of their characteristic energy (fortunately, the proton auroral energy flux is not in practice tremendously dependent on characteristic energy, but knowledge of this is of course preferable) [see Gérard et al., 2001]. In summary, the subtraction of the proton contribution to the LBH signal is based 1) on the knowledge of the LBH emission at the top of the atmosphere caused by an incident proton flux and 2) the knowledge of the LBH camera response to this signal. To accomplish this for IMAGE, the following was done:

<u>Knowledge of the LBH emission at the top of the atmosphere caused by an incident proton flux … requires:</u>

- An atmospheric model
- A method to generate the line profile at different energies. This can be done by a Monte Carlo method.
  Contact: Benoit Hubert, Jean-Claude Gérard
- Integration of the emission along the line of sight for different geometries of observation.
  Contact: Benoit Hubert, Jean-Claude Gérard

This work was done already by the Belgian group for the IMAGE mission. No need to do it again. For the UVAMC in the context of the Ravens imaging complement, this means we will need to carry out the following:

<u>Knowledge of the LBH camera response to this signal … requires:</u>

- The definition of the wavelength and the instrument response curve.
- Based on the outgoing emission (part 1) and the response curve, the last step is to simulate the response for this specific instrument.

A similar work was done already by Benoit Hubert for the IMAGE mission. This work is to be done again with the definition of the KuaFu (Ravens) instrument. The work time is estimated to ~ 2.5 months. Matthieu Meurant or an equivalent scientist would be the appropriate candidate for this.

Finally, knowledge of characteristic energy for previous efforts to estimate the contribution to the SI data from secondary electrons was obtained primarily from low altitude satellites with auroral energy ion detectors. These were DMSP and FAST. *There is a risk for the UVAMC in the context of the Ravens imaging complement because at present it is not clear that such detectors will be in flight in 2012.* Because of this, we are hoping that the Ravens SI instrument (from Belgium) can be designed in such a way that the images can be used to infer some information about the global distribution of characteristic energy. This is made possible by the fact that the Ravens SI will likely not capture images of 1356 Å emissions. Instead, we are arguing that the SI be redesigned so that images of two distinct parts of the Doppler shifted Lyman-alpha could be obtained, possibly simultaneously (if no moving parts), or possibly with alternating 30 second cadence frames (ie., 1 minute cadence in each bandpass). Such a scheme would allow crude tracking of the peak of the Doppler shift, and corresponding crude tracking of characteristic energy (please see discussion about the SI itself).

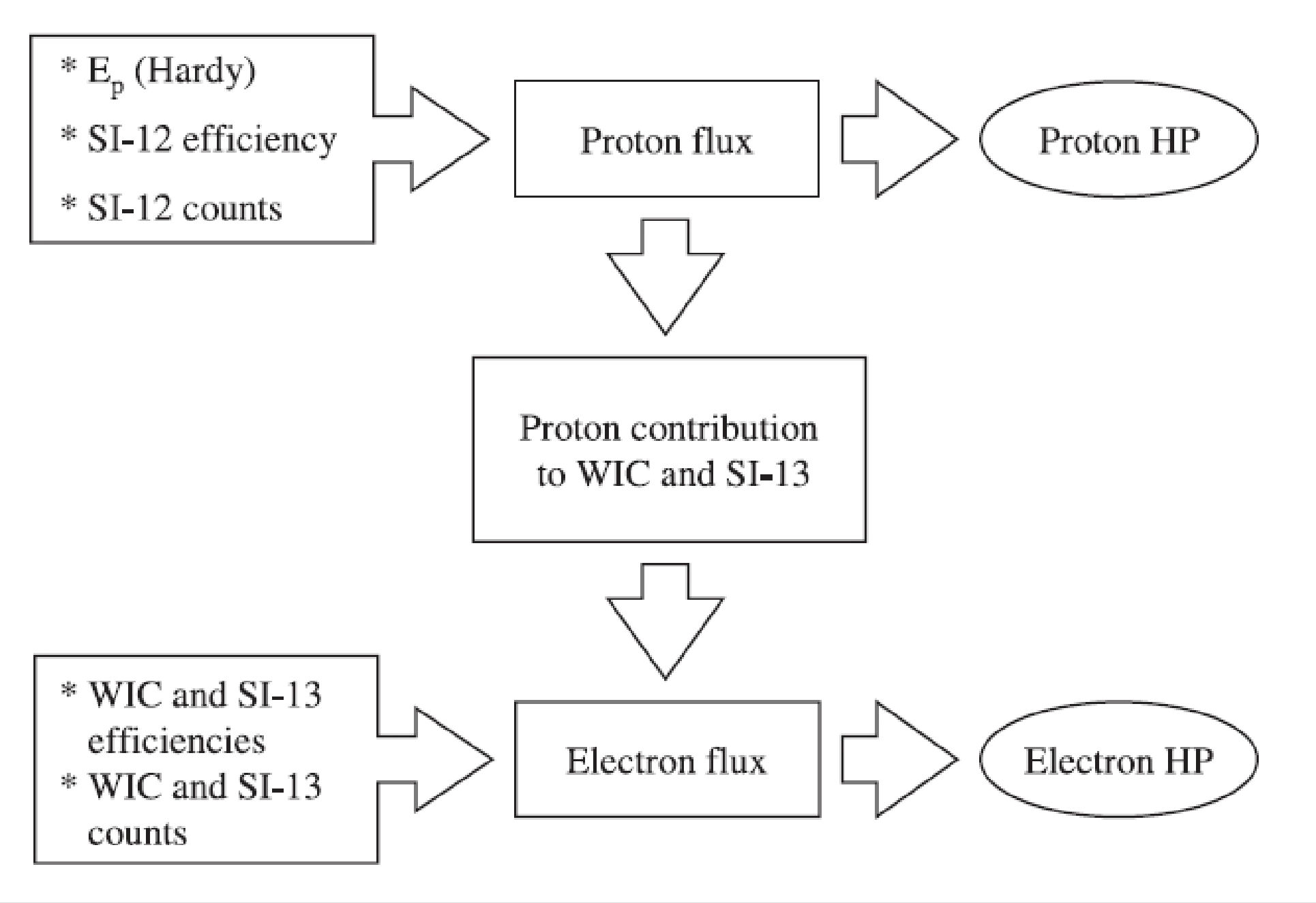

Figure 5.5: Schematic (provided by Benoit Hubert from the IMAGE SI team) indicating the steps necessary to subtract the proton contribution to the electron auroral signal

.

## 5.4 Separating LBH-L from LBH-S

Deriving the spatial distribution of the average energy and total energy flux of precipitating particles on a global scale is critical to determining ionospheric conductivity and quantifying the impact of solar forcing and magnetic storms to the neutral atmosphere. To quantitatively do this on a global scale is a difficult task. Several instruments have imaged the high latitude regions on a large scale both on the day and night side including AI (DE-1), Viking, UVI/VIS/Pixie (Polar), FUV-WIC/SI (IMAGE). Each of these instruments is responsible for significant strides toward understanding the dynamic nature of the global aurora and the first quantitative assessments of large scale auroral energetics. The most direct way of deriving the auroral energetics is to compare large scale images in two spectral regions of the far-ultraviolet Lyman-Birge-Hopfield (LBH) band-system; 140-160 nm (LBH low) where O2 absorption at auroral altitudes is significant and 160-190 (LBH high) where the emissions are relatively free of absorption (Strickland et. al., 1983; Germany et. al., 1994 a, b, 1990). Other approaches have been employed that are more model dependent. These use auroral observations that include 135.6, 130.4 nm OI lines, broadband LBH observations, 121.6 nm Doppler shifted hydrogen Lyman-α and auroral X-rays (Lummerzheim et. al., 1991, Hubert et. al., 2002, Østgaard et. al., 2002). There is evidence that the contribution of proton precipitation is not significant for magnetically disturbed events but should be taken into account for quiescent periods (Hubert et. al, 2002 and see previous section).

Dr. Dirk Lummerzheim of the Geophysical Institute in Fairbanks, Alaska has calculated the expected brightness of the $N_2$ LBH bands for a given electron energy flux and a range of average electron energies. He has kindly made his tables available for this concept study. His calculations correct for attenuation in the atmosphere and therefore provide the auroral surface brightness as viewed from above the atmosphere. We have summed the brightness values within each of the two spectral ranges (LBH-short and LBH-long). The numbers presented here therefore refer to the photons incident on the instrument without taking into account the instrumental characteristics such as the actual shape of the filter passbands. A more thorough analysis will be undertaken when the instrument characteristics are known.

The tables show that LBH-long brightness is in the range 100 to 250 R when the electron energy flux is 1 milliwatt per square meter ($mW/m^2$) and the average precipitating electron energy is in the range 1 to 10 keV. This energy flux gives rise to about 600 R of emission in the $N_2^+$1NG(0,0) band at 391.4 nm which corresponds to a relatively weak aurora. The instrument, with sensitivity characteristics as shown in Section 0 should therefore be capable of measuring 100 R in the LBH-long range. A 30 R/exposure noise contribution represents about 0.3 $mW/m^2$ uncertainty in the energy flux. For bright auroras (10 to 100 $mW/m^2$), the uncertainty due to the instrument noise becomes insignificant. This simple analysis shows that the cameras will have the necessary sensitivity to allow the electron energy flux to be extracted from the data. However, considerable further analysis is required. For example, it is known that proton precipitation, although weaker than electron precipitation also results in $N_2$ LBH photon emission; therefore, realistic models of global proton energy precipitation must be examined to ascertain the relative contribution of protons. The combination of a reliable precipitation model along with UVSI measurements should reduce this source of error. Another source of uncertainty is the fact that the LBH-long emission rate is not completely independent of the average electron energy, i.e. the emission rate is somewhat higher for incoming electrons at higher energy. Preliminary analysis indicates that this effect can introduce a variation of about 10 to 20% for the electron energy range 1 keV to 10 keV at all energy flux levels. This lack of complete independence was investigated by the Polar UVI team [Germany et al. 1998]. This dependence is illustrated in the figure below which was adapted from the paper of Germany et al. [1998] but is applicable to the Ravens UVAMC imagers.

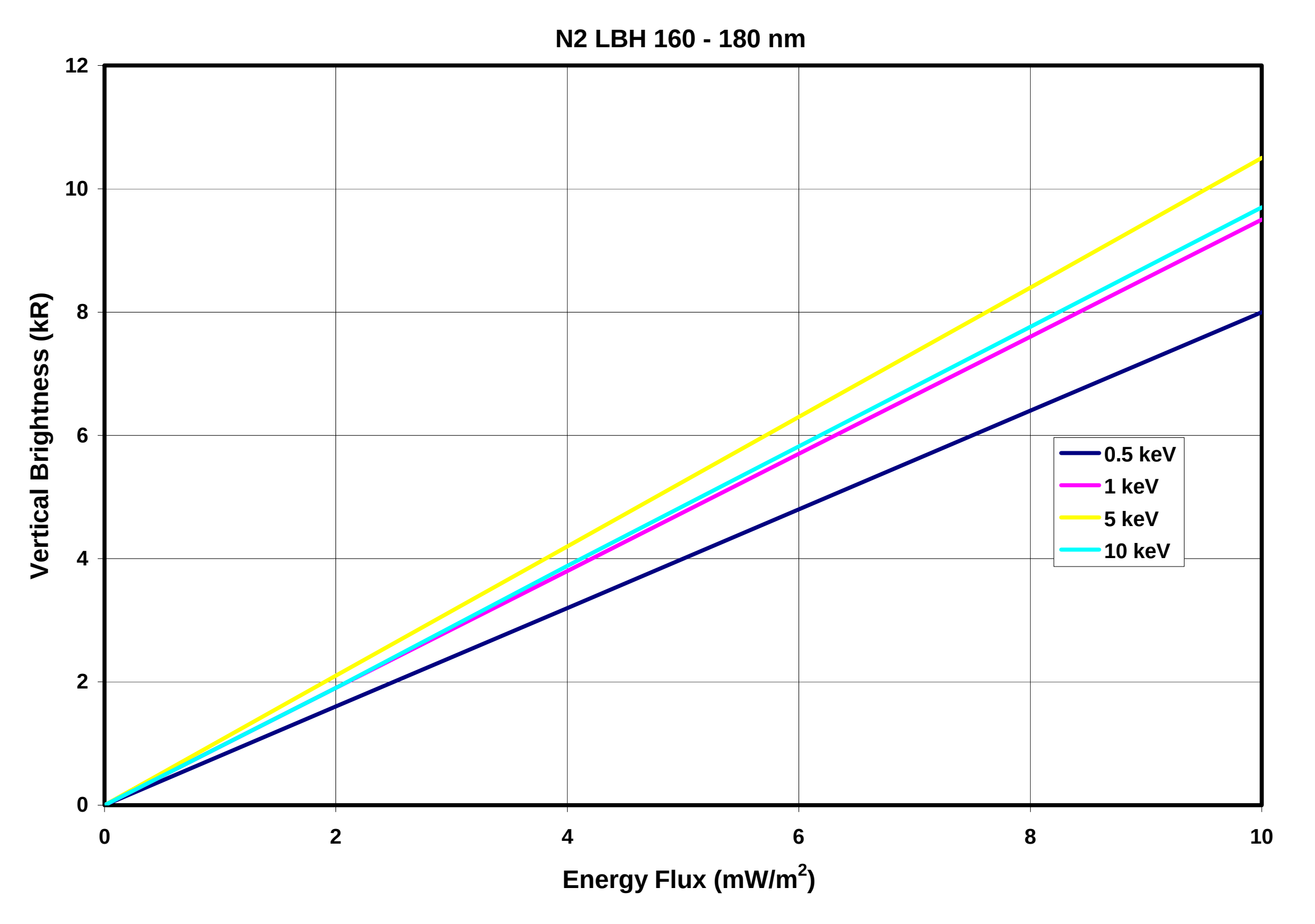

Figure 5.6: Dependence of LBH-long emission rate upon average electron energy.

A second necessary component of this preliminary analysis is to verify the feasibility of obtaining the average electron energy from the ratio of the LBH-short to LBH-long. The validity of this technique relies on the fact that the $O_2$ molecules absorb many more photons in the LBH-short wavelength range than in the longer LBH-long range. The Polar UVI team calculated the ratio for an electron energy range up to 10 keV based upon the UVI filter bandpasses. Using Lummerzheim's tables for the range 1 keV to 10 keV, we have calculated the expected LBH-short to LBH-long brightness ratio for the incoming electron flux in the two spectral ranges. These are shown in the figure below for an electron energy flux of 1 mW/m$^2$.

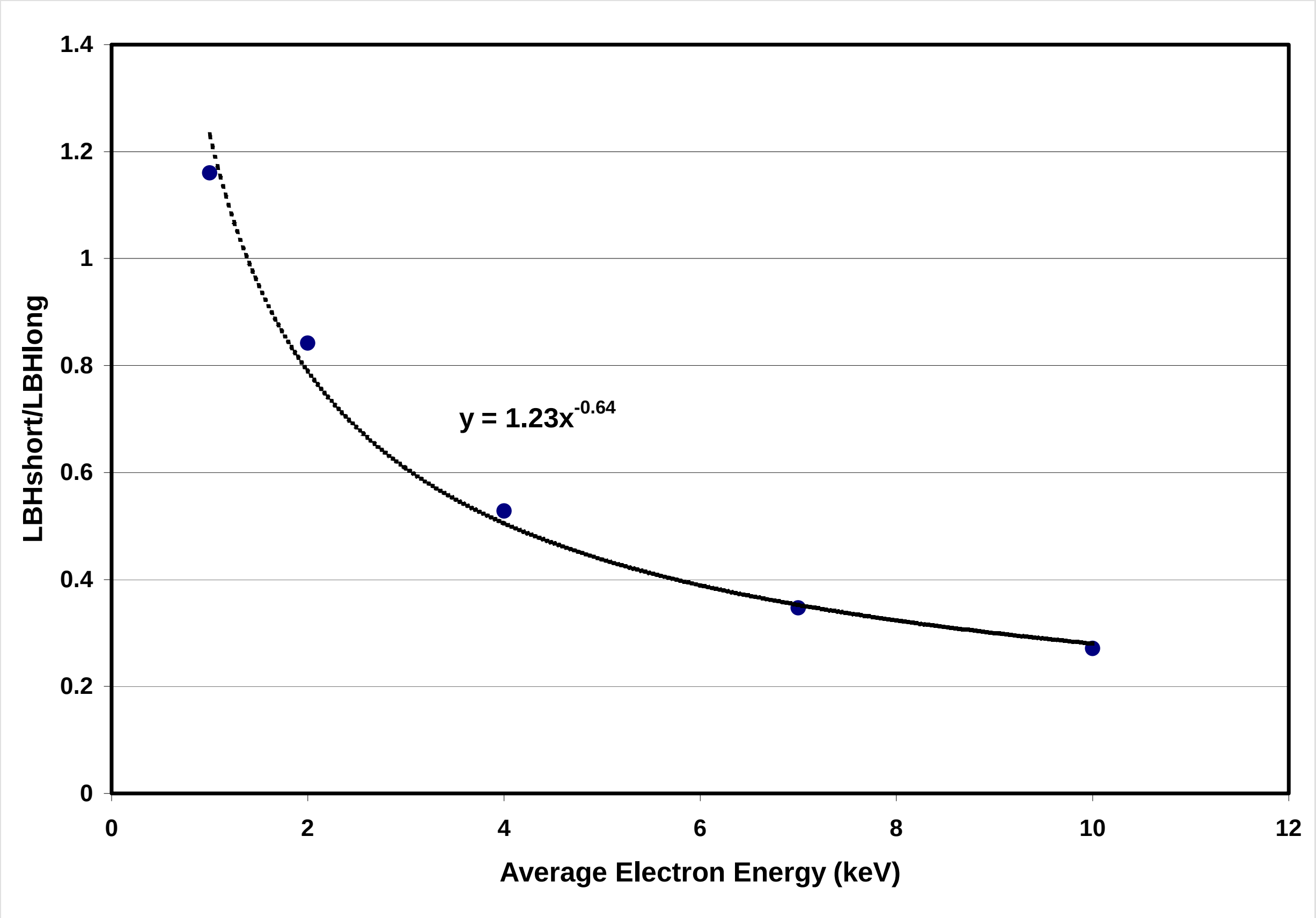

Figure 5.7: Expected LBH-short to LBH-long brightness ratios as a function of average electron energy, given an electron energy flux of 1 mW/m$^2$.

It is obvious that the expected uncertainty of this technique depends strongly on the auroral brightness. For a weak aurora (IBC I) that corresponds to an electron energy flux of around 1 mW/m$^2$ or less, the uncertainty is large for a single exposure with a noise equivalent signal of 30 R. However, for moderate auroras with energy flux near10 mW/m$^2$, the uncertainty introduced by the measurements is relatively small and is likely dominated by other sources of uncertainty such as variability in the $O_2$ density in the 100 to 150 km altitude range. It will therefore be important to utilize the best available model atmospheres for the analysis and to make the best estimates of uncertainty that can be generated by errors in the models. This topic will require considerable study during the pre-launch period. The good news is that this analysis technique is independent of the $N_2$ density because the same species emits in both spectral ranges.

In order to utilize these ratios and absolute values to infer characteristic energy and energy flux, the following technical requirements must be met. First, and most obviously, we must have simultaneous spectrally resolved images of the LBH high and low bands. Second, we must either characterize the out-of band contributions to these signals, or we need to sufficiently suppress them. The total signal due to scattered visible and IR sunlight is approximately 1E+11 R. If we wish to measure auroral signal as low as 10 R in the daylit side, the long-wavelength continuum Earth albedo contribution must be reduced by a factor of 1E+10. Although this is getting ahead of this stage of the development, this would be achieved by using a *very* pure CsI photocathode (1E+6 rejection) and radiation-hard FUV dielectric filters (1E+4 rejection). *It is important to note that driving down out of bandpass contributions will be expensive, complicated, and risky. This needs to be explored in great detail to make sure that we do not overstate our requirements unnecessarily. At the very least, we will likely back off on the 10 Rayleigh sensitivity requirement (we argue for 30 below), and that will allow us to back off on the suppression. Regardless this needs to be investigated – the cost implications are a risk from this particular point.*

At the present time, we estimate that our requirements can be met if we have sufficient signal to noise and we have 10 counts/kR/s sensitivity, and >1e10 out-of-band spectral rejection. As we describe in what follows, these requirements must be met on the spinning satellite (30 second spin period) which is in turn required by the UVSI. Separating the LBH into two spectrally resolved parts, and driving down the out-of-band contributions to the signals requires more light, and in turn a larger aperture, and ultimately a heavier, more complicated, and more expensive instrument. Dealing with those issues (or establishing whether we can back off on these stringent requirements and still meet our scientific objectives) will be the bulk of the Phase A UVAMC work (provided we go forward), and likely the single biggest factor in determining the cost of the mission to Canada. Third, we must also characterize the contributions to both the LBH-L and LBH-S signals due to secondary electrons created by primary auroral proton precipitation, as discussed in the previous section. This places a requirement on the UVSI that will be provided by Belgium.

## 5.5 UV Electron Auroral Intensities

There is a great tendency to quote rather stringent requirements (which often cannot be met) in proposing optical instrumentation in our field. In this section we present the results of an initial survey of the IMAGE WIC typical intensities. Our intention is to get a clear idea of what intensities we should be targeting with the UVAMC instrument. This is fundamentally important, because as can be seen from the technical budget tables in the following sections, achieving adequate signal to noise at low light levels is costly in terms of instrument mass (to the point that an unreasonably low intensity target could easily double, triple, or even make unfeasible the monetary budget for the instrument). As part of the Ravens Concept Study, we carried out some initial analysis on IMAGE WIC data, to try to get some idea of what kind of intensity variations in time and space, and absolute intensities we should be targeting. The objective of this kind of analysis, in no uncertain terms, is to see “what we can get away with”. If we can live with lower sensitivity than we might have thought, we can build a cheaper instrument that is more effective in other ways. In Phase A and beyond, should we proceed, our intention is to explore this issue more thoroughly (using models to estimate the breakdown of emission into the two proposed passbands, further exploration of the WIC data, similar analysis with the Polar UVI data, etc.), but in this subsection we present some of the work carried out under the Concept Study.

To begin with, IMAGE WIC was constructed using two Freja flight spares, with a detector provided by Ozzy Siegmund of UC Berkeley and a CCD from Lockheed. As such, the optical design is from the line of Canadian heritage instruments (Viking, Freja, InterBall). The bandpass for IMAGE WIC was wider than the bandpasses of either the LBH-L or LBH-S UVAMC, and in fact encompassed both. The CCD was 480X580, however the final images are 256X256 pixels, providing at best 100 km resolution. The intensifier gain had to be large in order to provide sufficient signal at an integration time of 33 msec (corresponding to TV rate readout). Although these factors conspired to limit the sensitivity of the instrument, it is an instrument members of our group understand. Further, the wide bandpass which encompasses effectively all of the LBH bands, makes for relatively straightforward interpretation of the numbers. From our perspective, we can take these numbers and divide them by 2 to get to initially explore the intensities in each of the UVAMC bandpasses.

The figure below shows a sequence of partial WIC images from January 8, 2001, highlighting the evening sector auroral oval showing in this instance a double oval. These images were collected at a cadence of 2 minutes. At this stage, we are interested in what the distribution of WIC intensities is, and the distributions of differences between adjacent pixels and differences between pixels in adjacent frames is. Given that UVAMC will operate at 4 times the cadence, with three times the spatial resolution, and that the LBH signal will be divided into two separate passbands, one could argue that to estimate what sensitivity would achieve what is necessary with the WIC parameters, and divide that required intensity by 4X3X2 or roughly 25. This is the limit that we should be thinking about for diffuse aurora, which actually have the smallest spatial and temporal variations in general.

On the other hand it is a fact that narrower features are in general the brightest features, and in general the features that change most rapidly in time. Thus, when we are trying to assess how reasonably we can push time and space resolution in a way that allows us to actually see differences we are interested in, we should be thinking about brighter features. Further, those bright features (arcs etc) are small compared to the typical size of a WIC pixel and comparable to or smaller than the size of the proposed UVAMC pixels. In this limit, it is not at all unreasonable that the smaller pixel size of UVAMC will better resolve the bright features, indicating even larger intensities for those features (the brightness measured in both cases is the average brightness in that pixel – for a smaller pixel with the same bright feature the measured brightness should be larger). So in this limit, UVAMC would only need to be able to track intensity differences as well as IMAGE could to achieve better results. *For the time being, we are asserting that this limit best represents the parameter space we are going after. We note that this means we are possibly underestimating the difficulty of going after better temporal and spatial resolution, but this will be revisited on an ongoing basis through the upcoming design phase.*

To assess the intensity requirements we should be going after, we decided to focus on the evening sector aurora, and more specifically taking data from 280 periods during January to April 2001 when IMAGE had a good view of the evening sector auroral oval (those keograms extracted from this WIC data subset are shown in the figure below). We wanted to make sure that this data reflected the typical range of magnetic activity. To do this, we compared SYM-H for the periods of interest with SYM-H for an entire solar cycle. The results of that comparison (shown in the histogram in Figure 5.9), indicate that we have an excellent cross-section of magnetic activity, according to SYM-H (which is the most appropriate index in terms of magnetic storms at least). The keograms from those periods are shown in the figure below.

We took that data, and focused our attention on typical auroral latitudes in the 2300 MLT sector. This in turn was used to create the histogram in Figure 5.10. Note the noise floor which we assert is due to a combination of readout noise and the exponential noise from the intensifier. Note also that the peak of intensity occurrence is roughly 200 Rayleighs above the noise floor. We limited our latitudes to just encompassing the typical auroral oval, and so we believe (though we cannot prove) that the decrease in frequency of occurrence left of the peak represents a true decrease of occurrence (ie., in the oval there is a peak in occurrence that is away from 0). Further, the decrease if it is real as we suggest is a convolution of the true occurrence statistics and the instrument response. In any event, we see that the peak intensity is ~200 Rayleighs above the noise floor, and most intensities are above that (in the evening sector auroral zone). Taking the rule of thumb discussed above, we assert that we will be able to do as well as IMAGE WIC in terms of intensities if we target a sensitivity of 100 Rayleighs (ie., as well in each of the two spectrally resolved channels). If we target lower intensities (which we intend to), we will do better. We also point out that our initial intention was to target 10 Rayleighs, but that meant an aperture that in turn meant a mass that was obviously prohibitively expensive. At the present time we believe (see below) we can get to 30 Rayleighs in each channel, which in each channel will be significantly better than IMAGE.

As might be expected, considering differences is more challenging in terms of interpretation. The histograms shown in Figures 5.11 and 5.12 show differences in space (between adjacent pixels separated north-south) and time (same pixel adjacent frames), broken down according to brightness. We are *most likely* most interested in accurate tracking of changes in bright features (see above). As well, subject to the caveats above, those differences are probably comparable in IMAGE WIC and UVAMC (each channel), at least within a factor of 2. Looking at the top two panels of each stack plot of histograms, again the number 100 Rayleighs could be viewed as a reasonable target. *We will target 30 Rayleigh sensitivity, and based on this cursory discussion we believe that we will be able to track bright quickly changing small features better than has ever been done before.*

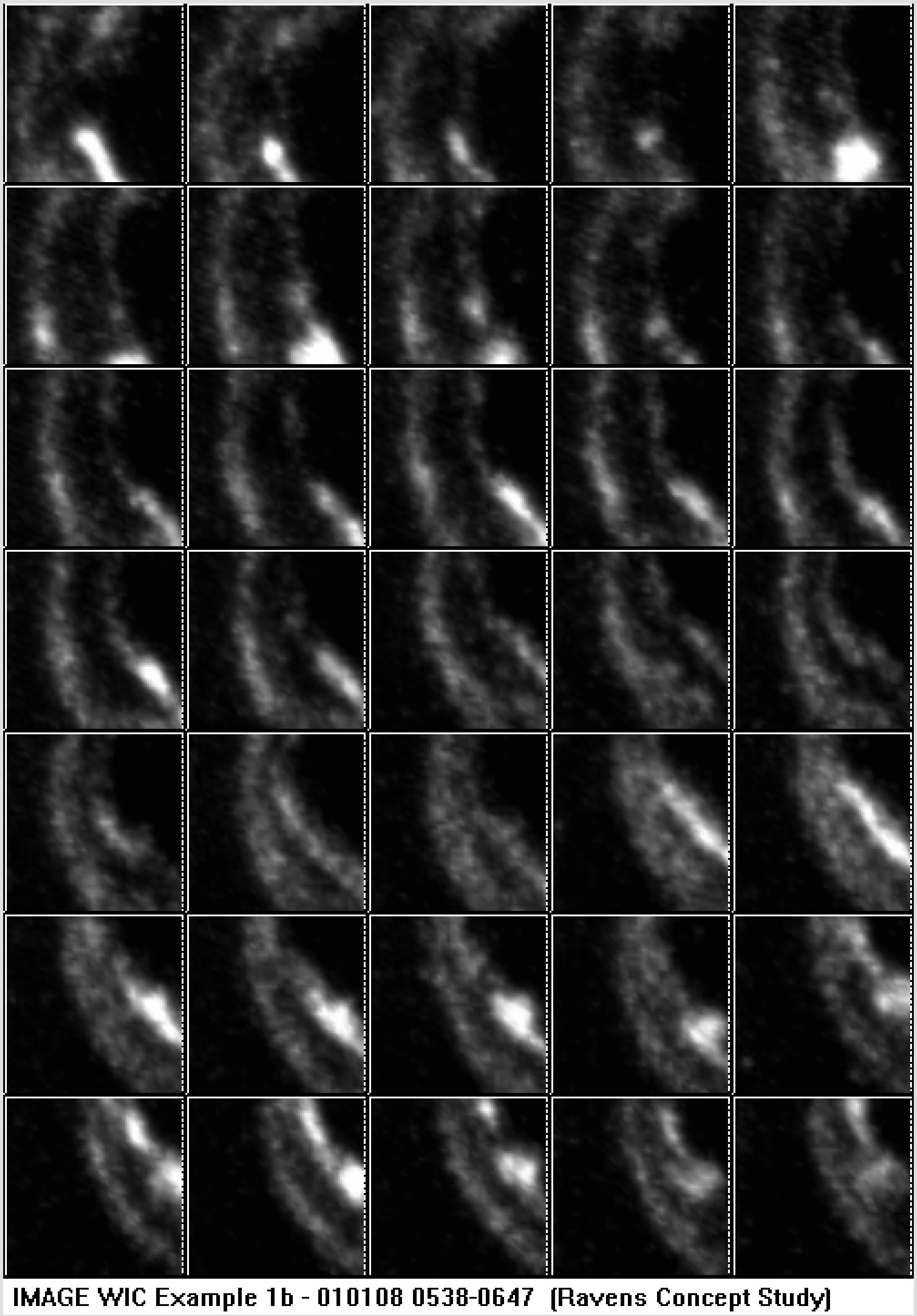

Figure 5.8: Detail from evening sector auroral oval from a sequence of WIC images.

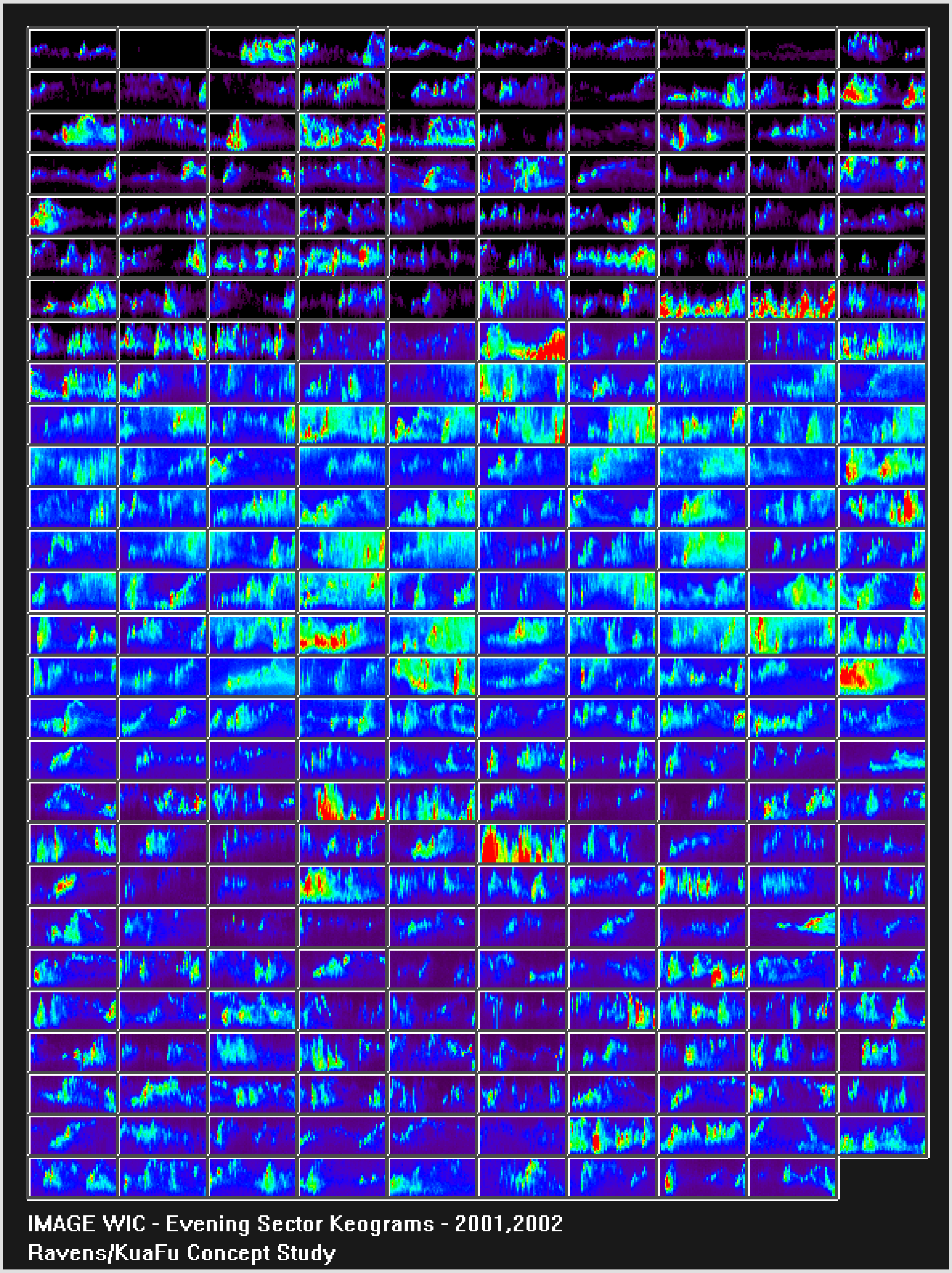

Figure 5.9: Keograms of the evening sector (MLT=2300) WIC data used in the analysis that produced the histograms in the following figures.

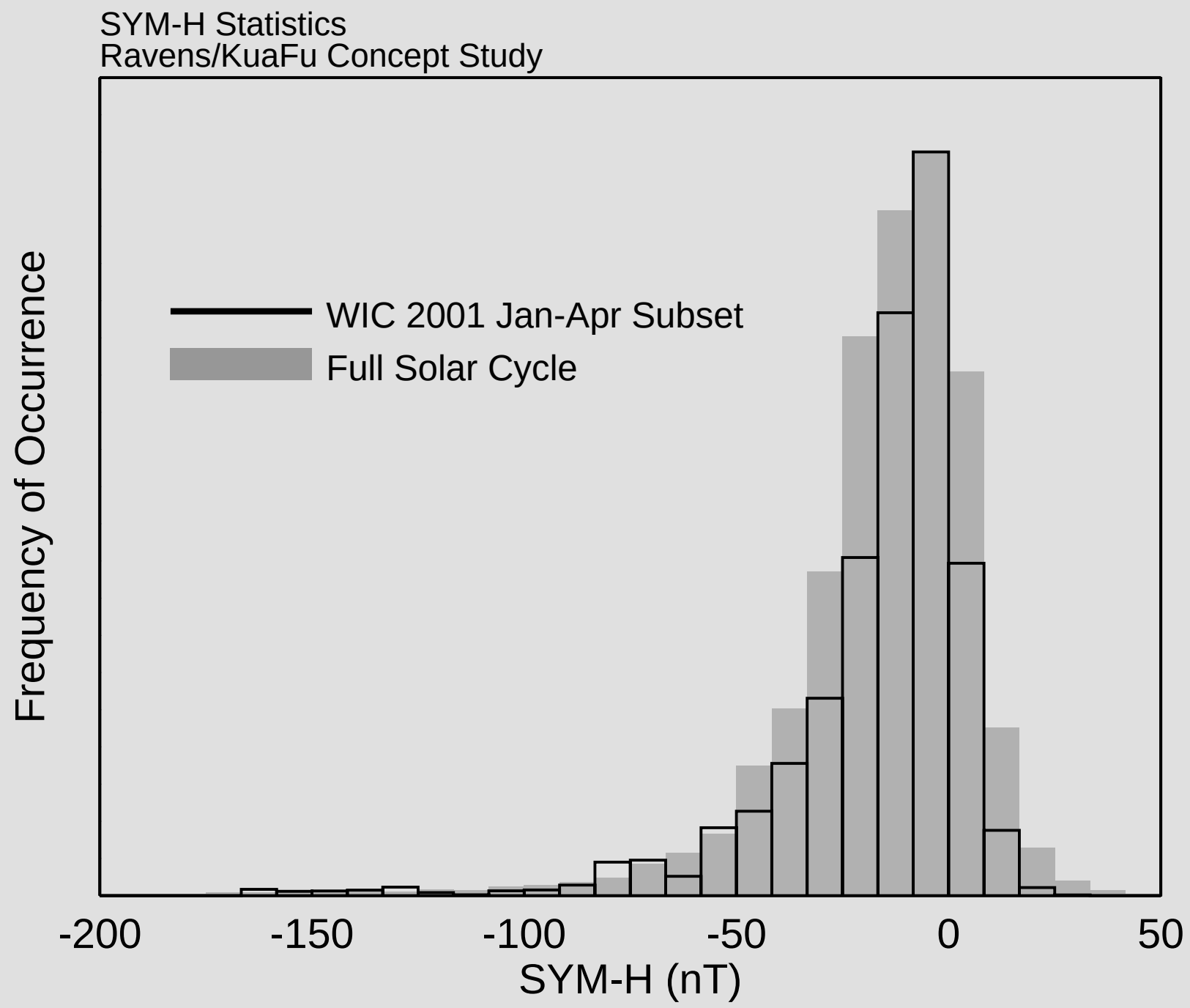

Figure 5.10: We wanted to be sure that our subset of WIC data was representative of appropriate levels of activity. The grey and black line histograms show SYM-H for an entire solar cycle, and the times for the WIC data, respectively. We have a good sampling of all levels of geomagnetic activity as measured by SYM-H.

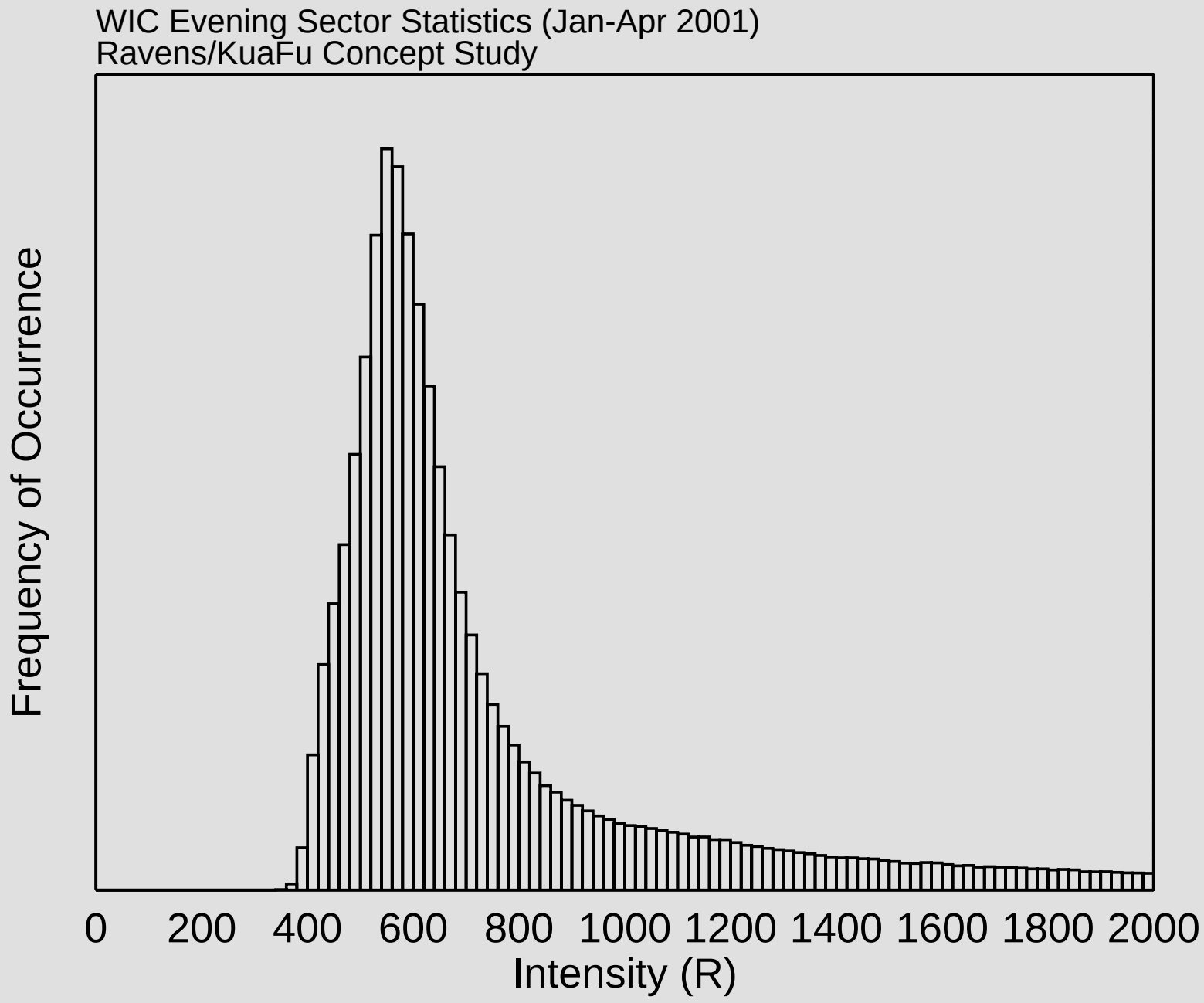

Figure 5.11: Histogram of WIC intensities for the 280 periods shown in the keograms above. Please see text for description.

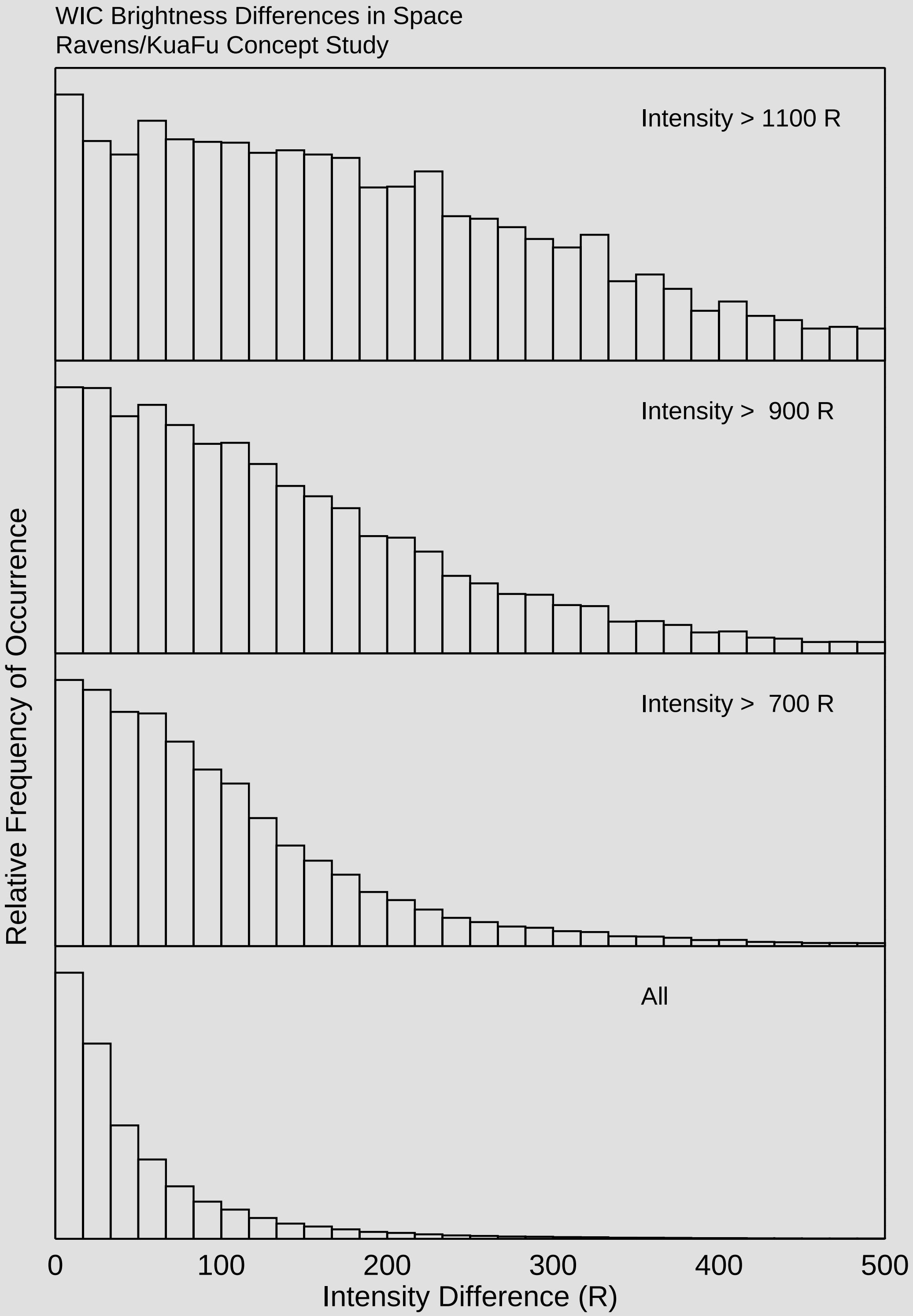

Figure 5.12: Histograms of differences in intensity between adjacent pixels in WIC that are roughly aligned north-south (in separation), broken down in terms of brightness of those pixels.

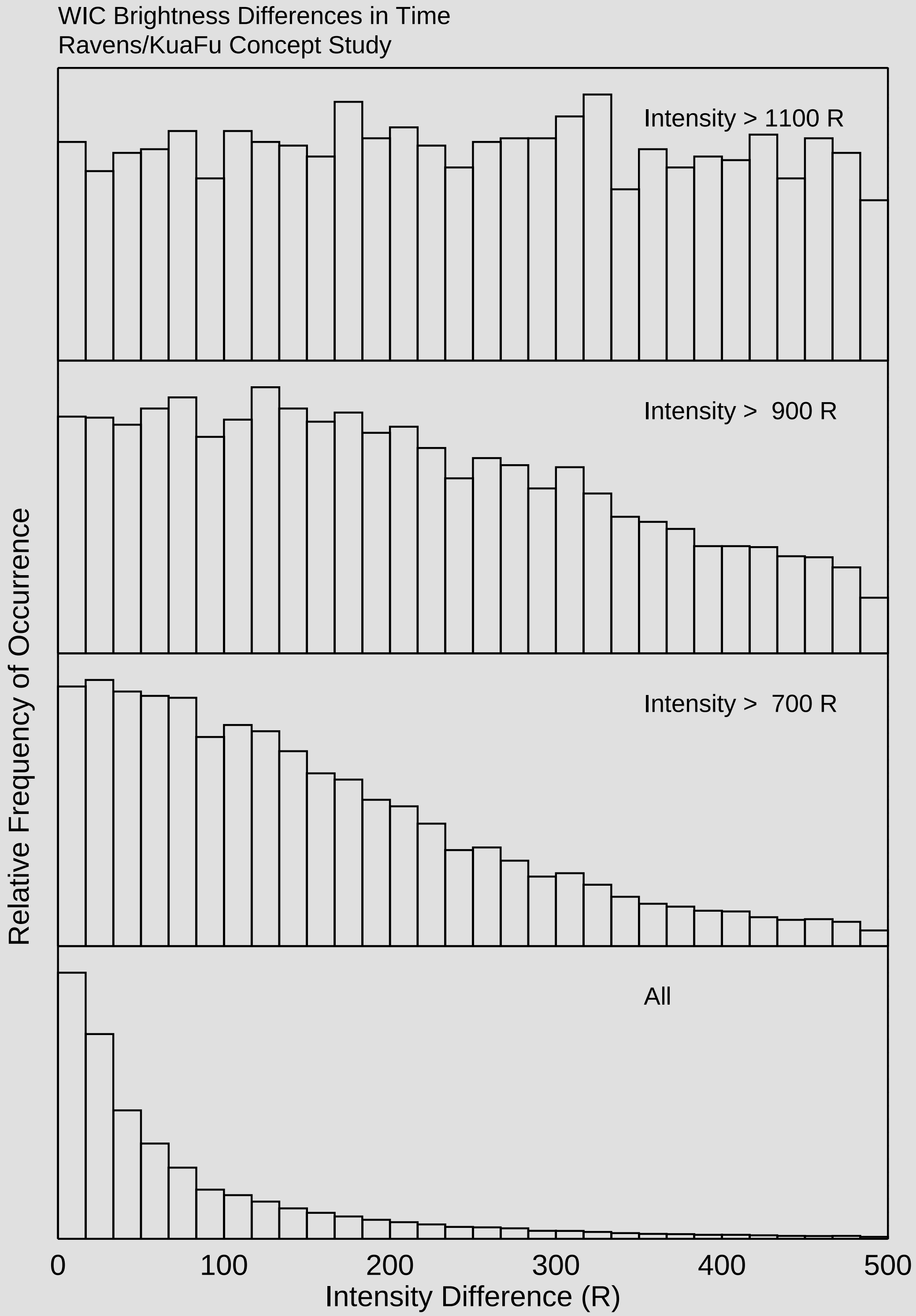

Figure 5.13: Same as the previous figure, but for differences from frame to frame.

## *5.6 Reasonable Spatial Resolution*

As can be seen from the table above, and straightforward consideration orbital parameters of satellites that have carried previous imagers and their CCDs, spatial resolution provided by global imagers, at nadir, has been at best ~70 km, and usually larger (ie., less well resolved). While these imagers have produced data that has clearly moved our understanding forward, it is equally clear that better resolution would be a welcome improvement. Resolution is limited by a number of factors, ranging from the point spread function which can be different in the direction of spin and across it, the pixilation of the CCD, the field of view, the distance from the emitting region, and the thickness of the emitting layer. We are confident that we can bring down data (for KuaFu) produced by a 512X512 effective CCD (this will be binned down to 512X512) at a frame rate of 1 image every 30 seconds. For length of mission, quality of view, and systematic conjugate imaging, we have settled on an orbit (that has been agreed to by the KuaFu engineering and science teams). At this point, there is not much more to do than to determine what the best possible spatial resolution would be.

To do this, we assumed an emitting layer with a thickness of 40 km. In reality, the layer is likely somewhat thinner (see Figure 8 in *Gérard et al.* [2001]), however there is variation in the height of the emission layer, and this is a reasonable compromise to get a first-look idea of achievable spatial resolution. We used a variation of programs developed for the Ravens Concept Study to use quality of view to explore the optimal orbital parameters. In this simulation, we placed two satellites on the proposed KuaFu orbits, each with a 512X512 25 degree field of view UVAMC. We restricted our attention to points above 40° geomagnetic (PACE 2000 epoch) latitude. We broke that region down into cells. Each cell is viewed by one, the other, or both UVAMC instruments, depending on where the satellites are on the orbit. The spatial resolution at the location of the cell depends on where the satellite is. So each cell, at any time, can be seen better by one satellite than the other (or occasionally the viewing quality is roughly equal but that does not matter). For each cell, we determine which satellite can see aurora at that location with better spatial resolution. The resolution provided by that satellite is the resolution we assign to that cell.

We did this for evenly spaced times during the orbital period of one of the satellites. The results are shown in the figure below. We found that, in general, nadir resolution us sometimes better than 30 km, and falls away with increasing distance from nadir. At first, that fall off is due to the increasing spatial footprint of the pixel width with increasing angle away from nadir. Inevitably, unless the thickness of the emitting layer is set to zero, the limiting factor becomes the fact that the line of sight from the pixel starts to look across too much of the thick emitting region. In other words, near nadir, spatial resolution is limited by pixel size. Well away from nadir, it is limited by thickness of the emitting region.

By carrying out this analysis for a number of different thicknesses and CCD density (ie., angular size of CCD pixels), we came to the conclusion that given the reality of a finite thickness of the emitting layer, which must be in reality exacerbated by variation in the height of the emitting layer within the field of view, a reasonable target resolution can be achieved by a 512X512 pixel CCD. As can be seen from the figure below, this translates to ~30 km resolution at nadir, falling away to more than 60 km resolution well away from nadir. With this, we will have most of the region over 40° magnetic latitude observed with ~35 km resolution, and the overall effect will be global images with much better spatial resolution than has previously been achieved. With KuaFu this will be achieved continuously for more than 2.5 years.

We must point out that in Phase A and Phase B, we will need to take instrument designs and attempt to assess the point spread function, taking into account both optical effects and potential uncertainties in pointing and nutation. At present we believe that with these additional factors, the spatial resolution will still be roughly what we are showing here.

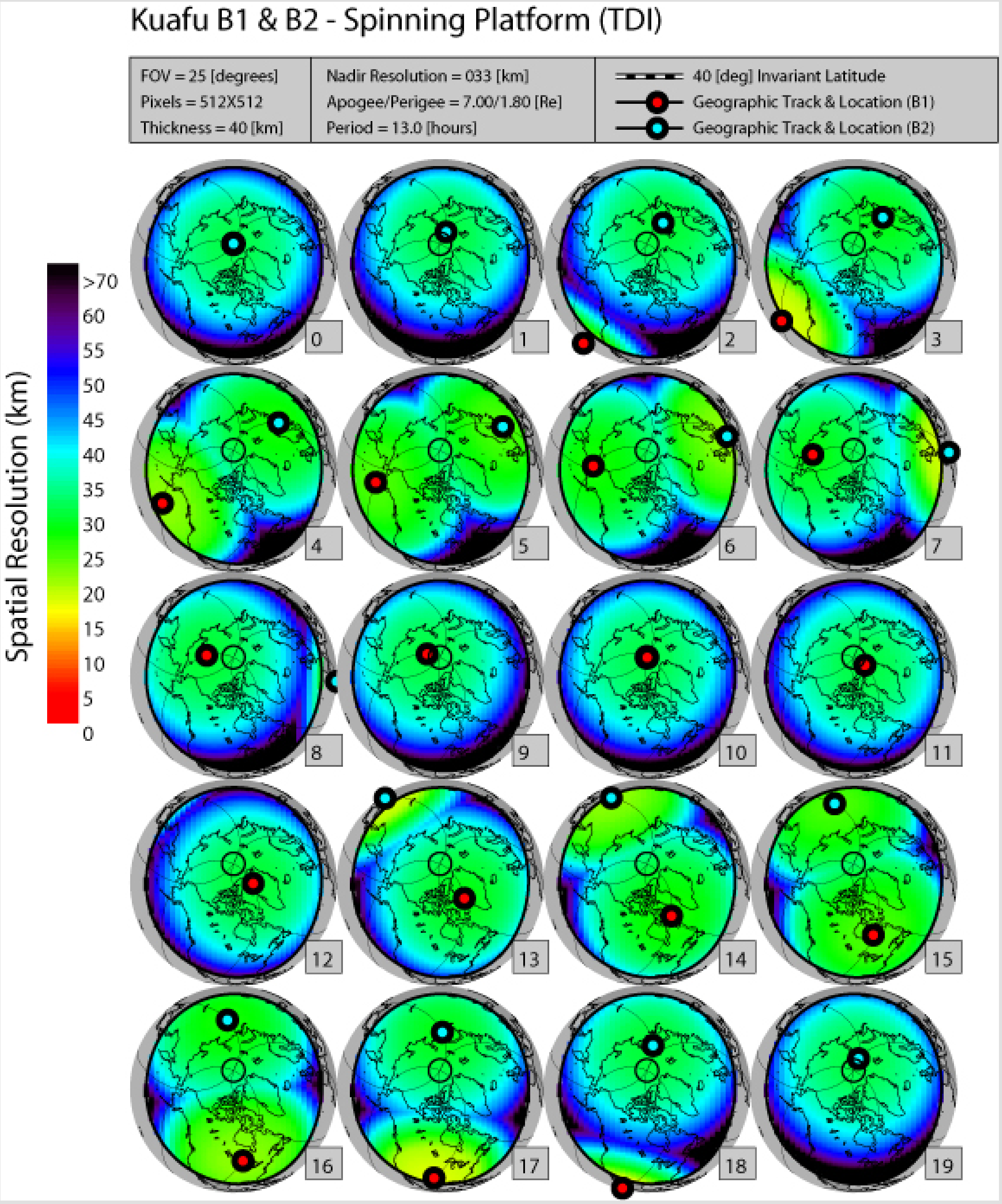

Figure 5.14: Distribution of spatial resolution at geomagnetic latitudes above 40° from 25 degree FOV UVAMC instruments on two satellites on the proposed KuaFu-B orbits at 19 evenly spaced times during one orbital period. We assume 512X512 pixel CCD, and a 40 kilometer thick emission layer. Note that the bulk of this region is viewed with spatial resolutions of ~30-35 km, but resolution inevitably increases (decreases in quality) to 60 km or more well away from nadir of either satellite. This is an inevitable consequence of the thickness of the emission layer (or variability in height of emission).

## 5.7 UVAMC Instrument Concept

### Scientific Requirements

The science requirements driving the design of the UVAMC are as follows:

1. The need for determination of characteristic energy and energy flux of precipitating electrons on a global scale.
2. The need for global images with on the order of 30 km spatial resolution to facilitate studies of multi-scale processes and natural complexity.
3. The need to image low light level auroral emissions, both on the nightside and sun-lit portions of the Earth (often simultaneously, within the same scene).

Deriving the spatial distribution of the average energy and total energy flux of precipitating particles on a global scale is critical to determining ionospheric conductivity and quantifying the impact of solar forcing and magnetic storms to the neutral atmosphere. To quantitatively do this on a global scale is a difficult task. Several instruments have imaged the high latitude regions on a large scale both on the day and night side including AI (DE-1), Viking, UVI/VIS/Pixie (Polar), FUV-WIC/SI (IMAGE). Each of these instruments is responsible for significant strides toward understanding the dynamic nature of the global aurora and the first quantitative assessments of large-scale auroral energetics.

Perhaps the most direct (model independent) way of deriving the auroral energetics is to compare large-scale images in two spectral regions of the far-ultraviolet Lyman-Birge-Hopfield (LBH) band system; 140-160 nm (LBH short) where O2 absorption at auroral altitudes is significant, and 160-180 (LBH long) where the emissions are relatively free of absorption. The UVAMC instrument (one instrument on each of two satellites) would thus each be comprised of *two FUV imagers* (UVAMC-0 and UVAMC-1, for LBH-short and LBH-long measurements, respectively). These would then deliver quantitative intensities from two well-separated parts of the LBH band, with the objective of using relative and absolute intensities for determination of characteristic energy and energy flux of precipitating electrons, on a global scale.

Instrument performance requirements are dictated by the scientific requirements. In the case of past and currently deployed FUV imaging missions, three primary obstacles to obtaining satisfactory quantitative results for the spatial distribution of auroral energetics can be identified:

- the lack of truly simultaneous observations of spectrally resolved LBH long and short bands,
- the lack of characterization of out-of-band contributions to the measurements,
- instrument performance that provides high spatial and temporal resolution images.

To quantify the spatial distribution of the average energy and total energy flux of precipitating particles on a global scale, and improve upon what is currently available, we set down the following measurement requirements:

- 30 s temporal resolution
- 30 km nadir spatial resolution
- 10-30 counts/kR/s sensitivity
- greater than $10^{10}$ out-of-band spectral rejection.

The requirement to maximize spatial and temporal resolution is derived from the need to determine the auroral energetics at the edges of the oval where the aurora is very dynamic. Such edges are particularly susceptible to effects of poor resolution as these regions are changing quickly from dark (no emissions) to bright emission, and vice-versa, during expansion. There is conjecture that at these edges, and also at substorm onset locations, emissions result from Alfvén driven processes (as opposed to the more static inverted-V acceleration process). In principle this can be verified by determining the auroral energetics using imaging. Current imagers on Polar and IMAGE are not capable of doing this. IMAGE has good image integration times (15 s every 120 s) but poor spatial resolution (100 km for WIC and worse for SI). Polar, even without the wobble problem, has no better than 50 km resolution and 37 seconds time integration, and appears to be suffering from leakage of a VIS-NIR contribution into its FUV bandpass.

The on-the-order-of 10 counts/kR/s sensitivity requirement is derived from the need to provide a significant advancement over what we have now. Currently, 50 R in 30 s is the best that is available in the FUV (Polar UVI). High sensitivity is needed to measure very weak polar cap emissions or dim remnants and precursor emissions of bright dynamic aurora that may help improve our understanding of the mapping to magnetospheric regions as well as auroral acceleration processes.

The requirement of more than $10^{10}$ out-of-band rejection is derived from the fact that the visible integrated dayglow (VIS-IR) is about $10^{11}$ R. Therefore, to measure a ~10 R signal on the daylit side of the Earth with an equal background, a rejection of at least $10^{10}$ is required.

Together, these requirements will push the envelope of current design, but not excessively so. That is, it is reasonable to expect that these requirements are achievable without revolutionary technological breakthroughs.

Consequently, there are at least three major instrumental issues that have to be resolved:

1. Acquiring a good solar blind photocathode (e.g., the purity of the CsI is critical)
2. Having the proper filter design/production. (Recently developed multilayer dielectric filters can provide an average measured reflectance longward of 200 nm of 3.6%, with maximum reflectance of greater than 85% at 155nm).
3. Making sure that the imaging performance and sensitivity meet the requirement. (Often, resolution has to be sacrificed for sensitivity or vice-versa. This is especially so in the case of allreflective systems, as they should not be over-designed by using too many optical components or they will be overly sensitive to component displacement. Spot size and throughput are critical to achieve the desired requirements. At least 4 filter reflections will be required in order to achieve the needed $10^{10}$ out-of-band rejection).

## Instrument Characteristics

The design of the UVAMC imagers is motivated by a desire to record snapshots of auroral dynamics with high temporal resolution under night-time as well as fully sunlit conditions. An intensified FUV CCD-based imager with fast optics is proposed. Such a design allows each image pixel to be exposed simultaneously, giving true snapshot ability as well as high temporal resolution. Given a spacecraft spin rate of 2 rpm, which is the current baseline, a too long exposure time would result in smearing of the image across the CCD. In fact, given the various parameters of the proposed mission, the allowable exposure time would be no more than a fraction of a second before smear would become significant. Even with a fast optical design and an image intensifier, the allowable exposure time is far too short to be able to image faint auroral emissions. To overcome this problem, the CCD will be run in a fast frame rate (snapshot) mode, with images being acquired at 30 frames per second and offset/summed in digital memory to simulate a longer effective exposure time. The advantages of this approach are an almost unlimited dynamic range, and the fact that any corrections for misalignment of CCD columns with respect to the spin plane, and other motion-induced distortions, may be corrected for digitally, on the fly. The MCP gain would have to be set accordingly, so as to make insignificant the increased CCD read-noise contribution associated with such fast readout.

**The following are some characteristics of the UVAMC imagers with a brief explanation of their rationale:**

**Field of view.** The instrument must have a field of view (FoV) wide enough to span the entire auroral zone and polar cap for a significant fraction of the orbit. A field of view of 25 degrees has therefore been chosen. As shown in Section 3.4, this ensures continuous global coverage of the northern auroral oval from the combined measurements from the two Ravens spacecraft, both on a 7 x 1.8 Re (geocentric) orbit.

**Image Format.** A 1024×1024 pixels frame-transfer CCD (tentatively the e2v CCD47-20) is baselined for this application. Images can be binned 2×2 to achieve a 512×512 format whenever spatial resolution can be sacrificed for sensitivity.

**Spatial Resolution.** The nadir apogee spatial resolution, employing 1×1 and 2×2 binning, is 16 km and 32 km, respectively, and therefore, to achieve the spatial resolution that we are aiming for from near apogee, the instrument must have at least 512×512 pixels. The 1.8×7 Re orbit (geocentric distances) proposed for the Ravens mission allows for continuous global imaging with excellent spatial resolution over much of the auroral and polar regions, and down to mid latitudes. To demonstrate that this is an effective orbit for UVAMC, see Figure 5.14 which is a panel plot of distributions of UVAMC spatial resolutions for pixels with geomagnetic latitudes greater than 40 degrees. In generating this figure, a 25-degree field of view and a 512×512 pixel image were assumed. There are 20 panels in the plot corresponding to equally spaced times during one orbit.

**Temporal Resolution.** Temporal resolution is determined by the spin rate of the satellite. The proposed spin rate is 2 revolutions per minute, yielding a maximum image cadence of 30 seconds. Higher image cadence would require a faster-spinning satellite, but this would impact imager sensitivity (next item).

**Exposure Time.** The effective image exposure time is determined by the imager's field of view and the satellite spin rate. For a 2 rpm spin rate (30 second spin period) and a 25 degrees field of view, it is readily shown that the maximum available exposure time is 2.1 seconds. Such a short exposure time severely impacts imager sensitivity threshold, and places high demands upon the optical design.

**Wavelength Separation.** The instrument shall provide simultaneous, wavelength separated LBH electron auroral observations over the global auroral distribution. Proposed passbands are 140 – 160 nm (LBH-short) and 160 – 180 nm (LBH-long). The latter provides a measure of energy flux, and the ratio of the two provides information on characteristic electron energy. Sufficient suppression of unwanted continuum contribution from shortward of 140 nm and longward of 200 nm must be provided. Figure **5.15**5.15 shows the LBH spectrum with LBH-short and -long passbands indicated.

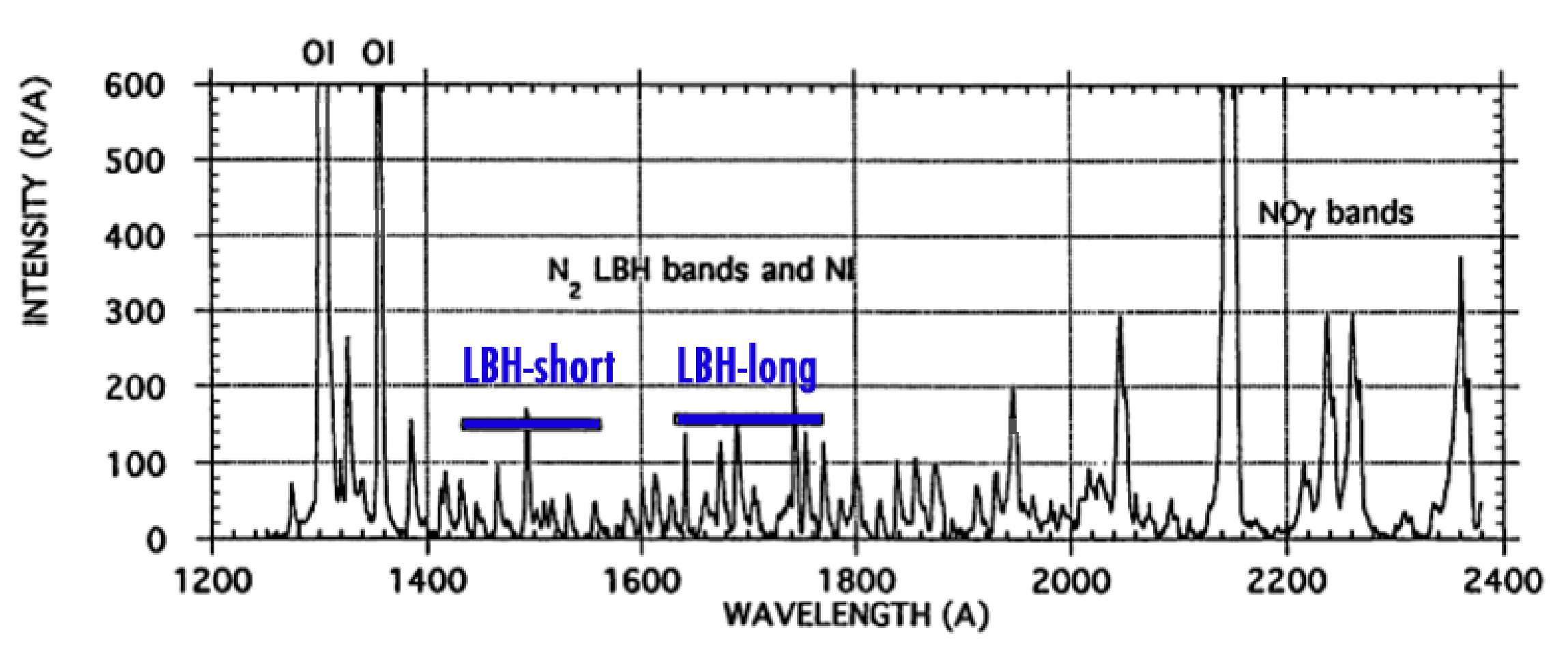

Figure 5.15: FUV spectrum, showing LBH bands and indicating the location of the desired "LBH-short" and "LBH-long" passbands.

**Sensitivity.** The imager's optical aperture must be as large as practically possible, with instrument mass and size restrictions in mind, so as to admit enough light for reasonable signal to noise ratio in the case of dim auroral features (tens of rayleighs, as integrated across the appropriate part of the spectrum) during the relatively short 2.1 s integration time. As shown below, a Noise Equivalent Signal (NES) of 30 R per exposure can be achieved with a 27.6 mm diameter aperture and careful selection of optical materials.

**Dynamic Range.** The instrument must be capable of measuring very bright signals (on the order of 100 kR) in the auroral oval as well as very weak signals (tens of rayleighs) in the polar cap. It may be necessary to measure weak regions simultaneously with bright regions within the same field of view. An intrascene dynamic range of 1000, and an interscene dynamic range of $10^4$ is desired. The interscene dynamic range can, as already indicated, be further extended by the use of image summation in memory during exposure.

Some important required instrument characteristics are summarized in Table 1.

**Table 1: Some required instrument characteristics.**

| | |
|---|---|
| *Field of View* | *25 degrees* |
| *Number of Cameras* | *2* |
| *Spectral Passband* | *140-160 nm (camera 0)*<br>*160-180 nm (camera 1)* |
| *Image Format* | *1024 x 1024 and 512x512 (selectable)* |
| *Nadir Spatial Resolution* | *16 km (1024x1024 pixels) and 32 km (512x512 pixels)* |
| *Temporal resolution* | *30 s (assuming 2rpm spin rate)* |
| *Exposure Time* | *2.1 s (assuming 2 rpm spin rate)* |
| *Sensitivity* | *30 R/exposure NES* |
| *Dynamic Range* | $10^4$ |
| *Rejection of VIS-NIR albedo* | $\sim 10^{10}$ |

The following is a somewhat more formal list of functional and performance requirements of the UVAMC imagers. Unless otherwise stated, requirements apply to both UVAMC-0 and UVAMC-1:

a. The UVAMC instruments on the two Ravens satellites shall together be capable of providing continuous global imaging of the auroral oval.

b. The UVAMC-0 instrument on the two Ravens satellites should provide images in the LBH-short band (140 nm to 160 nm) with a signal to noise ratio greater than unity for an auroral oval source brightness of 30 rayleighs per exposure.

c. The UVAMC-1 instrument on the two Ravens satellites should provide images in the LBH-long band (160 nm to 180 nm) with a signal to noise ratio greater than unity for an auroral oval source brightness of 30 rayleighs per exposure.

d. The UVAMC instrument shall provide auroral oval nadir spatial resolution from apogee of 30 kilometers or better in the final image, for an auroral altitude of TBD km.

e. The UVAMC instrument shall be capable of maintaining a square aspect ratio of an image pixel to within a tolerance that is to be determined.

f. The UVAMC instrument shall allow simultaneous imaging in two different wavelength bands using two co-aligned FUV intensified CCD (ICCD) camera units. The instantaneous field of view of each camera unit shall be 25 degrees across the spacecraft orbital track and 25 degrees along the spacecraft orbital track.

g. The CCD detectors shall be used in a fast frame rate snapshot mode, with a frame rate that is to be determined.

h. The UVAMC instrument shall be capable of providing images in a staring mode of arbitrarily long exposure times, to support laboratory calibration.

i. The design lifetime of the UVAMC instrument shall be not less than 2.5 years on orbit.

j. The UVAMC instrument shall be prevented from imaging the Sun.

k. The telemetry interface shall provide telemetry data to the Ravens spacecraft telemetry interface at either 0.1 Mbits/sec sustained, at the rate requested by the Ravens spacecraft, or as dictated by the pre-programmed imaging operations schedule. Other modes may be implemented as well.

l. The UVAMC instrument shall have the ability to store in buffer memory only one image pair, the latest image from both cameras, and make it available to the Ravens Data Handling Unit upon polling.

m. The UVAMC instrument shall be capable of utilizing limb sensor and/or other synchronization pulses from the Ravens spacecraft and/or other onboard experiments to initiate imaging sequences.

n. The UVAMC instrument shall operate from the spacecraft +28V ±6V input voltage.

o. The UVAMC instrument shall control all parameters which affect imaging. As a minimum these parameters include exposure time, binning size, image window size and location, high voltages and analog gains.

p. The UVAMC instrument shall provide a correlation of the UVAMC instrument clock to the spacecraft clock to within plus or minus 1 millisecond (TBD).

q. The UVAMC instrument shall have a low power standby mode which protects the integrity of any buffered image data, program data, and parameter lists, and minimizes the UVAMC instrument power dissipation.

r. The UVAMC instrument shall insert image headers into data memory contiguously with the associated image data.

s. The UVAMC instrument shall provide unique image time tags for a period greater than one orbit.

t. The UVAMC instrument shall be capable of imaging scenes when the satellite is out of range of a ground station. The images shall be preserved by the Ravens Data Handling Unit (DHU) data memory, if DHU so chooses, until the satellite is again within range of a ground receiving station, at which time the stored data will be transmitted.

u. The UVAMC instrument shall accept and execute commands through the Ravens spacecraft command interface.

v. The UVAMC instrument shall provide data telemetry to the Ravens spacecraft telemetry interface.

w. The image headers shall contain UVAMC instrument status parameters, UVAMC health parameters, and image parameters.

x. The UVAMC instrument shall execute three types of commands; instrument configuration commands, status and diagnostic commands, and imaging commands.

## Optical Design

### *Introduction*

The image-forming section of the UVAMC will be comprised of fast, all-reflecting telescopes. Two-, three-, and four-mirror systems have all been considered during this concept study, each having its own distinct advantages and disadvantages. The two-mirror system is based on the heritage University of Calgary line of imagers (Viking, Freja, InterBall, IMAGE-WIC), while the off-axis three- and four-mirror systems constitute a radical departure from the previous approach but provide superior visible-light rejection and isolation of the signal from the two separate parts of the LBH band (LBH-short and LBH-long). All optical designs considered are compact, fully corrected for spherical aberration, coma, and astigmatism. Moreover, they have excellent resolution over most of the field of view and have a high throughput in the far ultraviolet part of the spectrum. Each design provides a 25 degrees circular field of view with an optical axis perpendicular to the spacecraft spin axis. Adequate baffling will be implemented and a reusable aperture door is being considered.

In order to approach the selection of a suitable optical design, the required optical aperture and focal length were examined. Modeling (see Section 0, Sample Sensitivity Calculations) shows that in order to achieve the required sensitivity of 30 rayleighs per exposure, an aperture area of 6 $cm^2$ is required, translating into a diameter (unobscured) of 27.6 mm ($D_A$). The 25 degrees field-of-view projects onto the focal plane where an image intensifier photocathode is located. The maximum active diameter photocathode currently suitable for space applications is 40 mm. In order to project a FoV of 25 degrees onto a 40 mm diameter focal plane ($D_{FP}$), a telescope effective focal length (fl) of 90.2 mm is required. This is calculated from:

$$fl = D_{FP}/2 \, / \, \tan(FoV/2)$$

The *f*-ratio (fl/$D_A$) of the system is therefore *f*/3.3, a fast, but entirely realizable system.

Note that geometric distortion is a significant design issue in any fast optical system. For a spinning satellite, if the Time Delay and Integration (TDI) imaging mode is used, the geometric distortion must be sufficiently low to not cause blurring of the image during integration. The apparent speed of the image on the CCD must be constant. Geometric distortion in the optics will cause change in this speed, in turn causing a mismatch between CCD row number and the corresponding position at the target auroral feature. If the full 1024 pixels (or rows) over the FoV is used (high resolution mode), then the geometric distortion (deviation from zero error) must be less than about ±0.2 percent (1/512) peak, although this would need confirmation by simulation of the TDI process to determine how much distortion is acceptable before noticeable blurring occurs. Distortion would occur in both the telescope optics and the fibre-optic coupling between the intensifier phosphor screen and the CCD. Image deconvolution by digital post-processing may be possible, but this would have to be investigated by simulation in the next phase, to determine its benefits and how much distortion-induced blurring could be removed for the actual signal-to-noise ratios involved. Given the relatively low SNRs (e.g., 3:1), such deconvolution may indeed be useful, as deconvolution tends to add noise to high-SNR but blurred images, but may not noticeably add noise to already low SNR images with blurring. At this stage, however, a 30 fps fast-framerate (as opposed to TDI) readout mode is baselined, yielding significant flexibility when it comes to digital real-time correction of distortion and misalignment of the CCD array to the spacecraft and spacecraft spin plane. This mode is currently being successfully employed by the IMAGE WIC instrument.

If the optical *f*-ratio is further increased in an attempt to make the optics even more feasible, the focal length will become larger (keeping the aperture the same), and hence the focal plane area will increase, increasing instrument physical size. In theory, a fiber optic taper could be used in front of the image intensifier in order to improve the speed of the system while still keeping it compact, but there would be additional loss of light, which is highly undesirable. Further studies are needed.

## *Heritage two-mirror design*

The "classic" Canadian 2-mirror inverted Cassegrain camera flown successfully on Viking, Freja, Interball, and IMAGE-WIC is illustrated in Figure 5.16. This 22.4 mm, f/1.0, 25 degree field of view Burch-design is not considered suitable as-is for separately imaging LBH-long and LBH-short, as required by the Ravens mission objectives. The use of all-dielectric filters on the mirrors, as was done on the IMAGE WIC instrument, does not provide sufficient rejection of the longer visible wavelengths to enable quantitative analysis to be performed on the imager data: IMAGE WIC images are severely contaminated by out-of-band-contributions, while there is much less leakage on Polar UVI, which is a 3-mirror system [Jim Spann, personal communication, 2005].

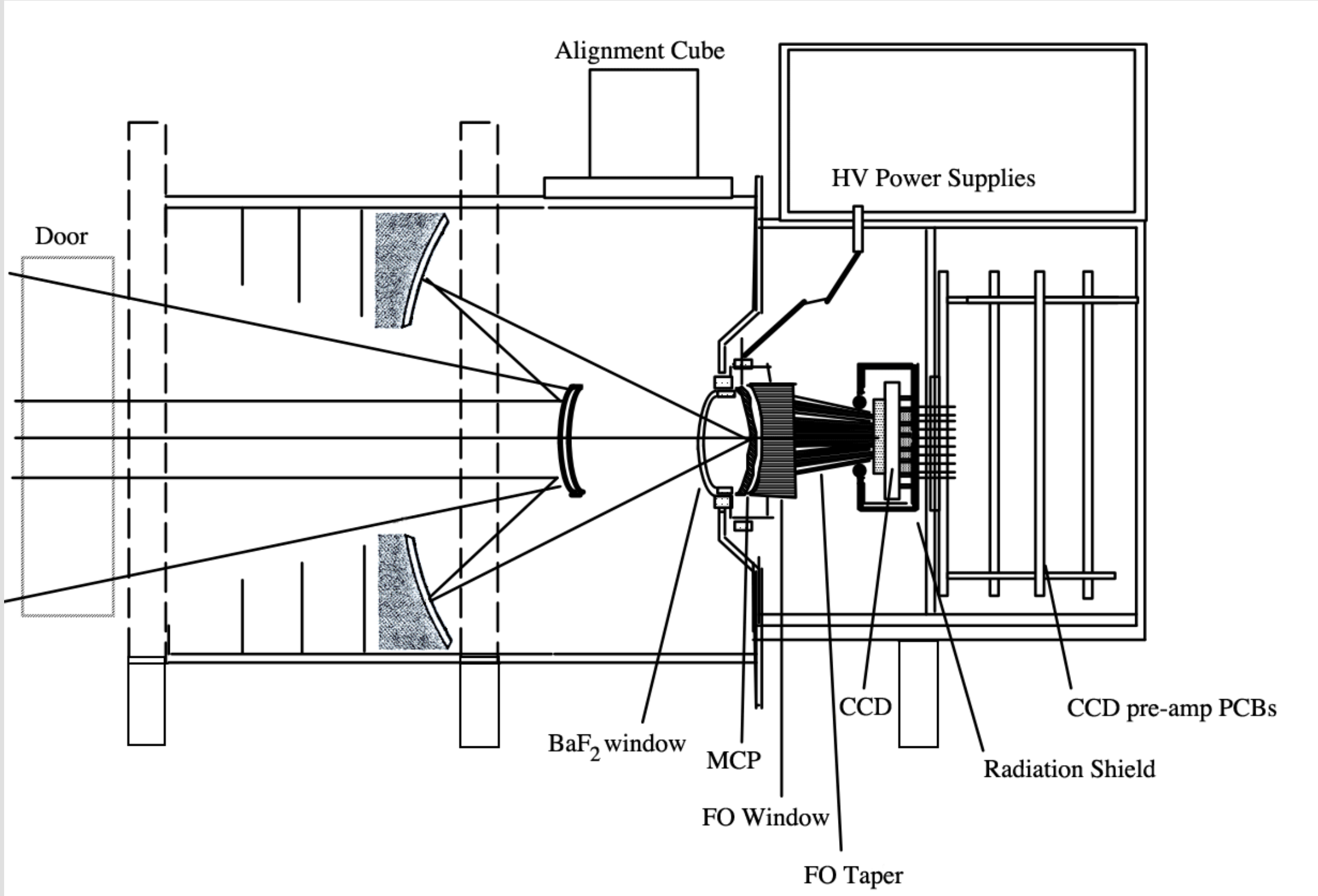

Figure 5.16: The University of Calgary 2-mirror Inverted Cassegrain design, used on Viking, Freja, Interball, and the NASA IMAGE mission.

The design may however still be considered useful for many kinds of non-quantitative applications. Flattening the curved focal plane would constitute a significant enhancement of this design, as will be touched upon in more detail in Section 0. A design study for the 2-mirror system is currently being conducted as a collaborative project between the University of Calgary and the Chinese Academy of Sciences. The outcome of this study will enable an assessment of the feasibility of flat focal plane designs. It is expected that the Chinese will then proceed to construct and test a model in anticipation of a future space mission.

One possibility for focal plane flattening in a 2-mirror camera is the introduction of an appropriate refractive element. The obvious problem with this approach is the high dispersion of materials used in the FUV. This problem goes away for cameras operated in a monochromatic mode. In the case of the present project, the first step in assessing this option is to look for suitable lens materials for which the refractive index does not change significantly over the passband of the instrument. If such a material can be found, then other properties would have to be investigated, such as resistance to radiation darkening and ease of fabrication. If such a material can be found, then there are two approaches to field-flattening: the detector window could be configured as a lens: i.e., the inside and outside surfaces would have different radii of curvature (in the simplest case); or a lens could serve as the entrance window to the imager. This lens could be coated to reflect all out-of-band light or could serve as a short-wavelength photon absorber.

### *Three-mirror design*

The UVAMC ultraviolet imagers are required to operate over what is an extremely wide field of view for most reflective telescope designs , ±12.5 degrees, with a speed of at least *f*/3.3 and an effective focal length of 90 mm. Since sensitivity is a driving factor, the design should be an off-axis, all-reflective system with no obscurations. Instrument size is driven by the resolution requirements, the required focal length, and field-of-view within which distortion must be minimized.

An initial survey of possible design solutions found that the more compact the off-axis design, the greater the distortion. There is thus a limit as to how compact the design can be and still have reasonable distortion performance. So, a maximum distortion of 10% is assumed as a starting point.

Given that this is an ultraviolet instrument, the surfaces of the mirrors must be optimized for smoothness. This is because surface scatter scales inversely with wavelength; the shorter the wavelength, the more significant the surface scatter. Such scatter is a significant factor in limiting the performance of instruments operating in the ultraviolet. The use of aspherical mirrors and mirrors with extreme conics should be limited as it is more difficult to manufacture extremely smooth surfaces with these types of mirrors than it is with spherical mirrors.

As will be explained in Section 0, UVAMC will operate in the "self filtered" mode, i.e., with reflective interference filter coatings deposited directly onto mirror surfaces. However, the larger the range of incident angles of the light a filter is to act upon, and the more the range of incident angles varies across the aperture of the filter, the more complex the filter design. The more complex the filter, the more thin film layers are required, potentially leading to manufacturing difficulties. Hence, for filter performance, surface complexity (shape as well as radius of curvature) must be limited.

One possible solution is the design depicted in Figure 5.17. This is a modified eccentric off-axis Schwarzschild telescope with an *f*-number of 3.1 and a focal length of 90 mm. Normally, in an Schwarzschild telescope, the stop is located just in front of the image plane, but given the large field-of-view of this instrument, the stop was moved further up in the system to reduce the size of the mirrors.  Given the limited degree of freedom in this design, both the primary and the tertiary mirrors are conics. The pupil is located at the secondary mirror to reduce the size of the other mirrors.  The tertiary mirror, located near where the stop is normally located, is a flat mirror.  Correction is adequate in this version of the design over the whole field-of-view (but further improvement is possible by aspherizing the mirror).  The distortion is of the order of 4%.  However, because of the large field-of-view and the location of the stop, the image plane and hence the micro-channel plate, must be curved.  Curving the tertiary mirror may reduce the curvature of image plane, at the cost of instrument performance and complexity.

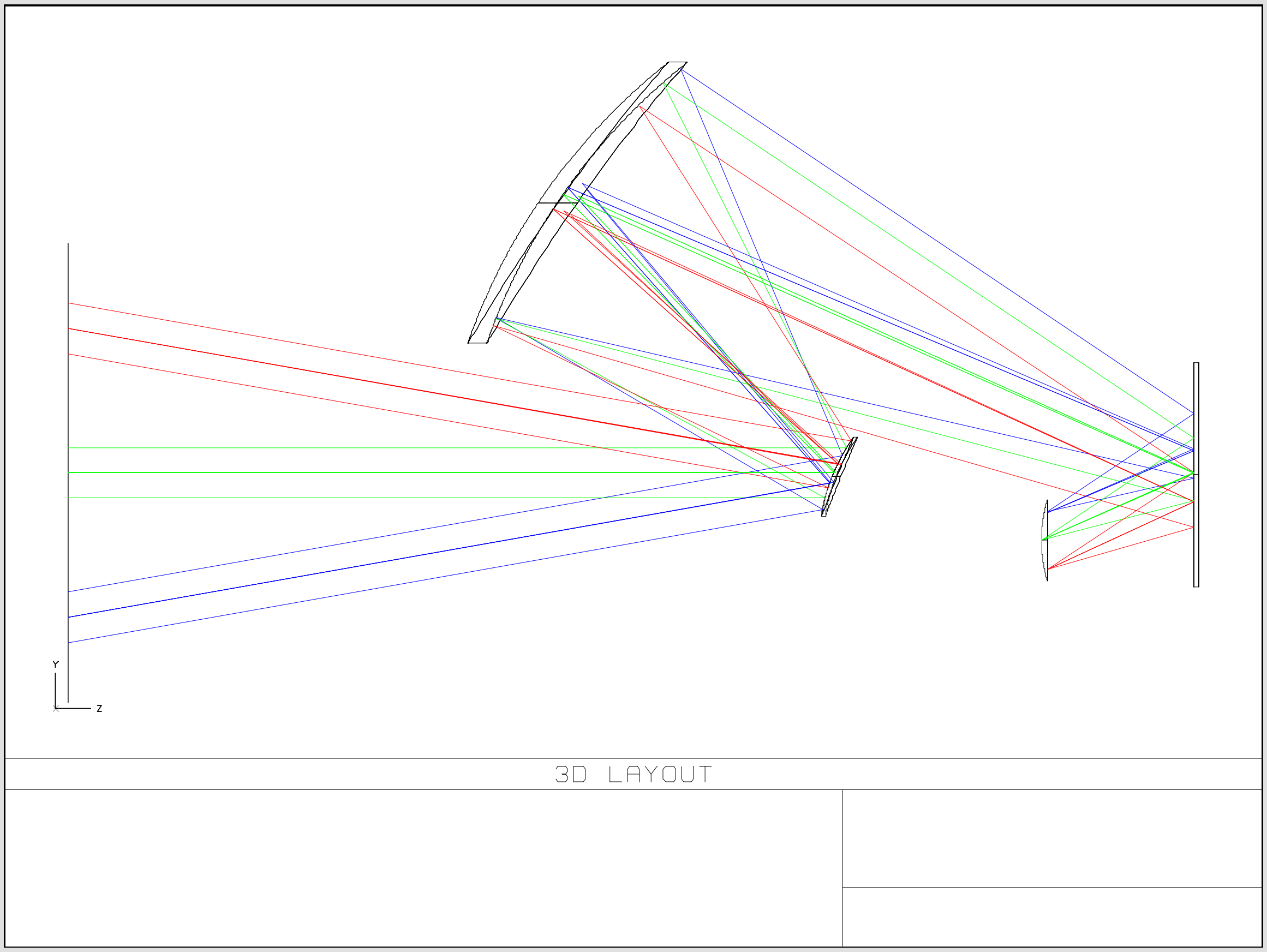

Figure 5.17: Off-Axis Eccentric Modified Schwarzschild

## *Four-mirror design*

Further modification to the off-axis design of Figure 5.17 results in the four-mirror wide-angle Wetherell-Womble design shown in Figure 1.18.  In this telescope, all mirrors are spherical, with the stop at the secondary mirror. The degrees of freedom required for correction comes from the fact that the mirrors do not have a common optic axis; they are all tilted with respect to each other.  Distortion performance is very good, considering the large field-of-view.  Distortion can be as low as 2%, depending on the tilt angle of the first mirror. There is also some latitude to adjust the curvature and tilt angles of the various mirrors in order to minimize the range of incident angles of light illuminating each optical component, for optimum coating performance. However, because the mirrors surfaces are less complex than those of the first design, the system is physically larger (longer) than the first design. It has the advantage of not employing conical surfaces making filter manufacturing easier. Importantly, the image plane is not curved, thereby greatly reducing the complexity and cost of the micro-channel plate image intensifier.

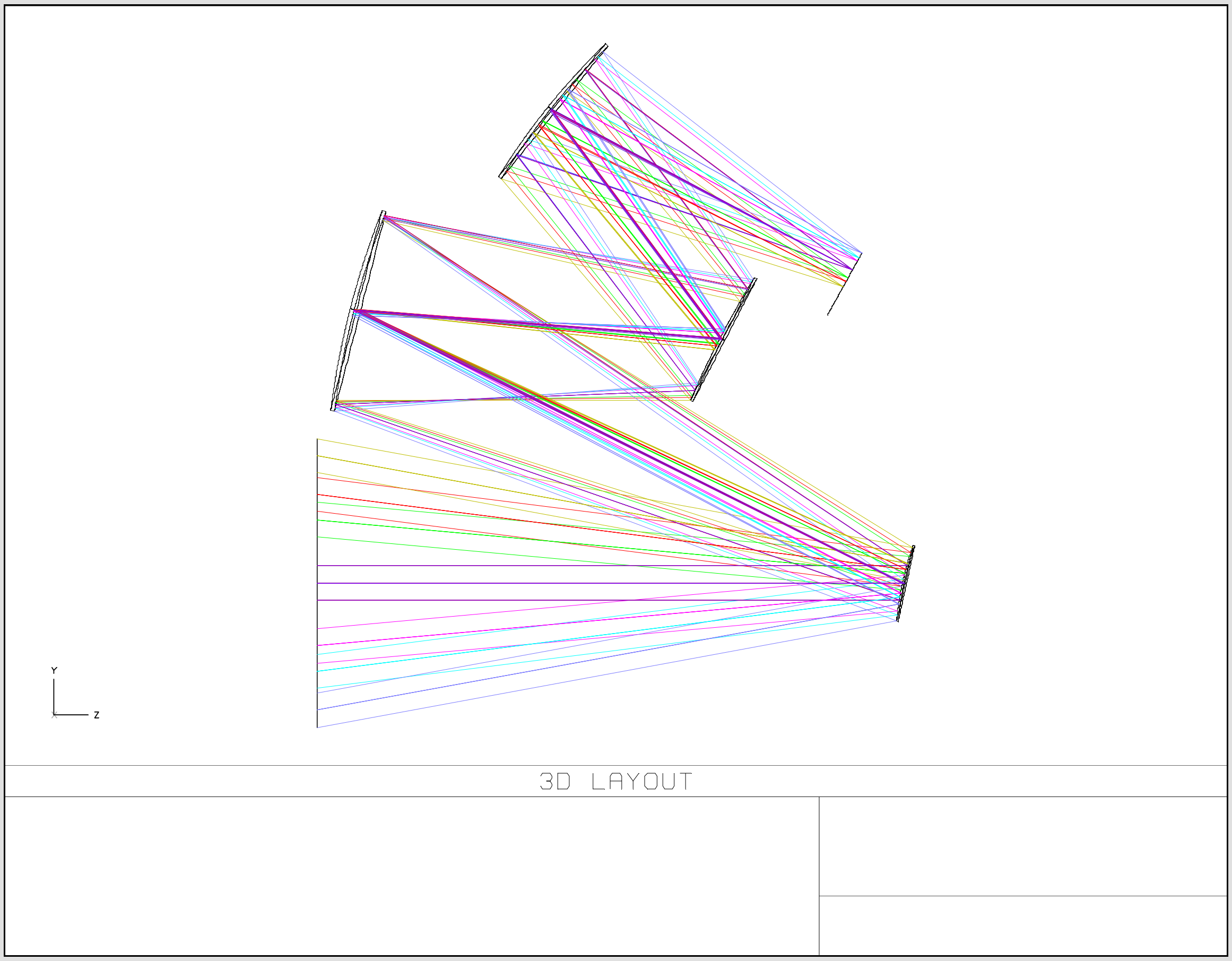

Figure 1.18: Wetherell-Womble Design

## *Discussion*

Three designs have been discussed. As will be shown in Section 0, the 4-mirror Wetherell-Womble is indeed the best choice if LBH wavelength separation and sufficient visible light rejection for quantitative analysis of data are required. A modified version of the 2-mirror Burch design may still be well suited in situations where mass and volume budgets are limited and LBH wavelength separation (and quantitative data analysis) can be sacrificed.

## 5.8 Selection of Optical Materials

### Filters

The UVAMC cameras will be operating in the "self filtering" mode, in which the imaging mirrors themselves are coated with all-dielectric multilayer filters which only reflect the desired wavelength. As such, the filtering and imaging properties are combined, helping reduce instrument size and mass while at the same time helping improve instrument performance. Detecting weak FUV emissions in the presence of bright visible light poses a significant design challenge. As mentioned earlier, and shown here in Figure 5.19, the integrated visible-to-infrared albedo is on the order of $10^{11}$ R.

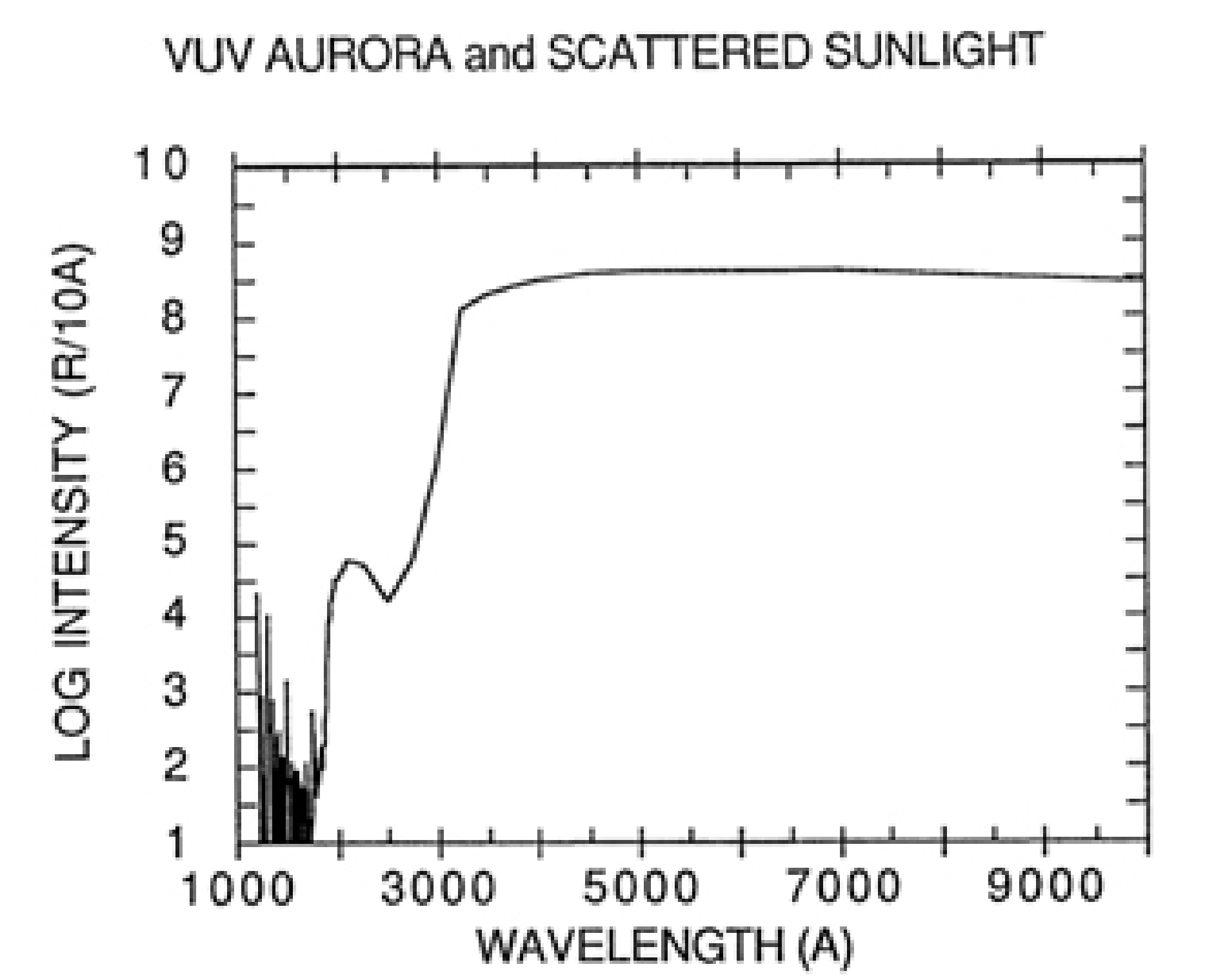

Figure 5.19: Brightness seen by nadir-viewing instrument looking at the sunlit Earth [Torr et. al, 1995].

To measure a signal on the order if 10 R, the long-wavelength continuum must be reduced by at least a factor of $10^{10}$. Such rejection has been attempted using various combinations of aluminized transmissive filters (DE-1), all-dielectric filters (Polar UVI, IMAGE WIC) and dispersive gratings (IMAGE SI). UVAMC will employ all-dielectric filters. Modern dielectric multilayer mirror coatings [Spann, 2005] will be used, having very low reflectivity in the Visible and Near-UV region and manufactured for high reflectivity in the LBH-short or LBH-long wavelength region, respectively. Most out-of-band wavelength photons impacting on the reflective surface of the mirrors (high quality fused silica or Zerodur) are either absorbed by the glass or pass through the mirror and are absorbed by absorptive coatings applied to the rear surface of the mirror. Using the latest filter technology, only 4% of out-of-band wavelength photons are reflected and eventually imaged with in-band photons. The reflectivity ratio is the ratio of mirror reflectivity of out-of-band photons to in-band photons. Typical peak reflectivity of these filters are 0.85, as shown in Figure 5.20. The reflectivity ratio rapidly decreases with multiple reflections, as $0.05^m$, where m is the number of reflective filters used. In the case of UVAMC, m=4, yielding an out-of-band rejection of about 6 x $10^{-6}$. Used in conjunction with a solar blind photocathode (see next item), a visible light rejection of $10^{10}$ can be accomplished.

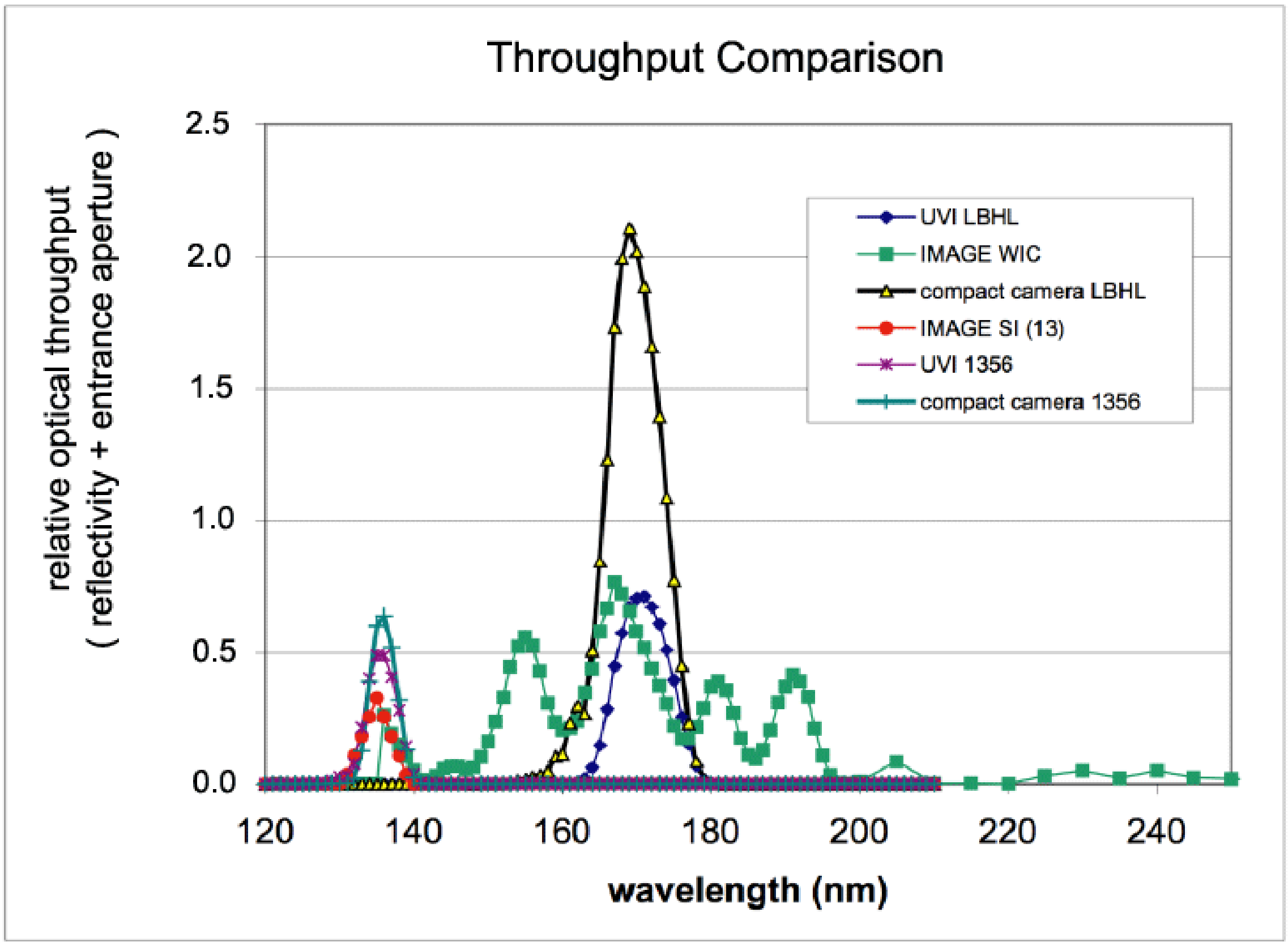

Figure 5.20: Traced in black is the passband of state-of-the-art LBH-long filter designed by Jim Spann, Marshal Space Flight Center. Average reflectance longward of 200 nm is 3.6%, reflectance at center of passband is greater than 80%. LBH-short filter has the same performance data. Similar technology will be employed for the Ravens UVAMC.

## Photocathode

A CsI (solar blind) MCP photocathode provides an additional factor of more than $10^4$ rejection longward of 200 nm. The CsI must be manufactured to the highest purity standards (~$10^6$ rejection has been shown to be achievable).

## Windows

The short-wavelength cutoff will be achieved by proper selection of detector window material. For LBH-short, a $BaF_2$ window will be used. For LBH-long, a UV-grade silica ($SiO_2$) is considered a suitable material.

## Sensor

A micro-channel plate (MCP) image intensifier with a solar-blind (CsI) photocathode placed at the focal plane of the cameras is illuminated by the FUV image formed by the all-reflecting optics. Rather than an open intensifier as was used in the VIKING UV Imager, a closed (i.e., sealed) design is employed in order to avoid contamination of sensitive elements (e.g., the photocathode) and saturation due to ambient electron fluxes. The sealing is accomplished by placing a $BaF_2$ and $SiO_2$ window in front of the photocathode for the LBH-short and LBH-long imagers, respectively. For structural reasons, the windows must be relatively thick; in fact a minimum of 2.5 mm of $BaF_2$ may be needed in order to provide both the structural strength and the short wavelength cut-off for the LBH-short imager. The required thickness of the $SiO_2$ window will be investigated in the next phase.

In order to maximize the resolution of the intensifier, the gap between the photocathode and the microchannel plate front needs to be minimized. However, if the gap is small, the photocathode "plate" must be stiff such that the launch vibration does not cause the photocathode to flex and touch the MCP surface. Appropriately designed 40 mm diameter flight intensifier units are currently being developed and qualified by Photek (U.K.) for the UVIT instrument on the India Astrosat mission [ROUTES AstroEngineering, personal communication, 2005].

Once a photon passes through the window, it impinges upon a photocathode material (CsI) designed to be photoemissive over a particular wavelength region. Once a photoelectron is confined to a MCP pore, it is accelerated by a potential of several kV to produce a cascade of electrons out from the rear of the MCP. This acceleration is necessary for the electrons to traverse an aluminum coating overlaid on a P43 phosphor. The purpose of the coating is to shield the subsequent optical chain from light which penetrates (without photoconversion) the MCP. A P43 anode screen is baselined in spite of its low efficiency (30% of P20) because of its particularly fast decay time (decay to 10% in 1.5 ms). This is required because of the relatively high spin rate of the spacecraft: since a pixel (a resolution element on the ground) sweeps through the field of view at a fairly high rate, it is required that the response of the phosphor due to one pixel must be sufficiently decayed before the next pixel arrives. P43 phosphor has a peak wavelength of 550 nm, which also happens to be the wavelength of maximum responsivity of the baselined CCD.

Once the electrons strike the phosphor, photons are produced. These photons are then fiber-optically coupled to the CCD. The proposed CCD is a 1024×1024 pixels scientific-grade frame transfer chip (e2v CCD47-20), which can be binned down to 512×512, yielding a ~32 km spatial resolution from apogee with a 25 degree field of view. The CCD is operated in fast readout mode and will not require a mechanical shutter. The CCD47-20 has 13-micron square pixels and therefore has an image dimension of 13.3 mm square. With a 40 mm diameter phosphor screen on the image intensifier, a fibre optic taper ratio of 3:1 would be required to reduce the image size of the screen to the size of the CCD. A 1:1 ratio is preferred to minimize geometric distortion. e2v does make 2k (H) x 4k (V) pixel CCDs that are side buttable to allow the use of a 1:1 taper, but this would likely be an expensive undertaking.

The spatial resolution of MCPs can often be the limiting factor in an intensified imaging system. In the case of Photek image intensifiers, these typically have a resolution of 22 line pairs (lp) per mm. If the active diameter of the intensifier is 40 mm, then there are a total of 880 line pairs resolvable across the image area. If a typical imaging system sampling factor of 2.4 pixels per line pair is used, then there are a total of 2112 pixels of "resolution" available in the Photek intensifier. The MCP is thus not a limiting factor in this case. Even a 2048 pixel CCD would be a good match to the intensifier, assuming the fibre optics taper between the intensifier and the CCD does not become the resolution-limiting factor (and, indeed, the taper will have to be chosen such that it does not become the limiting factor).

In regards to dark noise, in order to achieve 1 electron rms noise for an exposure time of 2.1 seconds, a dark signal of 0.48 electrons/pixel/s is required, corresponding to a temperature of about –15 deg C, from the datasheet graph for the CCD47-20 AIMO (advanced inverted mode operation) back-illuminated. This is not an unreasonable temperature for a passively cooled system. Adequate cooling may be accomplished by using a radiator that faces directly towards Earth, supplemented by low-power active cooling (or heating) from a TEC. This solution provides for best control of the CCD temperature at a fairly low power and mass penalty. Another solution is to use a radiator at the back of the spacecraft and having the spacecraft attitude control ensure that the radiator never sees the Sun. This is attractive in terms of required radiator area, but could be heavy, depending on final spacecraft topology. The spacecraft baseplate may be required to remain at a given specified temperature, from which a thermoelectric-cooler $\Delta T$ can be designed. These issues will have to be revisited at the appropriate time. The power figures given later, in Section 0 on Spacecraft Resource Requirements, include provision for some low-power active cooling. Temperature sensors will be mounted at strategic locations, and temperature will be continuously monitored.

## *5.9 Instrument Sensitivity Performance*

### Sample Sensitivity Calculation

A variety of modeled and calculated parameters pertaining to the presented 2, 3, and 4-Mirror designs are tabulated in Table 2 and Table 3, for LBH-S and –L, respectively. The tables are largely self-explanatory, but please see "Optical Efficiency" and "Signal to Noise (SNR)", below.

Table 2: Modeling of LBH-short sensitivity and calculation of some relevant parameters for 2-, 3- and 4-mirror all-reflective (self filtering mode) optical designs.

| ***LBH-SHORT CALCULATIONS (UVAMC-0)*** | | **2-Mirror** | **3-Mirror** | **4-mirror** |
|---|---|---|---|---|
| **Exposure** | CCD superpixels (2 x 2 binned) | 512 x 512 | 512 x 512 | 512 x 512 |
| | Imager Field of View [degrees] | 25 | 25 | 25 |
| | Angular superpixel FoV [degrees] | 0.049 | 0.049 | 0.049 |
| | Spin Period (Image Cadence) [s] | 30 | 30 | 30 |
| | Spin Rate [degrees/second] | 12.0 | 12.0 | 12.0 |
| | Effective Exposure time [seconds] | **2.1** | **2.1** | **2.1** |
| **Optical Efficiency** | Mirror reflectivity | 0.85 | 0.85 | 0.85 |
| | Resulting reflectivity | 0.72 | 0.61 | 0.52 |
| | Intens. window transmission ($SiO_2$) | 0.82 | 0.82 | 0.82 |
| | Photocathode QE (CsI) | 0.21 | 0.21 | 0.21 |
| | MCP efficiency | 0.60 | 0.60 | 0.60 |
| | Throughput | **0.075** | **0.063** | **0.054** |
| | Out of band rejection | **2.5E-03** | **1E-04** | **6E-06** |
| **Optical Params.** | Aperture [cm^2] | 2.6 | 6 | 6 |
| | Diameter [mm] | 22.4 | 27.6 | 27.6 |
| | Assumed MCP diameter [mm] | 13.4 | 40 | 40 |
| | f/# | 1.0 | 3.3 | 3.3 |
| | Pixel Omega [square pixels] | 7.26E-07 | 7.26E-07 | 7.26E-07 |
| **Signal to Noise (SNR)** | Radiance-kR conversion factor | 7.96E+07 | 7.96E+07 | 7.96E+07 |
| | Calculated Sensitivity [counts/kR/s] | 11.2 | 22.0 | 18.7 |
| | Counts/R/pixel/exposure | 0.023 | 0.046 | 0.039 |
| | **Noise Equivalent Signal [R/exp]** | **42.8** | **21.8** | **25.7** |
| | Desired SNR | 2.5 | 2.5 | 2.5 |
| | Desired SNR requires brightness [R/exp] | 270.0 | 135.0 | 160.0 |
| **Spatial Resolution** | 1 RE [m] | 6.38E+06 | 6.38E+06 | 6.38E+06 |
| | Satellite Apogee [Re] | 7 | 7 | 7 |
| | Derived Altitude [km] | 38280.0 | 38280.0 | 38280.0 |
| | Earth angle from above altitude [deg] | 16.43 | 16.43 | 16.43 |
| | Auroral Oval subtends from alt [km] | 1048 | 1048 | 1048 |
| | Nadir resolution at apogee [km] | **32.6** | **32.6** | **32.6** |
| **Data Rate** | Bits per pixel | 12 | 12 | 12 |
| | Total bits/image | 3.15E+06 | 3.15E+06 | 3.15E+16 |
| | Telemetry rate, no compr. [Mbits/s] | 0.10 | 0.10 | 0.10 |
| | Telemetry rate, compressed [kbits/s] | 30 | 30 | 30 |

Table 3: Modeling of LBH-long sensitivity and calculation of some relevant parameters for 2-, 3- and 4-mirror all-reflective (self filtering mode) optical designs.

| ***LBH-LONG CALCULATIONS (UVAMC-1)*** | | **2-Mirror** | **3-Mirror** | **4-Mirror** |
|---|---|---|---|---|
| **Exposure** | CCD superpixels (2 x 2 binned) | 512 x 512 | 512 x 512 | 512 x 512 |
| | Imager Field of View [degrees] | 25 | 25 | 25 |
| | Angular superpixel FoV [degrees] | 0.049 | 0.049 | 0.049 |
| | Spin Period (Image Cadence) [seconds] | 30 | 30 | 30 |
| | Spin Rate [degrees/second] | 12.0 | 12.0 | 12.0 |
| | Effective Exposure time [seconds] | **2.1** | **2.1** | **2.1** |
| **Optical Efficiency** | Mirror reflectivity | 0.85 | 0.85 | 0.85 |
| | Resulting reflectivity | 0.72 | 0.61 | 0.52 |
| | Intens. window transmission ($BaF_2$) | 0.76 | 0.76 | 0.76 |
| | Photocathode QE (CsI) | 0.15 | 0.15 | 0.15 |
| | MCP efficiency | 0.60 | 0.60 | 0.60 |
| | Throughput | **0.049** | **0.042** | **0.036** |
| | Out of band rejection | **2.5E-03** | **1E-04** | **6E-06** |
| **Optical Params.** | Aperture [cm^2] | 2.6 | 6 | 6 |
| | Diameter [mm] | 22.4 | 27.6 | 27.6 |
| | Assumed MCP diameter [mm] | 13.4 | 40 | 40 |
| | f/# | 1.0 | 3.3 | 3.3 |
| | Pixel Omega [square pixels] | 7.26E-07 | 7.26E-07 | 7.26E-07 |
| **Signal to Noise (SNR)** | Radiance-kR conversion factor | 7.96E+07 | 7.96E+07 | 7.96E+07 |
| | Calculated Sensitivity [counts/kR/s] | 7.4 | 14.6 | 12.4 |
| | Counts/R/pixel/exposure | 0.015 | 0.030 | 0.026 |
| | **Noise Equivalent Signal [R/exp]** | **64.6** | **33.0** | **38.8** |
| | Desired SNR | 2.5 | 2.5 | 2.5 |
| | Desired SNR requires brightness [R/exp] | 405.0 | 205.0 | 240.0 |
| **Spatial Resolution** | 1 RE [m] | 6.38E+06 | 6.38E+06 | 6.38E+06 |
| | Satellite Apogee [Re] | 7 | 7 | 7 |
| | Derived Altitude [km] | 38280.0 | 38280.0 | 38280.0 |
| | Earth angle from above altitude [deg] | 16.43 | 16.43 | 16.43 |
| | Auroral Oval subtends from alt [km] | 1048 | 1048 | 1048 |
| | Nadir resolution at apogee [km] | **32.6** | **32.6** | **32.6** |
| **Data Rate** | Bits per pixel | 12 | 12 | 12 |
| | Total bits/image | 3.15E+06 | 3.15E+06 | 3.15E+16 |
| | Telemetry rate, no compr. [Mbits/s] | 0.10 | 0.10 | 0.10 |
| | Telemetry rate, compressed [kbits/s] | 30 | 30 | 30 |

## *Optical Efficiency*

As described earlier, the instruments will be designed to operate in the self-filtering mode. Thus, the center-of-passband mirror reflectivity used here (0.85) is based on the properties of state-of-the-art multilayer dielectric filter coatings [Spann, 2005]. The number in the next row, labeled "Resulting Reflectivity," is calculated as $0.85^n$**,** where n is the number of reflecting surfaces used in the design (the three rightmost columns are for 2-mirror, 3-mirror, and 4-mirror design, respectively). The next row is the intensifier window transmission. In the case of LBH-short, a $SiO_2$, window material is used, with a 0.82 transmission at 150 nm ($BaF_2$ has a transmission of only 0.59 at 150 nm, for the required window thickness). In the case of LBH-long $BaF_2$ is baselined, with a transmission of 0.76 at 170 nm. The next row shows the CsI photocathode quantum efficiency at the center of the respective passbands (0.21 for LBH-short, 0.15 for LBH-long). The MCP FUV efficiency is listed next (approximately 0.6), and the total (effective) transmission is calculated in the next line as the product of values above. Finally, the out of band rejection is calculated. As already mentioned, about 4% of out-of-band photons are reflected and imaged with in-band photons. The reflectivity ratio is the ratio of mirror reflectivity of out-of-band photons (0.04) to in-band photons (0.85). The reflectivity ratio thus decreases rapidly with multiple reflections as $0.05^m$, where m is the number of reflective filters in the optical system (2, 3, and 4, respectively). Recall that an out-of-band rejection of better than $10^{10}$ is considered a requirement to meet science objectives. A solar blind photocathode (CsI) may provide a factor of $10^4$, which means that the rest of the optical system must provide the remaining $10^6$ of suppression. As can be seen, only the 4-mirror design satisfies this requirement. Note that the old 2-mirror design only provides $10^3$ rejection, which is still superior to that provided by the Acton "black coatings" originally used by the Viking, Freja, and Interball instruments (which did not operate in the self filtering mode).

## *Signal to Noise (SNR)*

Calculated sensitivity in counts per kilorayleigh per exposure is listed in this section of the table. The instrument's optical aperture was used in this calculation, and in the case of the 2-mirror design (the only on-axis design) the central obscuration was taken into account in order to arrive at an "equivalent" aperture. An important item here is the line tabulating Noise Equivalent Signal (NES), a figure often used for sensitivity comparison purposes. *As can be seen, in the case of the 4-mirror design (which is the only design with sufficient out-of-band rejection) the NES is in the neighborhood of 30 rayleighs per exposure, which is the original UVAMC sensitivity target (see Section 0).* This is the equivalent of one count per exposure and will require a very high gain image intensifier, operating in photon-counting mode. The required intensifier gain for UVAMC will need further investigation, with consideration being given to the intra-scene dynamic range required. It is possible that bright signal detection and automatic gain reduction are required.

## *Conclusion*

**We conclude that the 4-mirror design is the most viable candidate for the Ravens UVAMC instrument.**

## Instrument Calibration

It is clear from the above discussion of the auroral measurements that absolute calibration of the Ravens UVAMC is a requirement for the determination of the electron energy flux and the average electron energy. Previous Canadian FUV imagers have been characterized in the laboratory but not calibrated because they were developed for the purpose of providing morphological rather than physical measurements.

For the Ravens imagers the characterization and testing (point-spread functions, distortion, variations in responsivity with incident angle, etc.) will be carried out in a similar manner as done previously. This will require appropriate ground support equipment including a vacuum tank, positioning table, FUV light sources, monochromator etc. Where this will take place will be determined partly by the size requirement for the vacuum tank. A detailed verification and characterization plan will be developed as part of a later phase.

In addition to the above, there will be a need for a calibrated FUV source along with whatever is needed to ensure that the entrance aperture of the imager is completely filled. The absolute spectral responsivity will be measured by combining a monochromator and FUV source.

Once in orbit, variations in sensitivity can be monitored by means of star images.

## Performance Enhancement

Although the 4-mirror instrument design presented here appears to satisfy the criteria that will achieve the scientific objectives, there is potential within the estimated schedule to make substantial improvements, even breakthroughs, in FUV auroral imaging instrumentation. Consideration should therefore be given to spending effort on such potential enhancements. Some suggestions in this regard are outlined here. These suggestions have been prompted by a frank assessment of previous missions that have flown FUV imagers, whether Canadian or international.

**The overall sensitivity of previous and present designs is very low.** The major loss of sensitivity is in the detector system whose photocathode converts UV photons to electrons and whose microchannel plate intensifier accelerates them to excite a phosphor that is coupled to a CCD array detector. This arrangement results in an overall collection efficiency of well under 10% for FUV photons reaching the detector. Given this low quantum efficiency, there is an obvious challenge to look into possible radical departures from the traditional ICCD detectors that have been used for the Viking, Freja and Interball missions. One concept having potential merit is to consider a UV detector such as that sketched in Figure **5.21**5.21. The challenge is to provide a method of converting FUV photons to visible photons before they reach the photocathode. A window is selected for high reflectivity at wavelengths longer than 200 nm and high transmission at wavelengths shorter than 200 nm but longer than 130 nm; for example a magnesium fluoride ($MgF_2$) glass coated with appropriate materials to form a bandpass filter. The selected FUV photons then strike some type of photo fluorescent material that converts FUV photons to visible photons that are then transferred by means of a fibre optic block to a standard CCD. If the focal plane is not curved, the correcting fibre block may not be necessary. Depending on the efficiency of the photon converter, the transmission of the filter, etc., it seems feasible that the overall collection efficiency could be as high as 50%. The crucial test for this concept is whether appropriate photo fluorescent materials (phosphors, sodium salicylate etc.) that possess the necessary properties are available.

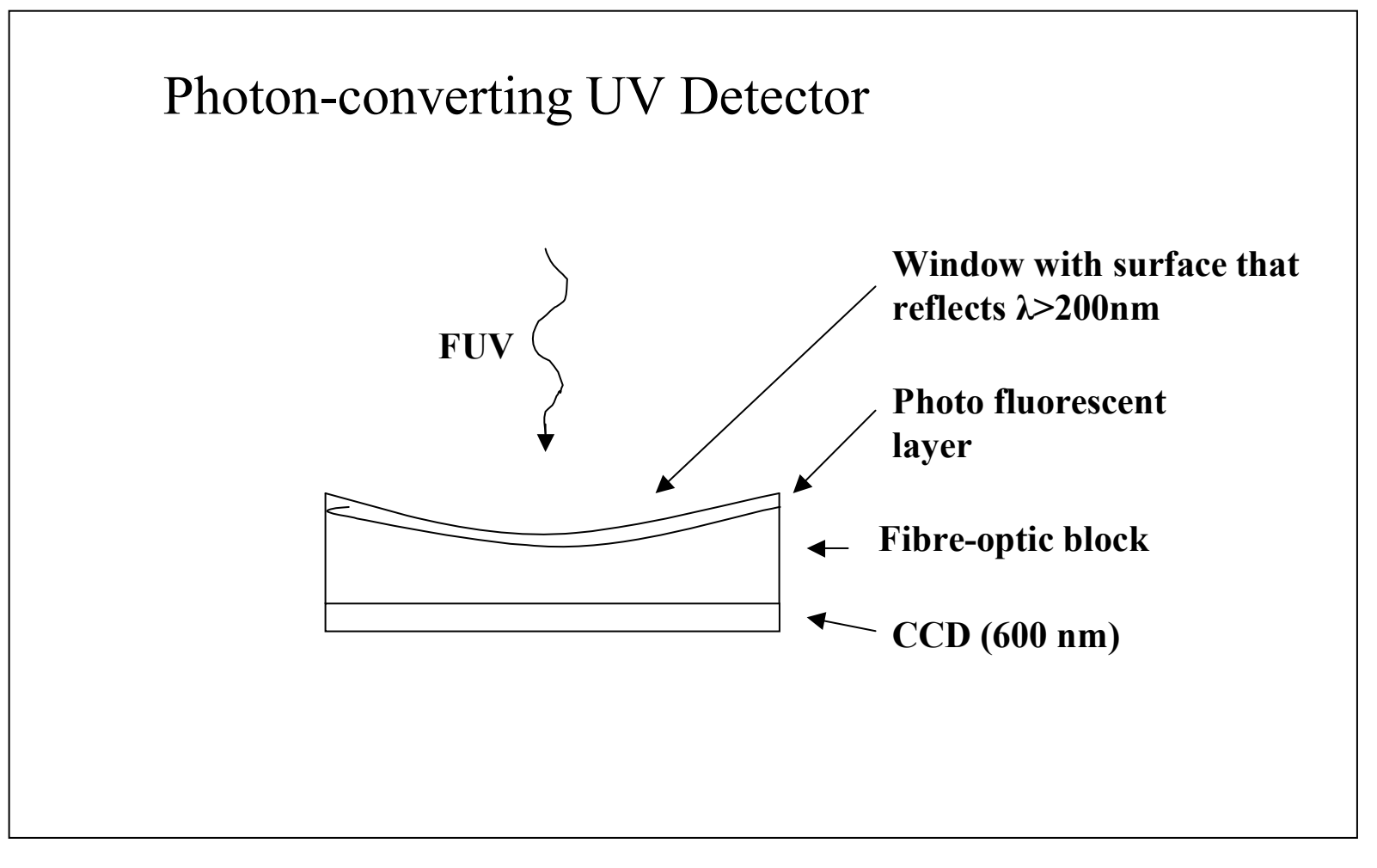

Figure 5.21: Alternate FUV detector concept.

Another possible remedy for the FUV sensitivity is to increase the FUV quantum efficiency of the photocathode of the conventional ICCD detector. Although the current plan is to use 2×2 binning on the CCD (resulting in 512×512 pixel images and a nadir spatial resolution of 32 km) in order to reach the desired sensitivity threshold, there is a very real possibility that soon-to-be-available GaN photocathode materials [Siegmund, 2005] would increase our optical efficiency by a factor of 3 to 5, resulting a NES sensitivity of 10 rayleighs per exposure in the case of the 4-mirror design. If this new technology reaches flight-readiness in time to be included in the UVAMC imaging systems, one might also try to avoid binning altogether, still achieving a better sensitivity than that baselined, and improving apogee nadir resolution to ~16 km.

**Mission lifetimes have been cut short due to failures in the high voltage system.** The presence of very high voltage (4 kV to 6 kV) constitutes a major risk, judging from experience with Canadian imagers. The novel concept outlined above (Figure 5.21) would eliminate the requirement for a microchannel plate intensifier and therefore for high voltage. Even if the above concept suffers from readout and thermal noise, it is feasible to introduce a standard MCP with a conventional visible-range photocathode that operates at a much lower and safer voltage (e.g. 1 kV).

**A curved focal plane is a complication.** Not only are curved microchannel plates expensive and of limited availability, but they also complicate the already difficult alignment procedure. It is therefore considered worth the effort to attempt to produce a camera design that has a flat focal plane. This was achieved in the Polar UVI 3-mirror design by combining elliptical and hyperbolic aspheric mirrors. However, the Polar camera field of view was only 8 degrees whereas the UVAMC requires at least 18 degrees. An initial design concept for UVAMC, utilizing four mirrors, has thus been presented in Section 0.

The above suggestions serve as examples of what pursuits could be carried out over the next few years in an effort to make major strides forward in the development of better FUV instruments***. It is only through improved technology that Canada can retain its prominent international position in the field of auroral imaging from space. It is also evident from history that advances in technology lead to advances in scientific knowledge.***

## *5.10 Future Studies related to the KuaFu Mission*

In Section 0 some suggestions were made regarding possible future studies that could lead to fairly radical changes in FUV instrumental design. This section deals with the current designs especially as they relate to the proposed Chinese KuaFu mission, possibly the first flight opportunity for the "Ravens" UVAMC imagers. There are a number of studies to be undertaken before a detailed design of the instrument can be initiated. The following list identifies some of these needed studies.

### Choice of basic optical configuration

Three concepts for the cameras have been identified, one utilizing 2 mirrors in the tradition of the previous Canadian FUV imagers, one using 3 mirrors, and finally one option utilizing 4 mirrors. At this point in time, only the 4-mirror design appears to be feasible for proper separation of LBH-long and LBH-short. It may however be wise to carry also the more compact 2-mirror design forward into the next phase when a selection will be made based upon more detailed considerations than have been possible so far. Examples of such considerations and tradeoffs are ease of manufacture, mass and volume, passband requirements, susceptibility to scattered light, ease of alignment, optical throughput, and thermal stability requirements.

There is considerable merit in addressing the trade-offs involved in altering the focal length of each design. Implications for the size of the detector and the field of view will necessarily be a part of this study.

## Choice of passbands

During this concept study the design team has chosen to measure the LBH brightness in two wavelength ranges. What is needed is an assessment of available filtering technologies to assist with the final selection of the camera passbands. Ideal passbands would transmit 100% of the radiation within the passband and 0% outside the passbands. Such filters do not exist and are not likely to be developed in the near future, so the best compromise must be selected. As discussed earlier, a very important consideration is the rejection of out-of-band radiation from the sunlit earth and other sources.

## Refinement of predicted performance

It will be necessary to improve the sensitivity calculations based on better information about the camera designs. The performance goals should be made consistent with the KuaFu mission; not just the scientific objectives, but also with the expected orbit and the various constraints imposed upon the instruments.

Related to the sensitivity prediction is the need to refine the anticipated uncertainty in the measurements. Of particular importance is the need to better estimate the various sources of noise. Some sources of noise are directly related to the instrument; however, as pointed out earlier, with the goal of determining the electron energy flux and the average electron energy comes the need to better understand the uncertainties in the analysis procedure. The sources of these uncertainties may be in the degree of validity of the necessary assumptions. They may be due to departures of the real molecular densities from the model atmosphere. They may be due to various sources of unwanted FUV photons such as from scattered light, proton precipitation, etc.

It is also important to be prepared to alter the chosen approach to determining the electron energy flux and average electron energy. Further study and developments in filtering techniques might indicate that one should measure OI 135.6 nm emission rather than LBH-short, for example. Such options are possible to adopt when the schedule permits a paced development, as is the case for a launch in 2012 as proposed for KuaFu.

## *5.11 Control Electronics*

The two Ravens imagers (UVAMC-0 and UVAMC-1) on each spacecraft will be controlled by one common electronics box, the UVAMC-E. The controller provides bias voltages, three phase clock signals to the CCD parallel and serial registers, signal processing to extract and digitize (16 bits, via a dual-slope integrator) the CCD output signal, and a fast serial interface (e.g., asynchronous RS-422) to the spacecraft main data handling unit. There will be built in a capability to command the imagers into a variety of operational modes (frame rates, binning, CCD and MCP gain). The design of the dual-slope integrator and clock drivers are optimized for fast settling time. As the electronics unit (and the preamplifier) will have more than sufficient speed, it will be possible to continuously clock the CCD and thus generate images in memory greater in extent than the 1024 rows of the CCD. This allows one to exceed the 25-degree field of view in the scan plane. Thus, a full 360-degree image could in principle be generated, useful for in-orbit sensitivity and spatial calibrations using stars. As mentioned, by utilizing on chip binning, signal to noise ratio can be increased during operational modes where some spatial resolution can be sacrificed. The instrument input power supply voltage will be +28VDC ± 6V. The instrument will survive indefinite application of input voltages between 0 and +40V, with no damage.

Instead of DSPs (as used on ePOP, MOST), one or more FPGAs will likely be used for the implementation of all digital logic and processing on the UVAMC instruments. FPGAs will generate all required CCD and video processing timing, receive data from the A/D converter(s), format and send the data to the satellite, receive and decode commands, store commands, parameters, and data (via external RAM if required). Real-time image data processing is also possible, if required.

Using a DSP may at first look seem like a quite attractive solution, especially as the more recent models are readily programmed using standard C-language and optimizing code generator tools. However there are currently few military-qualified DSP devices on the market. In addition, radiation dose and single event effects are not well known, if at all, for most DSPs. FPGAs on the other hand are readily available in military and space qualified versions, including radiation-hardened versions (notwithstanding possible ITAR restrictions). Even if non radiation-hardened FPGAs are used, single event mitigation techniques can likely be used to make such an FPGA suitable for the mission; the same cannot be said for a DSPs. Potential industrial partner Routes has considerable experience with FPGAs, including several in-orbit.

Regarding power dissipation, FPGAs are generally not very power efficient, however the power dissipation can be minimized using power efficient design techniques, e.g., local clock speed reduction and architecture optimization. DSPs are somewhat more power efficient in general than FPGAs, however there is usually little that can be done to reduce power in the DSP device, other than reducing memory cycles and hence power in external RAM connected to the DSP. It should also be noted that modern DSPs with their complex L1 and L2 caches are known to have caused program stalls, which might affect the generation of CCD- and other hardware-timing signals.

## *5.12 Spacecraft Resource Requirements*

Table 4.1 shows the initial resource requirements budget for UVAMC, assuming the 4-mirror Wetherell-Womble design.

**Table 4: Spacecraft resource requirements (preliminary) for new 4-mirror design.**

| | UVAMC-0 | UVAMC-1 | UVAMC-E |
|---|---|---|---|
| **Mass** [kg] | 9.0 | 9.0 | 3.0 |
| **Dimensions** [$cm^3$] | 25 x 24 x 15 | 25 x 24 x 15 | 20 x 20 x 10 |
| **Power** [W] | 4 | 4 | 3 |
| **Data Rate** [Mbits/s] | 0.1 | 0.1 | N/A |
| **Temperature Limits** | -40 to +60 C (surv.) -25 to +20 C (ops.) | -40 to +60 C (surv.) -25 to +20 C (ops.) | -40 to +60 C (surv.) -25 to +40 C (ops.) |

Table 5 shows resource requirements budget for UVAMC, assuming the traditional Canadian 2-mirror Burch design.

**Table 5: Spacecraft resource requirements for the "classic" 2-mirror design.**

| | UVAMC-0 | UVAMC-1 | UVAMC-E |
|---|---|---|---|
| **Mass** [kg] | 1.89 | 1.89 | 2.9 |
| **Dimensions** [$cm^3$] | 26 x 15 x 13 | 26 x 15 x 13 | 18 x 18 x 8.5 |
| **Power** [W, peak] | 2 | 2 | 3 |
| **Data Rate** [Mbits/s] | 0.1 | 0.1 | N/A |
| **Temperature Limits** | -40 to +60 C (surv.) -20 to +30 C (ops.) | -40 to +60 C (surv.) -20 to +30 C (ops.) | -40 to +60 C (surv.) -20 to +40 C (ops.) |

Note that these are preliminary ROM estimates that will need to be adjusted according to final design options that will be selected during a Phase A study. The spacecraft design must ensure that the combined conducted and radiated emissions from all sources will not adversely affect the correct scientific operation of the UVAMC instruments or degrade the performance of elements of the spacecraft bus subsystems.

## 5.13 Identifiable Risks and Risk Mitigation

At the completion of the Ravens Concept study, we have identified standard risks that must be dealt with in the design phases of the mission. As part of the study, a preliminary analysis of radiation dosage with implications (accounted for in the technical budgets above) for skin thickness was carried out. This is presented below, and in addition an expanded list of risks identified, which are (see expanded list which follows the Radiation subsection):

- Procurement Risks
- Cost Risks
- Schedule Risks
- High-voltage issues
- Exposure of detector to the Sun (orbit selection, reusable door)
- Stray light analysis (surface smoothness, dynamic range)
- Integrity of optical focus (through proper selection of materials and/or adequate thermal control)
- Cleanliness (dust, humidity issues)
- Drift of calibration baseline values due to gain issues, as experienced on IMAGE-WIC (one may use early-type FUV stars for characterizing long-term drifts)

### Radiation

The UVAMC radiation analysis was performed with ESA's SPENVIS analysis web based program. In the analysis, trapped proton and electron fluxes, trapped proton flux anisotropy, and solar proton fluences were considered as radiation sources. The total radiation presented in the results is the accumulated radiation after 2.6 years for a silicon material surrounded by aluminum sphere of given thickness. Using a starting RASC of 0° and 100° allowed the representation of a 4-day orbit that covered the entire globe once as demonstrated in 5.22 and Figure **5.23**5.23.

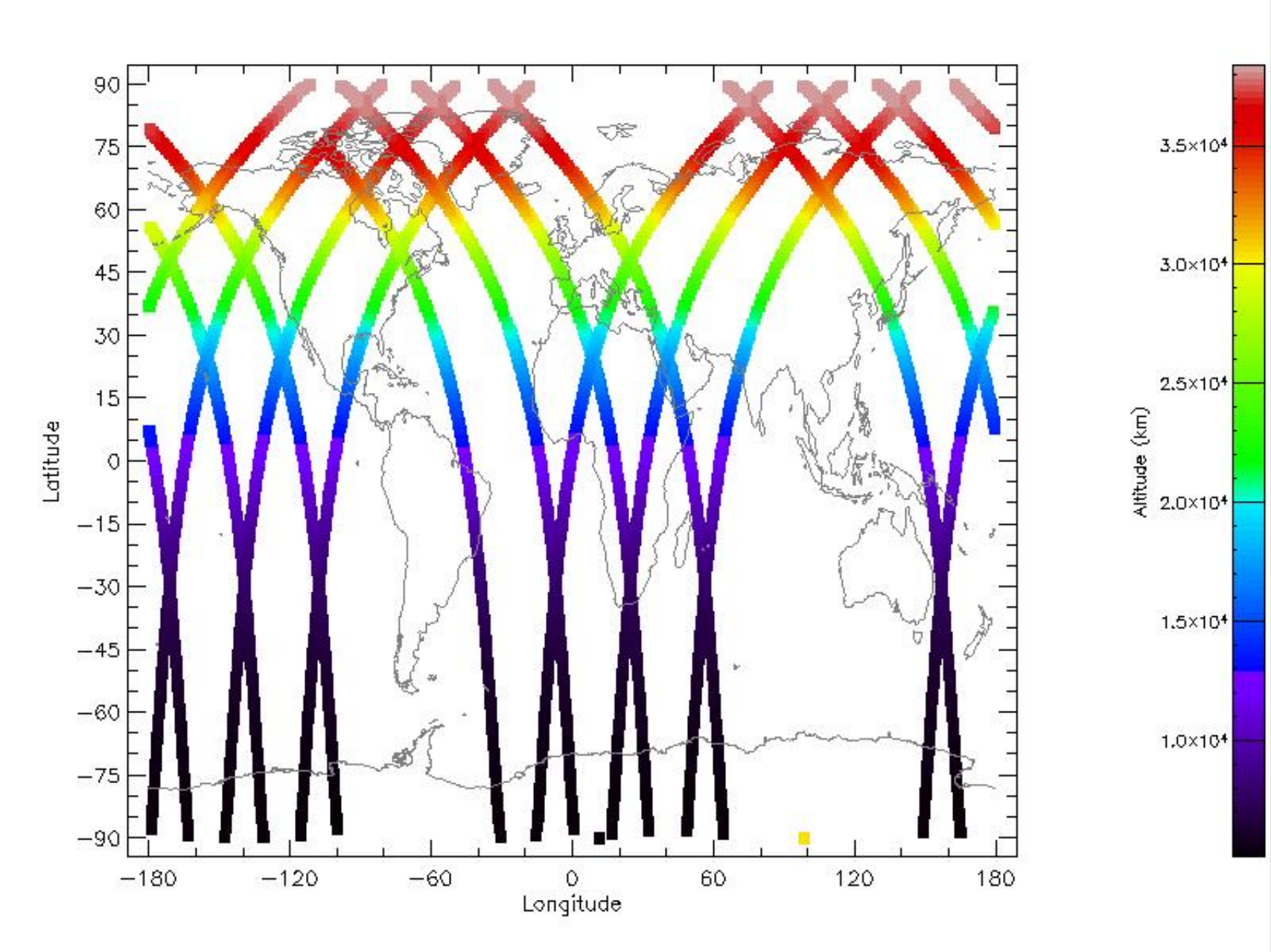

Figure 5.22: 4-day orbit spanning the globe (RAsc 0°). Altitude is represented by colors.

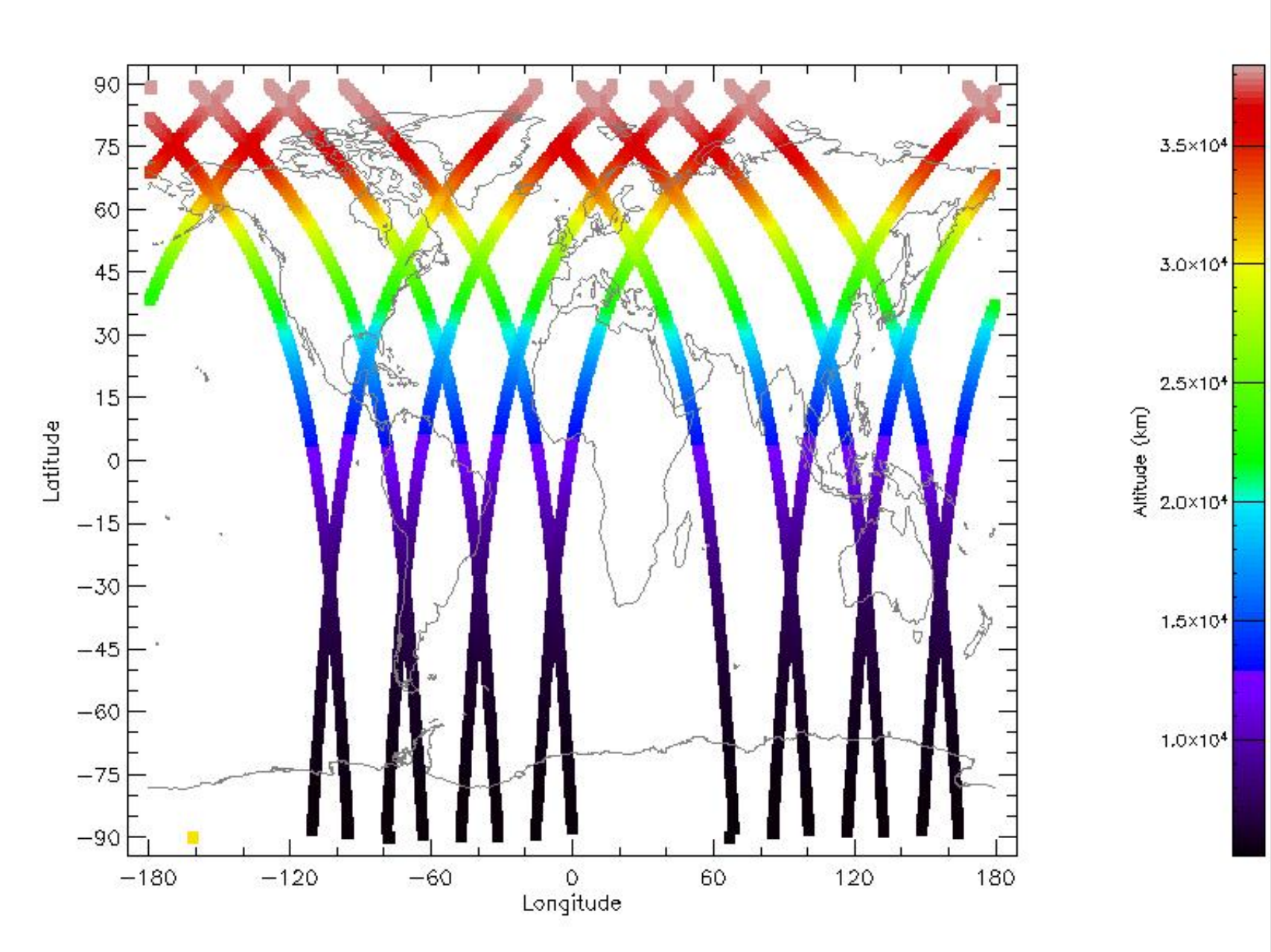

Figure 5.23: 4-day orbit spanning the globe (RAsc 100°).

The average total dose for these 4 days was then linearly scaled to represent the dose for 2.6 years for a particular argument of perigee. Three arguments of perigee were considered: 225°, 270°, and 315°. These three arguments were chosen as representative of possible spacecraft arguments during the 2.6 year mission. The average of these three arguments is what is represented as the total radiation shown in the figure below. Based on a total allowable dose of 5 krad over the 2.6 year mission, the model suggests that 8.5 mm of aluminum shielding would be needed. For the purposes of calculation of the mass of the instrument, one may assume that the spacecraft will provide approximately 2.5mm of shielding. Therefore, the walls of the instrument will be required to be ~6mm thick. This number was used in calculating the expected mass, as listed in Table 4.1

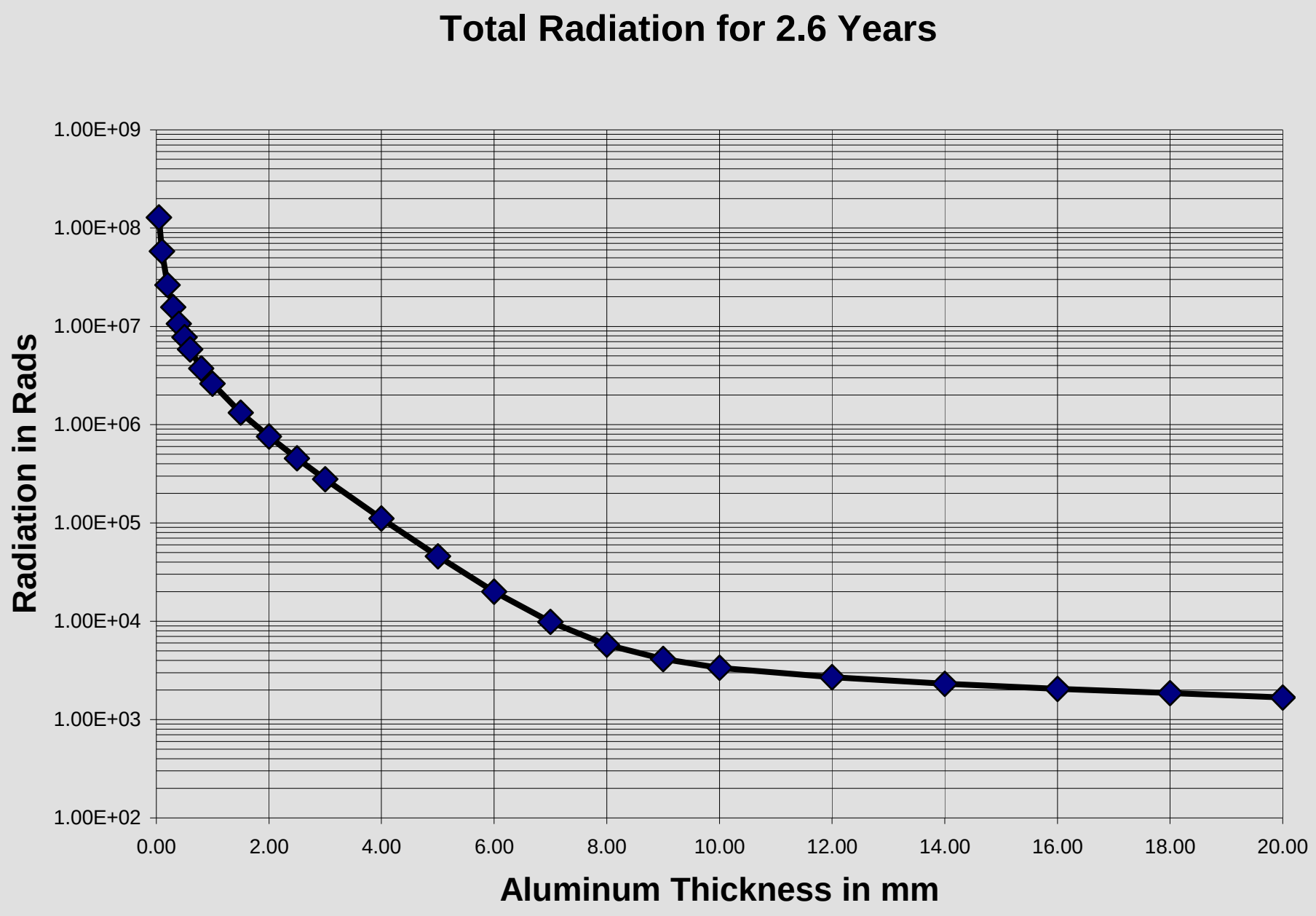

Figure 5.24: Estimated total radiation dose received during a 2.6 years mission on a Ravens orbit, as a function of aluminum shielding thickness**.**

### *Radiation induced upsets*

In case of radiation-induced upsets, internal program memory and registers can be reloaded from EDAC protected memory periodically, as signaled by a watchdog timer. More drastic failures may require a re-initialization from on-board EPROM.

*Damage of mirror coatings due to radiation*

All dielectric filter coating materials will be chosen for radiation resistance. Far-ultraviolet filters are rated at less than 5% change from exposure to up to 250 krad of high-energy radiation in space applications [Keffer, et al., 1994]. In the case of IMAGE-WIC: some degradation seen (but are likely due to gain issues). In the case of Polar UVI: no degradation seen to date (one is however starting to discern clouds in UVI imager, though this is thought to be a detector issue) [Jim Spann, personal communication, 2004].

## Procurement (availability, long-lead) Risks

A major risk is associated with whether or not a suitable detector is available. There is currently no Canadian expertise in manufacturing of multilayer dielectric optical coatings and microchannel plates. Due to ITAR restrictions, dealing with manufacturers in the USA may be a risky proposition.

## Cost Risks

None are anticipated. Note however that QA and Documentation practices that are commonly used in major, non-cost-capped, Space-Station-type projects, if allowed here, will represent a significant budgetary risk. The alternative is to follow the early "ePOP model".

## Schedule Risks

No significant risks identified at this time. However, if KuaFu moves forward, this will be a multi-instrument multi-national program with significant interdependencies between teams in different countries. There is expectable risk in schedule slip die to mismatch of work schedules in different countries. This must be avoided if at all possible, which will require effective management *at the mission level*.

## High Voltage

The use of on-board High-Voltage Power Supplies has proven to be risky. HV PSU failure is thought to have caused the end of both Viking and Freja imaging operations. High Voltage is only required if image intensified detectors (such as microchannel plates) are used. The associated risk can be reduced by reducing the actual voltage level required. Trade-off studies will be performed during Phase A.

## Detector Exposure to Sun

It is highly likely that the UVAMC will point at the Sun during certain periods throughout the instrument lifetime. This might also happen briefly during initial orbital maneuvers, as well as during any subsequent (temporary) malfunctions of the satellite platform stabilization system. Modeling will be performed to calculate the expected damage to detector from such a transient event. A lens cap or shutter with a sun-sensor would protect against this risk (and might also be useful in acquiring dark-frames for calibration purposes), albeit introducing the risk associated with employing movable parts. There are Canadian expertise in the area of Flight Qualified re-usable lens caps and shutters.

## Stray Light

The UVAMC is a highly sensitive instrument, due to the low light levels involved. Because of this, any stray light reaching the detector will have a potentially serious impact on data integrity. In addition to a stray light analysis of the optical system (including baffles), the instrument testing program should include measures to identify stray light problems. During early phase collaboration with the Chinese, one must ensure that no booms or other protruding features of other instruments co-inhabiting the platform are allowed to intrude into the FUVAMC field of view. Indeed, the layout and topology of the entire set of instruments on the satellite platform should be considered during stray light analysis.

## Integrity of Optical Focus

In the case of very fast optical systems, the cone of light impacting upon the detector is fairly steep, and the tolerance requirements for maintaining optical focus are severe. It is imperative that an optical tube assembly is designed that maintains strict focus across the variety of environmental condition expected during launch and while in-orbit. The Freja design was severely deficient in this respect, and, because of an almost non existent documentation trail, it subsequently came back to haunt the IMAGE WIC project (which essentially just used a Freja flight-spare – a rather well-kept secret). The use of graphite fiber epoxy composites (GFEC) will be considered (subject to mass restrictions), as it provides the necessary thermal expansion characteristics over the expected operating temperature range [TBC].

## Cleanliness

UVAMC contamination requirements must be carefully defined and communicated to the Canadian team as well as the Chinese. There must be requirements as to Instrument Purging, Handling, Cleaning, Storage, as well as Materials Selection, Outgassing, Surface Cleanliness, and Bake-out requirements. As for facility requirements, cleanliness levels must be specified (room Class 10,000 or better), monitoring/verification (e.g., witness samples), and personnel operations (enter/exit procedures, garments, etc.).

### Instrument Calibration

Drift of baseline calibration values (such as the detector gain) must be characterized and monitored continually. One may use Early-type FUV stars for characterizing long-term drifts. An on-board calibration source is also an option that will be considered in the next Phase.

### Operational Risks

As for communication (telemetry and upload of command sequences/patches), "Deep Space" capability will be required due to the particular KuaFu orbit characteristics and 24/7 operational modes. Suitable ground-station facilities must be identified. The fact that the Chinese are involved may in effect serve to reduce the number of options available. The various engineering issues associated with onboard data storage / store-and-forward will be included in this discussion, in Phase A, if awarded.

## *5.14 Work Sharing*

The UVAMC development will be a significant project. The cost is likely to approach $20,000,000, and the project will involve at least four and as many as six technical teams. The project is Canadian, and the bulk of the work must be done in Canada, however there are strategic and cost benefits to incorporating technical effort from other groups in other countries. That external to Canada work would not be funded by Canada.

The lead University team will be at the University of Calgary Institute for Space Research (ISR). The team will consist of Eric Donovan, Trond Trondsen, Leroy Cogger, Brian Jackel, and Mattieu Meurant during phase A. During later phases, this would possibly expand to include technical staff with expertise like that held by Greg Enno and Peter King, though the specific details of who those individuals would be is not at all clear right now (nor does it need to be).

The “Industry Prime” has not been established, however two companies (Routes AstroEngineering and COMDEV) have expressed strong interest in the project and have contributed to this concept study. The U. Calgary team is of the opinion that the expertise necessary for this undertaking is significant, and that the optimal solution is likely a coordination between *at least* COMDEV and Routes. We would hope that the CSA or (even better) the industries themselves could come up with a plan for which company can and should be prime.

In addition, the KuaFu team in China has requested, for a number of reasons, that every instrument team have a Chinese Co-PI and that the Co-PI's research and technical team be directly involved in the development of the instrument. Among other things, this protects the overall mission from criticism that it is not directly benefiting Chinese research (a criticism which is unfounded in our view but that there is no harm in addressing), and will help a great deal when it comes time for integration of instruments, and in the end data to accomplish science. In our case, the situation is very positive, for we were approached by Huygen Yang, the Deputy Director of the Polar Research Institute of China in Shanghai. Their group is a key research group in space science in China, and has many overlapping interests with Canada on the ground-based front. As well, their technical expertise seems strong, and there is every reason to believe that their group will contribute significantly to the UVAMC program. As of summer 2006, Huygen Yang is formally the Co-PI of UVAMC.

Three other research groups have approached us. We were first approached by James Spann of NASA who has offered considerable input in the form of discussions and criticism to the development of the project thus far. He has expressed a strong desire to be involved in Ravens and of course KuaFu. Given the realities of ITAR, this is problematic for obvious reasons. Nevertheless, Spann's extensive experience in global auroral imaging from space means that we will be including him in the development with continued discussion. Furthermore, Marshall Space Flight Center (MSFC) and Alabama Universities have created the National Space Science and Technology Center which may make involvement in a Chinese mission easier from a legal perspective than Spann as a NASA employee. In any case, should the situation change in terms of the difficulties imposed by ITAR, we would add him to the team, and identify tasks which he could seek funding for in the US system.

Early in the Concept Study, we were approached by Tuija Pulkkinen of the Finnish Meteorological Institute (FMI) who asked that her team play a role in UVAMC. There are significant advantages to this, for us. The FMI team, and Tuija in particular, have respected scientific capabilities in areas that are of importance to KuaFu, so the partnership in whatever form it takes will serve us well on the science end. More practically, the FMI team has asked to be able to contribute an isolated component of the Data Processing Unit (DPU) or other equivalent size isolated element. This would be funded by Finland. As well, there has been some discussion of Finland contributing a ground-station for deep space tracking, but that is much less likely due to the significant cost involved unless there is some other project that would motivate Finland to do so. In order to begin the process of deciding on appropriate work sharing between FMI and Canada, Tuija Pulkkinen and Hannu Koskinen of FMI, John Hackett (of COMDEV) and Eric Donovan from Calgary initiated a series of telecoms and face-to-face meetings. At the present stage, this has led to the production of the schematic included in the next figure, which is being used as the basis for identifying isolated elements of the instrument electronics that are potential candidates for a Finnish contribution.

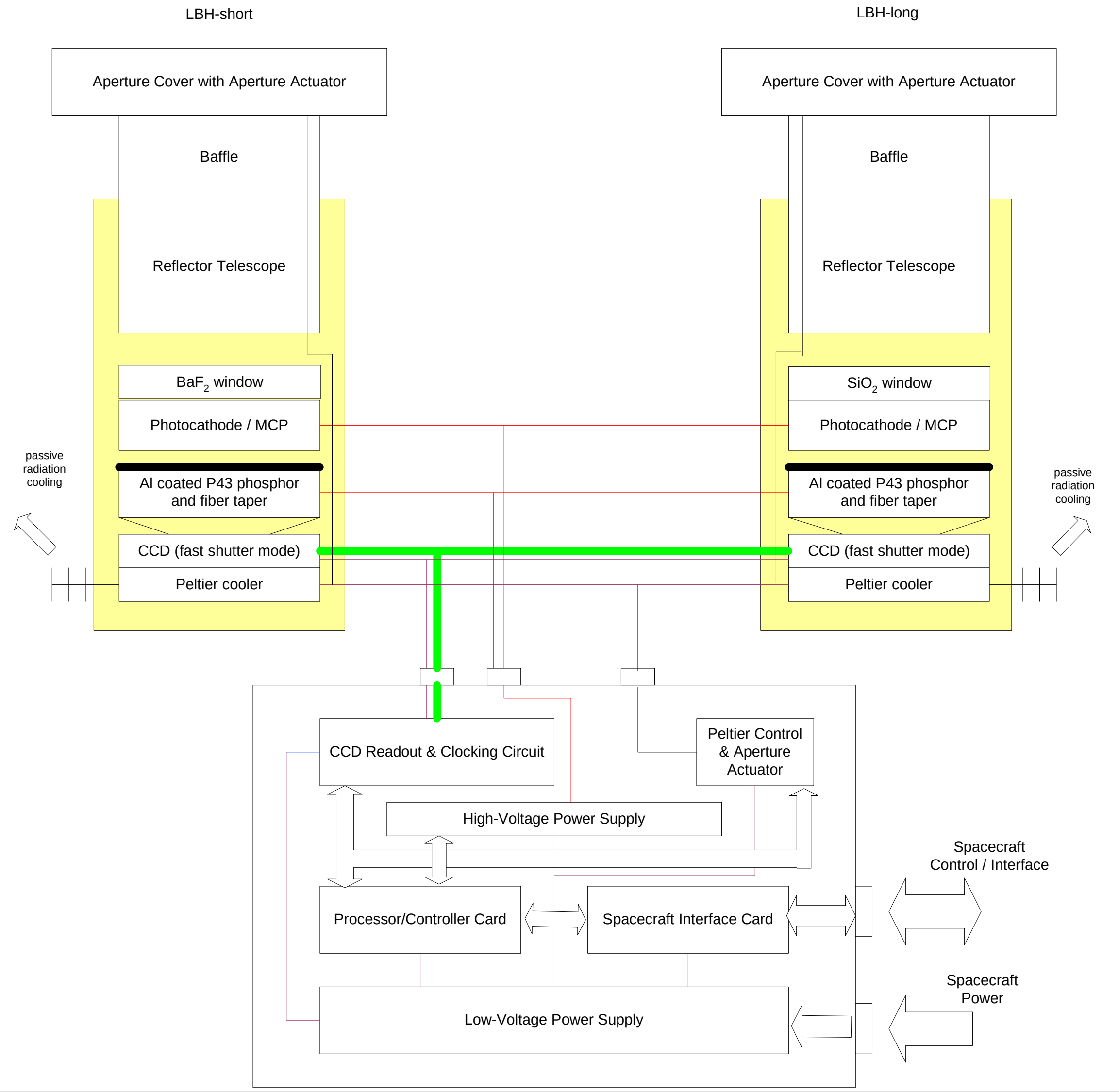

Figure 5.25: Schematic of proposed UVAMC instrument developed by COMDEV in coordination with the U. Calgary and FMI UVAMC teams.

This past summer, Nikolai Østgaard of the University of Bergen in Norway approached us and made a similar request to the one made by Tuija Pulkkinen. Again, there are significant advantages to Canada if such an arrangement could be worked out. Ostgarrd's group has been a leading light in X-Ray auroral imaging, and Østgaard himself is the PI of the proposed CASTRO mission. CASTRO would involve two satellites on elliptical polar orbits with their lines of apsides 180 degrees out of phase. In other words, when one satellite is at apogee in the Northern hemisphere, the other would be at apogee in the Southern hemisphere. The scientific objective is conjugate imaging. While KuaFu-B (or Ravens) would support ~40 minutes of conjugate images every orbit, CASTRO would support hours of conjugate imaging with a higher quality combined view every orbit. Conjugate imaging is an important and still largely unexplored scientific thrust. Provided

KuaFu goes forward, it is not at all unreasonable that CASTRO could evolve as a single satellite complement to KuaFu-B (see figure below). In short, we are interested in a hardware contribution from Østgaard's group of similar scope to that proposed by Pulkkinen. As well, we are excited by the prospect that KuaFu-B could evolve into a three satellite mission via CASTRO, and significantly improve our contribution to conjugate auroral studies.

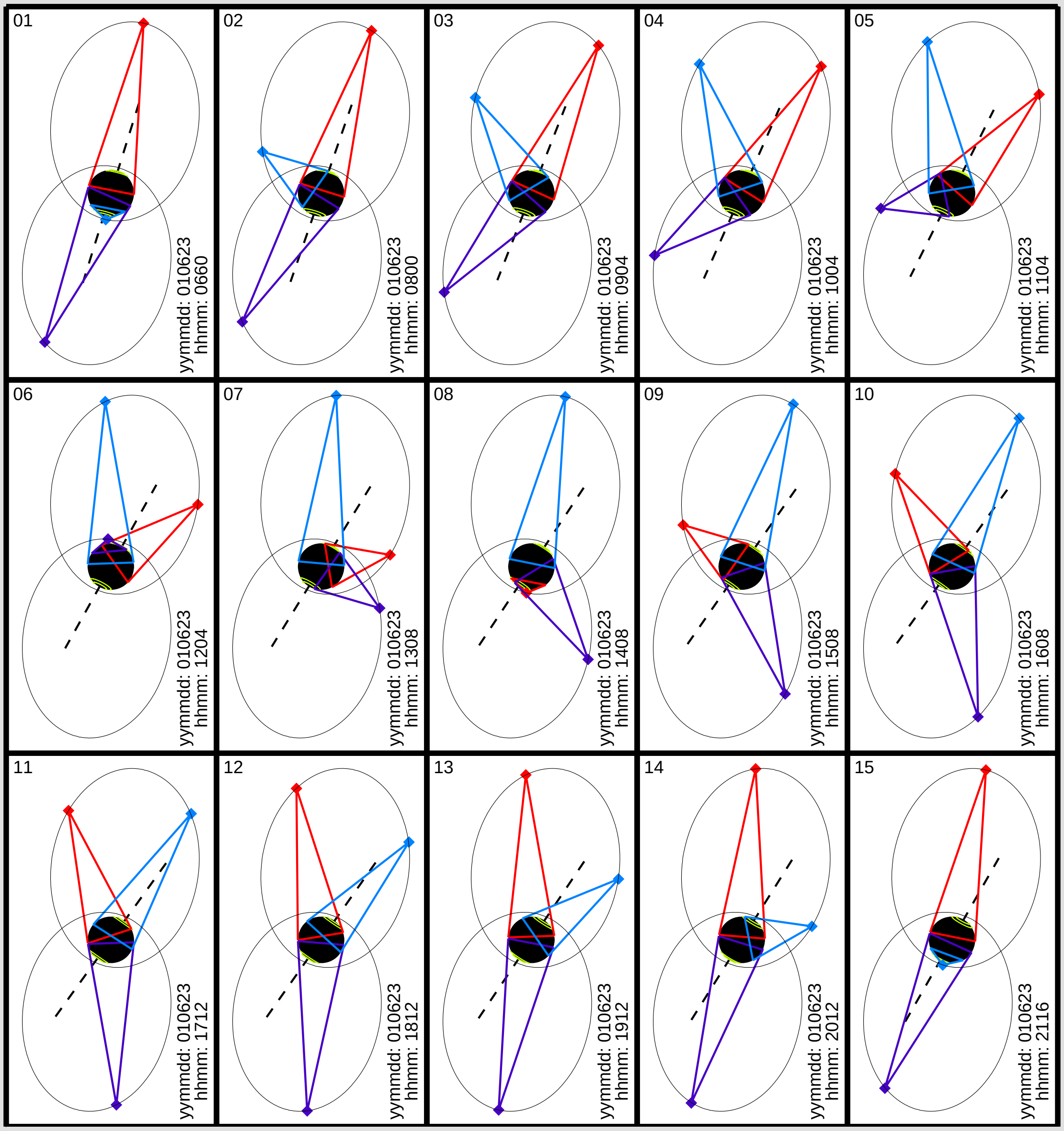

Figure 5.26: panel plot showing possible three satellite mission involving a Ravens-type pair providing 24/7 imaging of the northern hemisphere and a third satellite providing extended (8 hour or longer) periods of conjugate imaging. If KuaFu proceeds through to launch, and if CASTRO can be developed within the European sector as mission targeting conjugate observations, then CASTRO could be achieved with only one satellite, flown at the same time as KuaFu.

## 5.15 UVAMC In Depth - Conclusion

The objective of this study has been to perform a preliminary analysis in order to assess the feasibility of the proposed "Ravens" Far Ultraviolet Auroral Monitoring Camera (UVAMC) instrument package & mission concept. We have looked at a conceptual imager design, with an eye towards approaching an actual "proof of concept" in the next phase. Our emphasis has mainly been to lay the groundwork for determining the *technical* and *scientific* feasibility of the proposed project. As may be expected from a pre Phase A study, the level of detail reached in the study is not uniform. Rather, an attempt has been made to concentrate the effort on items the UVAMC project team has deemed to be critical. This work has been performed by the UVAMC team with support from an industrial subcontractor.

We have described the UVAMC camera concept, discussing several options. The following are some conclusions we have reached.

- **SPECTRAL PERFORMANCE.** The described instrument appears to be capable of providing both the necessary spectral *sensitivity* as well as spectral *selectivity* required in order to derive quantitative information on electron energy flux and average electron energy. This conclusion is based in part on a survey of IMAGE WIC data (Section 5.5) and appears to be valid for moderate and large electron energy fluxes, but more knowledge of the instrument will be required before a minimum usable electron energy flux can be estimated with confidence. These are issues that will be addressed during the next phase.

- **SPATIAL PERFORMANCE.** Spatial resolution is limited by the available exposure time and by the brightness (or, lack of such) of the features being imaged. We have found that a 30 km nadir spatial resolution appears feasible while at the same time meeting our proposed spectral and temporal requirements. However, physical realities of optical depth, emission layer thickness (itself on the order of 30 km), and intrascene spread in the emission layer height will all conspire to quickly and significantly decrease spatial resolution at off-nadir look-angles, regardless of how good is the resolution at nadir. Trade-off studies will be performed in the next phase, in order to determine whether a 30 km nadir resolution is defensible (in view of the associated cost of optics and satellite pointing stability requirements, and the very limited nadir patch of real-estate enjoying this level of resolution), or whether we could or should indeed back off from this number somewhat, and by how much.

- **TEMPORAL PERFORMANCE.** Assuming a satellite platform spinning at 2 rpm, we obtain a 30 second cadence, sufficient for studying the evolution of bright, narrow forms of 30 -50 km scale size (such as spirals, westward traveling surges). With a 25 degrees field of view, an effective exposure time of about 2 seconds is achieved. This appears to allow us to meet the spectral and spatial performance requirements. Further analysis will be conducted during the next phase.

In the next phase, we plan to – in collaboration with Industry – re-visit these design and performance aspects, and more fully flesh out a UVAMC instrument "proof of concept".

Our science objectives will also have to be further refined during that phase. As well, we intend to define spacecraft, launch, and orbit requirements, and to generate a preliminary operations plan, product assurance plan, design and implementation plan, as well as a management plan. Also to be included are preliminary costing data, with a discussion of build-versus-buy decisions.

# 6. Mission Level Considerations

The Chinese KuaFu Space Weather project consists of three spacecraft: KuaFu-A and KuaFu B1 and B2. KuaFu-A will be located around the L1 point. The two KuaFu-B satellites will be in identical elliptical polar Earth orbits relatively phased so that when one is at perigee the other is at apogee. KuaFu-A will be instrumented to continuously image the solar disk in EUV (19.5 nm and Lyman alpha) emission, to register Coronal Mass Ejections (CMEs) in Lyman alpha radiation and white light, to trace CME propagation by radio wave measurements, and to provide in situ measurements of the solar wind plasma, magnetic field, and solar energetic particles at L1. Another remote sensing instrument will observe the hard X-ray and Gamma-ray spectrum for timing of the origin of CME.

KuaFu-B1 and -B2 will provide continuous (24 hours per day, 7 days per week, or "24/7") FUV images of the northern hemisphere electron and proton aurora as well as ENA images of the ring current, systematic conjugate aurora observations and EUV image of the plasmasphere. KuaFu-B will also carry a suite of *in situ* instruments including a fluxgate magnetometer and charged particle detectors.

The Canadian *Ravens* Team was approached by the Chinese (Tu Chuan-Yi) in 2003, with a suggestion that the *Ravens* might be a perfect match for their proposed KuaFu Space Wather mission. The Canadian team was invited to fly their proposed *Ravens* instrument complement on each of the two KuaFu B satellites.

KuaFu project is currently in the Comprehensive Review phase supported by the Chinese National Space Administration (CNSA) and by the National Natural Science Foundation of China (NSFC) addressing scientific payload, spacecraft platforms, launching, tracking and control, and data transfer.

As the project is still in the Comprehensive Review Phase, only somewhat tentative data are available on spacecraft baseline design. During this project, the Chinese will draw upon the full expertise developed during their Double-Star Project.

The KuaFu mission is scheduled to start at the next solar maximum (a 2012 launch is aimed for), with an initial mission lifetime of two to three years.

The KuaFu-B pair will likely be double-launched into a parking orbit by one single Long March 3B (CZ-3B) rocket (as seen on the right hand side of the following figure). The parking orbit would be a small elliptical orbit with 0.8 RE perigee altitude and a 90-degree inclination.

**LM-3B (China) – taken from**
**http://www.fas.org/spp/guide/china/launch/cz-3b.htm**
**(without permission)**

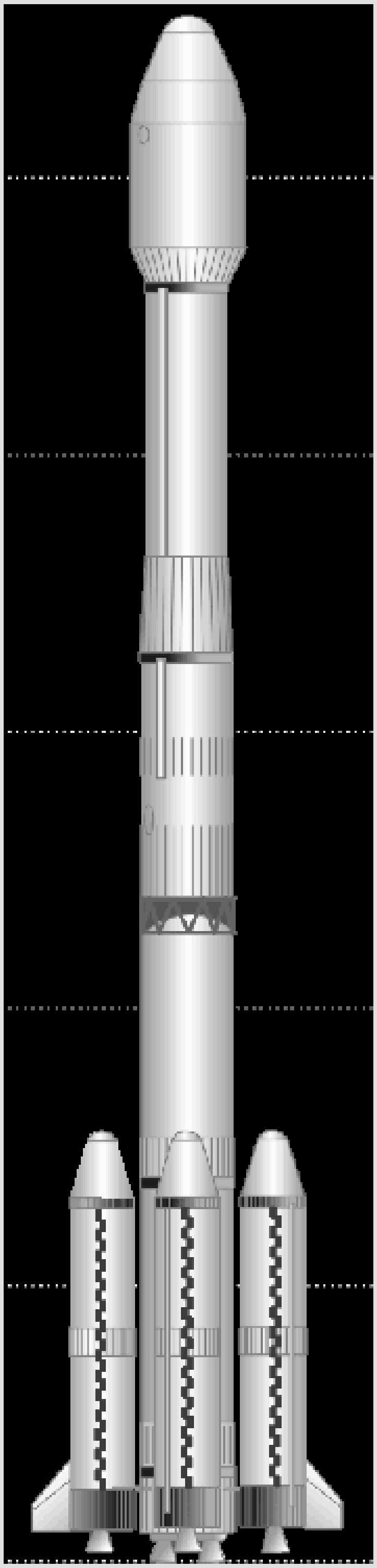

**Background Information**
**First Launch:** 14 February 1996
**Flight Rate:** 1 per year
**Launch Site:**
Xichang Space Launch Center, China
**Capability:**
29,900 lb to LEO; 9,900 lb to GTO
4,950 lb to Geosyschronous (with AKM)

**History**

- Chinese rocket program started in the late 1950s
- Evolved from Chinese surface-to-surface series IRBMs
- LM-3B is the same as LM-2E first stage with strap-ons, LM-3 second stage, and LM-3A LO2/LH2 third stage

**Description**

- Three-stage vehicle with four strap-on boosters
- Stage 1 consists of four YF-20 motors burning UDMH/N2O4 providing a total thrust of 664,000 lb
- Stage 2 uses one YF-22 engine and four YF-23 verniers burning UDMH/N2O4 generating a total thrust of 172,400 lb
- Stage 3 uses two YF-75 engines burning LO2/LH2 providing a thrust of 35,200 lb

**Length**: 190 ft
**Launch Weight:** 952,000 lb
**Diameter:** 11 ft
**Liftoff Thrust:** 1,328,000 lb
**Payload Fairing:** D>9.56m long X 4.0m diameter

The two satellites constituting the KuaFu-B pair will then independently maneuver into their respective orbits, carefully timed so as to achieve the required 180-degree phase difference. Within this launch scenario, the two satellites (KuaFu-B1 and –B2) will be accommodated into the fairing of the rocket as shown in the next figure.

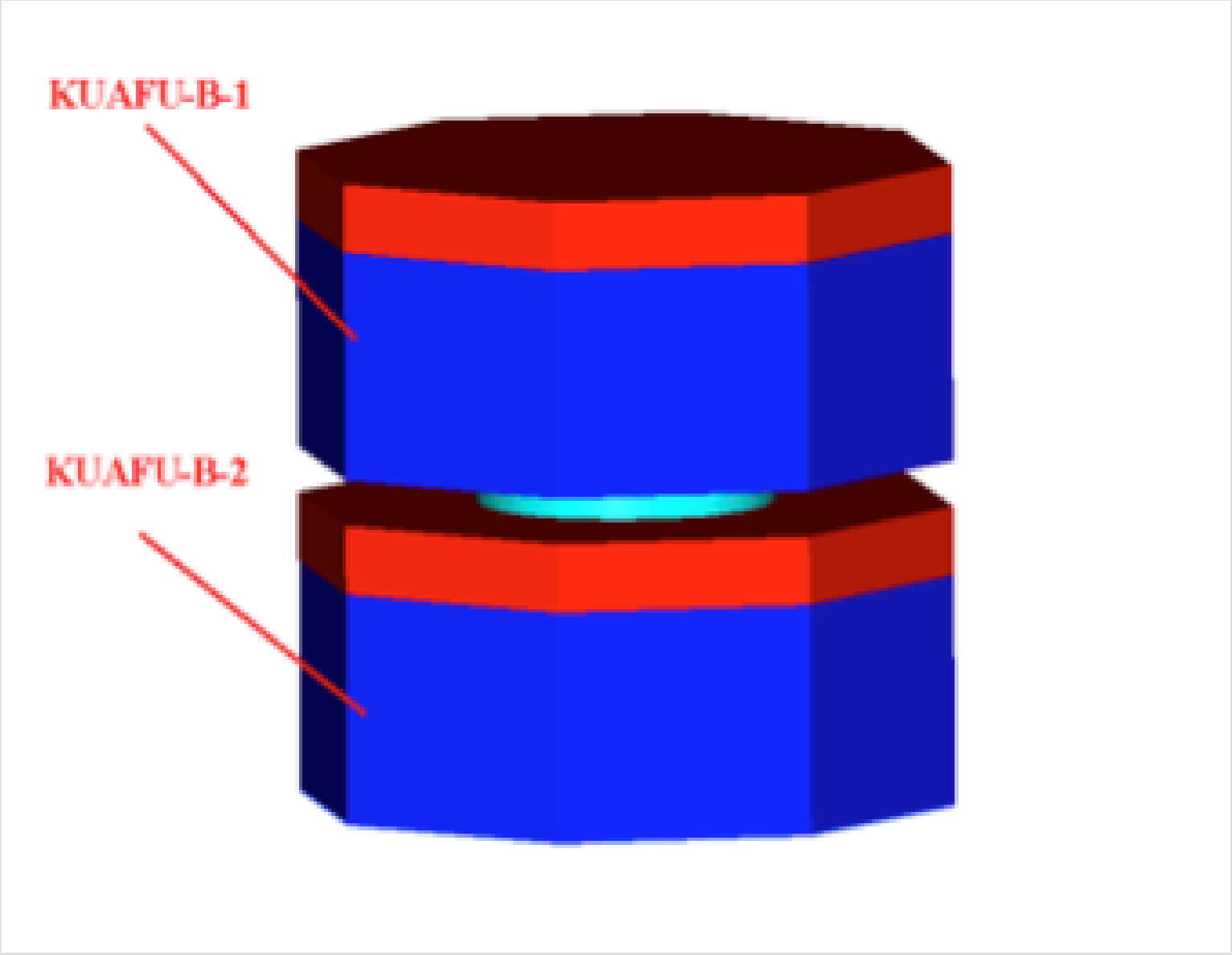

The concept drawing below depicts a fully deployed KuaFu–B satellite. Total mass is on the order of 400 kg (including propellant), and baselined dimensions are 2.9 m diameter and 1.3 m height. The octagon cylinder platform will be spin stabilized at a spin rate dictated in large part by *Ravens* science requirements. From an Engineering point of view, spin rate may be as low as 0.5 rpm and as high as 6 rpm. From a *Ravens* science perspective, a 2 rpm would be ideal. The Chinese team is therefore currently baselining a 2 rpm spin rate. The spin axis will be perpendicular to the orbit plane, in line with KuaFu-B mission science requirements.

Two rigid booms are installed, the length of which are 6 meters. These are stowed during launch. At the end of one boom, the Flux gate Magnetometer experiment is mounted, in order to meet the rigorous requirements as to electromagnetic magnetic cleanliness and to avoid interference from the spacecraft.

Solar panels are attached to the KuaFu-B's 8 side panels and 2 end panels, providing on the order of 250 W power, sufficient to power the science instrument complement and spacecraft subsystsems. When in eclipse, electrical power is provided by a battery.

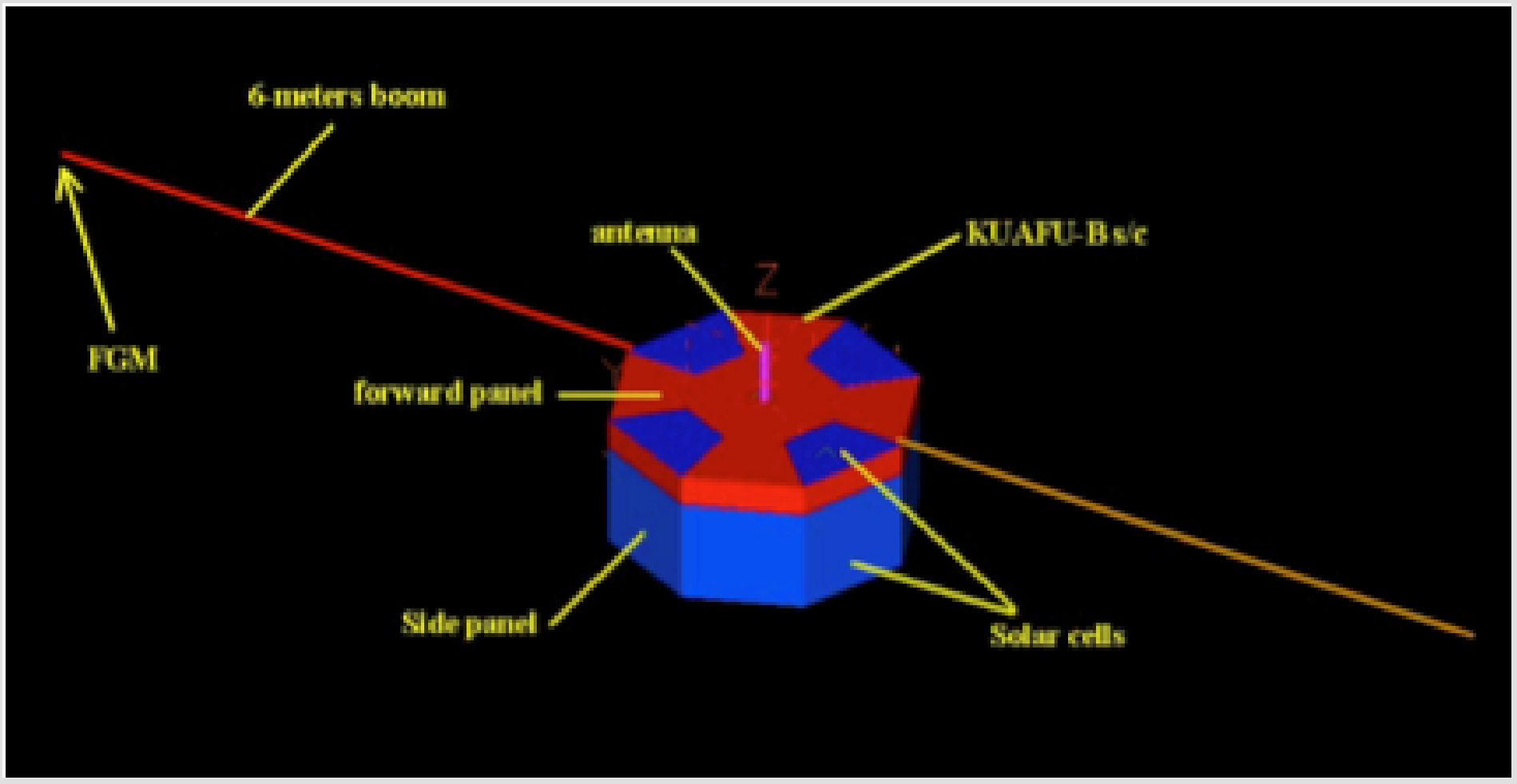

Currently, a mass of 65 kg and power of 50 W are allocated to science payloads on each of KuaFu-B1 and –B2, though at the time the table below was provided to us by the Chinese engineering team the payload limit was lower. This will likely change further by the end of the Comprehensive Review Phase.

Orbit parameters are largely dictated by Ravens science requirement of 24/7 coverage of the polar regions. Current parameters are 90 deg. inclination, 6 RE altitude, 0.8 RE perigee, with an orbital period of 17.7 hrs. The two spacecrafts co-inhabit the same orbit, but are spaces out by 180 degrees.

Table 7: KuaFu-B Spacecraft mass allocation is shown in the following table.

| | |
|---|---|
| Mechanical Sub-system (Deployable Booms): | 140 kg (35 kg) |
| Attitude Determination & Control Sub-system: | 40 kg |
| On-board Data Handling Sub-system: | 55 kg |
| Communications | 12 kg |
| Electrical Power Sub-system: | 70 kg |
| Thermal Sub-system: | 20 kg |
| Science Instruments: | 60 kg |
| | |
| Total (margin): | 400 kg (20 kg) |

Table 8: KuaFu Power Allocation.

| Sub-system | | Power (W) |
|---|---|---|
| KuaFu-B s/c | Attitude Determination & Control Sub-system | 17 |
| | Communications | 21 |
| | On-board Data Handling Sub-system | 25 |
| | Thermal Sub-system | 18 |
| | Electrical Power Sub-system | 12 |
| Science Payload | Science instruments | 50 |
| | Data Handling | 35 |
| | Transmission | 80 |
| Total (W) | | 250 |

# 7. Summary

This is the final report for the CSA funded Ravens Concept Study. During this study, the Ravens team did the following:

- Established a baseline instrument complement for Ravens of a two-channel LBH imager from Canada (UVAMC) and a spectroscopic imager from Belgium (FUVSI) for global imaging of the proton aurora in Lyman-alpha.
- Established a partnership with the Université de Liège and the Centre Spatial de Liège wherein they agreed to contribute the FUVSI to Ravens.
- Established that the 24/7 global imaging at the heart of the Ravens concept can be sustained for more than 2.5 years with feasible orbits.
- Developed a partnership with the Chinese KuaFu team, who agreed to modify KuaFu-B to accommodate the Ravens imaging payload and to place the KuaFu-B satellites on orbits specified by the Ravens Concept Study.
- Worked with the KuaFu Mission Definition team (which included E. Donovan, T. Trondsen, C. Jamar, and P. Rochus) to establish an imaging complement for KuaFu-B that would meet our desire to carry out system-level science as described in section 3.1, and that would support systematic conjugate imaging.
- Worked with the KuaFu Mission Development team to establish a set of *in situ* instruments for KuaFu-B that would fit within our mass budget and facilitate observations important to our overarching objectives.
- Interacted with the instrument teams for the other imagers to begin the process of technical budgeting (ie., splitting up the available payload, power, and telemetry), and to start assessing the interdependencies of the observations (for example, what does the UVAMC team need from the UVSI data to subtract the contribution to our signal from secondary electron emissions?).
- Contributed significantly to the KuaFu Assessment Study Report, and the writing of the KuaFu paper in Advances in Space Research, which are included as Appendices E and C, respectively.
- Interacted with two Canadian industries who are interested in being “Industry Prime” if UVAMC and KuaFu proceed.
- Established partners for UVAMC development in Finland, Norway, and China, and developed an initial work sharing plan.
- Carried out a feasibility study for building the UVAMC instruments in Canada.
- Wrote two papers on the results of this concept study (included as Appendix B and D).
- Identified another opportunity for flight of a Canadian UV auroral imager (the Molniya Mission – see Appendix G).
- Developed a strawman Phase A plan and ROM budget as the basis for negotiation of Phase A work on KuaFu and UVAMC in Canada.

On the basis of the work carried out under the Ravens Concept Study, we have demonstrated that the concept for the mission is sound and the instrument is feasible, though challenging. Further, we have identified an opportunity for the Ravens instruments and concept to become real as a part of the Chinese KuaFu mission. KuaFu partners in Europe and China have either started Phase A work or are negotiating Phase A or extended pre-Phase A contracts with their respective agencies.

In summary, the Ravens Concept Study has led directly to a flight opportunity for a Canadian UV auroral imager. Further, it helped shape the KuaFu-B mission to facilitate the never before achieved continuous 24/7 global imaging, with better spatial and spectral resolution than ever before achieved. KuaFu-B would provide the only global imaging during ILWS, and would enable system-level science. This is an exciting mission opportunity that is likely to proceed to launch, and in which Canada will have a high profile leading role.

# 8. References

## *Publications*

## Government Reports

## Ancillary References (not quoted in text – for future reference)

# 9. Overview of Appendices

The appendices to this document are included as separate attachments. The appendices are each documents that were produced either entirely or with significant contributions based on work carried out with financial support from and as a part of the Ravens Concept Study. In one case, that of the KuaFu Assessment Study (Appendix E), the appendix was submitted by the KuaFu team in China to the Chinese National Space Administration for a competition for prospective science missions led by China. Significant parts of that report were provided by Ravens team members who did the work under the Ravens Concept Study proposal. That material pertains to the science objectives of the KuaFu-B satellite pair (which is essentially and acknowledged to be derived from the Ravens concept), the FUVAMC instrument, and specifics of the KuaFu-B orbit design. The appendices are available via the web page on which the Final Report is staged. This URL is

http://aurora.phys.ucalgary.ca/doc/ravens/

**Appendix A:** *This is a strawman plan for Phase A activities for Canadian involvement in KuaFu and continued development of the FUVAMC by Canadian industry (prime to be determined on a competitive basis but likely to be one of Routes or COMDEV).*

**Appendix B:** *This paper was written for Advances in Space Research, outlining the motivations for future global imaging initiatives. This paper was written based on work that the PI presented at the November 2004 NASA Huntsville GeoSpace Imaging workshop. This was a brainstorming session during which the American imaging community was developing a "White Paper" on future directions. The PI was there as the only non-American scientist, by invitation of the host (James Spann), and specifically to present an invited talk on global imaging in the future. The paper is an attempt to highlight the motivations for global imaging, and to elucidate technical advancements (such as 24/7 global imaging) that would allow us to address new and exciting science. This refereed paper has been accepted and is in press. The current reference is*

> E. Donovan, T. Trondsen, J. Spann, W. Liu, E. Spanswick, M. Lester, C.-Y. Tu, A. Ridley, M. Henderson, T. Immel, S. Mende, J. Bonnell, M. Syrjäsuo, G. Sofko, L. Cogger, J. Murphree, P.T. Jayachandran, T. Pulkkinen, R. Rankin, and J. Sigwarth, **Global auroral imaging in the ILWS era**, Adv. Space Res. (2006), doi:10.1016/j.asr.2006.09.028.

*Please note that William Liu of the Canadian Space Agency contributed to the writing of this paper with a short section on system-level science. He was invited to do so by the author, and carried this out in his capacity as Chair of the International Living With a Star Steering Committee.*

**Appendix C:** *This paper was written for Advances in Space Research, outlining the science and technical objectives of the KuaFu satellite program. This paper was written based on work carried out in preparation of the KuaFu Assessment Study Report (Appendix E below) and reflects the coordinated effort of the KuaFu mission development team. Although Eric Donovan is the only Ravens team member on the author list, the paper contains descriptions of work carried out by Trond Trondsen and Leroy Cogger, both of whom were working under the Ravens Concept Study. This refereed paper is currently in review. The current reference is*

> C.-Y. Tu (a), R. Schwenn (b), E. Donovan (c), E. Marsch (b), J.-S. Wang (a), L.-D. Xia (d), and Y.-W. Zhang (e), **Space Weather Explorer – The KuaFu Mission**, Adv. Space Res. (2006), in review.

**Appendix D:** ***This paper was written as a follow on to an invited talk given by Trond Trondsen at the Canadian Aeronautics and Space Institute "ASTRO 2000" meeting in Montreal in 2006. The paper provides an initial technical overview of the UVAMC instrument. The current reference is of this paper is***

> Trondsen, T., E. Donovan, L. Cogger, J.-S. Wang, and C.-Y. Tu, "Ravens" – The far ultraviolet auroral monitoring imagers on the Chinese KuaFu mission, paper submitted in association with invited talk by Trondsen at ASTRO 2005, in press, 2006.

**Appendix E:** *This report describes the results of the KuaFu Assessment Study, and was submitted to the Chinese National Space Administration on July 6, 2005. This report was reviewed, along with roughly 20 others, in a competition for future science missions to be carried out by the CNSA. The KuaFu mission ranked first in that competition, and has currently moved forward to an Advanced Study. Sections of this report relating to the overall science objectives of KuaFu, the specific objectives of KuaFu B related to imaging, and the material describing the UVAMC were written in part or completely by Eric Donovan, Trond Trondsen, and Leroy Cogger.*

**Appendix F:** *Letters from members of the KuaFu team and the international GeoSpace team are included to corroborate statements about the relevance of KuaFu and the willingness of international partners to participate. A preliminary list is as follows.*

1. Chuanyi Tu
2. Mark Lester as International PI of SuperDARN
3. Mark Lester as PI of WFAI
4. Pierre Rochus representing CSL (builders of the KuaFu UVSI)
5. Jean-Claude Gerad (post launch KuaFu UVSI PI)
6. Susan McKenna-Lawlor (PI of NAIK)
7. Nikolai Østgaard as CoI on UVAMC
8. Tuija Pulkkinen as CoI on UVAMC
9. Heygen Yang as Co-PI of UVAMC
10. Jing Song Wang representing Chinese Meteorological Agency
11. Jan Sojka as Chair of the NSF CEDAR Steering Committee

*Additional letters will be added to this appendix over the few weeks following submission of the Ravens Concept Study Final Report.*

**Appendix G:** *Strawman proposal for Canadian involvement in the proposed NASA Molniya mission. This mission is proposed as a technology pathfinder for future Earth Imaging missions in Molniya orbits. The scientific motivation is nearly continuous (in the case of one satellite) to continuous (in the Ravens-like case of two satellites) imaging of the clouds in the Northern hemisphere polar regions. This complements GOES, and has the advantage of being a natural platform for global auroral imaging. As the proposed pathfinding satellite would not be spinning, this has the opportunity of allowing us to go after much better spatio-temporal resolution than we can achieve with the proposed Ravens platform. This proposal was solicited by the Molniya mission PI, who in turn got the idea of including a Canadian auroral imager from a member of the Ravens team (Tuija Pulkkinen). The Molniya mission could not compete with Ravens/KuaFu in that the global imaging from the pathfinder would not be continuous, and there would be no complementary proton aurora, ENA, conjugate or other observations that are required to advance system level science. On the other hand, the fact that the instrument can be Nadir pointing would allow us to build an inexpensive, single channel (likely only two mirrors and much smaller aperture than that for FUVAMC), nadir point instrument that could provide global images with ~15 kilometer spatial resolution at ten second temporal resolution. While we could not address system-level science, we could address specific and currently "hot topic" questions related to cross-scale coupling in geospace This would be an excellent lead in to KuaFu should Molniya and KuaFu both materialize, or an excellent stand-alone opportunity if KuaFu does not materialize. The Ravens Concept Study activity has raised the profile of Canadian space-based auroral imaging and has created at least two opportunities – namely Molniya and KuaFu. If both were to go at the same time, Molniya FUV imaging would complement the lower time resolution lower spatial resolution KuaFu-B auroral and ENA imaging.*

**Appendix H:** *This appendix cannot be attached with the document. It is a web tool for browsing results of the simulations we carried out to determine reasonable orbital parameters for the mission. This web tool is available via the URL*

http://aurora.phys.ucalgary.ca/doc/ravens/

*from which one clicks on "Appendix H". Once the web tool has been accessed, the user can step the satellites forward and backward on the orbit, as well as change the argument of perigee (the tool allows 11 choices), as well as apogee and perigee. By browsing through this tool, one can get a very good idea how apogee, perigee, and argument of perigee affects the orbit averaged quality of view. We attempted to do this with Satellite Tool Kit, but our particular needs warranted development of our own code. That code is available to the CSA and Public Works on request.*